\documentclass[longauth]{aa}       
\usepackage{natbib}
\usepackage{graphicx}
\usepackage{txfonts}

\newcommand{\Lsun}{{\hbox {$L_\odot$}}}
\newcommand{\Msun}{{\hbox {$M_\odot$}}}

\newcommand{\zgal}{{\hbox {$z$-GAL\ }}}

\begin{document} 

  \title{$z$-GAL - A NOEMA spectroscopic redshift survey of bright {\it Herschel} galaxies: [I] Overview}
  \author{P. Cox\inst{1}           \and
          R. Neri\inst{2}          \and
          S. Berta\inst{2}          \and
          D. Ismail\inst{3}         \and
          F. Stanley\inst{2}        \and
          A. Young\inst{4}          \and
          S. Jin\inst{5,6}             \and
          T. Bakx\inst{7,8,9}        \and 
          A. Beelen \inst{3}       \and 
          H. Dannerbauer\inst{10,11}  \and 
          M. Krips\inst{2}          \and
          M. Lehnert\inst{12}       \and 
          A. Omont\inst{1}          \and
          D.~A. Riechers\inst{13}    \and
          A.~J. Baker\inst{4,14}        \and 
          G. Bendo \inst{15}        \and 
          E. Borsato \inst{16}      \and
          V. Buat\inst{3}          \and 
          K. Butler \inst{2}       \and
          N. Chartab\inst{17}       \and
          A. Cooray\inst{17}       \and 
          S. Dye\inst{18}          \and
          S. Eales\inst{19}         \and
          R. Gavazzi\inst{3}       \and 
          D. Hughes\inst{20}       \and 
          R. Ivison\inst{21,22,23,24}     \and 
          B.~M. Jones  \inst{13}     \and 
          L. Marchetti\inst{25,26}     \and  
          H. Messias\inst{27,28}      \and  
          A. Nanni \inst{29,30}       \and
          M. Negrello\inst{19}      \and  
          I. Perez-Fournon\inst{10,11} \and    
          S. Serjeant\inst{31}      \and 
          S. Urquhart\inst{31}   \and
          C. Vlahakis\inst{32}     \and 
          A. Wei{\ss}\inst{33}     \and 
          P. van der Werf\inst{34} \and 
          C. Yang\inst{7}         
}         

  \institute{Sorbonne Universit{\'e}, UPMC Universit{\'e} Paris 6 and CNRS, UMR 7095, 
             Institut d'Astrophysique de Paris, 98bis boulevard Arago, 75014 Paris, France  \email{cox@iap.fr}   
        \and 
             Institut de Radioastronomie Millim\'etrique (IRAM), 300 rue de la Piscine, 38400 Saint-Martin-d'H{\`e}res, France 
        \and     
             Aix-Marseille Universit\'{e}, CNRS and CNES, Laboratoire d'Astrophysique de Marseille, 
             38, rue Frédéric Joliot-Curie 13388 Marseille, France 
        \and 
             Department of Physics and Astronomy, Rutgers, The State University of New Jersey, 
             136 Frelinghuysen Road, Piscataway, NJ 08854-8019, USA 
        \and
             Cosmic DAWN Center, Radmandsgade 62, 2200 Copenhagen N, Denmark 
        \and 
            DTU Space, Technical University of Denmark, Elektrovej 327, DK-2800 Kgs Lyngby, Denmark 
        \and
            Departement of Space, Earth and Environment, Chalmers University of Technology, Onsala Space Observatory, 439 92 Onsala, Sweden  
        \and 
            Division of Particle and Astrophysical Science, Graduate School of Science, 
            Nagoya University, Aichi 464-8602, Japan 
        \and 
            National Astronomical Observatory of Japan, 2-21-1, Osawa, Mitaka, Tokyo 181-8588, Japan 
        \and    
             Instituto Astrof{\'i}sica de Canarias (IAC), E-38205 La Laguna, Tenerife, Spain  
        \and
             Universidad de La Laguna, Dpto. Astrofísica, E-38206 La Laguna, Tenerife, Spain 
        \and 
             Centre de Recherche Astrophysique de Lyon - CRAL, CNRS UMR 5574, UCBL1, ENS Lyon, 9 avenue Charles Andr\'e, F-69230 Saint-Genis-Laval, France 
        \and 
              I. Physikalisches Institut, Universit\"at zu K\"oln, Z\"ulpicher Strasse 77, D-50937 K\"oln, Germany 
        \and
            Department of Physics and Astronomy, University of the Western Cape, Robert Sobukwe Road, Bellville 7535, Cape Town, South Africa 
        \and
             UK ALMA Regional Center Node, Jodrell Bank Center for Astrophysics, Department of Physics and Astronomy, The University of Manchester, Oxford Road, Manchester M13 9PL, United Kingdom 
        \and 
             Dipartimento di Fisica \& Astronomia "G. Galilei", Universit\`a di Padova, vicolo dell'Osservatorio 3, Padova I-35122, Italy 
        \and
             University of California Irvine, Department of Physics \& Astronomy, FRH 2174, Irvine CA 92697, USA  
        \and 
             School of Physics and Astronomy, University of Nottingham, University Park, 
             Notthingham NG7 2RD, UK 
        \and 
             School of Physics and Astronomy, Cardiff University, Queens Building, The Parade, Cardiff, CF24 3AA, UK 
        \and
             Instituto Nacional de Astrofísica, \'Optica y Electr\'onica, Astrophysics Department, Apdo 51 y 216, Tonantzintla, Puebla 72000 Mexico 
        \and 
             European Southern Observatory, Karl-Schwarzschild-Strasse 2, D-85748 Garching, Germany 
        \and 
             Department of Physics and Astronomy, Macquarie University, North Ryde, New South Wales, Australia 
        \and
             School of Cosmic Physics, Dublin Institute for Advanced Studies, 31 Fitzwilliam Place, Dublin D02 XF86, Ireland 
        \and
             Institut for Astronomy, University of Edinburgh, Blackford Hill, Edinburgh EH9 3HJ, UK 
        \and 
             Departement of Astronomy, University of Cape Town, 7701 Rondebosch, Cape Town, South Africa 
        \and  
             Istituto Nazionale di Astrofisica, Istituto di Radioastronomia - Italian ARC, Via Piero Gobetti 101, 40129 Bologna, Italy 
        \and
             Joint ALMA Observatory, Alonso de C\'ordova 3107, Vitacura 763-0355, Santiago de Chile, Chile 
        \and
             European Southern Observatory, Alonso de C\'ordova 3107,Vitacura, Casilla 19001, Santiago de Chile, Chile 
        \and
             National Centre for Nuclear Research, ul. Pasteura 7, 02-093 Warsaw, Poland 
        \and
             INAF - Osservatorio astronomico d'Abruzzo, Via Maggini, SNC 64100 Teramo, Italy 
        \and    
             Department of Physical Sciences, The Open University, Milton Keynes MK7 6AA, UK 
        \and 
            National Radio Astronomy Observatory, 520 Edgemont Road, Charlottesville VA 22903, USA 
        \and 
            Max-Planck-Institut f{\"u}r Radioastronomie, Auf dem H{\"u}gel 69, 53121 Bonn, Germany 
        \and
            Leiden University, Leiden Observatory, PO Box 9513, 2300 RA Leiden, The Netherlands 
         }

  \date{Received May 3, 2023/ accepted July 10, 2023}

 \abstract 
   {Using the IRAM NOrthern Extended Millimetre Array (NOEMA), we  conducted a Large Programme ($z$-GAL) to measure redshifts for 126 bright galaxies detected in the {\it Herschel} Astrophysical Large Area Survey (H-ATLAS),  the HerMES Large Mode Survey (HeLMS), and the {\it Herschel} Stripe 82 (HerS) Survey. We report reliable spectroscopic redshifts for a total of 124 of the {\it Herschel}-selected galaxies. The redshifts are estimated from scans of the 3 and 2-mm bands (and, for one source, the 1-mm band), covering up to 31~GHz in each band, and are based on the detection of at least two emission lines. Together with the Pilot Programme from \cite{Neri2020}, where 11 sources had their spectroscopic redshifts measured, our survey has derived precise redshifts for 135 bright {\it Herschel}-selected galaxies, making it the largest sample of high-$z$ galaxies with robust redshifts to date. Most emission lines detected are from $^{12}$CO (mainly from $J$=2-1 to 5-4), with some sources seen  in [C{\small I}] and $\rm H_2O$ emission lines. The spectroscopic redshifts are in the range $0.8<z<6.55$ with a median value of $z=2.56\pm0.10$, centred on the peak epoch of galaxy formation. The linewidths of the sources are large, with a mean value for the full width at half maximum $\rm \Delta V$ of $\rm 590 \pm 25 \, km \, s^{-1}$ and with 35\% of the sources having widths of $\rm 700 \, km \, s^{-1} < \Delta V < 1800 \, km \, s^{-1}$. Most of the sources are unresolved or barely resolved on scales of $\sim 2$ to $3\arcsec$ (or linear sizes of $\sim 15-25$~kpc, unlensed). Some fields reveal double or multiple sources in line emission and the underlying dust continuum and, in some cases, sources at different redshifts. Taking these sources into account, there are, in total, 165 individual sources with robust spectroscopic redshifts, including lensed galaxies, binary systems, and over-densities. This paper presents an overview of the $z$-GAL survey and provides the observed properties of the emission lines, the derived spectroscopic redshifts, and a catalogue of the entire sample. The catalogue includes, for each source, the combined continuum and emission lines' maps together with the spectra for each of the detected emission lines. The data presented here will serve as a foundation for the other $z$-GAL papers in this series reporting on the dust emission, the molecular and atomic gas properties, and a detailed analysis of the nature of the sources. Comparisons are made with other spectroscopic surveys of high-$z$ galaxies and future prospects, including dedicated follow-up observations based on these redshift measurements, are outlined.}


\authorrunning{P. Cox et al.}

\titlerunning{z-GAL Redshift Survey of Bright {\it Herschel} Galaxies}

\maketitle

\section{Introduction}
Over the last two decades, surveys in the far-infrared and sub-millimetre wavebands have opened up a new window into our understanding of the formation and evolution of galaxies, revealing a population of massive, dust-enshrouded galaxies forming stars at enormous rates in the early Universe \citep[see, e.g. reviews by][]{Blain2002, Carilli-Walter2013, Casey2014, Hodge&DaCunha2020}. In particular, the wide-field extragalactic imaging surveys carried out with the {\it Herschel Space Observatory} have increased the number of known dust-obscured star-forming galaxies (DSFGs) from hundreds to several hundreds of thousands \citep{Eales2010, Oliver2012, Lutz2011, Elbaz2011, Ward2022}. Including the all-sky Planck-HFI \citep[High Frequency Instrument;][]{Planck2015} and the South Pole Telescope (SPT) surveys \citep{Vieira2010}, today, we have large samples of luminous DSFGs that are among the brightest in the Universe, whose rest-frame 8-1000\,$\mu$m luminosities ($L_{\rm IR}$) exceed a few $10^{12} \, \Lsun$, and include numerous examples of strongly lensed systems \citep[][]{Negrello2010, Wardlow2013, Bussmann2013, Bussmann2015, Planck2015, Spilker2016, Nayyeri2016, Negrello2017, Bakx2018} and rare cases of extreme galaxies such as hyper-luminous infrared galaxies (HyLIRGs) with $L_{\rm IR} > 10^{13} \, L_\odot$ reaching star formation rates of $\rm 1000 \, \Msun yr^{-1}$ or more  \citep[see, e.g.][]{Ivison2013, Fu2013, Oteo2016, Riechers2013, Riechers2017, Ivison2019, Wang2021}.
  
The exact nature of these luminous DSFGs is still debated \citep{Narayanan2015}, although many of them are mergers \citep{Tacconi2008, Engel2010}. Compared to local ultra-luminous infrared galaxies, DSFGs are more luminous and several orders of magnitude more numerous by volume. DSFGs are most commonly found around $z \sim2-4$ \citep[e.g.][]{Danielson2017}, where the cosmic densities of star formation and black hole accretion peak \citep{Madau2014} and the gas accretion onto galaxies reaches its maximum \citep{Walter2020}. DSFGs therefore play a critical role in the history of cosmic star formation as the locus of the physical processes driving the most extreme phases of galaxy formation and evolution. 

Detailed follow-up studies of these galaxies require precise estimates of their distances to investigate their nature and physical properties. Photometric redshifts, which are based on spectral energy distribution templates designed for infrared-bright high-$z$ galaxies, provide useful indications of their redshift range; however, in most cases, this method remains too imprecise for efficient dedicated follow-up observations \citep[see, e.g.][]{Casey2012, Chakrabarti2013, Ivison2016, Bendo2023}. The dusty nature of these luminous galaxies makes optical and near-infrared spectroscopy using even the largest ground-based facilities difficult, if not impossible \citep{Chapman2005}, and, in the case of gravitational lensing, the optical spectra detect the foreground lensing galaxies rather than the background high-$z$ lensed object.  

Measuring molecular or atomic emission lines via sub-millimetre spectroscopy, which are unambiguously related to the (sub-)millimetre source and are not affected by dust extinction, has proved to be the most reliable method to derive secure redshifts for high-$z$ dust-obscured objects \citep[e.g.][]{Weiss2009, Swinbank2010}. The increase in the instantaneous frequency range of receivers operating at millimetre and sub-millimetre facilities has made this method more efficient over time and recent dedicated redshift surveys have provided crucial information that enable targeted studies to explore the nature of DSFGs in the early Universe. From the first sub-millimetre galaxies (SMGs) detected by the Submillimetre Common-User Bolometer Array (SCUBA) in the continuum \citep{Smail1997, Hughes1998}, whose redshifts were only measured a decade later \citep[e.g.][]{Chapman2005, Weiss2009, Walter2012}, to subsequent observations of tens of bright sources from the {\it Herschel} wide surveys that enabled the measurement of their redshifts  \citep[e.g.][]{Harris2012, Lupu2012, Riechers2011}, sub-millimetre spectroscopy was established as the prime method to precisely derive unambiguous distances to high-$z$ DSFGs. 

Due to their increased sensitivity, millimetre and sub-millimetre interferometers, in particular the IRAM Northern Extended Millimetre Array (NOEMA) and the Atacama Large Millimetre/submillimetre Array (ALMA), play a central role in measuring redshifts of high-$z$ DSFGs selected from wide field surveys. Using ALMA, a spectroscopic redshift survey of 81 high-$z$ strongly gravitationally lensed galaxies selected in the SPT 2500~deg$^2$ cosmological survey was presented by \cite{Reuter2020} - see, also, \cite{Weiss2013} and \cite{Strandet2016}. Initially selected at 1.4~mm, the SPT-selected DSFGs have redshifts spanning the range $1.9 < z <6.9$ with a median of $z=3.9\pm0.2$. This survey contains roughly half of the spectroscopically confirmed $z>5$ DSFGs, making it the most complete and largest high-$z$ sample of DSFGs prior to the sample we present herein. A survey of molecular gas in SMGs selected from 870~$\rm \mu m$ continuum surveys of the COSMOS, UDS, and ECDFS fields was performed using ALMA and NOEMA by \citet{Birkin2021} resulting in spectroscopic redshifts for 45 SMGs (from the 61 originally selected) with a median redshift of $z=3.0\pm0.2$. Spectroscopic redshift measurements of 17 bright SMGs selected from the AS2COSMOS survey are reported in \citet{Chen2022} with redshifts in the range $z=2-5$ with a median of $z=3.3\pm0.3$. For a sample of 24 strongly lensed high-$z$ galaxies, identified in the Planck survey, measurements of their spectroscopic redshifts, followed by detailed studies of the physical gas properties through multi spectral line $^{12}$CO and $\rm [C{\small I}]$ observations, were performed using a suite of single dish telescopes (e.g. IRAM 30m, GBT, APEX, and LMT) by \citet{Canameras2015} and \citet{Harrington2016,Harrington2021}.

 \begin{figure*}
   \centering
\includegraphics[width=1.0\textwidth]{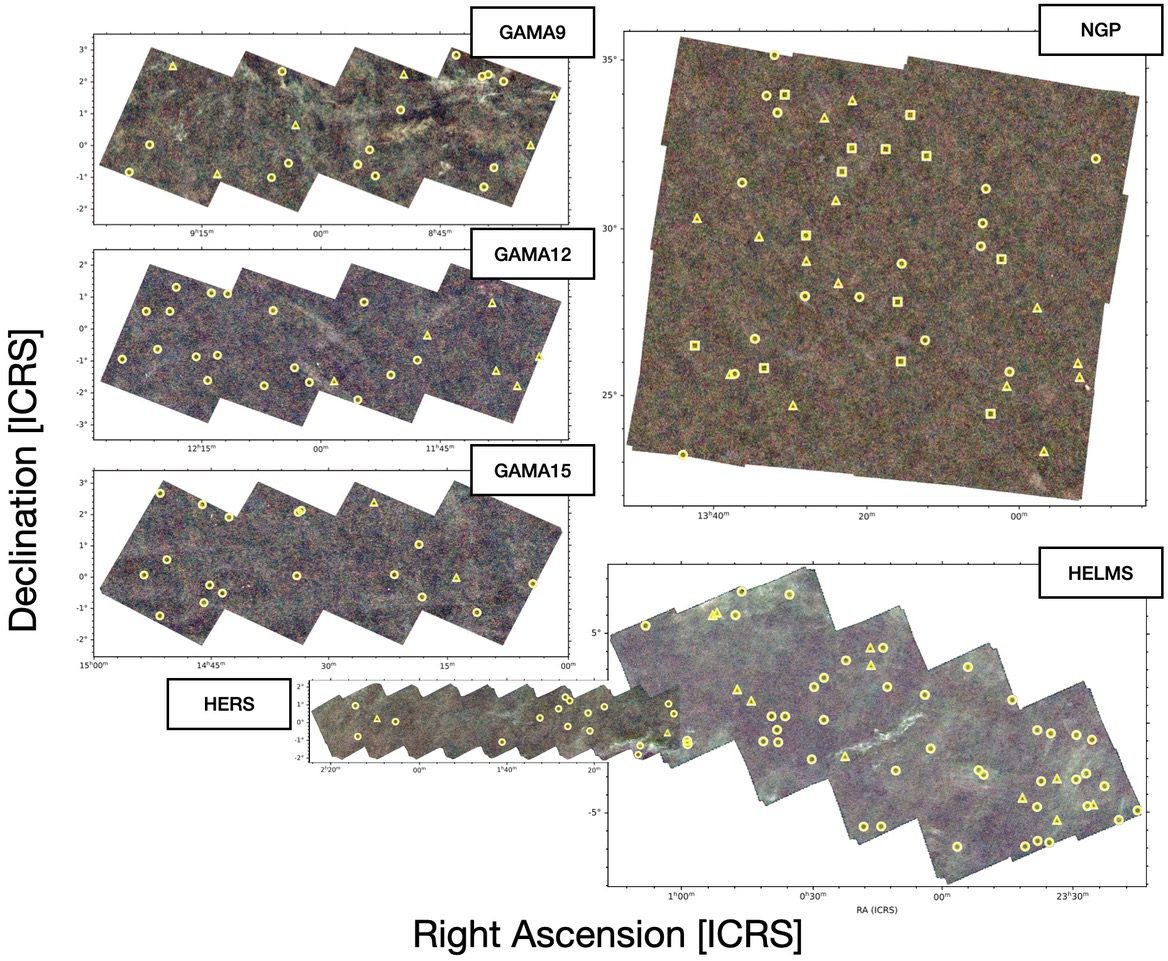}
\caption{{\it Herschel}/SPIRE colour maps of the H-ATLAS (GAMA9, GAMA12, GAMA15, and NGP), HeLMS and HerS fields with the positions of the 137 sources that were observed with NOEMA in the \zgal project and the Pilot Programme \citep{Neri2020}, highlighted with yellow circles and squares, respectively. The yellow triangles indicate the sources for which spectroscopic redshifts were already available (see Table~\ref{Appendix_table:sources_with_redshifts}). The maps display the distribution of the \zgal sources across the sky and show the proximity of some of the sources that enabled for them to be grouped together during the NOEMA observations, resulting in a more efficient use of telescope time (Sect.\ref{subsection:observations}).}
   \label{figure:sources-fields}%
    \end{figure*}

A number of very bright and/or very red {\it Herschel}-selected high-$z$ sources have been observed using various facilities, including the GBT, the IRAM 30m telescope, CARMA, NOEMA, and ALMA, yielding secure redshifts for about $\sim$70 sources \citep[][and references therein]{Cox2011, Harris2012, Wardlow2013, Ivison2016, Riechers2013, Riechers2017, Negrello2017, Fudamoto2017, Bakx2018, Nayyeri2016, Bakx2020b, Riechers2021}. Recently, \citet{Urquhart2022} presented a spectroscopic redshift survey of 85 high-$z$ {\it Herschel}-selected galaxies from the southern fields of the H-ATLAS survey \citep[{\it Herschel} Astrophysical Terahertz Large Area Survey - e.g.][]{Eales2010, Valiante2016}, listed in \citet{Bakx2018, Bakx2020b}, using ALMA and the Atacama Compact Array (ACA, also known as the Morita Array), and they derived robust redshifts for 71 DSFGs that were identified in the fields of 62 of the {\it Herschel}-selected sources. 

The correlator PolyFiX on NOEMA, with its capability to process a total, instantaneous single polarisation bandwidth of 31~GHz, alleviates one of the main problems related to the measurement of redshifts of high-$z$ DSFGs, namely the large overheads that are currently required in spectral-scan mode. This was demonstrated by \citet{Neri2020} who conducted a Pilot Programme to measure redshifts for 13 bright galaxies detected in the northern fields of the H-ATLAS survey with $\rm S_{500 \, \mu m} \geq 80 \, mJy$ and reported reliable spectroscopic redshifts for 11 of the selected sources. In one field, two galaxies with different redshifts were detected and, in two cases, the sources were found to be binary galaxies.  The spectroscopic redshifts were derived from scans of the 3 and 2~mm bands, and based on the detection of at least two emission lines. The spectroscopic redshifts are in the range  $2.08<z<4.05$ with a median value of $z=2.9\pm0.6$. The linewidths of the sources were large, with a median value for the full width at half maximum (FWHM) of $\rm 800 \, km \, s^{-1}$, potentially indicating cases of HyLIRGs and extreme mergers. In addition to the detailed information obtained on their structure and kinematics (mainly through emission from mid-$J$ $^{12}$CO transitions), the large instantaneous bandwidth of NOEMA provides, for each source, an exquisite sampling of the underlying dust continuum emission in the 3 and 2~mm (i.e. 90 and 150~GHz) bands. Together with the studies mentioned above, the number of {\it Herschel}-selected high-$z$ galaxies with precise redshifts amounts to $\sim$150 sources in total. 

Building upon this successful NOEMA Pilot Programme and taking advantage of the new capabilities of NOEMA \citep[ten to 12 antennas and PolyFiX; see, e.g.][]{Neri2022}, here we present the NOEMA $z$-GAL project, a comprehensive redshift survey of 126 of the brightest ($\rm S_{\rm 500 \mu m} \ge 80 mJy$) DSFGs with photometric redshifts $z_{\rm phot} \gtrsim2$ selected from the {\it Herschel} surveys in northern and equatorial fields and for which no reliable redshift measurements were previously available. Secure redshifts, based on at least two emission lines, were derived for all but two of the selected sources. This makes the NOEMA $z$-GAL project, together with the results from the Pilot Programme, the largest of these spectroscopic redshift surveys to date. 

The results of this comprehensive redshift survey are presented in a series of three papers that report on various aspects of the available data: (i) the current paper, the first in the series, describes the sample of selected {\it Herschel} sources, outlines the observational strategy, reviews the data reduction, and presents the basic results of the spectroscopic data, including the measurement of the spectroscopic redshifts, which was the main goal of this study; (ii) Paper~II \citep{Ismail2023} discusses the continuum properties of the sources in the sample and presents a detailed study of the derived dust properties and their evolution with redshift, based on the available spectral energy distributions including the {\it Herschel}, SCUBA-2, and NOEMA data; and (iii) Paper~III \citep{Berta2023} reports on the physical properties of the sources derived from the molecular and atomic gas ($^{12}$CO, [C{\small I}], and $\rm H_2O$) and dust, including the excitation conditions, the depletion timescales and first estimates of the stellar masses. In a subsequent study (Paper~IV), Bakx et al. (in prep.) will report on the nature of the sources and identify, using ancillary optical, near-infrared and radio data, gravitationally lensed sources, buried Active Galactic Nuclei (AGN) starbursts, proto-clusters, and hyper-luminous galaxies. 
  
The structure of this paper, the first of the $z$-GAL series, is as follows. In Section~\ref{section:obs}, we describe the sample selection, the observations, and the data reduction. In Section~\ref{section:results}, we present the main results including the redshift measurements, the spectral properties of the sources, and their morphology. A catalogue of the entire sample is presented in Appendix~\ref{Appendix: Appendix A} including, for each source, combined continuum and emission line maps together with spectra of all the emission lines that were detected. In Section~\ref{section:discussion}, we compare the spectroscopic and photometric redshifts, discuss the widths of the CO emission lines, and present the general properties of the $z$-GAL sources. Finally, in Section~\ref{section:conclusions}, we summarise the main conclusions and outline future prospects.   
 
Throughout this paper, we adopt a spatially flat $\Lambda$CDM cosmology with $H_{0}=67.4\,{\rm km\,s^{-1}\,Mpc^{-1}}$ and $\Omega_\mathrm{M}=0.315$ \citep{Planck2020}.

  \begin{figure}
   \centering
\includegraphics[width=0.5\textwidth]{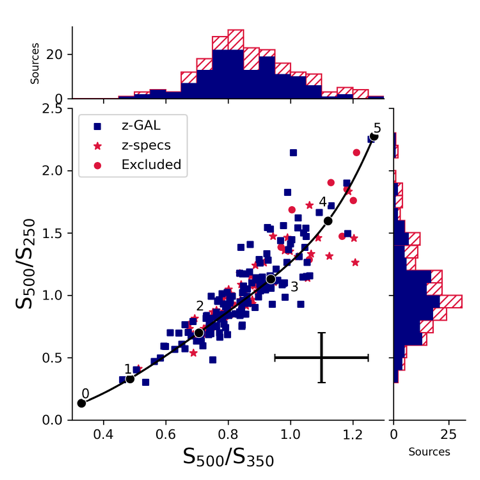}
\caption{{\it Herschel}/SPIRE colour-colour plot ($\rm S_{500 \, \mu m}/S_{250 \, \mu m}$ vs $\rm S_{500 \, \mu m}/S_{350 \, \mu m}$) for the $z$-GAL and Pilot Programme sources (shown as blue squares). Typical uncertainties on the colours are shown in the bottom right corner of the figure. The sources listed in \cite{Bakx2018} and \cite{Nayyeri2016} with spectroscopic redshifts (labelled 'z-spec') and listed in Table~\ref{Appendix_table:sources_with_redshifts}) are included for comparison and indicated with a red star. The sources that were excluded from the $z$-GAL selection (labelled "Excluded") are indicated with a red dot (see Table~\ref{Appendix_table:dropped_sources}). The side histograms display the distribution of the values as a function of the number of sources. The overlaid curve shows the redshift track expected for a galaxy with a spectrum energy distribution comparable to the Cosmic Eyelash galaxy; the photometric redshifts are indicated along the curve \citep[see][]{Ivison2016}.}
   \label{figure:sample_selection_plots}
    \end{figure}

\begin{figure*}
   \centering
\includegraphics[width=1.0\textwidth]{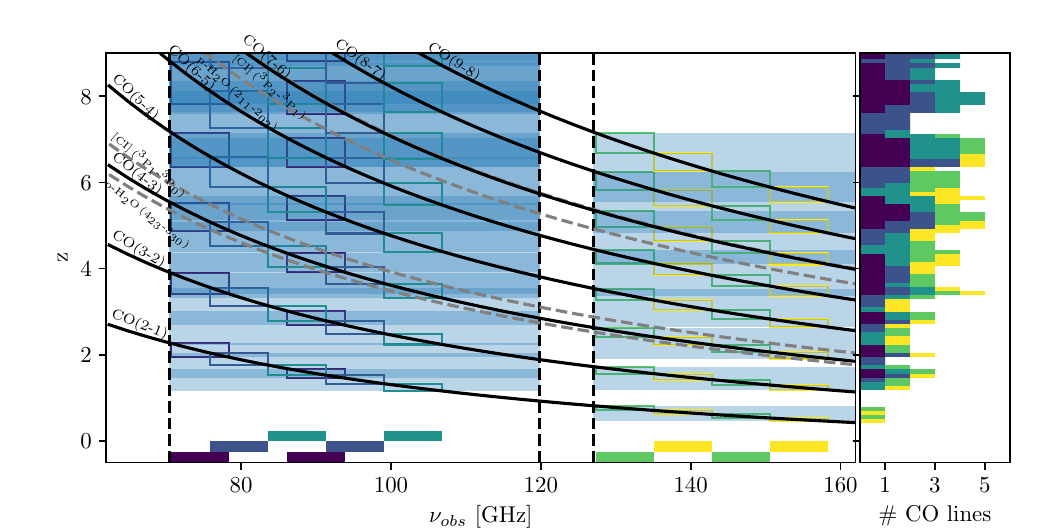}
   \caption{Spectral coverage of all the molecular and atomic lines searched for in the $z$-GAL survey
   as a function of redshift in the 3 and 2~mm atmospheric windows covering the frequency ranges 70.689-106.868~GHz and 127.377-158.380~GHz. The boxes at the bottom show the LSB and USB frequency settings used in the \zgal Large Programme for the 3-mm band (blue and dark green, respectively) and the 2-mm band (light green and yellow, respectively); those that were selected for the additional observations in the 3-mm band are highlighted in indigo (the set-up at 1~mm used for HeLMS-17 is not displayed - see text for further details). The lines detected in the \zgal survey include the following: the $^{12}$CO emission lines (solid black) from $J$=2-1 to 7-6; the $\rm [C{\small I}]\,(^3P_1$-$\rm ^3P_0)$ and $\rm [C{\small I}]\,(^3P_2$-$\rm ^3P_1$) fine-structure lines (492 and 809\,GHz rest frequencies, respectively, shown as grey dotted lines); and two water transitions, namely, the ortho-H$\rm _2O\,(4_{23}$-$3_{30})$ and para-H$\rm _2O\,(2_{11}$-$2_{02})$ (rest frequencies at 448.001 and 752.033~GHz, respectively, also shown as grey dotted lines). To enhance readability, the right panel exclusively displays the $^{12}$CO emission lines that can be detected for a given redshift in each of the frequency settings (identified with their corresponding colour). The 2~mm frequency windows were selected to optimally cover the range of spectroscopic redshifts based on the 3~mm observations. The dark blue zones identify the redshift ranges where at least two emission lines are detected at 3 or 2~mm, while the light blue zones indicate the redshift ranges where only one line is present. The wide frequency coverage enabled by NOEMA enables the detection of at least one emission line in each receiver band, except for a few small redshift gaps (further details are provided in Sect.~\ref{subsection:observations}).}
   \label{figure:spectral-coverage}
\end{figure*}


\section{Sample selection and observations}
\label{section:obs}

\subsection{Sample selection}
\label{subsection:sample-selection}
The sources of the NOEMA $z$-GAL Large Programme were chosen from three {\it Herschel} surveys. 

Firstly, we selected galaxies from the HerMES Large Mode Survey (HeLMS) and the {\it Herschel} Stripe 82 (HerS) Survey based on the source selection reported by \cite{Nayyeri2016}. Together, these partially overlapping surveys cover $\rm 372 \, deg^2$ on the sky. From the list of 77 sources with $\rm S_{500 \mu m} \ge 100\,mJy$ listed in \cite{Nayyeri2016}, we retained 60 sources for which no spectroscopic redshift data was available (43 from HeLMS and 17 from HerS). One of these sources (HeLMS-33) turned out to be a blazar (J0030$-$021) and was subsequently dropped. Only one source for which no spectroscopic redshift was available in the original list was overlooked in the selection and consequently could not be included in the observations, namely HeLMS-53. In total, we observed 59 {\it Herschel}-selected sources from HeLMS and HerS.

Secondly, we selected sources from the {\it Herschel} Bright Sources (HerBS) sample, which contains the 209 galaxies detected in the H-ATLAS survey with $\rm 500 \, \mu m$ flux densities $\rm S_{500 \mu m} > 80\,mJy$ and photometric redshifts $z_{\rm phot} > 2$ \cite[]{Bakx2018}. Most of these galaxies (189 in total) have been observed at 850\,$\mu$m using SCUBA-2, and 152 were detected. The SCUBA-2 flux densities originally reported by \cite{Bakx2018} were revised by \cite{Bakx2020a} together with the photometric redshifts estimated for each of the sources. 

From the HerBS sample, we selected all the sources from the 121 high-redshift galaxies in the North Galactic Pole (NGP) and equatorial (GAMA) fields of H-ATLAS (covering 170.1 and 161.6~deg$^2$, respectively), which had no spectroscopic redshift measurements. In addition to the 11 sources of the Pilot Programme \citep{Neri2020}, only 22 sources in the HerBS sample had available spectroscopic redshifts based on previous multi-line detections and 4 sources are listed with spectroscopic redshifts based on tentative, single line detections \citep[see references in][and Table~\ref{Appendix_table:sources_with_redshifts}]{Bakx2018}. In a subsequent study, \cite{Bakx2020b} presented spectroscopic measurements of four sources selected from the HerBS catalogue; two of these sources had robust redshifts derived from multi-line detections and were not included in the $z$-GAL Large Programme (HerBS-52 and HerBS-64), while two other sources (HerBS-61 and HerBS-177) were kept in the final sample.  

In the selection process, we eliminated 14 sources: two blazars - HerBS-16 (J1410+020) and HerBS-112 (J1331+30) - and 12 weak sources ($\rm S_{500 \mu m} \le 95 \, mJy$) with low-quality SPIRE data and/or no SCUBA-2 detection, based upon the experience gained in the Pilot Programme \citep{Neri2020}. The majority of the selected sources are reliably detected in the SCUBA-2 maps with $>3 \sigma$ $\rm 850 \, \mu m$ flux densities and positionally coincident with the {\it Herschel} source, and have good-quality SPIRE data at $\rm 500 \, \mu m$ with a signal-to-noise ratio $\rm S/N \ge 10$. In addition, one source, that was part of the $z$-GAL Pilot Programme (HerBS-204), and for which no spectroscopic redshift could be derived by \cite{Neri2020}, was added to the sample of the $z$-GAL Large Programme; and one source, HerBS-173, for which no emission line was detected neither at 2 or 3-mm, was dropped in the final selection. In total, we observed 67 {\it Herschel}-selected galaxies from H-ATLAS.  

The number of HerBS, HeLMS and HerS-selected galaxies observed in the $z$-GAL Large Programme, that cover the northern and equatorial regions, is therefore 126. Together with the 11 HerBS-selected sources for which spectroscopic redshifts are available from the $z$-GAL Pilot Programme \citep{Neri2020}, the complete $z$-GAL survey amounts to a total of 137 {\it Herschel}-selected galaxies. Fig.~\ref{figure:sources-fields} shows the distribution of the 137 \zgal-selected sources on {\it Herschel}/SPIRE colour maps of the H-ATLAS, HeLMS and HerS fields. 

The {\it Herschel} (and SCUBA-2) flux densities, as well as the photometric redshifts $z_{\rm phot}$ (Sect.~\ref{section:photometric-redshifts}), are listed in Table~\ref{Appendix_table:sources1} for the HeLMS and HerS sources and in Table~\ref{Appendix_table:sources2} for the HerBS sources that were selected for the $z$-GAL Large Programme\footnote{For the names of the sources, we here adopt HeLMS and HerS to be consistent with the names in the HerBS catalogue.}. The 11 sources of the $z$-GAL Pilot Programme are listed separately in Table~\ref{Appendix_table:sources3}. Finally, the sources for which spectroscopic redshifts were available are listed in Table~\ref{Appendix_table:sources_with_redshifts} and those that were not retained in the final selection or not observed are listed in Table~\ref{Appendix_table:dropped_sources}. 

Figure~\ref{figure:sample_selection_plots} displays the {\it Herschel}/SPIRE colour-colour plot ($\rm S_{500 \, \mu m}/S_{250 \, \mu m}$ vs. $\rm S_{500 \, \mu m}/S_{350 \, \mu m}$) for the $z$-GAL and Pilot Programme sources together with sources listed in \cite{Bakx2018} and \cite{Nayyeri2016} with spectroscopic redshifts and the sources that were excluded from the $z$-GAL selection. The colour-colour plot displays the selection function resulting from the flux limits in the samples together with the redshift track of a typical DSFG, the Cosmic Eyelash \citep[SMM J2135$-$010 - ][]{Ivison2010, Swinbank2010}, and indicates the expected photometric redshifts covered by the selected sources.

  \begin{figure*}
   \centering
\includegraphics[width=0.8\textwidth]{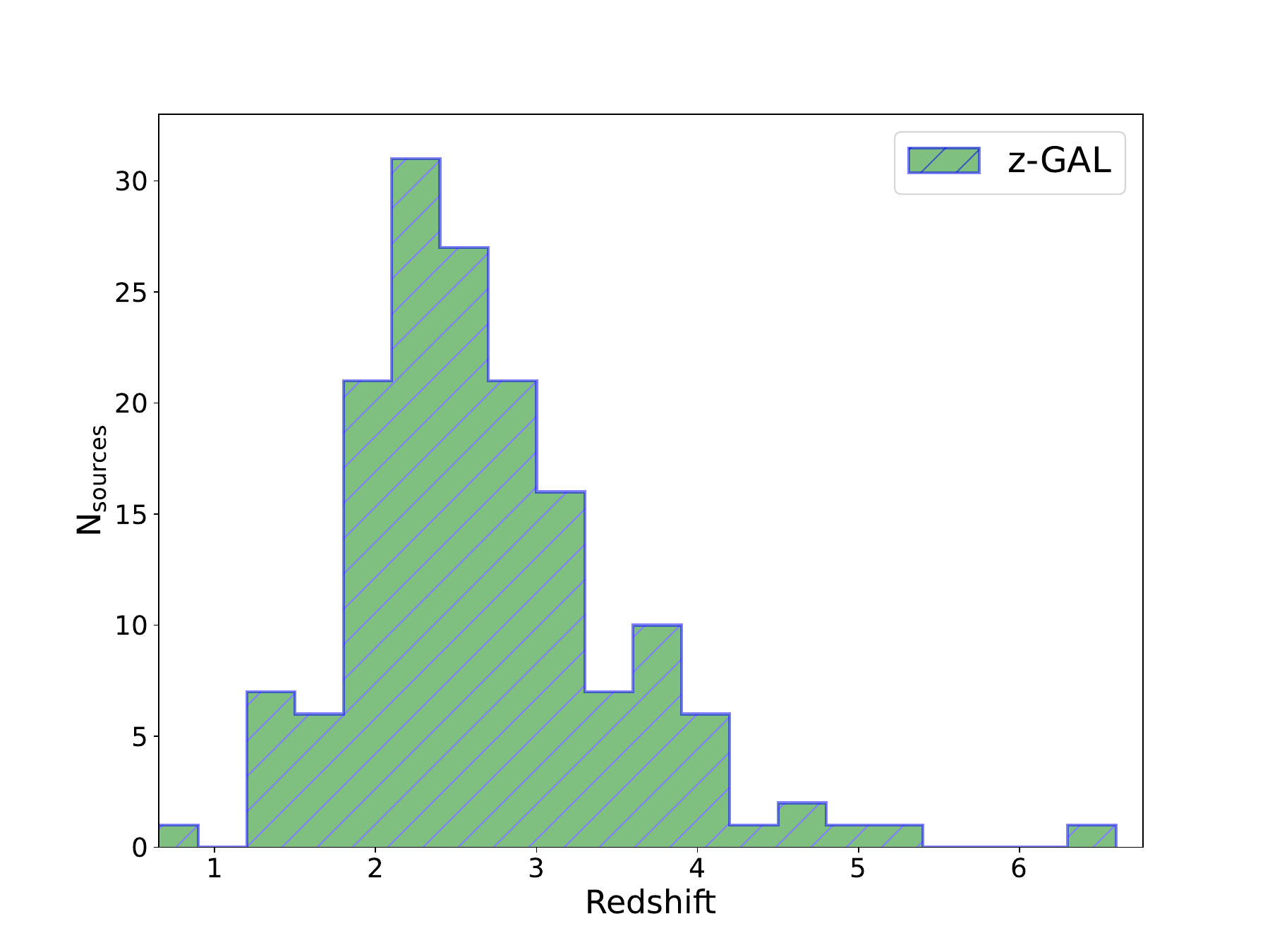}
\includegraphics[width=0.8\textwidth]{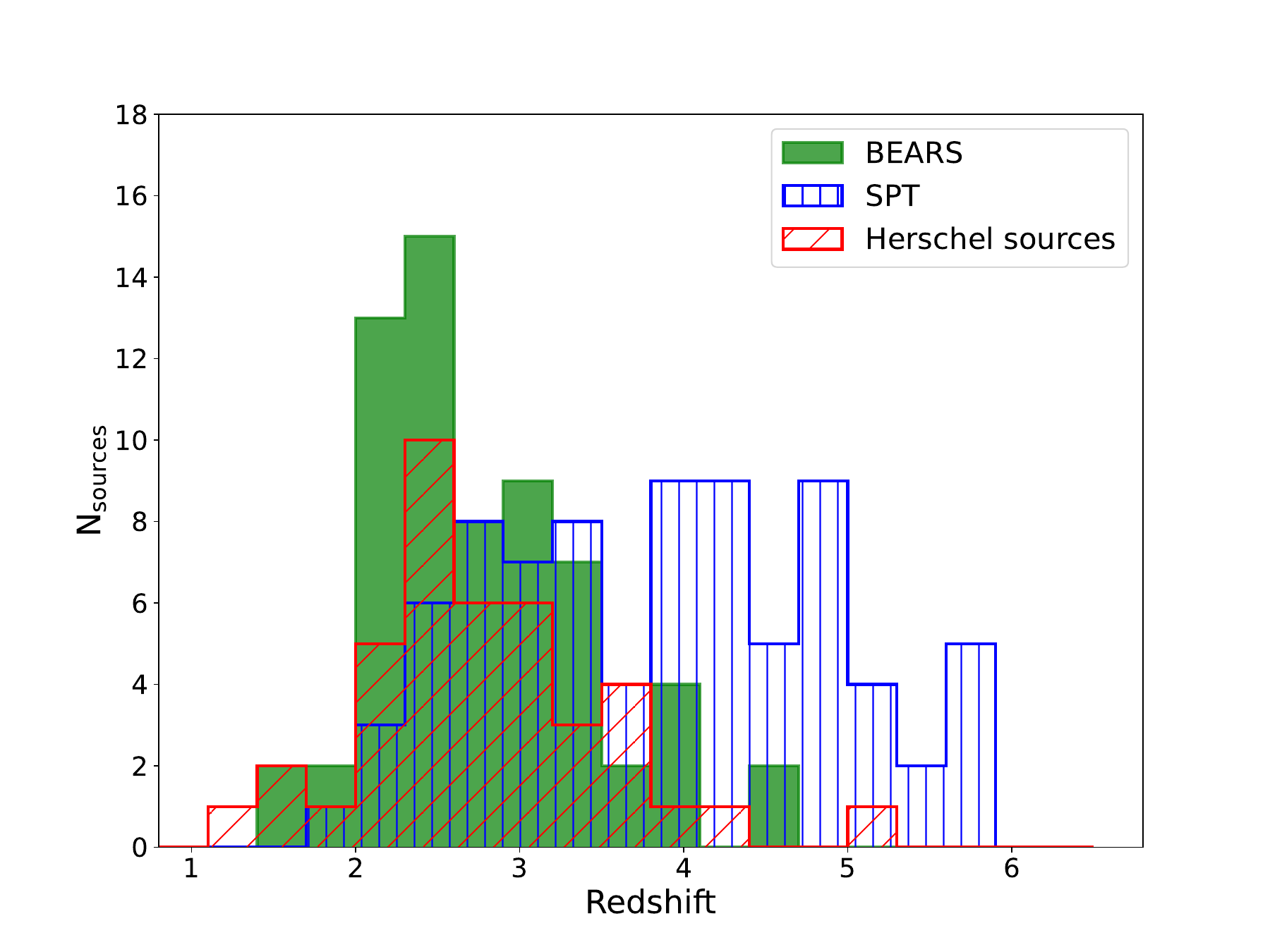}
   \caption{Spectroscopic redshift distribution of the \zgal survey compared to other high-$z$ galaxy samples. (Upper panel) Spectroscopic redshift distribution for the 165 individual sources identified in the fields of the 137 bright {\it Herschel}-selected galaxies of the {\it z}-GAL sample and the Pilot Programme \citep{Neri2020}. All the sources with robust spectroscopic redshifts were detected in at least two emission lines, mostly from CO (Table\,\ref{Appendix_table:emission-lines1}, \ref{Appendix_table:emission-lines2} and \ref{Appendix_table:emission-lines3}). The spectroscopic redshifts are in the range $0.8<z<6.5$ with a median value of $z=2.56\pm0.10$. (Bottom panel) Spectroscopic redshift distribution of the SPT sample \citep{Reuter2020}, the BEARS sample \citep{Urquhart2022}, and the {\it Herschel} sources for which $z_{\rm spec}$ values are listed in \cite{Bakx2018} and \cite{Nayyeri2016} with updates provided in Table~\ref{Appendix_table:sources_with_redshifts}.} 
   \label{figure:spectroscopic-redshifts}%
    \end{figure*}

  \begin{figure*}
   \centering
\includegraphics[width=0.8\textwidth]{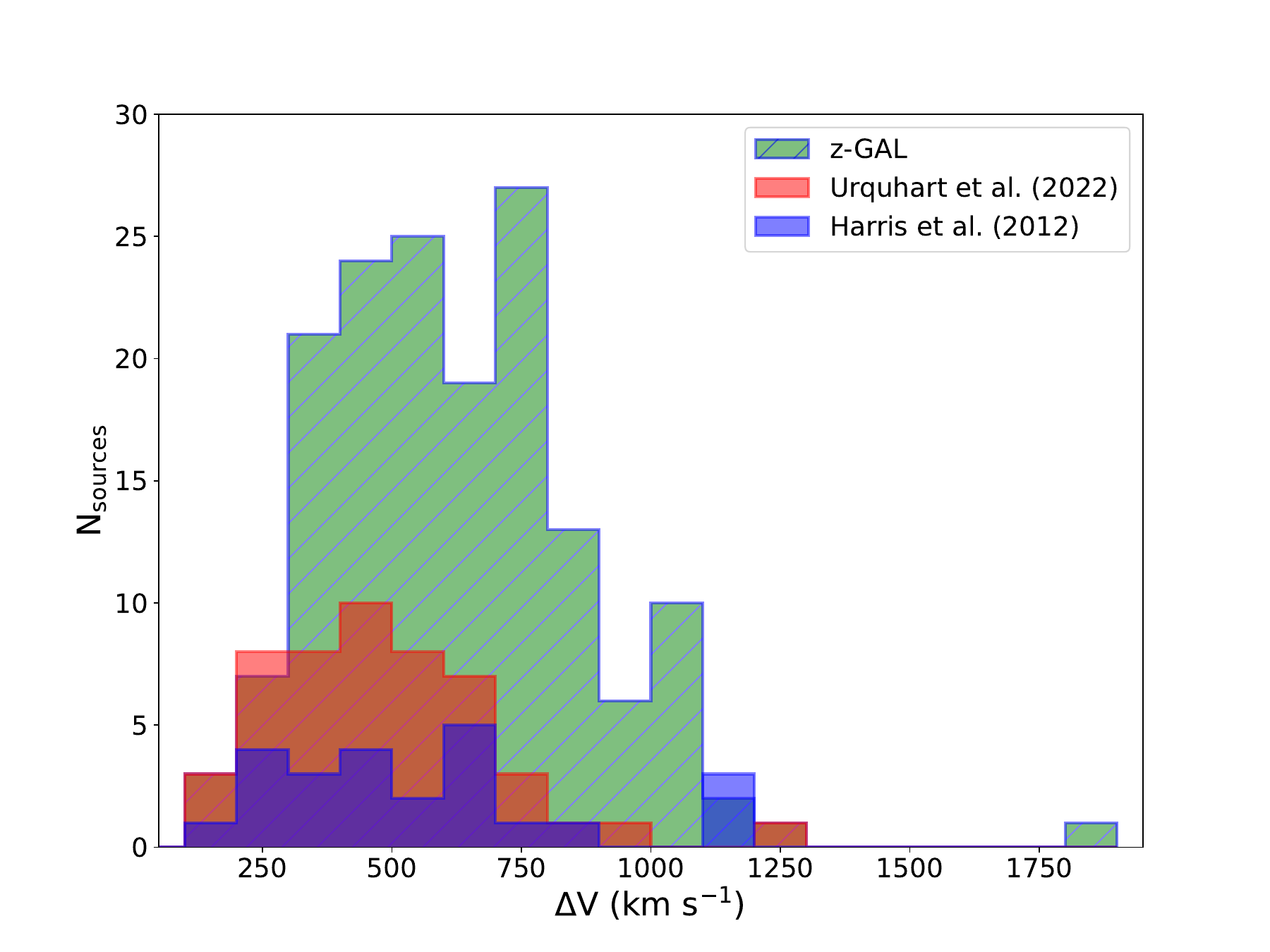}
   \caption{Distribution of the FWHM for all the individual sources identified in the fields of the 137 bright {\it Herschel}-selected galaxies of the $z$-GAL sample and the Pilot Programme \citep{Neri2020, Stanley2023} for which spectroscopic redshift were measured - see Tables\,\ref{Appendix_table:emission-lines1}, \ref{Appendix_table:emission-lines2} and \ref{Appendix_table:emission-lines3}. The linewidths of the sources are on average large with a median value for the FWHM of $\rm 590 \pm 25\, km s^{-1} $ (Sect.\ref{section:line widths}). For comparison, the distributions of the FWHM for the BEARS sample \citep{Urquhart2022} and the {\it Herschel}-selected galaxies reported in \citet[][and references therein]{Harris2012} are shown in the red and blue histograms, respectively. The median values of the FWHM are $\rm 465 \pm 37 \, km s^{-1}$ for the BEARS sample and $\rm 485 \pm 73 \, km s^{-1}$ for the sample of \citet{Harris2012}  - further details are provided in Sect.~\ref{section:line widths}.}  
   \label{figure:line_widths}%
    \end{figure*}

  \begin{figure}
   \centering
\includegraphics[width=0.5\textwidth]{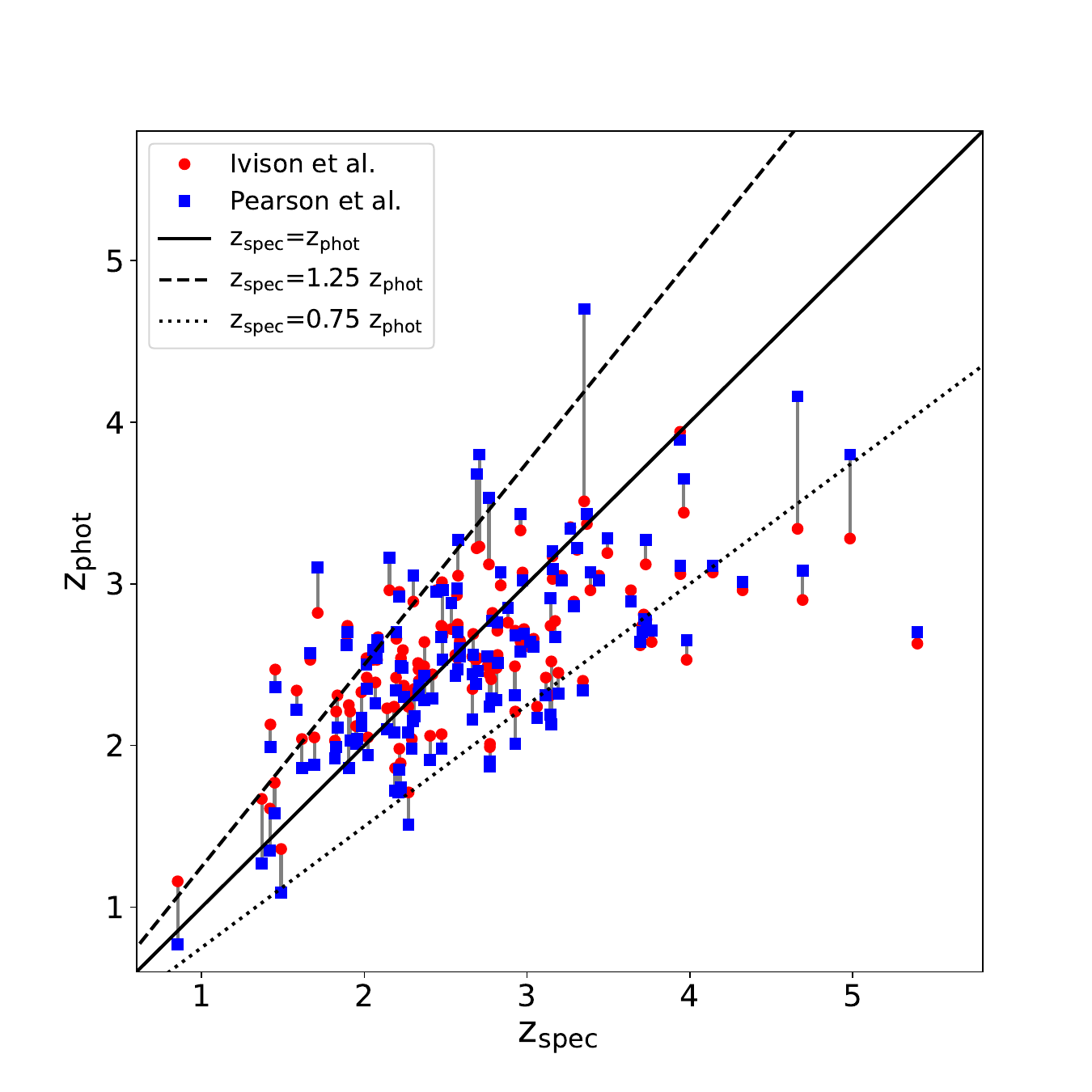}
   \caption{Comparison of the spectroscopic redshifts of the sources in the $z$-GAL sample that are single and multiple (where each source has a similar redshift) with the photometric redshifts derived using the methods of \cite{Ivison2016} and \cite{Pearson2013} - blue and red symbols, respectively (as listed in Tables~\ref{Appendix_table:emission-lines1}, \ref{Appendix_table:emission-lines2} and \ref{Appendix_table:emission-lines3} and Tables~\ref{Appendix_table:sources1} to \ref{Appendix_table:sources3}, respectively). For each source, the two photometric redshift values are linked by a grey vertical line positioned at the spectroscopic redshift value. The solid line shows where the photometric and spectroscopic reshifts are equal and the dashed lines indicate the $\pm$25\% deviations. The significant spread here displayed is consistent with previous findings \cite[e.g.][]{Ivison2016, Reuter2020, Urquhart2022} and demonstrates that photometric redshifts, based on {\it Herschel} and, when available, SCUBA-2 $\rm 850 \, \mu m$ continuum flux densities, are only indicative. 
   }
   \label{figure:z_phot-vs-z_spec}%
    \end{figure}

\subsection{Observations}
\label{subsection:observations}
We used NOEMA to observe the 126 {\it Herschel}-selected bright DSFGs (Table~\ref{Appendix_table:sources1} \& \ref{Appendix_table:sources2}) in the NGP and equatorial fields and derive their redshifts by scanning the 3 and 2~mm bands to search for at least two emission lines. The observations of the $z$-GAL Large Programme were carried out under project M18AB (PIs: P.\,Cox, T.\,Bakx \& H.\,Dannerbauer). Starting on December 10, 2018, NOEMA continuously and regularly observed the 126 {\it Herschel}-selected sources for a two-year period, using nine to ten antennas in two configurations (C and D) of the array. The $z$-GAL Large Programme was completed in July 26, 2020 with 190 hours of observing time used. A subsequent Director's Discretionary Time (DDT) project (D20AB) started soon thereafter with the aim of measuring an additional emission line for 13 sources that had an emission line falling outside the frequency settings originally defined for the Large Programme and secure for these sources unambiguous spectroscopic redshifts. On January 15 2021, all of the 126 {\it Herschel}-selected galaxies were reported in at least two emission lines and, henceforth, robust redshifts measurements were potentially acquired for the entire originally proposed sample. Following these observations and after a careful analysis of the data, a dozen sources were found to still have ambiguous redshift solutions (Sect.~\ref{section:spectroscopic-redshifts}). Additional observations were then performed under the same DDT project to measure a third emission line in these sources. This removed remaining ambiguities, giving robust spectroscopic redshifts for 124 sources of the 126 selected sources, thereby successfully achieving the original goal of this project. 

Observing conditions were on average excellent during the entire time frame of this project with an atmospheric phase stability of typically 10-40\,deg RMS and 2-5\,mm of precipitable water vapour. The correlator was operated in the low-resolution mode to provide spectral channels with a nominal resolution of 2\,MHz. The NOEMA antennas are equipped with 2SB receivers that cover a spectral window of 7.744\,GHz in each sideband and polarisation. Since the two sidebands are separated by 7.744~GHz, two frequency settings are necessary to span a contiguous spectral range of 31\,GHz. At 3~mm, we adjusted the spectral survey to cover a frequency range from 75.874 to 106.868~GHz. At 2~mm, we then selected two frequency windows that covered as well as possible the candidate redshifts allowed by the emission lines detected at 3~mm, across the frequency range from 127.377 to 158.380~GHz (Fig.~\ref{figure:spectral-coverage}). Additional observations to secure the redshifts for a dozen sources were done at 3~mm and covered the frequency ranges from 70.689 to 78.433~GHz and 86.177 to 93.921~GHz. For one source (HeLMS-17), we performed additional observations at 1~mm across the frequency ranges from 208.884 to 216.628 and 224.372 to 232.116~GHz to obtain a robust redshift. 

The wide spectral coverage of the NOEMA correlator ensures that a scan of both the 3 and 2~mm spectral windows can detect at least two $^{12}$CO emission lines for every $z \gtrsim 1.2$ source, between $\rm ^{12}CO$\,(2-1) and $\rm ^{12}CO$\,(9-8), in each band, with the exception of a few redshift gaps. The gaps most relevant to the present observations are at 3~mm for $1.733<z<1.997$, and at 2~mm for $1.668<z<1.835$ (Fig.~\ref{figure:spectral-coverage}). The redshift range $1.733<z<1.835$ was not covered by any of the 3 and 2~mm settings. The spectral coverage of these observations also includes the $\rm [C{\small I}]\,(^3P_1$-$\rm ^3P_0)$ and $\rm [C{\small I}]\,(^3P_2$-$\rm ^3P_1$) fine-structure lines, two transitions of water, namely ortho-H$_2$O\,(4$_{23}$-$3_{30})$ and para-H$_2$O\,(2$_{11}$-$2_{02})$, and HCO$^+$(3-2) and HCN(3-2), which were all detected in sources selected for this survey (Table~\ref{Appendix_table:emission-lines1}, \ref{Appendix_table:emission-lines2} and \ref{Appendix_table:emission-lines3}). In total, 358 emission lines were detected in the combined $z$-GAL survey and Pilot Programme. These include among the brightest lines in the sub-millimetre range and the most likely species \citep{vanderwerf2010} to be picked up in redshift searches \citep[see also, e.g.][]{Bakx2022}. Exploring both the 3 and 2~mm spectral bands is therefore a prerequisite for detecting, in most cases, at least two $^{12}$CO emission lines in the $2<z<6$ {\it Herschel} bright galaxies selected for this survey and deriving reliable spectroscopic redshifts. 
 
All observations were carried out in track-sharing mode by cyclically switching between groups of galaxies within a track, as was made possible due to the proximity of many sources to each other (Fig.~\ref{figure:sources-fields}). Track-sharing allowed us to observe efficiently by not repeating pointing and some calibration observations for sources belonging to a group of five to seven galaxies in a small field (within 10$\rm ^o$ to 15$\rm ^o$). The sky positions of the 126 sources were distributed in five large fields, namely: H-ATLAS NGP, GAMA9, GAMA12 and GAMA15, and HerMES HeLMS/HerS (H Field) shown in Fig~\ref{figure:sources-fields}. In summary, the sources were ordered in 21 groups of five to seven sources. On average, we spent $\sim$1.5~hour of total observing time per source, including overheads and calibration. Two different configurations (C and D) of the array were used, yielding angular resolutions between $1\farcs2$ and $3\farcs5$ at 2~mm, and $1\farcs7$ and $\sim$6$''$ at 3~mm. Observations were started by first observing sources in one track in the lower 3~mm frequency setting. Galaxies that did not show a robust line detection were then observed again in the upper 3~mm frequency setting. For every line detection, the most probable redshift was estimated taking into account the photometric redshift. The galaxies were subsequently observed in one of the two 2~mm frequency settings, and when no line was detected, observed again in the second setting. To resolve the cases of ambiguous redshift determinations, additional frequency windows were observed according to inclusion or exclusion criteria based on photometric redshifts and relative intensities of the detected lines (Sect.~\ref{section:spectroscopic-redshifts}). 

\subsection{Data reduction}
Phase and amplitude calibrators were chosen in close proximity to the track-shared sources in order to reduce uncertainties in astrometric and relative amplitude calibration. MWC349 and LkH$\alpha$101 were used as absolute flux calibrators. The positions of the source correspond to the barycenter of the emission region in the pseudo-continuum maps (Appendix~\ref{Appendix: Presentation of Catalogue}). The precision of the position relies on the S/N, and we typically achieved an absolute astrometric accuracy of $< 0\farcs2$ (Tables~\ref{Appendix_table:emission-lines1} and \ref{Appendix_table:emission-lines2}). The absolute flux calibration was estimated to be accurate to within 10\%. The data were calibrated, averaged in polarisation, imaged, and analysed in the GILDAS software package\footnote{http://www.iram.fr/IRAMFR/GILDAS/}.

The {\it uv}-tables were produced from calibrated visibilities in the standard way and cleaned using natural weighting and support masks defined on each of the detected sources. Continuum maps were produced for each sideband and setup, excluding the spectral ranges that include emission lines. Continuum extraction was performed on cleaned continuum maps, after correcting for primary beam attenuation, using polygonal apertures of variable size and shape, and a pseudo-continuum map as guidance. This was achieved independently for each source according to the procedure described in Appendix~\ref{Appendix: Presentation of Catalogue}.

In case multiple sources were detected in the field of view, the fluxes of each individual object were extracted. Source positions were computed as the barycenter of the flux distribution in the sky of the given object, within the extraction polygon. The spectra of each source were extracted from cleaned cubes within polygonal apertures defined by hand on the emission line channels to enhance the S/N of the spectral detection. The emission lines of the majority of the 165 sources revealed in the {\it Herschel} fields were detected with good signal-to-noise ratios.  Only in a few cases, the sources (10 in total) displayed at least one emission line with $\rm S/N \lesssim 3$. 

The simultaneous fitting of all the detected lines of a particular source, utilising single or double Gaussian profiles, offers the advantage of improved S/N compared to fitting the lines individually (Appendix~\ref{Appendix: Presentation of Catalogue}). This approach benefits from the combined information on the line properties and thereby enhances the accuracy of the fitting process. Additionally, by assuming, for all the detected lines, common velocity widths and velocity separations for double Gaussian profiles, the simultaneous fitting method optimises the precision of the measurements. For the majority of the sources, the line widths have been estimated with $\rm S/N \gtrsim 5$, with the exception of three sources, namely: HerBS-72, HerBS-124 E and HerBS-194 S (for which $\rm 3.5 < S/N < 4.5$). The errors on the derived redshift values, which are estimated from the differences in centroid positions of the fitted profiles, are typically of the order of a few $10^{-5}$ to $10^{-4}$.

The spectroscopic redshifts and the extracted properties of the spectral lines are listed in Tables~\ref{Appendix_table:emission-lines1} and \ref{Appendix_table:emission-lines2}, together with the measurements of the sources of the $z$-GAL Pilot Programme from \citet{Neri2020} in Table~\ref{Appendix_table:emission-lines3}. The combined continuum and emission line images of all the $z$-GAL sources are displayed, together with the detected molecular and atomic emission lines, in the series of figures in the catalogue presented in Appendix~\ref{Appendix: Appendix A}. Detailed comments on individual sources are given in Appendix~\ref{Appendix: Comments on sources}.

In addition to the emission lines, the continuum flux densities of the sources were extracted from up to ten available polarisation-averaged 7.744\,GHz wide sidebands over the frequency range from 76.3 to 206.5\,GHz. All the $z$-GAL sources were detected in the continuum in at least four sidebands. The NOEMA continuum measurements are presented and discussed in detail in Paper~II, including tables listing the continuum flux densities for the entire $z$-GAL sample\footnote{All $z$-GAL data, including line emission and continuum data products, are available from the $z$-GAL collaboration and can be viewed on the IRAM Large Programme Archive web page: https://iram-institute.org/science-portal/proposals/lp/completed/a-noema-spectroscopic-redshift-survey-of-bright-herschel-galaxies/}.

\section{Results}
\label{section:results}

The observations from the $z$-GAL project resulted in a large data set containing a wealth of information about the {\it Herschel}-selected sources. In addition to unambiguous spectroscopic redshifts measurements, the NOEMA data provide key information on the morphology and the multiplicity of the sources, and high-quality measurements of the properties of the emission lines (widths and line fluxes) as well as on the underlying dust continuum. In order to display the richness of the data and present in the clearest possible way an overview of the $z$-GAL survey, we built a catalogue  (Appendix~\ref{Appendix: Appendix A}) that shows, for each of the $z$-GAL sources, the combined continuum and emission lines' maps together with all the molecular and atomic emission lines that were detected in the HeLMS, HerS and HerBS sources (Fig.~\ref{figure:spectra_continuum_HeLMS}, \ref{figure:spectra_continuum_HerS} and \ref{figure:spectra_continuum_HerBS}, respectively). The continuum maps and emission lines detected in the sources from the Pilot Programme have been presented in \citet{Neri2020}. The sources' positions, the spectroscopic redshifts and emission lines properties are tabulated in Table~\ref{Appendix_table:emission-lines1} for the HeLMS and HerS sources and in Table~\ref{Appendix_table:emission-lines2} and \ref{Appendix_table:emission-lines3} for the HerBS sources. 

In this section, we summarise the main results on the spectroscopic redshifts (Sect.~\ref{section:spectroscopic-redshifts}) and the line widths of the sources (Sect.\ref{section:line widths}) and present a comparison between the spectroscopic and photometric redshifts in Sect.~\ref{section:photometric-redshifts}. The morphology and general properties of the sources will be discussed in Sect.~\ref{section:discussion}.

\subsection{Spectroscopic redshifts}
\label{section:spectroscopic-redshifts}
The main goal of the $z$-GAL project has been achieved by measuring, in the originally selected sample of 137 {\it Herschel}-selected sources, robust spectroscopic redshifts for 135 sources based on the detection of at least two emission lines in the 3 and 2~mm spectral bands. Taking into account the individual sources that were identified in some of the fields as double or multiple, as well as sources with different redshifts in the same field (e.g. HerBS-43), there are, in total, 165 individual sources with robust redshifts, including lensed galaxies, binary systems (such as HerBS-70 and HerBS-95) and clusters (e.g. HerBS-150) in the $z$-GAL sample (Sect.~\ref{section:morphology}). The sources span the redshift range $0.8<z<6.5$, with a median of $z_{\rm median} = 2.56 \pm 0.10$ and are centred around the peak of cosmic evolution (Fig.~\ref{figure:spectroscopic-redshifts}), which is consistent with the original selection of sources with $z_{\rm phot}\gtrsim2$. The majority of the $z$-GAL sources are at $z<3.5$ (i.e. $86\%$) and only 21 sources are at $z>3.5$. 

As shown in the bottom panel of Fig.\ref{figure:spectroscopic-redshifts}, the spectroscopic redshift distribution of the $z$-GAL sources is comparable to both the distributions of the {\it Herschel}-selected sample of southern H-ATLAS sources, the BEARS survey \citep{Urquhart2022}, and the {\it Herschel}-selected sources for which spectroscopic redshifts were measured in previous studies \citep[as listed in][]{Nayyeri2016, Bakx2018}. This similitude reflects the fact that the wavelengths and the source selection criteria of these surveys are comparable to those of the $z$-GAL survey.  

This is in contrast with the spectroscopic redshift distribution of the SPT-selected galaxies reported in \cite{Reuter2020}, which is clearly different than all the samples based on {\it Herschel}-selected galaxies (Fig.\,\ref{figure:spectroscopic-redshifts}), with the SPT sources peaking at significantly higher redshifts. The SPT-selected galaxies show a flat distribution between $z=2.5$ and $z=5.0$, with a major fraction ($76\%$) of the sample at $z>3$ and a median redshift of $z_{\rm median}=3.9\pm0.2$. The SPT galaxies were selected from a survey performed at a longer wavelength than the {\it Herschel} surveys and the difference in redshifts between the SPT and {\it Herschel}-selected galaxies is consistent with expectations for the selected wavelengths of these surveys \citep[see, e.g.][and references therein]{Reuter2020, Bethermin2015}. The systematic study of galaxies selected from the {\it Herschel} and SPT surveys thus offers an opportunity to gather critical complementary information on galaxy populations at different epochs of cosmic evolution, with {\it Herschel}-selected sources probing the peak of star formation activity around $2<z<3$, while the SPT-selected galaxies provide crucial information about the nature of star formation at earlier epochs. 

For all but two (HerBS-82 and HerS-19) of the selected sources, we detected at least two emission lines, mostly from $^{12}$CO ranging from the $J$=2$-$1 to the 7$-$6 transition and, in some case, from $\rm H_2O$ emission lines, $\rm [C {\small I}]$ and, in one case, $\rm HCO^+$ and HCN (Table~\ref{Appendix_table:emission-lines1} and \ref{Appendix_table:emission-lines2}). The majority of the sources, that is 85\%, were detected in two transitions of $^{12}$CO emission, only 8\% were detected in three $^{12}$CO emission lines and 21\% in $^{12}$CO and lines of other species (water or atomic carbon). In total, 40 sources were detected in $^{12}$CO(2-1) and 109 sources in $^{12}$CO(3-2). The CO emission lines are all relatively strong, resulting in high signal-to-noise ratios ($\rm S/N>5$) that provide the necessary quality to derive precise and reliable redshifts. In addition, significant information is available about the properties of the molecular gas, including the morphology and dynamics for sources that are resolved, and the physical conditions of the molecular gas (Paper~III).

Based on the successful outcome of $z$-GAL, the following remarks can be made regarding the derivation of robust redshifts: (i) reliable spectroscopic redshift need most of the time the detection of  at least two lines and one line is never enough; (ii) photometric redshifts are only indicative and sometimes ambiguous. Their values depend on the SED wavelength coverage and adopted template for the SED used in the analysis and are found to be, on average, within 20-30\% of the $z_{\rm spec}$ values (Fig.~\ref{figure:z_phot-vs-z_spec} and Sect.~\ref{section:photometric-redshifts}); (iii) most emission lines detected are from $^{12}$CO (mostly $J$=2-1 to 5-4), with some sources seen in $\rm H_2O$ and $\rm [C{\small I}]$ emission lines; for one source at $z=0.85$, the two detected lines are HCN(3-2) and $\rm HCO^+$(3-2); and (iv) some targets required a third line to lift any ambiguity, as different redshift solutions were compatible with the available photometry. This is mainly due to the degeneracy of the CO ladder, resulting in frequencies of the emission lines that are equally spaced. As an example, two emission lines were detected at 97.75 and 145.60~GHz in the source HerBS-108 (Appendix~\ref{Appendix_table:emission-lines2}). These lines could be identified either with $^{12}$CO(2-1) and $^{12}$CO(3-2) or with $^{12}$CO(4-3) and $^{12}$CO(6-5) indicating spectroscopic redshifts of $z_{\rm spec}=1.25$ or $z_{\rm spec}=3.71$, respectively. The photometric redshift for HerBS-108 ($z_{\rm phot}=2.4$; Table~\ref{Appendix_table:sources2}) lies in between these two values. The measurement of a third line, detected at 73.3~GHz and identified with $^{12}$CO(3-2), provided the required information to derive a robust spectroscopic redshift of $z_{\rm spec}=3.7166$ (Table~\ref{Appendix_table:emission-lines2}). It should be noted that, in many of those cases, the relative intensities of the CO emission lines can provide additional useful constraints, as some solutions seem to be incompatible with the expected excitation conditions in high-$z$ dusty star-forming galaxies. Although, differential lensing could result in unusual CO line ratios (see, e.g. the case of HeLMS-19 described in Appendix~\ref{Appendix: Comments on sources}), for all the \zgal sources where a third line was measured to disentangle between two spectroscopic redshift solutions, the CO line fluxes were always found to be compatible with an excitation akin to dusty starburst galaxies. 

\subsection{Line widths}
\label{section:line widths}
The $z$-GAL sources display an exceptionally wide distribution of line widths, as shown in the histogram in Fig.\ref{figure:line_widths}, with a first ensemble of sources ($\sim 65\%$ of the sample) with a FWHM of $\rm \Delta V < 700  \, km \, s^{-1}$ and a second one (representing 35\% of the sample) displaying broad lines, extending from $\rm 700  \, km \, s^{-1}$ to $\rm 1800 \, km \, s^{-1}$ for the most extreme case (HerBS-147 - see Appendix~\ref{Appendix: Comments on sources}). The mean value for the FWHM of the \zgal sample is $\rm 590\pm 25 \, km \, s^{-1}$. Many of the broad line sources show profiles that are asymmetrical or double-peaked, with separation between the peaks of up to a few $\rm 100 \, km \, s^{-1}$, indicative of merger systems and/or rotating disks (as illustrated in the catalogue in Appendix~\ref{Appendix: Appendix A}). 

Compared to other samples of high-$z$ dusty-star forming galaxies, the $z$-GAL sample stands out by the large fraction of sources with very large line widths. This is illustrated in Fig.~\ref{figure:line_widths} where the \zgal sample is compared to the BEARS sample \cite{Urquhart2022} and the sources reported in \citet{Harris2012}, that have mean values of the FWHM of $\rm 465\pm 37 \, km \, s^{-1}$ and $\rm 485\pm73 \, km \, s^{-1}$, respectively, both distinctly smaller than the mean of the \zgal sample. In the case of the {\it Herschel}-selected H-ATLAS sources in the BEARS sample, only ten sources display line widths in excess of $\rm 700 \, km \, s^{-1}$ (Fig.~\ref{figure:line_widths}). The unlensed DSFGs from \cite{Bothwell2013} have a mean value of the FWHM of $\rm 510 \pm 80 \, km \, s^{-1}$, with individual values ranging from 200 to $\rm 1400 \, km \, s^{-1}$, very similar to the mean value of $\rm 540\pm40 \, km s^{-1}$ derived for a sample of 45 high-$z$ DSFGs presented by \citet{Birkin2021}. For the SPT sources, the weighted-average FWHM is $\rm 370 \pm 130 \, km \, s^{-1}$, a value that is comparable to the average of lensed DSFGs and Main Sequence $z<2$ galaxies \citep[][and references therein]{Aravena2016}.

\begin{table*}[!htbp]
\caption{Multiplicity of the $z$-GAL sources.}
\begin{tabular}{lcl}
\hline
\hline
\vspace{0.3cm}
Number of sources detected within the NOEMA field & Nb. of fields & Source names \\
\hspace{0.7cm} Emission lines in fields with multiple sources   &     & \\
\hline
1   &  102  &    \\ 
    &      &     \\
2   &      &    \\
\hspace{0.7cm} Emission lines for both sources (same $z_{\rm spec}$)   & 6 & HeLMS-1 \& 19; HerS-3; HerBS-61, 92 \& 116   \\
\hspace{0.7cm} Emission lines for both sources (different $z_{\rm spec}$) & 12 & HeLMS-26, 40; HerS-18; HerBS-43 \& 70 \\
     &    & HerBS-76, 95, 124, 162, 194, 199 \& 204 \\
\hspace{0.7cm} Emission lines for one source & & \\
\hspace{0.7cm} with second source only detected in the continuum &  5 & HeLMS-32, 35 \& 43;  HerBS-74 \& 89  \\  
    &      &   \\
3   &      &   \\
\hspace{0.7cm} Emission lines for all sources (different $z_{\rm spec}$) & 5 & HeLMS-17; HerBS-38, 109, 150 \& 205 \\
\hspace{0.7cm} Emission lines for two sources (different $z_{\rm spec}$) &  & \\
\hspace{0.7cm} with the third source only detected in the continuum & 4 & HerS-19; HerBS-53, 91 \& 187\\
\hspace{0.7cm} Emission lines for one source &  & \\
\hspace{0.7cm} with two other sources only detected in the continuum & 3 & HerS-7; HerBS-105 \& 180 \\

\hline
\end{tabular}
\tablefoot{All the 137 NOEMA fields are centred on the coordinates of the {\it Herschel}-selected sources. The names of the {\it Herschel} sources are only given for the fields where two or three sources were detected. Detailed comments on these sources are given in Appendix~\ref{Appendix: Comments on sources}.}
\label{table:source-morphology}
\end{table*}

\subsection{Comparison to photometric redshifts}
\label{section:photometric-redshifts}

As has been shown in previous studies, the spectroscopic redshifts of bright high-$z$ dusty galaxies can significantly differ from the photometric redshifts based on the available measurements of the dust continuum flux densities \citep[e.g.][]{Neri2020, Reuter2020, Urquhart2022}. Deriving redshifts using sub-millimetre SEDs of galaxies with known redshifts and dust temperatures as templates remains uncertain \citep[e.g.][]{Jin2019, Bendo2023}. This is particularly true when using SPIRE data alone because the 250, 350, and 500~$\rm \mu m$ bands are close to the peak of the observed SED for $2<z<4$ galaxies. In order to homogenise the photometric redshifts of all three HerBS, HerMES and HeRS samples and show differences when using different analysis, we here adopted the methods recommended by \cite{Ivison2016} and \cite{Pearson2013}. The former method is based on functions fitted to the SED of the Cosmic Eyelash (J2135$-$0102) and two composite SEDs of multiple sub-millimetre galaxies; this method has been shown to perform better than other methods to estimate photometric redshifts \citep[see discussion and references in][]{Blain1999, Bendo2023}. The latter method is based on the sums of two blackbody SEDs (based only on {\it Herschel} flux densities) and derived from H-ATLAS sources with known redshifts. The $z_{\rm phot}$ values derived for the sources of $z$-GAL sample using both methods are listed in Table~\ref{Appendix_table:sources1} to \ref{Appendix_table:sources3}. 

The $z_{\rm phot}$ values are on average consistent within 20-30\% of the $z_{\rm spec}$ values, with deviations that become larger for redshifts $z>4.5$; only for a few sources are the two values comparable within 10\% (Fig.~\ref{figure:z_phot-vs-z_spec}). This is in agreement with previous studies that have shown that the average uncertainty in a photometric redshift estimate is $\Delta(z)/(1+z) = 0.12$ for sources for which prior information indicates that they are at $z>1$ \citep{Pearson2013}. The poor accuracy and reliability of redshifts derived from sub-millimetre continuum photometry is due to the degeneracy between temperature, the spectral index $\beta$, and redshifts \citep[e.g.][]{Bendo2023}. In addition, the presence of multiple sources in the field, particularly when the spectroscopic redshifts of the sources are different, introduces another level of inaccuracy in the photometric redshift derived from the {\it Herschel} and SCUBA-2 flux densities (e.g. HerBS-109 or HerBS-53). The derived values of the redshifts based on continuum measurements alone remain therefore indicative and not precise enough to follow up efficiently with targeted observations of molecular or atomic gas.

\begin{figure*}
   \centering
\includegraphics[width=0.85\textwidth]{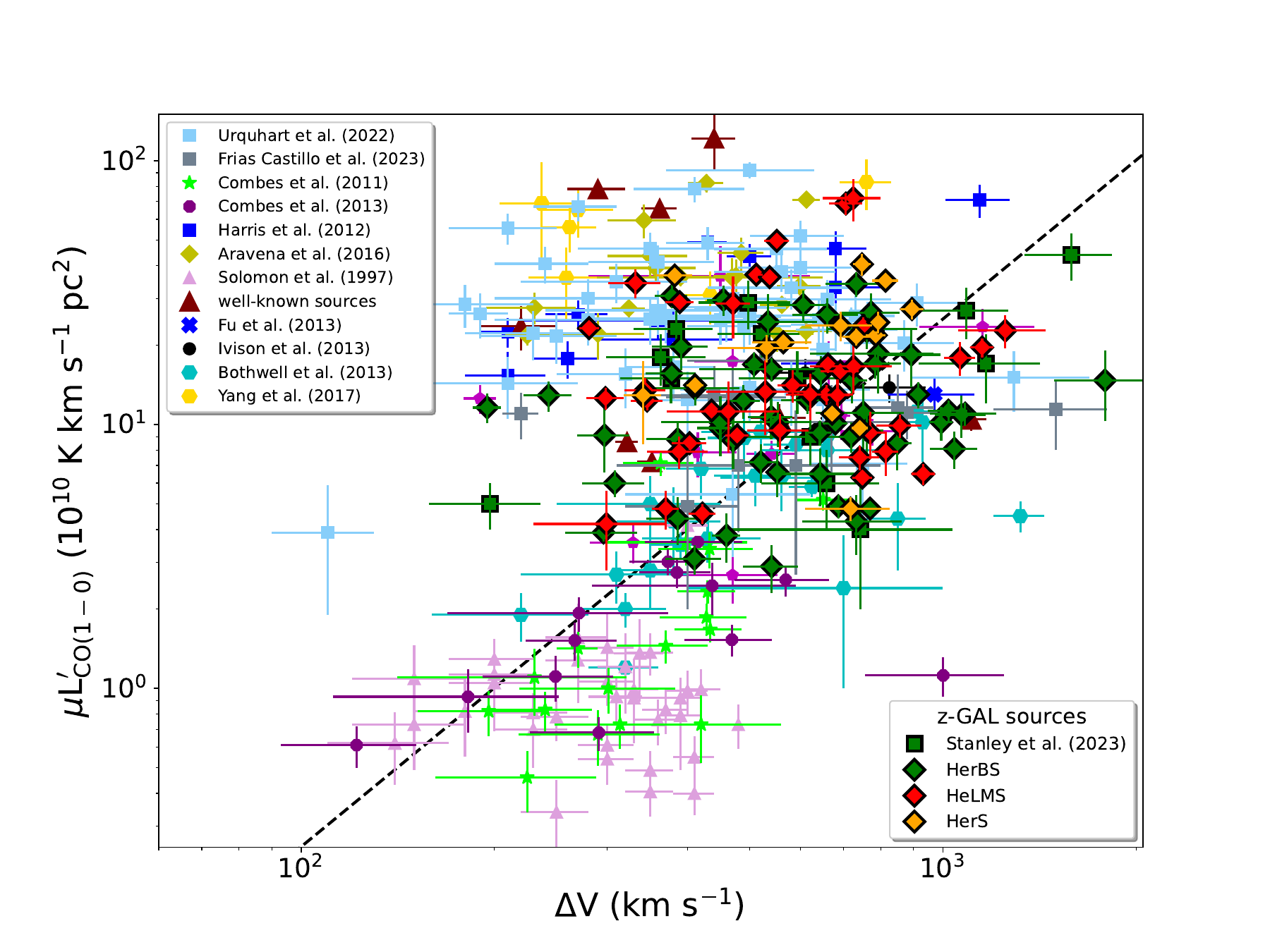}
\caption{$^{12}$CO luminosity ($L'_{\rm CO(1-0)}$) plotted against the linewidth ($\Delta V$) of the $^{12}$CO emission line for the $z$-GAL sources. The HerS, HeLMS and HerBS samples are identified with orange, red and green diamond shaped symbols, respectively, where the values for $L'_{\rm CO(1-0)}$ are from Paper~III. The sources of the Pilot Programme from \citet{Neri2020} are shown as green squares with the $^{12}$CO(1-0) luminosity values from \cite{Stanley2023}. The $z$-GAL sample is compared to high-$z$ lensed and unlensed galaxy samples \citep{Harris2012, Bothwell2013, Yang2017, Urquhart2022, FriasCastillo2023}, individual galaxies including the two binary hyper-luminous systems HATLAS~J084933 at $z=2.41$ \citep{Ivison2013} and HXMM01 at $z=2.308$ \citep{Fu2013}, well-known high-$z$ sources, both lensed (IRAS~F10214, Eyelash, Cloverleaf, APM~08279 and the Cosmic Eyebrow) and unlensed (BR1202N and S, and BR1335), from \citet[][and references therein]{Carilli-Walter2013} and \cite{Dannerbauer2019}, and local ultra-luminous infrared galaxies (ULIRGs) \citep{Combes2011, Combes2013, Solomon1997} with corresponding symbols as indicated in the upper left panel. The sources of \cite{Harris2012} are the ones reported in Table~1 of that paper, except for HATLAS~J084933 where we used the follow-up measurements of \cite{Ivison2013}. No correction for amplification was applied to the $^{12}$CO luminosities. With the exception of the sources of this paper and the sources selected from \cite{Yang2017}, \cite{Bothwell2013} and \cite{Urquhart2022}, all the other sources used in this plot have been measured in $\rm ^{12}CO$(1-0) or in $\rm ^{12}CO$(2-1). Corrections for excitation were applied for sources for which only $^{12}$CO transitions higher than $J$=2-1 are available, using the ratios listed in \cite{Carilli-Walter2013}. The dashed line shows the best-fitting power-law fit derived from the data for the unlensed DSFGs, $L'_{\rm CO(1-0)} = 10^{5.4} \times \Delta V^2$ \cite[e.g.][]{Bothwell2013, Zavala2015}. 
}
\label{fig:Deltav-Lco1-0}
\end{figure*}

\begin{figure*}
   \centering
\includegraphics[width=0.85\textwidth]{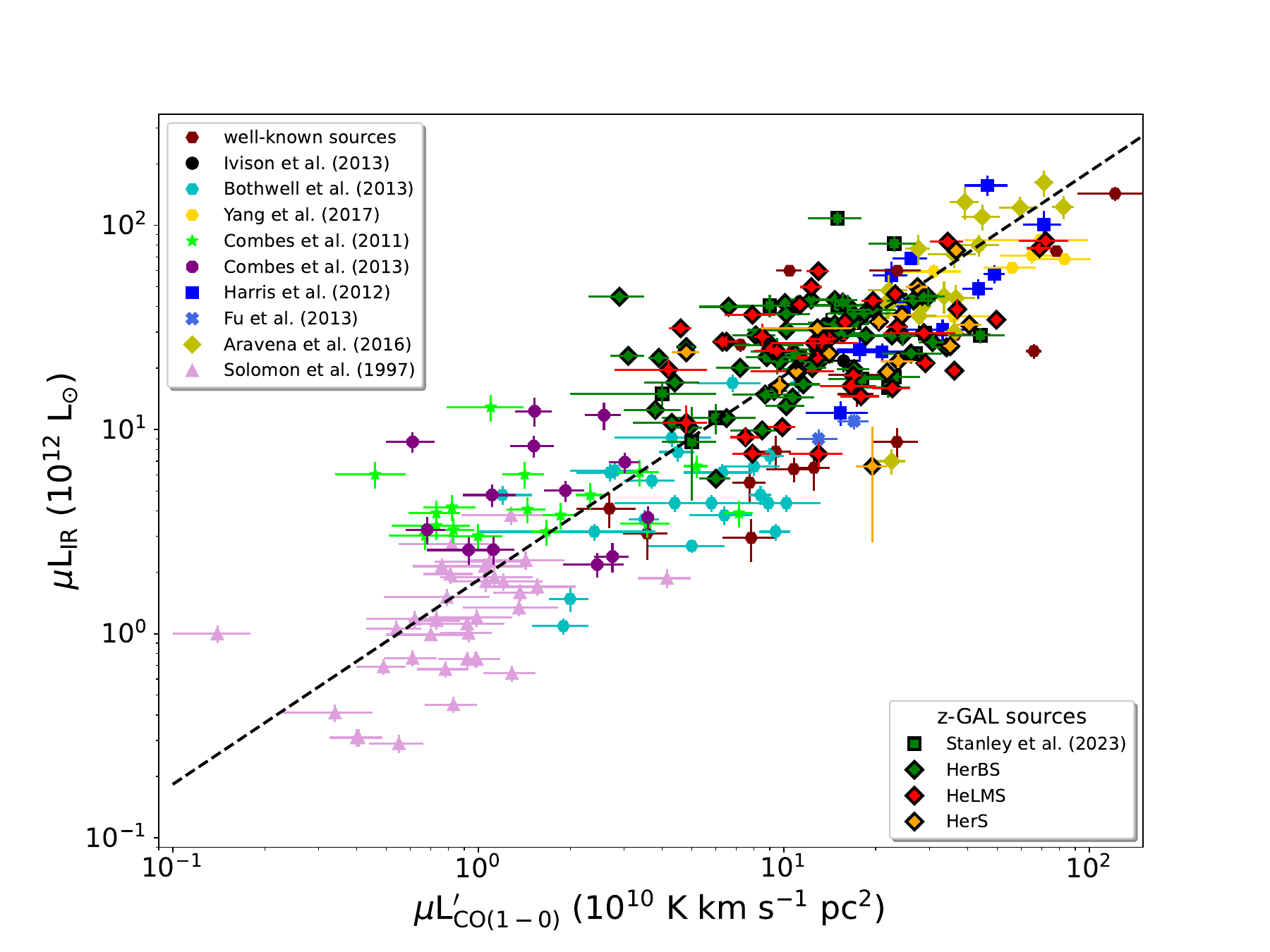}
\caption{Infrared luminosity ($L_{\rm IR}$; 8-1000~$\mu$m) as a function of $L'_{\rm CO(1-0)}$ for the $z$-GAL sample, including the sources from the Pilot Programme \citep{Neri2020, Stanley2023}. The sources are identified using the same symbols as in Fig.~\ref{fig:Deltav-Lco1-0}. The infrared luminosities are derived from Paper~II (Sect.~\ref{section:general_properties}). For comparison, we also plot high-$z$ lensed and unlensed galaxies, as well as local ultra-luminous infrared galaxies (ULIRGs) from the literature (also shown in Fig.~\ref{fig:Deltav-Lco1-0}) with references and corresponding symbols given in the upper left panel. No correction for amplification was applied to the $^{12}$CO and infrared luminosities. The dashed line shows the linear relationship between $L_{\rm IR}$ and $L'_{\rm CO(1-0)}$ for a gas depletion timescale of $\tau_{\rm dep}=0.2 \, \rm Gyr$ (Paper~III).}
\label{fig:Lir-L'co(1-0)}
\end{figure*}

\section{Discussion} 
\label{section:discussion}

In this section, we describe general properties of the sample of high-$z$ bright {\it Herschel}-selected galaxies observed in $z$-GAL. We first comment on the morphology of the sources and highlight some of their properties as a guide to the catalogue presented in Appendix~\ref{Appendix: Appendix A} and then focus on global characteristics of the sample, based on relationships between the infrared luminosity, the CO luminosity and the line widths of the sources. Finally, we place the main findings of this overview in the context of the sample selection and will discuss the implications for probing the activity of DSFGs at the peak of cosmic evolution.   

\subsection{Morphology of the $z$-GAL sources}
\label{section:morphology}
The main goal of this section is to highlight the diversity of the galaxies observed in the \zgal project, comment on the multiplicity of the sources and derive first conclusions on the sources' properties, that will be further analysed in Bakx et al. (in prep.) based on multi-wavelength ancillary data and a more in-depth exploitation of the NOEMA data. Detailed comments on individual sources are given in Appendix~\ref{Appendix: Comments on sources} and the combined continuum and emission line maps are shown in Figs.~\ref{figure:spectra_continuum_HeLMS}, \ref{figure:spectra_continuum_HerS} and \ref{figure:spectra_continuum_HerBS} for the HeLMS, HerS and HerBS sources, respectively. 

All the sub-millimetre bright galaxies in the $z$-GAL sample remained unresolved in the $\rm 17\arcsec$-$35\arcsec$ beam of the {\it Herschel}-SPIRE and $13\arcsec$ beam of the SCUBA-2 observations. The angular resolution of NOEMA, between $1\farcs2$ to $3\farcs5$ at 2~mm, enabled us to measure with high precision the location of each of the {\it Herschel}-selected sources and, in some cases, to resolve them or to reveal multiple counterparts in the NOEMA field. A summary of the sources' morphology is provided in Table~\ref{table:source-morphology}, which highlights the number of double and triple systems found in the \zgal survey.      

In the majority of the cases (102 in total), the {\it Herschel} field contains only one source in the NOEMA data. In a few cases, unrelated sources are detected in the field of view of these high-$z$ galaxies (Appendix~\ref{Appendix: Comments on sources}). For instance, about $15\arcsec$ to the east of HerBS-74, a galaxy at $z=2.559$, a strong continuum emission source is detected that corresponds to a foreground galaxy seen in the {\it HST} near-infrared image (Borsato et al. subm.). In the case of HerBS-89a, a starburst at $z=2.95$, a weak source is detected in the continuum about $6\arcsec$ to the east \citep{Neri2020, Berta2021}; no emission lines are seen in the NOEMA data and there is no corresponding source in the SDSS catalogue. 

Some of these unresolved high-$z$ galaxies are likely lensed (Sect.~\ref{section:general_properties}) and higher angular resolution follow-up observations are needed to confirm their precise morphology. Examples of such gravitationally lensed sources, that have been recently confirmed, include HerBS-89a and HerBS-154 from the Pilot Programme \citep{Neri2020} where $0\farcs3$ resolution follow-up observations at 1-mm with NOEMA revealed in these sources $1\farcs0$ and $2\farcs0$ diameter Einstein rings \citep[][and Butler et al. in prep.]{Berta2021}.

A few $z$-GAL sources were resolved in the NOEMA observations. Remarkable cases include HeLMS-19 that displays two close components at the same redshift ($z_{\rm spec}=4.688$) tracing the arcs of a large Einstein ring seen in the {\it HST} near-infrared image (Borsato et al. subm.), and HerBS-204, a pair of interacting galaxies at $z_{\rm spec}=3.293$ linked by extended emission, which is reminiscent of the Antennae galaxy (NGC~4038/NGC~4039). 

Other fields contain multiple sources separated by $5\arcsec$ to $10\arcsec$ with redshifts that differ slightly and separations that are too large to be consistent with gravitational amplification. These sources point to genuine binary or triple systems. Examples are the binary systems HerBS-70 and HerBS-95 at $z_{\rm spec}=2.31$ and $z_{\rm spec}=2.97$ \citep[][the former is part of a proto-cluster; Bakx et al. subm.]{Neri2020}, and the triple systems HerBS-150 and HerBS-205 at $z_{\rm spec}=2.960$ and 3.67, respectively. 
 
In total, we identified 184 individual sources in the 137 fields centred on the {\it Herschel}-selected sources (Table~\ref{table:source-morphology}), 169 of which are detected in emission lines. Of these sources, precise redshifts were measured for 165 of them; the sources for which only tentative redshifts were derived are HerBS-82, the west and south-east sources of HerS-19, and the source to the west of HerS-18, likely tracing a foreground galaxy (further detailes are provided in Appendix~\ref{Appendix: Comments on sources}). 

\subsection{General properties of the $z$-GAL sample}
\label{section:general_properties}
We present in this sub-section the global properties of the $z$-GAL sample and place the findings of this survey in the context of other studies of high-$z$ DSFGs and local ultra-luminous infrared galaxies (ULIRGs). The two relationships that will be used are $L'_{\rm CO(1-0)}$  vs. $\rm \Delta V$ and $L_{\rm IR}$ vs. $L'_{\rm CO(1-0)}$  (Figs.\ref{fig:Deltav-Lco1-0} and \ref{fig:Lir-L'co(1-0)}). Detailed analysis of these relationships are presented in Paper~III for the latter to derive the depletion timescales and in Bakx et al. (in prep.) for the former to constrain the nature of the sources.   

We compute the $^{12}$CO luminosities of the sources (in $\rm K \, km \, s^{-1} \, pc^2$) 
using the standard relation given by \cite{Solomon-VandenBout2005}, 
\begin{equation}\label{eq:CO-Luminosity}
L'_{\rm CO} = 3.25 \times 10^7 \, S_{\rm CO} \, {\Delta V \, \nu_{\rm CO}^{-2}} \, D_{\rm L}^2 \, (1+z)^{-1}
,\end{equation}
where $S_{\rm CO}$\,$\Delta V$ is the velocity-integrated $^{12}$CO line flux in $\rm Jy \, km s^{-1}$, $\rm \nu_{CO}$ the rest frequency of the $^{12}$CO emission line in GHz, and $D_{\rm L}$ the luminosity distance in Mpc. The $L'_{\rm CO(1-0)}$ luminosities were derived using the lowest available $J\rightarrow{}(J-1)$ $^{12}$CO transition for each \zgal source after applying a correction for the excitation adopting the median brightness temperature ratios values listed in \cite{Carilli-Walter2013}, and applying similar corrections, where needed, for sources taken from the literature (see Berta et al. 2023 for the tabulated values). For the sources of the Pilot Programme, we used the $L'_{\rm CO(1-0)}$ values and the line widths from \cite{Stanley2023}. All the $L'_{\rm CO(1-0)}$ values were recalculated for the cosmology adopted in this study. 

The total infrared luminosities $L_{\rm IR}$(8-1000$\rm \mu m$) of the \zgal sources are based on the 50-1000~$\rm \mu m$ luminosities derived and listed in Paper~II after applying a correction of 0.7 following \citet{Berta2013} - readers can refer to Paper~III for more details.  

\subsubsection{The $L'_{\rm CO(1-0)}$-$\Delta V$ relationship}
\label{section: Tully-Fisher}

Figure~\ref{fig:Deltav-Lco1-0} shows the relation between the apparent $^{12}$CO luminosities, $\mu L'_{\rm CO(1-0)}$, and the FWHM ($\Delta V$) of the $^{12}$CO emission lines for all the \zgal sources detected in $^{12}$CO(1-0) up to (5-4) emission lines (corresponding to a total of 122 sources). The \zgal sample is compared to high-$z$ lensed and unlensed DSFGs, as well as local ULIRGs from the literature (see figure caption for details and references). With the exception of the sources selected from \cite{Urquhart2022}, \citet{Yang2017} and \citet{Bothwell2013}, all the other sources selected from the literature have been observed in $^{12}$CO(1-0) or in $^{12}$CO(2-1). In the case of the BEARS sample of \citet{Urquhart2022}, we retained, to be consistent with the approach adopted for $z$-GAL, only those sources for which two emission lines were detected and for which unambiguous spectroscopic redshifts could be derived (50 sources in total).  

This relationship, also known as the Tully-Fischer relationship, has first been based on observations of the 21~cm line of atomic hydrogen in nearby galaxies \citep{Tully&Fisher1977} and was proposed to be used for CO observations by \cite{Dickey&Kazes1992} - see also \cite{Schoniger&Sofue1994} and references therein. The $^{12}$CO line emission traces the kinematics of the potential well in which a galaxy's molecular gas lies, and can therefore provide a measure of the dynamical mass of the galaxy, modulo any inclination or dispersion effects. The CO Tully-Fisher relationship therefore provides a rough estimate of the correlation between the gas mass and the galaxy dynamics. As was first pointed out by \cite{Harris2012}, strong lensing boosts the apparent CO luminosity without affecting the line width, placing the gravitationally amplified sources above the roughly quadratic relationship between $L'_{\rm CO(1-0)}$ and $\Delta V$ that is observed for the unlensed sources. Since then, this relationship has therefore been used in various studies of the molecular gas traced by CO in high-$z$ DSFGs \cite[e.g.][]{Bothwell2013, Carilli-Walter2013, Aravena2016, Yang2017, Neri2020, Birkin2021, Urquhart2022, Stanley2023}. However, the usefulness of this effect for measuring the lensing magnification has been debated in the literature, and the consensus is that deriving exact values for the magnification factors, based on the Tully-Fischer relationship and using unresolved CO data, remains unreliable \cite[e.g.][]{Aravena2016, Stanley2023}. Notwithstanding this caution, the separation between strongly lensed and unlensed (or weakly lensed) sources is clearly present in the $L'_{\rm CO(1-0)}$ versus $\Delta V$ relationship including high-$z$ galaxies and, in principle, could be used to distinguish sources that are strongly lensed from intrinsically hyper-luminous sources. 

The most obvious feature in the relationship shown in Fig.~\ref{fig:Deltav-Lco1-0} is indeed the difference in position between the unlensed or weakly lensed galaxies that lie along the power-law fit derived from the data for the unlensed DSFGs \citep{Bothwell2013, Zavala2015} and the sources that are strongly lensed (see references in the figure caption). Interestingly, the majority of the \zgal sources are located in the upper part of the relationship indicating that several of them are strongly lensed, similarly to what is observed for the sources from the BEARS sample. Many of the sources that are farthest away from the relationship, that is with the largest apparent $L^\prime_{\rm CO(1-0)}$ luminosities, have line widths $\rm \lesssim 700 \, km s^{-1}$, and include sources that are located amongst the most magnified high-$z$ galaxies known to date. In contrast, the sources with the larger line widths (that are dominated by the \zgal sample) tend to be distributed along or close to the quadratic relationship for the unlensed sources. These sources could correspond to hyper-luminous galaxies ($L_{\rm IR} \gtrsim 10^{13} \, L_\odot$) akin to the binary sources HATLAS~J084933 \citep{Ivison2013} and HXMM01 \citep{Fu2013}. The number of these potential HyLIRGs is remarkable amongst the sample of \zgal bright {\it Herschel}-selected galaxies and more detailed observations are required to explore in greater detail their nature and derive their intrinsic properties. 

\subsubsection{The $L_{\rm IR}$-$L'_{\rm CO(1-0)}$ relationship}
\label{section: Lir-Lco}

Fig.~\ref{fig:Lir-L'co(1-0)} shows the relationship between $L^\prime_{\rm CO(1-0)}$ and $L_{\rm IR}$ for the \zgal sample, compared to other samples of high-$z$ DFSGs and local ULIRGs from the literature (references are provied in the figure caption). The infrared luminosity is a measure of the on-going star formation activity, while the CO emission is a tracer of the molecular gas available to form stars. The $L_{\rm IR}$-$L^\prime_{\rm CO(1-0)}$ relation is therefore an indication of the fraction of the molecular gas ($M_{\rm mol}$) that is converted to new stars, measuring the star formation efficiency, and equivalent to the integrated Kennicutt-Schmidt relation \citep{Kennicutt1998}. There is a tight linear correlation between $L_{\rm IR}$ and $L^\prime_{\rm CO(1-0)}$ with the \zgal sources populating the high end of the infrared luminosities ($6\times10^{12} < L_{\rm IR}/L_{\odot} < 8\times10^{13}$) and the region corresponding to the most massive galaxies ($4 < L^{\prime}_{\rm CO(1-0)}/{\rm 10^{10} K \, km \, s^{-1} \, pc^2} < 80$), with $M_{\rm mol}>10^{11} M_\odot$, intermediate between the high-$z$ strongly lensed galaxies \citep[e.g. the SPT sample;][]{Aravena2016} and the unlensed high-$z$ and local ULIRGs \citep{Bothwell2013, Combes2011, Solomon1997}. It should be noted that this relationship will hold when correcting the lensed sources for the magnification, as the amplification will affect similarly the $^{12}$CO and the dust continuum emission (provided that they are co-spatial) and shift the data points along the relationship. The gas depletion timescales for the \zgal sample are estimated and discussed in Paper~III and are shown to be in the range between 0.1 and 1.0~Gyr, populating the part of the main sequence of star formation and the locus of starburst outliers.           
 
\subsection{Selection effects}
Based on the previous sections, we subsequently place the results in the context of the sources' selection criteria and highlight some of the main findings with respect to model predictions \citep[e.g.][]{Negrello2010}. 

The \zgal sources were selected based on a flux density threshold at 500~$\mu$m, using $S_{\rm 500 \mu m} > 100 \, {\rm mJy}$ for the HeLMS and HerS sources \citep{Nayyeri2016} and $S_{\rm 500 \mu m} > 80 \, {\rm mJy}$ for the HerBS sources \citep{Bakx2018}. This selection resulted in a sample containing amongst the brightest high-$z$ galaxies, including potentially gravitationally lensed galaxies. Indeed, \cite{Negrello2010} demonstrated that strongly lensed DSFGs dominate over unlensed galaxies for $S_{\rm 500 \mu m} > 100 \, \rm mJy$, a flux density region where the count of unlensed DSFGs decreases steeply. Using a galaxy evolution model for the HerBS fields specifically, \cite{Bakx2020c} showed that $\sim$70\% of the HerBS sources are predicted to be lensed. 

From the \zgal sample selection criteria, aimed at identifying infrared bright high-$z$ galaxies that are potentially lensed \citep[see, e.g.][]{Negrello2017}, the majority of the selected sources are expected to be single and unresolved in the NOEMA observations. The number of fields for which single sources were detected amounts to 102 (Table~\ref{table:source-morphology}), representing 75\% of the \zgal sample, which is in line with this expectation. The results from the \zgal survey further support the predictions from \cite{Negrello2010} and \cite{Bakx2020c} by the large number of sources of the sample that are located well-above the power-law fit derived from the data for the unlensed DSGFs in the $L'_{\rm CO(1-0)}$ vs $\rm \Delta V$ relationship. It is interesting to note that this effect is more pronounced for the sources selected from the HeLMS and HerS samples than those from the HerBS sample, with about half of the HerBS sources distributed along or below the linear relation (Fig.~\ref{fig:Deltav-Lco1-0}). In addition, there are more HerBS sources with line widths in excess of $\rm 700 \, km s^{-1}$ (33 in total) than for the HeLMS and HERS sources (22 in total), with nearly 4$\times$ as much HerBS sources with $\rm \Delta V > 900 \, km s^{-1}$ than HeLMS and HerS sources (15 and 4, respectively). As noted in \cite{Aravena2016}, unlensed DSFGs tend to display larger linewidths and these differences could indicate that the sources selected from the HerBS sample contain a significant number of potential hyper-luminous galaxies. 

Another notable difference between the HeLMS and HerS samples and the HerBS sample is the number of NOEMA fields centred on the {\it Herschel} positions where multiple sources are revealed. Not taking into account resolved lensed sources (e.g. HeLMS-19 and HerBS-91) and only retaining the binary and triple systems (with spectroscopic redshifts), there are 16 HerBS sources that turned out to be multiple systems, compared to only 4 HeLMS and 2 HerS sources (Table~\ref{table:source-morphology} and Appendix~\ref{Appendix: Comments on sources}). 

Follow-up observations at higher angular resolution than the current NOEMA data are essential to fully explore the nature of the \zgal sources, further probe their properties and assess their true nature. The \zgal spectroscopic survey has allowed for significant progress in our understanding and classification of these bright {\it Herschel}-selected galaxies, centred on the peak epoch of cosmic galaxy formation, revealing lensed galaxies, multiple systems, interacting galaxies and potential HyLIRGs. The sources' characteristics extracted from the currently available data will hopefully serve as a foundation to study with better spatial resolution and sensitiviy some of these exceptional DSFGs as well as critical aspects of galaxy evolution.

\section{Conclusions}
\label{section:conclusions}
In this paper we introduce and present an overview of a Large Programme, {\it z}-GAL, using NOEMA to measure reliable redshifts for a sample of 126 bright {\it Herschel}-selected galaxies. From this original sample, spectroscopic redshifts were measured for 124 sources. Together with the $z$-GAL Pilot Programme \citep{Neri2020}, where 11 sources had their spectroscopic redshifts measured, this amounts to 135 dusty, bright {\it Herschel}-selected galaxies with precise redshifts, making it the largest sample of high-$z$ galaxies with unambiguous redshifts to date and more than doubling the number of {\it Herschel}-selected galaxies with robust spectroscopic redshifts. 

The results and conclusions are summarised as follows:    

\begin{itemize}
  \item Precise spectroscopic redshifts ($z_{\rm spec}$) were established for 135 of the {\it Herschel}-selected galaxies based on the detection of at least two emission lines, mostly from $\rm ^{12}CO$ ranging from the $J$=2--1 to the 7--6 transition. In addition, for 28 sources, we also report the detection of the emission of the atomic carbon fine-structure line $\rm [C{\small I}]\, (^3P_1$-$\rm ^3P_0)$, and for eight sources, the detection of water emission lines (mostly in the para-H$_2$O\,($2_{11}$-$2_{02}$) transition and for one source in the ortho-H$_2$O\,($4_{23}$-$3_{30}$) transition). For one source, the redshift was derived based on the detection of the HCN(3-2) and HCO$^+$(3-2) emission lines. The derived spectroscopic redshifts are in the range $0.8 <z< 6.5$ with a median value of $z_{\rm spec}=2.56\pm0.10$ and a tail in the distribution to $z>3.5$. 
  \item The majority of the \zgal targets are individual sources (102 in total), which are unresolved or barely resolved on scales of 15 to 25\,kpc. In total, 35 fields display multiplicity, including double and triple sources with lensed systems (e.g. HeLMS-19), mergers (e.g. HerBS-204), protoclusters (e.g. HerBS-150 and HerS-19), and fields with sources at different redshifts (e.g. HerBS-38). Taking these sources into account, the total number of individual sources in the $z$-GAL sample amounts to 184, of which 165 have robust spectroscopic redshifts. 
  \item The photometric redshifts ($z_{\rm phot}$), which depend on the wavelength coverage and the adopted template for the SED, are only indicative and are, on average for the sources studied in this paper, within 20-30\% of the $z_{\rm spec}$ values here reported. 
  \item The linewidths of the sources are large, with a mean value for the CO FWHM of $\rm 590 \pm 25 \, km \, s^{-1}$ and with $\sim 35\%$ of the sources having widths of $\rm 700 \, km \, s^{-1} < \Delta V < 1800 \, km \, s^{-1}$. Compared to other high-$z$ dusty star-forming galaxies, the $z$-GAL sample stands out by the large fraction of sources with large line widths. About half of the sources display double-peaked profiles indicative of merger systems and/or rotating disks. 
  \item The wealth of information on the molecular and atomic emission lines in the $z$-GAL offers the possibility to analyse their physical properties in detail and study their evolution across cosmic time. This analysis is presented in Paper~III.
  \item In addition to the emission lines, the measurements of the underlying dust continuum at different frequencies across the 3 and 2-mm bands provides a solid foundation to explore the dust properties of the \zgal sample and compare it to other available samples.  This study is presented in Paper~II. 
  \item Based on the location of the $z$-GAL sources in the $L'_{\rm CO(1-0)}$ versus $\Delta V$ relationship, we conclude that a majority of the selected high-$z$ galaxies are gravitationally amplified and that the sample includes HyLIRGs. Precise measurements of the amplification factors and the derivation of the properties of these sources will require higher resolution follow-up observations in the sub-millimetre and at near-infrared wavelengths to study the characteristics of the lensed high-$z$ galaxies and the foreground amplifying galaxies. 
\end{itemize}

The observations presented in this study have enabled one of the largest comprehensive measurement of redshifts of high-$z$ dusty star-forming galaxies, providing a useful database for exploring in detail the properties of these sources and, using follow-up observations, the properties of the lensing systems in the case of gravitational amplification. This extended sample provides, together with other already available redshift measurements, a sizeable and homogeneous ensemble of $\sim$300 bright {\it Herschel} selected galaxies with reliable redshifts, which will allow us to increase the number of known lensed galaxies at the peak of cosmic evolution, to provide a new sample of HyLIRGs, and to identify additional rare objects.

\begin{acknowledgements}
      This work is based on observations carried out under project numbers M18AB and, subsequently D20AB, with the IRAM NOEMA Interferometer. IRAM is supported by INSU/CNRS (France), MPG (Germany) and IGN (Spain). The authors are grateful to IRAM for making this work possible and for the continuous support that they received over the past four years, in particular, from the staff at the NOEMA observatory, to make this Large Programme a success. The authors are also grateful to the IRAM director for approving the DDT proposal that enabled to complete the survey. The anonymous referee is thanked for providing useful comments that helped to improve the contents of this paper. 
      We would like to thank I. Cortzen and C. Herrera for their contributions in the early stages of this project and recognise the essential work of the \zgal Cat-Team and Tiger-Team, who performed the reduction, calibration and delivery of the \zgal data. We are grateful to S. Wombat for continuous support and many discussions during this project. 
      This work benefited from the support of the project Z-GAL ANR-AAPG2019 of the French National Research Agency (ANR). 
      S.S. was partly supported by the ESCAPE project. ESCAPE - The European Science Cluster of Astronomy \& Particle Physics ESFRI Research Infrastructures has received funding from the European Union’s Horizon 2020 research and innovation programme under Grant Agreement no.~824064.  
      A.J.B. and A.J.Y acknowledge support from the National Science Foundation grant AST-1716585.  
      T. B. acknowledges support from NAOJ ALMA Scientific Research Grant No. 2018-09B and JSPS KAKENHI No.~17H06130, 22H04939, and 22J21948.
      S.J. is supported by the European Union's Horizon Europe research and innovation program under the Marie Sk\l{}odowska-Curie grant agreement No. 101060888.
      H.D. acknowledges financial support from the Agencia Estatal de Investigación del Ministerio de Ciencia e Innovación (AEI-MCINN) under the grant agreement PID2019-105776GB-I00/DOI:10.13039/501100011033 and acknowledge support from the ACIISI, Consejería de Economía, Conocimiento y Empleo del Gobierno de Canarias and the European Regional Development Fund (ERDF) under the grant agreement PROID2020010107.
      A.N. acknowledges support from the Narodowe Centrum Nauki (UMO-2020/38/E/ST9/00077).
      R.J.I acknowledges funding by the Deutsche Forschungsgemeinschaft (DFG, German Research Foundation) under Germany's Excellence Strategy -- EXC-2094 -- 390783311.
      C.Y. acknowledges support from ERC Advanced Grant 789410.
      D.A.R. acknowledges support from the National Science Foundation under grant numbers AST-1614213 and AST-1910107 and from the Alexander von Humboldt Foundation through a Humboldt Research Fellowship for Experienced Researchers as well as from the Deutsche Forschungsgemeinschaft (DFG) through SFB~956.
\end{acknowledgements}

\bibliographystyle{aa} 
\bibliography{references} 

\begin{appendix}

\section{Catalogue of the $z$-GAL sources} 
\label{Appendix: Appendix A}

The following catalogue presents, for all the \zgal sources, the combined continuum and emission line maps together with the molecular and atomic lines that were detected in each galaxy identified in the NOEMA field around the {\it Herschel} position. A series of tables lists the derived spectroscopic redshifts for each of the sources and the properties (widths and line fluxes) of all the emission lines that were detected.  

The {\it Herschel} bright high-$z$ galaxies that were selected for the \zgal project are listed in Tables~\ref{Appendix_table:sources1}, \ref{Appendix_table:sources2} and \ref{Appendix_table:sources3}, that include the {\it Herschel} coordinates, the SPIRE and, when available, the SCUBA-2 flux densities, together with the estimated photometric redshifts. The sources without $z_{\rm spec}$, that were not selected in (or dropped from) $z$-GAL, are presented in Table~\ref{Appendix_table:dropped_sources} and the sources with previously available $z_{\rm spec}$ measurements are listed in Table~\ref{Appendix_table:sources_with_redshifts}. 

The main results of the \zgal survey are given in Tables~\ref{Appendix_table:emission-lines1}, \ref{Appendix_table:emission-lines2} and \ref{Appendix_table:emission-lines3} that include, for all the sources identified in the NOEMA fields, the spectroscopic redshifts and the emission lines properties for the HeLMS and HerS sources, and the HerBS sources from the Large and Pilot Programmes, respectively.

\subsection{Presentation of the catalogue} 
\label{Appendix: Presentation of Catalogue}

For all the sources, the continuum and line emission observed at 3 and 2~mm (including for one source, HeLMS-17, the emission lines detected at 1-mm) were combined to increase the signal-to-noise of the maps and reveal the morphology and the multiplicity of the galaxies detected in each of the fields around the position of the {\it Herschel}-selected source. Thus, as a first approximation, it was assumed that the line and the continuum emission originate from the same region. 

The maps of the \zgal sources were created using a multi-step procedure. First, pure continuum $uv$-tables were generated for each sideband and each source using their line-free frequency ranges. Every $uv$-table was then Fourier-transformed adapting natural weighting to produce a dirty map and cleaned to measure the continuum flux of each source in all sidebands. The spectral index of each source was then determined over the frequency range covered by the \zgal observations, by fitting a single power-law function ($S_{\nu}\sim\nu^{\alpha}$) to the continuum fluxes with weighting factors taking into account the sensitivity achieved in each sideband. In a second step, pseudo-continuum $uv$-tables were generated by taking both the continuum and line emission into account. To preserve source sizes and maximise sensitivity, we proceeded as follow: (i) the $uv$-coordinates of the visibilities of each sideband were scaled to the highest frequency sideband by a factor $\nu_S/\nu_S^H$, where $\nu_S^H$ and $\nu_S$ were the respective central frequencies of the highest frequency sideband and of the one to scale; (ii) the real and imaginary parts of the visibilities of the individual $uv$-tables were scaled by a factor $(\nu_S/\nu_S^H)^{-\alpha}$; and (iii) the respective visibility weights were corrected by a factor  $(\nu_S/\nu_S^H)^{2\alpha}$. By concatenating the scaled and re-weighted $uv$-tables of each sideband, a single $uv$-table was then created for each source and Fourier-transformed to generate a dirty map using natural weighting. Finally, the mapped emission regions were cleaned down to the 1-$\sigma$ noise level. Given the compactness of the pseudo-continuum emission regions, no corrections were made for the primary beam attenuation in the various sidebands.

The basic strategy that was adopted to measure the redshifts of the \zgal sources was to perform least squares Gaussian fits to the identified spectral lines. The algorithm used derives the spectroscopic redshift of a source and its associated error by fitting simultaneously up to four spectral lines. This has the significant advantage that the redshift of a source is not measured independently on different spectral lines and that spectral profiles that are noisier can also be taken into account. All spectral profiles of the \zgal sources were fitted with double-peaked Gaussians when the single Gaussian profile appeared unsuitable. To reduce the number of degrees of freedom, the algorithm assumes for each source the same velocity width, $\Delta w$, for the set of detected spectral lines. For a source with double-peaked profiles, the algorithm also assumes that the velocity spacing, $\Delta s$, between the fitted Gaussian profiles is the same for all detected lines. The line widths of the double-peaked profiles were estimated as $\Delta v = \Delta w + \Delta s$. This method enabled to derive with great precision the spectroscopic redshifts for each source, yielding typical errors of a few $\rm \sim 10^{-4}$. The determination of the redshifts, line intensities, line widths, velocity separations, and their respective errors, was based solely on the rest frequencies of the lines and the signal-to-noise with which they were detected. The results of this analysis are presented in Tables~\ref{Appendix_table:emission-lines1} and \ref{Appendix_table:emission-lines2}.  

The catalogue is organised following the order adopted in the tables of this paper with the HeLMS sources shown in Fig.~\ref{figure:spectra_continuum_HeLMS}, the HerS sources in Fig.~\ref{figure:spectra_continuum_HerS} and the HerBS sources in Fig.~\ref{figure:spectra_continuum_HerBS}. The continuum maps and the emission lines detected in the sources from the Pilot Programme can be found in \citet{Neri2020}, while their spectroscopic redshifts and line properties are listed in Table~\ref{Appendix_table:emission-lines3}. The source name and the derived spectroscopic redshift are indicated along the top of each series of panels displaying the emission lines of that source. The emission lines, detected in the 2~mm (top) and 3~mm (bottom) bands, are identified in the upper (left or right) corner of each spectrum; for the triple sources, the emission lines at 2 and 3~mm of the third source (e.g. HerBS-38) are shown in the bottom left and right panels, respectively. In the case of HeLMS-17, the $^{12}$CO emission lines detected in the 1~mm and 2~mm bands are displayed in the top and bottom (left and right) panels, respectively, while the $\rm H_2O(2_{11}$-$2_{02})$ emission line, which is detected at 1~mm in the Western source, is shown in the lower centre panel. The spectra are shown with the continuum subtracted and each emission line is centred at the zero velocity corresponding to its rest frequency. Fits to the emission line profiles are displayed as solid red lines. The contours of the combined continuum and emission line maps are plotted starting at $3\sigma$. The synthesised beam is shown in the lower left corner of each map.

\newpage

\subsection{Comments on individual sources}
\label{Appendix: Comments on sources}

In the following, we comment on individual sources highlighting their specificity and underlining some of the issues found in the fields where they are located. The sources here discussed are listed in their order of appearance in the catalogue.

\begin{itemize}

\item {\bf HeLMS-1} - This is an extended source with a second peak $\sim 8\arcsec$ to the south that is resolved in the 2~mm dust continuum and the $^{12}$CO(4-3) line emission. A detailed study of this source, which is amplified by a foreground lensing galaxy cluster, will be presented in Cox et al. (in prep.). \\

\item {\bf HeLMS-17} - To derive a robust spectroscopic redshift of this multiple source, additional observations were made at 1~mm, which enabled the detection of the $^{12}$CO(6-5) and $\rm H_2O(2_{11}$-$2_{02})$ emission lines. Together with the $^{12}$CO(4-3) emission line detected at 2~mm, spectroscopic redshifts of $z_{\rm spec}$=2.2983 for the single source to the East and of $z_{\rm spec}$=2.2972 for the double source to the west were derived. It should be noted that the two components of the western source have the same redshift and that both are detected in the $^{12}$CO(6-5) and the water emission lines. Observations performed at 3~mm covered the redshifted $^{12}$CO(3-2) emission line (at $\rm \sim$104.8~GHz), although this line remained barely detected and deeper observations will be needed to extract the line properties of this transition in HeLMS-17. \\

\item {\bf HeLMS-19} - This source is resolved in two components to the east and west, separated by $\sim 5\arcsec$, that have been detected in the $^{12}$CO(7-6), (5-4) and (4-3) emission lines, yielding spectroscopic redshifts that are nearly identical, namely $z_{\rm spec}$=4.6871 and 4.6882, respectively. The [C{\small I}](1-0) and [C{\small I}](2-1) emission lines shown in the catalogue have been integrated over the east and west components of this source to enhance the S/N of these relatively weak lines. The two components trace an Einstein ring that has been detected with the {\it HST} (Borsato et al. subm.). It is interesting to note that the line fluxes of the CO emission lines in HeLMS-19 have unusual ratios with the $^{12}$CO(5-4) being weaker than both the $^{12}$CO(3-2) and (7-6) emission lines (Table~\ref{Appendix_table:emission-lines1}). This could be a result of differential amplification and higher angular resolution observations are needed to further explore the spatial distributions of the CO emission lines and reveal if and how they differ. \\

\item {\bf HeLMS-26} - Two sources are detected to the east and the west in the field around the {\it Herschel} position. They have slightly different redshifts ($z_{\rm spec}$=2.6899 and 2.6875, respectively) and are separated by $\sim 12\arcsec$ corresponding to a linear distance of $\sim 100$~kpc.  \\

\item {\bf HeLMS-32} - A second source, tentatively detected in the continuum ($3 \sigma)$, is seen $\sim 7\arcsec$ to the east of the $z_{\rm spec}$=1.7153 source. \\

\item {\bf HeLMS-35} - A weak source, only detected in the continuum, is present $\sim 6\arcsec$ to the east of the main $z_{\rm spec}$=1.6684 source. \\

\item {\bf HeLMS-40} - Two unresolved sources are revealed in the NOEMA field, HeLMS-40~W1 \& W2, with slightly different redshifts of $z_{\rm spec}$=3.1445 and 3.1395, respectively, and offset to the west with respect to the {\it Herschel} position. The two sources are separated by $\sim 4\arcsec$, corresponding to a linear distance of $\sim 30$~kpc. \\

\item {\bf HeLMS-43} - East to the main source at $z_{\rm spec}$=2.2912, there is a weaker source that is only detected in the continuum. \\

\item {\bf HeLMS-57} - This source displays an arc-like structure extending towards the south-west with a size of $\approx 6\arcsec$. \\

\item {\bf HerS-3} - HerS-3 is extended over $\sim 10\arcsec$ in the north-east and south-west direction with two distinct peaks detected in the $^{12}$CO(3-2) and (5-4) transitions, yielding a similar redshift of $z_{\rm spec}$=3.0607 for both components.  \\

\item {\bf HerS-7} - This source, close in position to the {\it Herschel} coordinates, is detected in the continuum and the $^{12}$CO(2-1) and $^{12}$CO(4-3) emission lines, yielding a spectroscopic redshift of $z_{\rm spec}$=1.9838. A second weaker component, $\sim 3.5\arcsec$ to the east, is revealed in the higher angular resolution 2~mm data; the detection of an emission line at the same frequency (154.5~GHz) as the main component indicates that both are at the same redshift. Two other sources in the field are only detected in the continuum emission, one $\sim 11\arcsec$ to the north and a weaker source $\sim 12\arcsec$ to the West. \\

\item {\bf HerS-8} - The galaxy in the field, at $z_{\rm spec} = 2.2431$ based on the detection of the $^{12}$CO(4-3) and (3-2) emission lines, is offset by $\sim 8 \arcsec$ to the north with respect to the {\it Herschel} position. \\  

\item {\bf HerS-9} - This source is the only one detected in the HCN(3-2) and $\rm HCO^+$(3-2) emission lines yielding an unambiguous spectroscopic redshift of $z_{\rm spec} = 0.8530\pm0.0001$. HerS-9 coincides in position with a bright galaxy seen in the {\it HST} (Borsato et al. subm.). The SDSS DR12 spectroscopic redshift of this galaxy agrees precisely with the NOEMA derived value, namely $z_{\rm phot}=0.853$ \citep{Nayyeri2016}. \\ 

\item {\bf HerS-18} - The field around HerS-18 reveals two distinct sources to the east and the west. The source to the East is detected in the $^{12}$CO(2-1) and (3-2) emission lines, yielding a robust redshift of $z_{\rm spec}$=1.6926. The source to the west is only seen in one emission line that is detected at a frequency of 147.4~GHz and is therefore unrelated to the east component for which the $^{12}$CO(2-1) and (3-2) emission lines are detected at frequencies of 182.4 and 85.6~GHz, respectively. In the field of HerS-18 there is a foreground galaxy seen in the SDSS at $z_{\rm phot}=0.495\pm0.075$ \citep{Nayyeri2016}. The identification of this emission line is therefore tentatively associated with the $^{12}$CO(2-1) transition that would indicate a spectroscopic redshift of $z_{\rm spec}$=0.56, in agreement with the value from the SDSS. However, additional observations are needed to confirm this result.  \\ 

\item {\bf HerS-19} - The NOEMA observations reveal a triple, possibly quadruple system with an unresolved source to the west, an extended structure with a strong component in the south-east, a weak source in the north-east and a tentatively detected source in the centre of the field. In the frequency range covered by the 3~mm observations (75.894 to 83.638 GHz and 91.382 to 99.126~GHz), well defined emission lines are detected in the south-east and west components at 82.9 and 83.6~GHz; the latter line is found at the border of the bandwidth and, as a result, its profile is only partially covered. At 2~mm, HerS-19 was observed in the frequency ranges 127.377 to 135.135~GHz and 142.865 to 150.623~GHz. The continuum emission of the west and south-east components are dominating the spectra at 2~mm (with total flux densities of 2.1 and 1.5~mJy at 146 and 131~GHz, respectively - see Paper II). However, despite a detailed analysis of the available data, no emission lines were found at 2~mm that could unambiguously be identified with $^{12}$CO transitions (or with other species) compatible in frequency with the emission lines detected at 82.9 and 83.6~GHz, thereby preventing us to derive a solid redshift solution for HerS-19. The lack of any clear emission line(s) in the frequency ranges that were covered from 91.382 to 150.623~GHz could be compatible with a redshift of $z\sim3.16$, if the emission line detected around 83~GHz is identified with the CO(3-2) transition.  However, additional observations are needed to definitely confirm this redshift solution, which is close to the photometric redshift of HerS-19, ($z_{\rm phot}$=3.03 --Table~\ref{Appendix_table:sources1}), by covering the 2-mm frequency ranges 135.134-142.892 and 150.622-158.380~GHz, that would enable to detect the $^{12}$CO(5-4) emission line (Fig.~\ref{figure:spectral-coverage}).  \\             

\item {\bf HerBS-38} - In the field of HerBS-38, there are three well-separated sources. The south-eastern and western sources have redshifts of $z_{\rm spec}$=2.4775 and 2.4158, respectively. These redshifts are based on the detection of the CO(3-2) emission line at 99.4 and 101.3~GHz and the [C{\small I}](1-0) and $^{12}$CO(4-3) emission lines at 141.5 and 136.5~GHz in the south-east and west sources (with only half of the $^{12}$CO(4-3) profile detected in the west source). The south-eastern and western sources are separated by $\sim 20 \arcsec$, corresponding at a linear distance of $\rm \sim$160~kpc, with the difference in redshift indicating a comoving radial distance of $\rm \sim 70\,Mpc$. The third source to the north-east is detected in two emission lines at frequencies of 99.4 and 106~GHz, with no other line seen in the frequency ranges that were covered in the observations. Both lines have relatively low S/N. The derived spectroscopic redshift of $z_{\rm spec}$=6.5678, based on the identification of the above lines with the $\rm H_2O(2_{11}$-$2_{02})$ and CO(7-6) emission lines, appears to be the only solution and therefore the most likely. The detection of a third CO emission line in this source would, however, definitely confirm this result.  \\   

\item {\bf HerBS-53} - Two sources are revealed to the east and west in the NOEMA field with $^{12}$CO(2-1) and (3-2) emission lines yielding spectroscopic redshifts of $z_{\rm spec}$=1.4219 and 1.4236, respectively. The distance of the sources is $\sim 5\arcsec$, corresponding to a projected linear distance of $\sim$40~kpc. The closeness of their redshifts indicates that these sources are most likely related. To the north, another source is detected in the continuum only, with no emission lines found in the frequency ranges covered by the observations (i.e. 75.8-106.8~GHz and 127.3-158.3~GHz). \\  

\item {\bf HerBS-61} - This source is resolved at 2~mm into two components separated by $\sim$2$\arcsec$-3$\arcsec$ that are both at the same redshift ($z_{\rm spec}$=3.7293). Only the combined spectra are shown to improve the S/N of the $^{12}$CO(6-5), (4-3) and (3-2) emission lines and reveal the details of the line profiles. This spectroscopic redshift is in excellent agreement with the value of $z_{\rm spec}=3.7275$ derived by \cite{Bakx2020b}. \\ 

\item {\bf HerBS-74} - The NOEMA field is dominated by the strong continuum emission of another source, $\sim 10\arcsec$ to the East, that coincides with a bright nearby galaxy seen in the {\it HST} (Borsato et al. subm.). No emission lines were detected with NOEMA in this foreground galaxy. The source at high redshift is centred near the {\it Herschel} coordinates. It is weak in the continuum and detected in the emission lines of $^{12}$CO(4-3) and (3-2), yielding a secure spectroscopic redshift of $z_{\rm spec}=2.5610$.  \\  

\item {\bf HerBS-76} - This source is extended in the east-west direction. To the east, a main component displays strong, narrow ($\Delta(V)=200\pm15 \, \rm km s^{-1}$) $^{12}$CO(4-3) and (3-2) emission lines. The western component shows weaker $^{12}$CO(4-3) and (3-2) emission lines with significantly larger line widths ($\Delta(V)=742\pm149 \, \rm km s^{-1}$). The derived spectroscopic redshifts are slightly different with $z_{\rm spec}$=2.3302 and 2.3319 for the eastern and western components, respectively. \\  

\item {\bf HerBS-82} - This source is unresolved and shows a strong, double-peaked emission line centred at a frequency of 150.7~GHz with a line width of $\rm 181\pm15 \, km \, s^{-1}$ and a line flux of $\rm 4.07\pm0.38 \, Jy \, km \, s^{-1}$. A second broader ($\rm 1584\pm338 \, km \, s^{-1}$) emission line (with a line flux of $\rm 1.43\pm0.38 \, Jy \, km s^{-1}$ and a 20x times weaker peak flux density) is detected at 97.7~GHz. By identifying the 150.7~GHz emission line with the $^{12}$CO(6-5) transition, the lower frequency emission could correspond to the $\rm H_2O$($4_{23}$-$3_{30}$) transition resulting in a spectroscopic redshift of $z_{\rm spec}$=3.5871. However, the notable difference in the profiles of the two lines makes this identification doubtful; in addition, this solution is far off from the estimated photometric redshift  of $z_{\rm phot}\sim2.49$ (Table~\ref{Appendix_table:emission-lines2}).  

Another possibility is that the 150.7~GHz emission line traces the $^{12}$CO(5-4) transition, yielding a spectroscopic redshift of $z_{\rm spec}=2.8277$. In this case, the [C{\small I}] emission line should have been detected at 128.5~GHz; however, nothing is seen at that frequency in the available NOEMA data. Therefore the identification with the $^{12}$CO(4-3) is favoured, which would correspond to a spectroscopic redshift of $z_{\rm spec}=2.0583$ and would be compatible with the fact that no other line, with a similar profile as the 150.7~GHz emission line, is detected in the frequency ranges that were covered for this high-$z$ source (Fig.~\ref{figure:spectral-coverage}). This result should be confirmed by measuring another emission line, for example of a lower CO transition. 

Finally, the emission line detected at 97.7~GHz is likely unrelated to the {\it Herschel}-detected source. HerBS-82 is indeed centred on a bright, foreground lensing galaxy seen in the {\it HST} (Borsato et al. subm.) and the emission line could therefore trace the molecular gas of this nearby galaxy. Identifying this emission line with the $^{12}$CO(1-0) transition would result in a spectroscopic redshift of $z_{\rm spec}=0.1788$ for the foreground lensing galaxy. \\

\item {\bf HerBS-91} - The combined continuum and emission lines map of this source reveals three components. The eastern and centre components are detected in the $^{12}$CO(4-3) and (3-2) emission lines and in [C{\small I}(1-0)] (the catalogue shows the combined spectrum) yielding a similar redshift for both, namely $z_{\rm spec}$=2.4047. The western source is only detected in the continuum. \\

\item {\bf HerBS-92} - The field of HerBS-92 reveals two main components that are separated by $\sim 15\arcsec$ and have the same redshift ($z_{\rm spec}=3.2644$). The weak $^{12}$CO(5-4) and (3-2) emission lines have been integrated over both components in order to increase the S/N of the spectra. South of the {\it Herschel} position, a weaker source is detected in the continuum only.  \\ 

\item {\bf HerBS-105} - The field of HerBS-105 reveals three components, two of which are only detected in the continuum. The bright component to the south-west is detected in the continuum and in the $^{12}$CO(5-4) and (4-3) emission lines yielding a redshift of $z_{\rm spec}=2.6684$. The two other sources are detected to the north-east in the continuum: the strongest of the two is located $\sim 6\arcsec$ from the south-western component, while the weaker ($4\sigma$) component is $\sim 9 \arcsec$ further to the east. \\

\item {\bf HerBS-109} - Three distinct components are detected in the field of this {\it Herschel} source. The north-western and southern components have comparable spectroscopic redshifts of $z_{\rm spec}$=1.5850 and 1.5843, respectively, based on the detection of the $^{12}$CO(3-2) and CO(2-1) emission lines. They are separated by $12\arcsec$ corresponding to a projected distance of $\rm \sim 100 \, kpc$. The component to the north-east is unrelated to this binary system with a derived spectroscopic redshift of $z_{\rm spec}$=2.8385 based on the detection of the $^{12}$CO(5-4) and (3-2) emission lines. \\

\item {\bf HerBS-116} - Two components to the east and west, separated by $\sim 4\arcsec$, are clearly identified at 2~mm but remain unresolved in the lower angular resolution 3~mm observations. The spectra of both components have therefore been combined. Two emission lines are detected in each component, corresponding to the $^{12}$CO(5-4) and (3-2) transitions, yielding a spectroscopic redshift of $z_{\rm spec}$=3.1573 for this system. \\  

\item {\bf HerBS-124} - Within the field of this source, there are two sources separated by $\sim 4\arcsec$ that are both detected in the continuum and the emission lines of $^{12}$CO(4-3) and CO(3-2), resulting in redshifts of $z_{\rm spec}=2.2781$ and 2.2772 for the eastern and western components, respectively. The eastern source is weak in both the continuum and line emission. \\ 

\item {\bf HerBS-127} - This source is detected in the continuum and in the emission lines of $^{12}$CO(5-4) and (3-2) and displays weak extended emission to the south. The spectroscopic redshift is $z_{\rm spec}$=3.1958. \\ 

\item {\bf HerBS-136} - The field of HerBS-136 reveals one main component detected in the continuum and in the $^{12}$CO(5-4) and (3-2) emission lines, yielding a spectroscopic redshift of $z_{\rm spec}$=3.2884. About $4 \arcsec$  to the east, two weaker components are detected in the continuum, forming a $\sim 5 \arcsec$ extended structure. \\

\item {\bf HerBS-147} - This galaxy, at $z_{\rm spec}$=3.1150, displays the largest line width of the entire \zgal sample with $\rm  \Delta V = 1789\pm300 \, km s^{-1}$ for the $^{12}$CO(5-4) and (3-2) emission lines. \\ 

\item {\bf HerBS-150} - Three distinct sources are revealed in the east, centre and west in the combined continuum and emission line map of this field separated by $\sim$10$\arcsec$ and $5\arcsec$. The eastern source is detected in the $^{12}$CO(6-5) and (4-3) emission lines, yielding a robust spectroscopic redshift of $z_{\rm spec}$=3.6732. The centre and western components are each detected in only one line seen around the same frequency as the CO(4-3) emission line in the eastern component ($98.6 < \nu_{\rm obs} < 98.8$~GHz) so that the identification of this line with the $^{12}$CO(4-3) is most likely. The derived spectroscopic redshifts are close to the eastern component with values of $z_{\rm spec}$=3.6787 and 3.6682, respectively. It should be noted that the emission line of the western source is only partially covered as it was detected at the edge of the bandwidth. Further observations will be needed to fully investigate the properties of this cluster of galaxies. \\     

\item {\bf HerBS-162} - The combined continuum and emission line map shows a strong component (to the south-west) centred near the {\it Herschel} coordinates and a second weaker component to the north-east. Both components are detected in the $^{12}$CO(4-3) and (3-2) emission lines, yielding spectroscopic redshifts that are slightly different of $z_{\rm spec}$=2.4739 and 2.4742, respectively. The sources show distinct profiles with double peaked CO lines for the south-west component and single profiles for the north-east component, suggesting that these sources are distinct. \\ 

\item {\bf HerBS-177} - The spectroscopic redshift of $z_{\rm spec}=3.9625$ is in excellent agreement with the value of $z_{\rm spec}=3.9633$ that was derived for this source by \cite{Bakx2020b}. \\ 

\item {\bf HerBS-180} - This field is dominated by a strong double peaked source to the north-east that is only detected in the continuum emission. No emission lines were found in this source in the frequency range that was covered (75.89-83.6~GHz, 91.38-99.12~GHz and 127.37-158.38~GHz). A weaker source to the south-west is detected in the $^{12}$CO(3-2) and (2-1) emission lines, resulting in a spectroscopic redshift of $z_{\rm spec}$=1.4527. \\ 

\item {\bf HerBS-187} - Three sources are detected in this field. The eastern and western sources, which are separated by $\sim 4\arcsec$, are both detected in the $^{12}$CO(2-1) and (4-3) emission lines, yielding spectroscopic redshifts that are close to each other, namely $z_{\rm spec}$=1.8285 and 1.8274, respectively. The weaker southern source is only detected in the continuum and additional observations are needed to measure its redshift. \\ 

\item {\bf HerBS-194} - Two sources are detected in the field of HerBS-194 to the north and the south. They are separated by $\sim 15\arcsec$ and equidistant from the {\it Herschel}-source coordinates. Both sources are seen in the $^{12}$CO(4-3) and (3-2) emission lines and have spectroscopic redshifts that are close to each other, namely $z_{\rm spec}$=2.3335 and 2.3316, respectively. The projected distance between the sources is $\sim$120~kpc. \\ 

\item {\bf HerBS-199} - Two sources are revealed in this field to the East (close to the {\it Herschel} postion) and to the West with a separation of $\sim$12~arcsec. They are both detected in the emission lines of $^{12}$CO(4-3) and $^{12}$CO(2-1) resulting in spectroscopic redshifts that are comparable, namely $z_{\rm spec}$=1.9248 and 1.9197, respectively. The projected distance between the sources is $\sim$100~kpc. \\ 

\item {\bf HerBS-204} - This source is the only case in the \zgal sample that shows two sources that are separated by $\sim 6\arcsec$ and are linked by a bridge of matter, with all components detected in the dust continuum and the emission lines of $^{12}$CO(5-4) and $^{12}$CO(4-3). The derived spectroscopic redshift is $z_{\rm spec}$=3.493 and the projected separation between the two main components, and hence the extent of this interacting system is $\sim$45~kpc. \\ 

\item {\bf HerBS-205} - The field of HerBS-205 reveals a triple system dominated by a source to the east that is resolved into two components (north-east and south-east) and another, weaker component $\sim 8\arcsec$ to the west. In all three sources, the $^{12}$CO(3-2) and $^{12}$(5-4) emission lines are clearly detected, yielding spectroscopic redshifts that are nearly identical with $z_{\rm spec}$=2.9600 for the Eastern double source and $z_{\rm spec}$=2.9630 for the western source, respectively. The projected distance between the Eastern and Western components is $\sim$60~kpc. \\  

\end{itemize}

\begin{table*}[!htbp]
\tiny
\caption{\zgal Sample: HeLMS and HerS sources.}
\begin{tabular}{lcccccccc}
\hline\hline
\multicolumn{2}{c}{Source name} & \multicolumn{2}{c}{{\it Herschel} Coordinates} & \multicolumn{2}{c}{$z_{\rm phot}$} & \multicolumn{3}{c}{Flux density}  \\
 &   &  \multicolumn{2}{c}{(J2000.0)}   &  &  & \multicolumn{3}{c}{{\it Herschel}} \\
{\it Herschel} Name  & Other &  &  & (a) & (b)  & $\rm S_{250 \, \mu m}$ & $\rm S_{350 \, \mu m}$ & $\rm S_{500 \, \mu m}$  \\
& & & & & \multicolumn{3}{c}{(mJy)}  \\
\hline
\multicolumn{8}{c}{HeLMS Sources} \\
\hline
HERMES J233441.0$-$065220 & HeLMS-1  & 23:34:41.0   & $-$06:52:20       & 2.25 & 1.86 & 431$\pm$6 & 381$\pm$7     & 272$\pm$7 \\
HERMES J000215.9$-$012819 & HeLMS-3  & 00:02:15.9       & $-$01:28:29   & 1.61 & 1.35 &   643$\pm$7       & 510$\pm$6     & 258$\pm$7 \\ 
HERMES J003929.6+002426   & HeLMS-11 & 00:39:29.6       & +00:24:26         & 2.73 & 2.96 & 140$\pm$7    & 157$\pm$7     & 154$\pm$8 \\
HERMES J235601.5$-$071142 & HeLMS-12 & 23:56:01.5       & $-$07:11:42   & 2.49 & 2.43 & 178$\pm$7 & 184$\pm$6     & 154$\pm$7 \\
HERMES J003619.8+002420   & HeLMS-14 & 00:36:19.8       & +00:24:20         & 2.04 & 1.86 & 251$\pm$6    & 247$\pm$6     & 148$\pm$7 \\   
HERMES J231857.2$-$053035 & HeLMS-16 & 23:18:57.2       & $-$05:30:35   & 2.71 & 2.76 & 143$\pm$7 & 183$\pm$7     & 146$\pm$8 \\
HERMES J232558.3$-$044525 & HeLMS-17 & 23:25:58.3       & $-$04:45:25   & 2.33 & 2.15 & 190$\pm$6 & 189$\pm$6     & 142$\pm$8 \\ 
HERMES J232210.3$-$033559 & HeLMS-19 & 23:22:10.3       & $-$03:35:59   & 2.90 & 3.08 & 114$\pm$6 & 160$\pm$7     & 134$\pm$8 \\
HERMES J233728.8$-$045106 & HeLMS-20 & 23:37:28.8       & $-$04:51:06   & 2.42 & 2.34 & 162$\pm$6 & 178$\pm$7     & 132$\pm$8 \\  
HERMES J001800.1$-$060235 & HeLMS-21 & 00:18:00.1       & $-$06:02:35   & 1.99 & 1.87 & 206$\pm$6 & 186$\pm$7     & 130$\pm$7 \\   
HERMES J005841.2$-$011149 & HeLMS-23 & 00:58:41.2       & $-$01:11:49   & 1.36 & 1.09 & 391$\pm$7 & 273$\pm$6     & 126$\pm$8 \\  
HERMES J003814.1$-$002252 & HeLMS-24 & 00:38:14.1       & $-$00:22:52   & 3.28 & 3.80 &  82$\pm$6 & 120$\pm$6     & 126$\pm$7 \\  
HERMES J004124.0$-$010307 & HeLMS-25 & 00:41:24.0       & $-$01:03:07   & 2.23 & 2.10 & 178$\pm$6 & 186$\pm$7     & 125$\pm$8 \\  
HERMES J004747.1+061444   & HeLMS-26 & 00:47:47.1       & +06:14:44         & 3.22 & 3.68 & 85$\pm$7     & 119$\pm$6     & 125$\pm$8 \\   
HERMES J003758.0$-$010622 & HeLMS-27 & 00:37:58.0       & $-$01:06:22   & 2.64 & 2.71 & 125$\pm$7 & 144$\pm$6     & 124$\pm$8 \\
HERMES J003009.1$-$020625 & HeLMS-28 & 00:30:09.1       & $-$02:06:25   & 2.72 & 2.88 & 114$\pm$6 & 135$\pm$6     & 122$\pm$7 \\   
HERMES J001027.1$-$024624 & HeLMS-30 & 00:10:27.1       & $-$02:46:24   & 2.03 & 1.92 & 185$\pm$6   & 170$\pm$6   & 121$\pm$7 \\
HERMES J001353.5$-$060200 & HeLMS-31 & 00:13:53.5       & $-$06:02:00   & 2.12 & 2.01 & 178$\pm$7 & 176$\pm$6     & 120$\pm$7 \\   
HERMES J000336.9+014013   & HeLMS-32 & 00:03:36.9       & +01:40:13         & 2.82 & 3.10 & 103$\pm$6    & 112$\pm$6     & 119$\pm$7 \\
HERMES J002719.5+001204   & HeLMS-34 & 00:27:19.5       & +00:12:04         & 1.71 & 1.51 & 248$\pm$6    &  20$\pm$7     & 116$\pm$8 \\
HERMES J232500.1$-$005643 & HeLMS-35 & 23:25:00.1       & $-$00:56:43   & 2.53 & 2.57 & 122$\pm$6 & 132$\pm$7     & 114$\pm$8 \\   
HERMES J234314.0+012152   & HeLMS-36 & 23:43:14.0       & +01:21:52         & 2.53 & 2.65 & 115$\pm$6    & 115$\pm$6     & 113$\pm$8 \\   
HERMES J010801.8+053201   & HeLMS-37 & 01:08:01.8       & +05:32:01         & 2.49 & 2.55 & 122$\pm$6    & 120$\pm$6     & 113$\pm$7 \\  
HERMES J002208.1+034044   & HeLMS-38 & 00:22:08.1       & +03:40:44         & 1.86 & 1.72 & 190$\pm$6    & 157$\pm$6     & 113$\pm$7 \\
HERMES J002936.3+020710   & HeLMS-39 & 00:29:36.3       & +02:07:10         & 3.12 & 3.53 &  81$\pm$6    & 107$\pm$6     & 112$\pm$7 \\
HERMES J235332.0+031718   & HeLMS-40 & 23:53:32.0       & +03:17:18         & 2.74 & 2.91 & 102$\pm$6    & 123$\pm$7     & 111$\pm$7 \\
HERMES J233633.5$-$032119 & HeLMS-41 & 23:36:33.5       & $-$03:21:19   & 2.40 & 2.37 & 130$\pm$6 & 131$\pm$6     & 110$\pm$7 \\
HERMES J234014.6$-$070738 & HeLMS-42 & 23:40:14.6       & $-$07:07:38   & 2.12 & 2.04 & 158$\pm$6 & 154$\pm$6     & 110$\pm$8 \\  
HERMES J233420.4$-$003458 & HeLMS-43 & 23:34:20.4       & $-$00:34:58   & 2.04 & 1.98 & 156$\pm$7 & 141$\pm$5     & 109$\pm$8 \\   
HERMES J231447.5$-$045658 & HeLMS-44 & 23:14:47.5       & $-$04:56:58   & 1.67 & 1.27 & 220$\pm$8 & 141$\pm$7     & 106$\pm$8 \\
HERMES J001226.9+020810   & HeLMS-45 & 00:12:26.9       & +02:08:10         & 2.63 & 2.70 & 107$\pm$6    & 142$\pm$6     & 106$\pm$7 \\   
HERMES J004622.3+073509   & HeLMS-46 & 00:46:22.3       & +07:35:09         & 3.05 & 3.27 &  82$\pm$9   & 113$\pm$9      & 105$\pm$10 \\
HERMES J234951.6$-$030019 & HeLMS-47 & 23:49:51.6       & $-$03:00:19   & 1.89 & 1.74 & 186$\pm$7 & 167$\pm$6     & 105$\pm$8 \\
HERMES J232833.6$-$031416 & HeLMS-48 & 23:28:33.6       & $-$03:14:16   & 3.51 & 4.70 &  49$\pm$6 & 104$\pm$6     & 105$\pm$8 \\  
HERMES J233721.9$-$064740 & HeLMS-49 & 23:37:21.9       & $-$06:47:40   & 1.98 & 1.85 & 173$\pm$6 & 161$\pm$7     & 105$\pm$8 \\
HERMES J235101.7$-$024426 & HeLMS-50 & 23:51:01.7       & $-$02:44:26   & 2.53 & 2.59 &   112$\pm$6       & 124$\pm$6     & 105$\pm$7 \\
HERMES J232617.5$-$025319 & HeLMS-51 & 23:26:17.5       & $-$02:53:19   & 2.96 & 3.16 &  86$\pm$6 & 109$\pm$6     & 104$\pm$7 \\
HERMES J233727.1$-$002343 & HeLMS-52 & 23:37:27.1       & $-$00:23:43   & 1.86 & 1.71 & 182$\pm$6 & 157$\pm$6     & 104$\pm$8 \\  
HERMES J002718.1+023943   & HeLMS-54 & 00:27:18.1       & +02:39:43         & 3.23 & 3.80 &  69$\pm$6    &  87$\pm$6     & 103$\pm$7 \\  
HERMES J232831.8$-$004035 & HeLMS-55 & 23:28:31.8       & $-$00:40:35   & 2.76 & 2.85 &  95$\pm$7 & 120$\pm$6     & 102$\pm$7 \\
HERMES J001325.7+042509   & HeLMS-56 & 00:13:25.7       & +04:25:09         & 2.96 & 3.07 &  89$\pm$6    &  98$\pm$6     & 102$\pm$7 \\   
HERMES J003519.7+072806   & HeLMS-57 & 00:35:19.7       & +07:28:06         & 2.33 & 2.17 & 134$\pm$6    & 135$\pm$7     & 101$\pm$8 \\   \hline
\multicolumn{8}{c}{HerS Sources} \\
\hline
HERS J012041.6$-$002705 & HerS-2  &     01:20:41.6 & $-$00:27:05        & 2.42 & 2.35 &   240$\pm$6 &     260$\pm$6 &     198$\pm$7   \\  
HERS J012751.1+004940   & HerS-3  &     01:27:54.1 & +00:49:40      & 2.24 & 2.17 &        253$\pm$6 &     250$\pm$6 &     191$\pm$7   \\   
HERS J012620.5+012950   & HerS-5  &     01:26:20.5 & +01:29:50      & 1.77 & 1.58 &        268$\pm$8 &     228$\pm$7 &     133$\pm$9   \\
HERS J010133.8+003157   & HerS-7  &     01:01:33.8 & +00:31:57      & 2.33 & 2.12 &        165$\pm$7 &     154$\pm$6 &     122$\pm$7   \\   
HERS J010911.7$-$014830 & HerS-8  &     01:09:38.9 & $-$01:48:30        & 2.37 & 2.30 &   146$\pm$8 &     152$\pm$7 &     118$\pm$10  \\  
HERS J010911.7$-$011733 & HerS-9  &     01:09:11.7 & $-$01:17:33        & 1.16 & 0.77 &   393$\pm$8 &     220$\pm$8 &     118$\pm$9   \\   
HERS J011722.3+005624   & HerS-10 &     01:17:22.3 & +00:56:24      & 2.93 & 2.97 &        105$\pm$6 &     125$\pm$6 &     117$\pm$7   \\   
HERS J005847.3$-$010017 & HerS-11 &     00:58:47.3 & $-$01:00:17        & 3.34 & 4.16 &    63$\pm$8 &     116$\pm$7 &     115$\pm$9   \\   
HERS J012546.3$-$001143 & HerS-12 &     01:25:46.3 & $-$00:11:43        & 2.24 & 2.08 & 152$\pm$8 &       135$\pm$7 &     113$\pm$9   \\   
HERS J012521.0+011724   & HerS-13 &     01:25:21.0 & +01:17:24      & 2.07 & 1.98 &        165$\pm$8 &     153$\pm$7 &     114$\pm$10  \\
HERS J014057.3$-$010547 & HerS-14 &     01:40:57.3 & $-$01:05:47        & 2.40 & 2.34 &   136$\pm$8 &     143$\pm$8 &     112$\pm$9   \\
HERS J012106.9+003457   & HerS-15 &     01:21:06.9 & +00:34:57      & 2.89 & 3.05 &        94$\pm$6  &     130$\pm$7 &     110$\pm$7   \\          
HERS J021434.4+005923   & HerS-16 &     02:14:34.4 & +00:59:23      & 2.66 & 2.70 &        110$\pm$9 &     134$\pm$8 &     109$\pm$10  \\  
HERS J021402.6$-$004612 & HerS-17 &     02:14:02.6 & $-$00:46:12        & 2.61 & 2.64 &   110$\pm$8 &     130$\pm$8 &     105$\pm$9   \\  
HERS J013212.2+001754   & HerS-18 &     01:32:12.2 &  +00:17:54     & 2.05 & 1.88 &        176$\pm$7 &     175$\pm$6 &     104$\pm$8   \\   
HERS J020529.1+000501   & HerS-19 &     02:05:29.1 & +00:05:01      & 2.83 & 3.03 &         89$\pm$6 &     112$\pm$6 & 102$\pm$8   \\      
HERS J010246.1+010543   & HerS-20 & 01:02:46.1 & +01:05:43          & 2.63 & 2.65 &        107$\pm$8 &     133$\pm$8 &     102$\pm$11  \\
\hline
\end{tabular}
 \tablefoot{The source names and {\it Herschel} flux densities are from \citet{Nayyeri2016}, adopting the names HeLMS-* and HerS-* for consistency with the names of the HerBS sources (Table~\ref{Appendix_table:sources2}). The sources are listed in order of decreasing $\rm S_{\rm 500 \mu m}$ flux density. The photometric redshifts ($z_{\rm phot}$) have been re-evaluated using the methods described by \citet{Ivison2016} and \citet{Pearson2013} listed as (a) and (b), respectively.}
  \label{Appendix_table:sources1}
\end{table*}

\begin{table*}[!htbp]
\tiny
\caption{\zgal Sample: HerBS sources.}
\begin{tabular}{lccccccccc}
\hline\hline
\multicolumn{2}{c}{Source name} & \multicolumn{2}{c}{{\it Herschel} Coordinates} & \multicolumn{2}{c}{$z_{\rm phot}$} & \multicolumn{4}{c}{Flux density} \\
 &   &  \multicolumn{2}{c}{(J2000.0)}  & &  &  \multicolumn{3}{c}{{\it Herschel}} & SCUBA-2 \\
{\it Herschel} Name  & Other &  &  & (a)  & (b) & $\rm S_{250 \, \mu m}$ & $\rm S_{350 \, \mu m}$ & $\rm S_{500 \, \mu m}$ & $\rm S_{850 \, \mu m}$  \\
& & & & & & \multicolumn{4}{c}{(mJy)}  \\
\hline
H-ATLAS 144608.6+021927    & HerBS-38   & 14:46:08.6 & +02:19:27    & 3.01 &  2.96 & 73.4$\pm$7.1 & 111.7$\pm$8.1 & 122.1$\pm$8.7 & 13.6$\pm$8.2 \\
H-ATLAS 144556.1$-$004853  & HerBS-46   & 14:45:56.1 & $-$00:48:53  & 2.31 & 2.11 & 126.7$\pm$7.3 & 132.6$\pm$8.4 & 111.8$\pm$8.7 & 20.3$\pm$5.2 \\
H-ATLAS 121301.5$-$004922  & HerBS-48   & 12:13:01.6 & $-$00:49:22      & 2.31 & 2.19 & 136.6$\pm$6.6 & 142.6$\pm$7.4 & 110.9$\pm$7.7 & 25.3$\pm$5.0 \\
H-ATLAS 120319.1$-$011253  & HerBS-50   & 12:03:19.1 & $-$01:12:54      & 2.71 & 2.68 & 114.3$\pm$5.3 & 142.8$\pm$8.2 & 110.2$\pm$8.6 & 47.2$\pm$5.4  \\
H-ATLAS 120709.2$-$014702  & HerBS-51   & 12:07:09.2 & $-$01:47:03      & 2.24 & 2.08 & 143.2$\pm$7.4 & 149.2$\pm$8.1 & 110.3$\pm$8.7 & 23.8$\pm$4.9 \\
H-ATLAS 115112.2$-$012637  & HerBS-53   & 11:51:12.3 & $-$01:26:37      & 2.13 & 1.99 & 141.2$\pm$7.4 & 137.7$\pm$8.2 & 108.4$\pm$8.8 & 11.6$\pm$5.5 \\
H-ATLAS 120127.6$-$014043  & HerBS-61   & 12:01:27.7 & $-$01:40:44      & 3.12 & 3.27 & 67.4$\pm$6.5 & 112.1$\pm$7.4 & 103.9$\pm$7.7 & 38.1$\pm$4.8 \\
H-ATLAS 121542.7$-$005220  & HerBS-62   & 12:15:42.6 & $-$00:52:20      & 2.54 & 2.47 & 119.7$\pm$7.4 & 135.5$\pm$8.2 & 103.4$\pm$8.6 & 32.4$\pm$5.1 \\
H-ATLAS 134422.6+231952    & HerBS-65   & 13:44:22.6 & +23:19:52        & 2.53 & 2.38 & 110.6$\pm$6.4 & 98.3$\pm$7.2  & 101.6$\pm$7.7 & 24.4$\pm$5.8 \\
H-ATLAS 144512.1$-$001510  & HerBS-72   & 14:45:12.1 & $-$00:15:11      & 2.96 &  2.89 & 78.8$\pm$6.5 & 100.7$\pm$7.4 & 99.6$\pm$7.7 & 34.2$\pm$4.7 \\
H-ATLAS 120600.7+003459    & HerBS-74   & 12:06:00.7 & +00:34:59        & 2.56 & 2.43 &  88.7$\pm$7.4 & 104.1$\pm$8.1 &  98.8$\pm$8.7 & 13.2$\pm$4.8 \\
H-ATLAS 133534.1+341835    & HerBS-76   & 13:35:34.1 & +34:18:35        & 2.51 & 2.31 & 108.5$\pm$5.9 & 124.3$\pm$6.0 &  98.5$\pm$7.0 & 27.4$\pm$4.4 \\
H-ATLAS 143352.4+020417    & HerBS-78   & 14:33:52.4 & +02:04:17        & 2.78 &  2.77 & 87.7$\pm$7.3 & 102.4$\pm$8.1 &  98.2$\pm$8.8  & 33.9$\pm$5.7 \\
H-ATLAS 121144.8+010638    & HerBS-82   & 12:11:44.8 & +01:06:38        & 2.49 & 2.41 & 114.5$\pm$6.7  & 123.2$\pm$7.6  &  96.8$\pm$8.0  & 29.4$\pm$6.0  \\
H-ATLAS 121812.8+011841    & HerBS-83   & 12:18:12.8 & +01:18:42        & 3.94 & 3.89 &  49.5$\pm$7.2  & 79.7$\pm$8.1   &  94.1$\pm$8.8  & 45.2$\pm$5.5  \\
H-ATLAS 114752.7$-$005831  & HerBS-85   & 11:47:52.8 & $-$00:58:31      & 2.56 & 2.51 & 92.1$\pm$6.6  & 104.2$\pm$7.4  & 96.0$\pm$7.7  &  8.0$\pm$6.1  \\
H-ATLAS 092135.6+000131    & HerBS-91   & 09:21:35.7 & +00:01:32        & 2.06 & 1.91 & 139.2$\pm$7.3  & 128.8$\pm$8.1  &  95.1$\pm$8.6  & 21.4$\pm$6.5  \\
H-ATLAS 133808.9+255153    & HerBS-92   & 13:38:08.9 & +25:51:53        & 3.35 &  3.34 & 42.2$\pm$5.7  &  75.3$\pm$6.0  &  94.9$\pm$7.2  & 17.9$\pm$6.2  \\
H-ATLAS 083932.2$-$011758  & HerBS-105  & 08:39:32.2 & $-$01:17:58      & 2.69 & 2.56 & 73.8$\pm$7.4  &  88.5$\pm$8.1  &  93.2$\pm$8.7  & 19.0$\pm$5.4  \\
H-ATLAS 083817.4$-$004134  & HerBS-108  & 08:38:17.4 & $-$00:41:35      & 2.81 & 2.78 & 84.5$\pm$7.4  & 106.1$\pm$8.2  &  93.0$\pm$8.8  & 34.7$\pm$6.0 \\
H-ATLAS 132900.4+281914    & HerBS-109  & 13:29:00.4 & +28:19:14        & 2.34 & 2.22 & 121.7$\pm$5.4  & 140.1$\pm$5.9  &  92.8$\pm$7.6  & 29.3$\pm$5.6 \\
H-ATLAS 141832.9+010212    & HerBS-110  & 14:18:33.0 & +01:02:13        & 2.99 &  3.07 & 66.0$\pm$6.6  & 106.5$\pm$7.5  &  92.8$\pm$7.8  & 30.1$\pm$4.6  \\
H-ATLAS 133538.3+265742    & HerBS-115  & 13:35:38.3 & +26:57:42        & 2.64 & 2.28 & 116.2$\pm$5  & 133.5$\pm$6.0  &  91.8$\pm$6.9  & 9.6$\pm$6.4  \\
H-ATLAS 121348.0+010812    & HerBS-116  & 12:13:48.0 & +01:08:13        & 3.17 &  3.20 & 65.1$\pm$7.4  &  96.6$\pm$8.2  &  93.6$\pm$8.5  & 37.6$\pm$5.4  \\
H-ATLAS 122158.5+003326    & HerBS-124  & 12:21:58.5 & +00:33:26        & 2.01 & 1.90 & 135.7$\pm$8.2  & 116.1$\pm$8.2  &  89.9$\pm$8.6  & 21$\pm$6  \\
H-ATLAS 130432.2+295338    & HerBS-125  & 13:04:32.2 & +29:53:38        & 2.75 & 2.70 & 75.7$\pm$5.8  & 103.4$\pm$5.7  & 89.8$\pm$7.1  & 24.0$\pm$4.1  \\     
H-ATLAS 145135.2$-$011418  & HerBS-126  & 14:51:35.3 & $-$01:14:18      & 2.65 &  2.55 & 81.9$\pm$7.2  &  95.9$\pm$8.2  & 89.8$\pm$8.8  & 22.3$\pm$5.7  \\     
H-ATLAS 132128.6+282020    & HerBS-127  & 13:21:28.6 & +28:20:20        & 2.45 & 2.32 & 110.0$\pm$5.5  & 122.7$\pm$6.1  & 89.5$\pm$6.9  & 30.0$\pm$6.1  \\     
H-ATLAS 130414.6+303538    & HerBS-128  & 13:04:14.6 & +30:35:37        & 2.39 & 2.26 & 106.4$\pm$5.7  & 111.2$\pm$5.9  & 89.2$\pm$7.1  & 24.9$\pm$5.1  \\     
H-ATLAS 130053.8+260303    & HerBS-129  & 13:00:53.8 & +26:03:03        & 3.21 & 3.22 & 59.4$\pm$5.9  &  85.4$\pm$5.9  & 89.0$\pm$7.0  & 35.2$\pm$4.4  \\     
H-ATLAS 133440.4+353141    & HerBS-134  & 13:34:40.4 & +35:31:41        & 2.77 &  2.67 & 69.9$\pm$5.9  &  97.3$\pm$6.2  & 87.9$\pm$7.3  & 20.4$\pm$4.9  \\
H-ATLAS 085308.5$-$005728  & HerBS-136  & 08:53:08.6 & $-$00:57:28      & 2.89 &  2.86 & 68.3$\pm$7.5  &  97.5$\pm$8.2  & 87.7$\pm$8.6  & 28.1$\pm$5.9  \\
H-ATLAS 145337.2+000407    & HerBS-137  & 14:53:37.2 & +00:04:08        & 2.66 & 2.61 &  86.0$\pm$7.2  & 103.6$\pm$8.0  & 87.7$\pm$8.6  & 29.4$\pm$6.6  \\
H-ATLAS 142140.3+000447    & HerBS-140  & 14:21:40.4 & +00:04:48        & 2.41 &  2.29 & 96.8$\pm$7.2  &  98.5$\pm$8.2  & 87.4$\pm$8.7  & 20.9$\pm$5.4  \\     
H-ATLAS 141810.0$-$003747  & HerBS-143  & 14:18:10.1 & $-$00:37:47      & 2.59 &  2.48 & 77.7$\pm$6.5  &  97.3$\pm$7.4  & 87.1$\pm$7.9  & 16.4$\pm$3.8 \\      
H-ATLAS 143403.5+000234    & HerBS-147  & 14:34:03.6 & +00:02:34        & 2.42 & 2.31 & 103.3$\pm$7.4  & 103.3$\pm$8.1  & 86.6$\pm$8.5  & 28.2$\pm$4.9 \\      
H-ATLAS 133827.6+313956    & HerBS-149  & 13:38:27.6 & +31:39:55        & 2.35 & 2.16 & 101.5$\pm$5.5  & 103.3$\pm$6.0  & 86.0$\pm$7.0  & 18.4$\pm$5.4 \\      
H-ATLAS 122459.1$-$005647  & HerBS-150  & 12:24:59.0 & $-$00:56:46      & 3.37 &  3.43 & 53.6$\pm$5  &  81.3$\pm$8.3  & 92.0$\pm$8.9  & 35.8$\pm$4.4 \\      
H-ATLAS 144243.4+015504    & HerBS-153  & 14:42:43.4 & +01:55:04        & 2.52 & 2.13 & 123.2$\pm$7.2  & 133.4$\pm$8.1  & 85.7$\pm$8.8  & 31.1$\pm$5.6 \\ 
H-ATLAS 084957.7+010713    & HerBS-157  & 08:49:57.7 & +01:07:13        & 2.74 & 2.70 & 81.2$\pm$7.3  &  98.9$\pm$8.2  & 85.2$\pm$8.7  & 30.9$\pm$5.8 \\      
H-ATLAS 144334.3$-$003034  & HerBS-162  & 14:43:34.3 & $-$00:30:34      & 2.74 &  2.67 & 76.1$\pm$6.5  &  92.5$\pm$7.3  & 84.6$\pm$7.7  & 26.7$\pm$5.1 \\      
H-ATLAS 121416.3$-$013704  & HerBS-164  & 12:14:16.3 & $-$01:37:04      & 2.54 & 2.50 & 88.0$\pm$6.4  &  99.3$\pm$7.4  & 84.3$\pm$7.7  & 19.4$\pm$5.6 \\      
H-ATLAS 090613.8$-$010042  & HerBS-165  & 09:06:13.8 & $-$01:00:42      & 2.54 & 2.49 &  73.4$\pm$7.4  &  80.2$\pm$8.0  & 84.3$\pm$8.7  & 15.1$\pm$6.2 \\
H-ATLAS 130341.5+313754    & HerBS-167  & 13:03:41.5 & +31:37:53        & 2.95 &  2.92 & 52.1$\pm$5.6  &  82.2$\pm$6.0  & 84.3$\pm$7.2  & 12.4$\pm$5.5 \\      
H-ATLAS 083859.3+021325    & HerBS-169  & 08:38:59.3 & +02:13:26        & 2.54 & 2.46 &  95.2$\pm$7.5  & 105.2$\pm$8.2  & 84.0$\pm$8.7  & 31.2$\pm$5.4 \\
H-ATLAS 083945.0+021021    & HerBS-171  & 08:39:45.1 & +02:10:22        & 2.68 & 2.53 & 71.3$\pm$7.3  &  97.4$\pm$8.1  & 83.4$\pm$8.6  & 19.6$\pm$4.5 \\
H-ATLAS 145040.5+003333    & HerBS-172  & 14:50:40.5 & +00:33:34        & 2.49 &  2.31 & 76.1$\pm$7.4  &  85.1$\pm$8.1  & 83.3$\pm$8.9  & 12.8$\pm$5.3    \\ 
H-ATLAS 121900.8+003326    & HerBS-175  & 12:19:00.8 & +00:33:28        & 3.03 & 3.09 & 56.7$\pm$7.4  &  81.5$\pm$8.0  & 81.9$\pm$8.8  & 20.6$\pm$6.0 \\
H-ATLAS 131222.2+270219    & HerBS-176  & 13:12:22.2 & +27:02:19        & 2.72 & 2.69 & 76.7$\pm$5.5  &  90.1$\pm$5.8  & 82.9$\pm$6.9  & 27.2$\pm$5.1     \\
H-ATLAS 115433.6+005042    & HerBS-177  & 11:54:33.6 & +00:50:42        & 3.44 & 3.65 & 53.9$\pm$7.4  & 85.8$\pm$8.1  & 83.9$\pm$8.6  & 51.4$\pm$4.9      \\         
H-ATLAS 115521.0$-$021329  & HerBS-179  & 11:55:21.0 & $-$02:13:30      & 3.06 & 3.11 & 62.9$\pm$7.3  &  79.9$\pm$8.2  & 82.2$\pm$8.5  & 32.6$\pm$5.3 \\
H-ATLAS 131539.2+292219    & HerBS-180  & 13:15:39.2 & +29:22:19        & 2.47 & 2.36 &  88.2$\pm$5.4  & 102.6$\pm$5.8  & 82.6$\pm$7.1  & 19.0$\pm$4.4    \\
H-ATLAS 090453.2+022017    & HerBS-183  & 09:04:53.3 & +02:20:18        & 2.64 & 2.62 & 87.0$\pm$7.2  &  98.2$\pm$8.0  & 82.3$\pm$8.8  & 33.2$\pm$6.2     \\
H-ATLAS 092408.8$-$005017  & HerBS-185  & 09:24:08.8 & $-$00:50:18      & 2.96 & 3.01 &  71.8$\pm$7.4  &  87.7$\pm$8.2  & 82.2$\pm$8.5  & 37.0$\pm$6.0    \\
H-ATLAS 083705.2+020033    & HerBS-187  & 08:37:05.2 & +02:00:33        & 2.21 & 1.99 & 108.0$\pm$7.2  &  97.0$\pm$8.1  & 82.0$\pm$8.6  & 18.5$\pm$4.7    \\      
H-ATLAS 084259.9+024959    & HerBS-188  & 08:42:59.9 & +02:49:59        & 2.44 & 2.24 &  84.2$\pm$7.4  & 101.5$\pm$8.1  & 81.8$\pm$8.6  & 15.1$\pm$4.7    \\      
H-ATLAS 090405.3$-$003332  & HerBS-190  & 09:04:05.3 & $-$00:33:32      & 2.64 & 2.60 & 82.7$\pm$7.3  &  90.8$\pm$8.2  & 81.6$\pm$8.7  & 28.7$\pm$5.6 \\
H-ATLAS 124753.3+322448    & HerBS-191  & 12:47:53.3 & +32:24:48        & 3.05 & 3.02 &  57.7$\pm$5.8  &  81.5$\pm$5.8  & 81.5$\pm$7.5  & 26.7$\pm$5.2 \\
H-ATLAS 085352.0$-$000804  & HerBS-193  & 08:53:52.1 & $-$00:08:05      & 2.62 & 2.64 &  96.0$\pm$7.3  &  95.0$\pm$8.1  & 81.4$\pm$8.9  & 35.8$\pm$5.5    \\      
H-ATLAS 085521.1$-$003603  & HerBS-194  & 08:55:21.2 & $-$00:36:04      & 2.47 & 2.32 &  95.6$\pm$7.5  &  98.8$\pm$8.1  & 81.3$\pm$8.5  & 26.0$\pm$6.1    \\      
H-ATLAS 122034.2$-$003805  & HerBS-197  & 12:20:34.1 & $-$00:38:05      & 2.44 & 2.29 & 81.9$\pm$7.5  &  93.8$\pm$8.2  & 84.8$\pm$8.7  & 11.1$\pm$5.1     \\
H-ATLAS 133352.2+334913    & HerBS-199  & 13:33:52.2 & +33:49:13        & 2.21 & 2.03 & 112.4$\pm$5.4  & 108.8$\pm$5.9  & 80.6$\pm$7.0  & 20.3$\pm$5.8    \\
H-ATLAS 141117.8$-$010655  & HerBS-201  & 14:11:17.9 & $-$01:06:56      & 3.07 &  3.11 & 52.2$\pm$7.2  &  78.6$\pm$8.2  & 80.5$\pm$8.7  & 24.8$\pm$4.7 \\
H-ATLAS 143328.4+020811    & HerBS-202  & 14:33:28.4 & +02:08:11        & 2.05 & 1.94 & 117.5$\pm$7.3  & 100.7$\pm$8.3  & 80.4$\pm$8.5  & 23.7$\pm$5.1 \\      
H-ATLAS 132909.5+300957    & HerBS-204  & 13:29:09.5 & +30:09:57        & 3.19 & 3.28 & 57.9$\pm$5.5  &  95.3$\pm$6.1  & 80.1$\pm$7.1  & 40.0$\pm$6.6     \\
H-ATLAS 145132.7+024101    & HerBS-205  & 14:51:32.8 & +02:41:01        & 2.64 &  2.58 & 84.5$\pm$7.2  & 104.4$\pm$8.3  & 80.2$\pm$8.9  & 31.1$\pm$7.2    \\      
H-ATLAS 140421.7$-$001217  & HerBS-206  & 14:04:21.8 & $-$00:12:17      & 2.48 &  2.28 & 79.3$\pm$7.4  & 102.6$\pm$8.4  & 80.2$\pm$8.8  & 12.9$\pm$4.3    \\
\hline
\end{tabular}
 \tablefoot{The source names and {\it Herschel} flux densities are from \citet{Bakx2018}. The SCUBA-2 flux densities of the HerBS sources are the revised values from \citet{Bakx2020a}. The sources are listed in order of decreasing $\rm S_{\rm 500 \mu m}$ flux density. For consistency with the HerS and HerMES sources, the photometric redshifts ($z_{\rm phot}$) for the HerBS sources have been re-evaluated using the methods described by \citet{Ivison2016} and \citet{Pearson2013} listed as (a) and (b), respectively.}
   \label{Appendix_table:sources2}
\end{table*}
\normalsize

\begin{table*}[!htbp]
\tiny
\caption{HerBS sources from the Pilot Programme \citep{Neri2020}.}
\begin{tabular}{lccccccccc}
\hline\hline
\multicolumn{2}{c}{Source name} & \multicolumn{2}{c}{{\it Herschel} Coordinates} & \multicolumn{2}{c}{$z_{\rm phot}$} & \multicolumn{4}{c}{Flux density} \\
 &   &  \multicolumn{2}{c}{(J2000.0)}  & &  &  \multicolumn{3}{c}{{\it Herschel}} & SCUBA-2 \\
{\it Herschel} Name  & Other &  &  & (a)  & (b) & $\rm S_{250 \, \mu m}$ & $\rm S_{350 \, \mu m}$ & $\rm S_{500 \, \mu m}$ & $\rm S_{850 \, \mu m}$  \\
& & & & & & \multicolumn{4}{c}{(mJy)}  \\
\hline
H-ATLAS 133413.8+260458    & HerBS-34   & 13:34:13.8 & +26:04:58    & 2.55 &  2.44 & 136.1$\pm$5.4 & 161.0$\pm$5.5 & 126.5$\pm$6.8 & 34.0$\pm$5.7 \\
H-ATLAS 132419.0+320752    & HerBS-43   & 13:24:19.0 & +32:07:52    & 3.05 &  3.02 &  84.4$\pm$4.9 & 116.0$\pm$5.2 & 115.4$\pm$6.3 & 37.0$\pm$5.1 \\
H-ATLAS 133255.8+342208    & HerBS-44   & 13:32:55.8 & +34:22:08        & 2.21 &  2.01 & 164.3$\pm$5.8 & 186.8$\pm$5.8 & 114.9$\pm$7.2 & 25.3$\pm$4.5 \\
H-ATLAS 131540.6+262322    & HerBS-54   & 13:15:40.6 & +26:23:22        & 2.96 &  2.95 &  94.0$\pm$5.7 & 116.1$\pm$6.1 & 108.6$\pm$7.1 & 44.7$\pm$4.6 \\
H-ATLAS 130333.1+244643    & HerBS-58   & 13:03:33.2 & +24:46:43        & 2.67 &  2.61 &  99.0$\pm$5.5 & 111.5$\pm$5.9 & 104.5$\pm$7.1 & 30.5$\pm$5.0 \\
H-ATLAS 130140.2+292918    & HerBS-70   & 13:01:40.2 & +29:29:18        & 2.35 &  2.18 & 119.6$\pm$5.8 & 136.8$\pm$5.8 & 100.0$\pm$7.1 & 21.9$\pm$5.5 \\
H-ATLAS 131434.1+335219    & HerBS-79   & 13:14:34.1 & +33:52:19        & 2.53 &  2.54 & 103.4$\pm$6 & 115.3$\pm$6.0 &  97.9$\pm$7.3 & 28.5$\pm$5.0 \\
H-ATLAS 131611.5+281219    & HerBS-89   & 13:16:11.5 & +28:12:19        & 3.33 &  3.43 &  71.8$\pm$5.7 & 103.4$\pm$5.7 &  95.7$\pm$7.0 & 52.8$\pm$4.3 \\
H-ATLAS 134342.5+263919    & HerBS-95   & 13:43:42.5 & +26:39:19        & 3.07 &  3.02 &  61.9$\pm$5.7 & 101.3$\pm$5.7  & 94.7$\pm$7.6 & 27.4$\pm$6.2 \\
H-ATLAS 131211.5+323837    & HerBS-113  & 13:12:11.5 & +32:38:37        & 2.82 &  2.77 &  80.7$\pm$5.9 & 103.4$\pm$6.0 &  92.0$\pm$7.0 & 32.0$\pm$5.2 \\
H-ATLAS 132258.2+325050    & HerBS-154  & 13:22:58.2 & +32:50:50        & 2.74 &  2.70 &  79.1$\pm$5.6 &  87.9$\pm$5.9 &  85.6$\pm$7.2 & 28.8$\pm$4.2 \\
\hline
\end{tabular}
\tablefoot{The source names and {\it Herschel} flux densities are from \citet{Bakx2018}. The SCUBA-2 flux densities of the HerBS sources are the revised values from \citet{Bakx2020a}. The sources are listed in order of decreasing $\rm S_{\rm 500 \mu m}$ flux density. For consistency with the HerS and HerMES sources, the photometric redshifts ($z_{\rm phot}$) for the HerBS sources have been re-evaluated using the methods described by \citet{Ivison2016} and \citet{Pearson2013} listed as (a) and (b), respectively.}
   \label{Appendix_table:sources3}
\end{table*}
\normalsize

\begin{table*}[!htbp]
\tiny
\caption{Northern galactic plane and equatorial sources, without spectroscopic redshift measurements, not selected in (or dropped from) $z$-GAL from the HerBS, HeLMS, and HerS catalogues.}
\begin{tabular}{lcccccccl}
\hline\hline
\multicolumn{2}{c}{Source name} & \multicolumn{2}{c}{{\it Herschel} Coordinates} & \multicolumn{4}{c}{Flux density} \\
 &   &  \multicolumn{2}{c}{(J2000.0)}  &  \multicolumn{3}{c}{{\it Herschel}} & SCUBA-2 & \\
{\it Herschel} Name  & Other &  &  & $\rm S_{250 \, \mu m}$ & $\rm S_{350 \, \mu m}$ & $\rm S_{500 \, \mu m}$ & $\rm S_{850 \, \mu m}$  & \\
& & & & \multicolumn{4}{c}{(mJy)}  & \\
\hline
\multicolumn{9}{c}{HerBS sources} \\
\hline
H-ATLAS 141004.7+020306   &  HerBS-16 & 14:10:04.7 &  +02:03:06 & 119.4$\pm$7.3 & 151.0$\pm$8.4 & 176.0$\pm$8.7 & 86.0$\pm$4.7 & Blazar \\ 
H-ATLAS 091809.5+001929   &  HerBS-99 & 09:18:09.5 &   +00:19:29 &  93.2$\pm$7.4 &  116.6$\pm$8.2 &  94.3$\pm$8.7 & 23.9$\pm$5.0 & \\ 
H-ATLAS 133108.4+303034   & HerBS-112 & 13:31:08.4 &   +30:30:34 &  71.8$\pm$5.8 & 87.0$\pm$5.8 & 92.2$\pm$7.0 & 220.3$\pm$6.2 & Blazar \\
H-ATLAS 113833.8$-$014655 & HerBS-119 & 11:38:33.8 & $-$01:46:55 &  68.5$\pm$7.2 & 85.6$\pm$8.1 & 91.2$\pm$8.6 & {\it 12.5$\pm$5.5} & \\
H-ATLAS 142706.4+002258   & HerBS-130 & 14.27:06.4 &   +00:22:58 & 119.4$\pm$7.3 & 118.7$\pm$8.1 & 88.8$\pm$8.6 & 15.5$\pm$5.1 & \\
H-ATLAS 134441.5+240345   & HerBS-133 & 13:44:41.5 &   +24:03:45 &  85.4$\pm$5.5 & 98.5$\pm$6.1 & 88.1$\pm$7.3 & {\it 17.4$\pm$5.9} & \\  
H-ATLAS 134855.6+240745   & HerBS-139 & 13:48:55.6 &   +24:07:45 &  76.9$\pm$5.9 & 82.9$\pm$5.9 & 87.4$\pm$6.8 & {\it 15.2$\pm$6.0} & \\
H-ATLAS 091454.0$-$010358 & HerBS-142 & 09:14:54.0 & $-$01:03:58 &  69.0$\pm$7.3 & 72.2$\pm$8.1 & 87.2$\pm$8.5 & 22.4$\pm$5.7 & \\
H-ATLAS 133057.5+311734   & HerBS-152 & 13:30:57.5 &   +31:17:34 &  47.7$\pm$5.6 & 53.4$\pm$6.0 & 85.8$\pm$6.9 & 14.1$\pm$5.0 & \\
H-ATLAS 132329.9+311528   & HerBS-158 & 13:23:29.9 &   +31:15:28 &  64.7$\pm$5.4 & 75.7$\pm$6.2 & 85.1$\pm$7.2 & 24.4$\pm$4.1 & \\
H-ATLAS 122407.4$-$003247   & HerBS-161 & 12:24:40.7 & $-$00:32:47 &  56.5$\pm$7.3 & 75.7$\pm$8.1 & 82.4$\pm$8.8 & {\it 9.8$\pm$ 4.3} & \\
H-ATLAS 145754.2+000018   & HerBS-195 & 14:57:54.2 &   +00:00:18 &  70.3$\pm$7.3 & 92.7$\pm$8.1 & 81.0$\pm$8.8 & 17.3$\pm$5.3 & \\
H-ATLAS 134403.1+242628   & HerBS-196 & 13:44:03.1 &   +24:26:28 &  86.9$\pm$5.7 & 92.3$\pm$6.3 & 81.0$\pm$7.1 & {\it 13.9$\pm$6.3} & \\ 
H-ATLAS 141827.4$-$001703   & HerBS-203 & 14:18:27.4 & $-$00:17:03 & 117.2$\pm$6.5 & 116.4$\pm$7.4 & 80.2$\pm$7.6 & 15.0$\pm$4.8 &  \\ 
\hline
\multicolumn{8}{c}{HeLMS and HerS sources} \\
\hline
HERMES 003032.1$-$021153 & HeLMS-33 & 00:30:32.1 & $-$02:11:53 & 81$\pm$7  & 98$\pm$6 &  118$\pm$8 &  & Blazar \\
HERMES 004532.6$-$000123 & HeLMS-53  & 00:45:32.6 & $-$00:01:23 &  48$\pm$7     &  85$\pm$6  & 103$\pm$8 &  &  \\
\hline
\end{tabular}
 \tablefoot{The source names and {\it Herschel} flux densities are from \citet{Bakx2018} and \citet{Nayyeri2016}. The SCUBA-2 flux densities of the HerBS sources are the revised values from \citet{Bakx2020a}. The SCUBA-2 flux densities indicated in italics are classified as non-detections, as discussed in Sect. 3 of \citet{Bakx2018}. The sources are listed in order of decreasing $\rm S_{\rm 500 \mu }$ flux density. For the sources that are identified as blazars the reader is referred to Sect.\,\ref{section:obs}.} 
   \label{Appendix_table:dropped_sources}
\end{table*}
\normalsize

\begin{table*}[!htbp]
\tiny
\caption{Northern galactic plane and equatorial sources with previously available spectroscopic redshift measurements from the HerBS, HeLMS, and HerS catalogues.}
\begin{tabular}{lcccccccc}
\hline\hline
\multicolumn{2}{c}{Source name} & \multicolumn{2}{c}{{\it Herschel} Coordinates} & \multicolumn{4}{c}{Flux density} & $z_{\rm spec}$ \\
 &   &  \multicolumn{2}{c}{(J2000.0)}  &  \multicolumn{3}{c}{{\it Herschel}} & SCUBA-2 & \\
{\it Herschel} Name  & Other &  &  & $\rm S_{250 \, \mu m}$ & $\rm S_{350 \, \mu m}$ & $\rm S_{500 \, \mu m}$ & $\rm S_{850 \, \mu m}$  & \\
& & & & \multicolumn{4}{c}{(mJy)}  & \\
\hline
\multicolumn{9}{c}{HerBS Sources} \\
\hline
H-ATLAS 134429.5+303034 & HerBS-1 & 13:44:29.5 & +30:30:34 & 461.9$\pm$5.8 & 465.7$\pm$6.5 & 343.3$\pm$7.1 & 90.3$\pm$5.3 & 2.30 \\
H-ATLAS 114637.9$-$001132 & HerBS-2 & 11:46:37.9 & $-$00:11:32 &        316.0$\pm$6.6 &       357.9$\pm$7.4 & 291.8$\pm$7.7 & 98.2$\pm$5.4 &  3.26 \\ 
H-ATLAS 132630.1+334408 & HerBS-3 & 13:26:63.0 & +33:44:08 & 190.5$\pm$5.6 &       281.3$\pm$5.9 & 278.6$\pm$7.5 & 73.8$\pm$5.0 &  2.95 \\
H-ATLAS 083051.0+013225 & HerBS-4       & 08:30:51.0 & +01:32:25 & 248.5$\pm$7.5 &       305.3$\pm$8.1 & 269.1$\pm$8.7 & 98.7$\pm$5.4 & 3.63 \\
H-ATLAS 125632.5+233627 & HerBS-5 & 12:56:32.5 & +23:36:27 & 209.3$\pm$5.6 & 288.5$\pm$6.0 & 264.0$\pm$7.0 & 97.6$\pm$5.0 & 3.56 \\
H-ATLAS 132427.0+284450 & HerBS-6 & 13:24:27.0 & +28:44:50 & 342.3$\pm$5.6 & 371.0$\pm$5.9 & 250.9$\pm$6.9 & 38.1$\pm$5.6 &        1.68 \\
H-ATLAS 132859.2+292327 & HerBS-7 & 13:28:59.2 & +29:23:27      & 268.4$\pm$4.4 & 296.3$\pm$4.8 & 248.9$\pm$5.9 & 77.4$\pm$6.5 &        2.78 \\
H-ATLAS 084933.4+021442 & HerBS-8 & 08:49:33.4 & +02:14:42      & 216.7$\pm$7.5 & 248.5$\pm$8.2 & 208.6$\pm$8.6 & 36.3$\pm$5.8 &        2.41 \\
H-ATLAS 125135.3+261458 & HerBS-9 & 12:51:35.3 & +26:14:58 &    157.9$\pm$5.9 &       202.2$\pm$6.0 & 206.8$\pm$6.9 &  83.5$\pm$5.5 & 3.68 \\
H-ATLAS 113526.2$-$014606 & HerBS-10 & 11:35:26.2 & $-$01:46:06 & 278.8$\pm$7.4 & 282.9$\pm$8.2 & 204.0$\pm$8.6 & 73.2$\pm$5.3  & 3.13 \\       
H-ATLAS 133008.6+245900 & HerBS-12 & 13:30:08.6 & +24:59:00 & 271.2$\pm$5.4 &       278.2$\pm$5.9 & 203.5$\pm$6.9 & 67.7$\pm$6.4 &  3.11    \\
H-ATLAS 142413.9+022303 & HerBS-13 & 14:24:13.9 & +02:23:03 & 112.2$\pm$7.3 &       182.2$\pm$8.2 & 193.3$\pm$8.5 & 89.7$\pm$5.4 &  4.24 \\ 
H-ATLAS 141351.9$-$000026 & HerBS-15 & 14:13:51.9 & $-$00:00:26 & 188.6$\pm$7.4 &217.0$\pm$8.1 &        176.4$\pm$8.7 & 44.0$\pm$4.9 &  2.48 \\
H-ATLAS 090311.6+003907 & HerBS-19 & 09:03:11.6 & +00:39:07 & 133.2$\pm$7.4 &       186.1$\pm$8.2 & 165.2$\pm$8.8 & 87.7$\pm$6.3 &  3.04 \\
H-ATLAS 132504.4+311534 & HerBS-20 & 13:25:04.4 & +31:15:34 & 240.6$\pm$5.4 &       226.6$\pm$6.0 & 164.9$\pm$7.3 & 26.2$\pm$4.9 &  1.84    \\
H-ATLAS 133846.5+255055 & HerBS-29 & 13:38:46.5 & +25:50:55 & 159.0$\pm$5.8 &       183.1$\pm$6.0 & 137.6$\pm$7.5 & 36.9$\pm$5.8 &  2.34    \\
H-ATLAS 132301.7+341649 & HerBS-30 & 13:23:01.7 & +34:16:49 &   124.1$\pm$5.6 &       144.5$\pm$6.0 & 137.0$\pm$7.2 & 48.2$\pm$4.7 &  2.19 \\
H-ATLAS 125652.5+275900 & HerBS-31      & 12:56:52.5 & +27:59:00 & 133.9$\pm$5.8 & 164.1$\pm$6.0 & 131.8$\pm$7.4 & 41.5$\pm$4.9 &        {\it 2.79}      \\
H-ATLAS 091840.8+023048 & HerBS-32 & 09:18:40.8 & +02:30:48 & 125.7$\pm$7.2 &       150.7$\pm$8.2 & 128.4$\pm$8.7 & 32.9$\pm$5.7 &  2.58    \\
H-ATLAS 133543.0+300402 & HerBS-35 & 13:35:43.0 & +30:04:02 & 136.6$\pm$5.4 &       145.7$\pm$5.8 & 125.0$\pm$6.9 & 35.8$\pm$4.7 &  2.68    \\
H-ATLAS 125125.8+254930 & HerBS-52 & 12:51:25.8 & +25:49:30 & 57.4$\pm$5.8 & 96.8$\pm$5.9 & 109.4$\pm$7.2 & 39.4$\pm$5.8 & 3.44 \\
H-ATLAS 091304.9$-$005344 & HerBS-59 & 09:13:04.9 & $-$00:53:44 & 118.2$\pm$6.4 &       136.8$\pm$7.4 & 104.3$\pm$7.7 & 42.0$\pm$5.4 &  2.63 \\
H-ATLAS 130118.0+253708 & HerBS-64 & 13:01:18.0 & +25:37:08 & 60.2$\pm$4.8 &       101.1$\pm$5.3 & 101.5$\pm$6.4 & 54.3$\pm$4.6 & 4.04 \\
H-ATLAS 115820.1$-$013752 & HerBS-66 & 11:58:20.1 & $-$01:37:52 & 119.8$\pm$6.8 &       123.7$\pm$7.7 & 101.5$\pm$7.9 & 25.8$\pm$4.2 &  2.19 \\
H-ATLAS 113243.0$-$005108 & HerBS-71    & 11:32:43.0 & $-$00:51:08 & 67.8$\pm$7.3 &       105.8$\pm$8.2 & 99.8$\pm$8.8 &  {\it 10.8$\it \pm$5.1} &        2.58 \\
H-ATLAS 083344.9+000109  & HerBS-88 &   08:33:44.9 & +00:01:09 & 71.0$\pm$7.6 &       96.0$\pm$8.1 &  95.9$\pm$8.8 & 16.3$\pm$5.8 &   {\it 3.10} \\ 
H-ATLAS 113803.6$-$011737 & HerBS-96 & 11:38:03.6 & $-$01:17:37 & 85.1$\pm$7.3 &       98.4$\pm$8.2 &  94.8$\pm$8.8 & 11.3$\pm$5.1 &   {\it 3.15}      \\
H-ATLAS 113833.3+004909 & HerBS-100 & 11:38:33.3 & +00:49:09 &  96.8$\pm$7.3 &       106.4$\pm$8.1 & 93.4$\pm$8.7 & 6.3$\pm$5.9 &    {\it 2.22}      \\
\hline
\multicolumn{9}{c}{HeLMS and HeRS Sources} \\
\hline
HERMES J233255.5$-$031134 & HeLMS-2 & 23:32:55.4 & $-$03:11:34 &        271$\pm$6 & 336$\pm$6 & 263$\pm$8 & &     2.6895 \\
HERMES J004410.2+011821   & HeLMS-4 & 00:44:10.2 & +01:18:21   & 113$\pm$7 & 177$\pm$6 & 209$\pm$8 &  &  4.1625 \\
HERMES J234051.3$-$041937 & HeLMS-5 & 23:40:51.3 & $-$04:19:37 & 151$\pm$6 & 209$\pm$6 & 205$\pm$8 & & 3.5027 \\
HERMES J233620.7$-$060826 & HeLMS-6 & 23:40:51.5 & $-$04:19:38 & 151$\pm$6 & 209$\pm$6 & 205$\pm$8 & & 3.4346 \\
HERMES J232439.4$-$043934 & HeLMS-7 & 23:24:39.4 & $-$04:39:36 & 214$\pm$7 & 218$\pm$7 & 172$\pm$9 & & 2.4726 \\
HERMES J004714.1+032453   & HeLMS-8 & 00:47:23.6 & +01:57:51 & 398$\pm$6 & 320$\pm$6 & 164$\pm$8 & & 2.2919 \\ 
HERMES J004723.3+015749   & HeLMS-9 & 00:47:23.6 & +01:57:51 & 398$\pm$6 & 230$\pm$6 & 164$\pm$8 & & 1.441 \\
HERMES J005258.4+061319   & HeLMS-10 & 00:52:58.6 & +06:13:19 &  88$\pm$6 & 129$\pm$6 & 155$\pm$7 & & 4.3726 \\
HERMES J001615.8+032433   & HeLMS-13 & 00:16:15.7 & +03:24:35 & 195$\pm$6 & 221$\pm$6 & 149$\pm$7 & & 2.765 \\    
HERMES J233255.7$-$053424 & HeLMS-15 & 23:32:55.7 & $-$05:34:26 & 148$\pm$6 & 187$\pm$6 & 147$\pm$9 & & 2.4022 \\
HERMES J005159.4+062240   & HeLMS-18 & 00:51:59.5 & +06:22:41 & 166$\pm$6 & 195$\pm$6 & 135$\pm$7 & & 2.392 \\    
HERMES J001626.0+042613   & HeLMS-22 & 00:16:26.0 & +04:26:13 & 117$\pm$7 & 151$\pm$6 & 127$\pm$7 & & 2.5093 \\ 
HERMES J002220.9-015520   & HeLMS-29 & 00:22:20.9 & $-$01:55:20 & 66$\pm$6 & 102$\pm$6 & 121$\pm$7 & & 5.1614 \\
HERS J020941.1+001557     & HerS-1 & 02:09:41.1 & +00:15:57 & 826$\pm$7 & 912$\pm$7 & 718$\pm$8 & & 2.5534 \\     
HERS 011640.1$-$000454    & HerS-4    & 01:16:40.1 & $-$00:04:54 & 137$\pm$7     & 196$\pm$7  & 190$\pm$8 &  & 3.791$\rm ^a$ \\ 
HERS J010301.2-003300     & HerS-6 & 01:03:01.2 & $-$00:33:00 & 121$\pm$7 & 147$\pm$6 & 130$\pm$8 & & 2.2153 \\
\hline
\end{tabular}
 \tablefoot{The source names and {\it Herschel} flux densities are from \citet{Bakx2018} and \citet{Nayyeri2016}. The SCUBA-2 flux densities of the HerBS sources are the revised values from \citet{Bakx2020a}. The SCUBA-2 flux densities indicated in italics are classified as non-detections, as discussed in Sect. 3 of \citet{Bakx2018}. The sources are listed in order of decreasing $\rm S_{\rm 500 \mu m}$ flux density. The references to the spectroscopic redshifts measurements are given in \citet{Bakx2018} and \citet{Nayyeri2016}; for the HeLMS and HerS sources, the $z_{\rm spec}$ values have been udpated for the sources included in \citet{Riechers2021} and \citet{Maresca2022} (and references therein); for HerBS-52 and HerBS-64, the spectroscopic redshifts are from \citet{Bakx2020b}. Spectroscopic redshifts in italics are derived from on a single line detection and are therefore considered tentative. (a) The spectroscopic redshift of HerS-4 was measured with the LMT (A. Baker, priv. comm.).} 
   \label{Appendix_table:sources_with_redshifts}
\end{table*}
\normalsize

\begin{table*}[!htbp]
\tiny
\caption{Spectroscopic redshifts and emission line properties: HeLMS and HerS sources.}
\begin{tabular}{lcrcccccccc}
\hline
\hline
Source & \multicolumn{2}{c}{Observed Position} & $z_{\rm spec}$ &  \multicolumn{7}{c}{Emission Lines} \\
 & RA & Dec. & & $\Delta V$ & Line & Line Flux &  Line & Line Flux  & Line & Line Flux \\ 
 & \multicolumn{2}{c}{(J2000)} & & $\rm (km \, s^{-1})$ & & $\rm (Jy \, km \, s^{-1})$ & & $\rm (Jy \, km \, s^{-1})$ & & $\rm (Jy \, km \, s^{-1})$ \\ 
\hline
\multicolumn{11}{c}{HeLMS Sources} \\
\hline
HeLMS-1 & 23:34:40.97 & $-$06:52:20.2 & 1.9047(1) & 281$\pm$10 & CO(2-1) &  4.11$\pm$0.25  & CO(4-3) & 12.19$\pm$0.57  & &  \\
HeLMS-3  & 00:02:15.96 & $-$01:28:30.5 & 1.4202(1) & 512$\pm$12$^*$     & CO(2-1) & 11.24$\pm$0.86 & CO(3-2) & 20.72$\pm$1.55  & &  \\
HeLMS-11 & 00:39:29.45 &  00:24:25.9 & 2.4829(1) & 751$\pm$18$^*$ & CO(3-2) & 4.71$\pm$0.41  & CO(4-3) &  6.81$\pm$0.55  & &  \\
HeLMS-12 & 23:56:01.47 & $-$07:11:43.1 & 2.3699(2) & 551$\pm$25  & CO(3-2) & 10.70$\pm$0.75 & CO(4-3) & 17.81$\pm$1.15  & & \\     
HeLMS-14 & 00:36:19.77 &  00:24:17.7 & 1.6169(1) & 537$\pm$11$^*$ & CO(2-1) & 8.62$\pm$0.52 & CO(3-2) & 13.99$\pm$0.86  & & \\
HeLMS-16 & 23:18:57.18 & $-$05:30:34.7 & 2.8187(1) & 1152$\pm$57$^*$ & CO(3-2) & 5.31$\pm$0.76  &       &    & [CI](1-0) & 6.47$\pm$1.38 \\
HeLMS-17 E & 23:25:58.56 & $-$04:45:25.6 & 2.2983(1) &  881$\pm$140 & CO(4-3) & 0.72$\pm$0.16 &  CO(6-5) & 1.27$\pm$0.31 &   &   \\
HeLMS-17 W & 23:25:57.92 &  $-$04:45:25.5 & 2.2972(1) & 744$\pm$27 & CO(4-3) & 3.12$\pm$0.24 &  CO(6-5) & 6.81$\pm$0.38  & $\rm H_2O(2_{11}$-2$_{02}$) & 1.80$\pm$0.19 \\
HeLMS-19 E & 23:22:10.20 & $-$03:35:59.8 & 4.6871(1) & 326$\pm$45  & CO(4-3) & 1.49$\pm$0.40 & CO(5-4) & 1.01$\pm$0.25 & CO(7-6) & 1.67$\pm$0.40 \\
HeLMS-19 W & 23:22:09.96 & $-$03:36:01.4 & 4.6882(1) & 250$\pm$24  & CO(4-3) & 1.76$\pm$0.38  & CO(5-4) & 1.01$\pm$0.25  & CO(7-6) & 2.71$\pm$0.45 \\
HeLMS-19 EW & 23:22:10.05 & $-$03:36:00.8 & 4.6885(1) & 332$\pm$31 & CO(4-3) & 3.25$\pm$0.40 & CO(5-4) &  1.76$\pm$0.16 &    CO(7-6)  & 4.38$\pm$0.42 \\
HeLMS-19 EW  &            &               &  &          &                         &              &          [CI](1-0) & 0.78$\pm$0.35    & [CI](2-1) & 2.00$\pm$0.88 \\
HeLMS-20 & 23:37:28.83 & $-$04:51:06.3 & 2.1947(1) & 298$\pm$20  & CO(4-3) & 3.83$\pm$0.85  &       &   &  [CI](1-0) & 2.05$\pm$0.49 \\
HeLMS-21 & 00:18:00.21 & $-$06:02:35.3 & 2.7710(2) &  346$\pm$20 & CO(3-2) & 2.02$\pm$0.18  & CO(5-4) & 6.41$\pm$0.52   &  &    \\ 
HeLMS-23 & 00:58:41.09 & $-$01:11:49.0 & 1.4888(1) & 403$\pm$25  & CO(2-1) & 2.35$\pm$0.29  & CO(3-2) &  5.16$\pm$0.42   &  & \\   
HeLMS-24 & 00:38:14.00 & $-$00:22:52.5 & 4.9841(1) & 706$\pm$42 & CO(4-3) & 6.08$\pm$0.67  & CO(5-4) & 6.33$\pm$0.58  & [CI](1-0) & 1.49$\pm$0.45 \\
HeLMS-25 & 00:41:24.12 & $-$01:03:07.8 & 2.1404(2) & 932$\pm$39$^*$ & CO(2-1) & 0.94$\pm$0.17   &  CO(4-3) & 3.36$\pm$0.46  & &  \\   
HeLMS-26 E & 00:47:47.56 & 06:14:39.4 & 2.6899(1) &  328$\pm$53  & CO(3-2) & 0.35$\pm$0.14  & CO(5-4) &  1.65$\pm$0.44   & [CI](1-0)  & 1.19$\pm$0.36 \\
HeLMS-26 W & 00:47:46.76 & 06:14:45.3 & 2.6875(1) & 815$\pm$159 & CO(3-2) & 0.89$\pm$0.30  & CO(5-4) & 2.82$\pm$0.75   & & \\
HeLMS-27  & 00:37:58.07 & $-$01:06:20.1 & 3.7652(1) & 658$\pm$67 & CO(4-3) & 1.61$\pm$0.28  & CO(6-5) & 5.63$\pm$0.78   & &  \\
HeLMS-28  & 00:30:09.28  & $-$02:06:25.1 & 2.5322(3) & 749$\pm$27$^*$ & CO(3-2) & 1.20$\pm$0.13  &          &  & [CI](1-0) & 1.64$\pm$0.13 \\
HeLMS-30  & 00:10:27.16  & $-$02:46:26.4 & 1.8197(1) & 662$\pm$15$^*$ & CO(2-1) & 3.26$\pm$0.26  & CO(4-3) &  8.58$\pm$0.67   & & \\    
HeLMS-31  & 00:13:53.53  & -06:02:00.2 & 1.9494(1) & 389$\pm$20$^*$ & CO(2-1) & 4.96$\pm$0.48  & CO(4-3) & 10.46$\pm$0.99  & & \\     
HeLMS-32  & 00:03:37.03  &  01:40:12.2 & 1.7149(3) & 742$\pm$53$^*$ & CO(2-1) & 1.63$\pm$0.38  & CO(3-2) & 1.63$\pm$0.43   & & \\     
HeLMS-34  & 00:27:19.60  &  00:12:02.5 & 2.2714(1) & 436$\pm$7$^*$       & CO(3-2) & 2.61$\pm$0.15  & CO(4-3) & 4.10$\pm$0.22 & & \\     
HeLMS-35  & 23:24:59.96  & $-$00:56:44.1 & 1.6684(2) & 857$\pm$70 & CO(2-1) & 2.25$\pm$0.37  & CO(3-2) & 7.91$\pm$0.86   & &  \\    
HeLMS-36 & 23:43:14.11  &  01:21:55.4 & 3.9802(1) & 388$\pm$42 & CO(4-3) & 1.09$\pm$0.24  & CO(5-4) & 1.68$\pm$0.36 & CO(6-5) & 1.69$\pm$0.31  \\
HeLMS-37  & 01:08:01.83  &  05:32:01.2 & 2.7576(2) & 422$\pm$20$^*$ & CO(3-2) & 0.78$\pm$0.13  & CO(5-4) & 2.00$\pm$0.31   & & \\     
HeLMS-38  & 00:22:08.10  &  03:40:42.0 & 2.1898(1) & 480$\pm$28$^*$ & CO(2-1) & 1.95$\pm$0.28 & CO(3-2)  & 4.27$\pm$0.66 & CO(4-3) & 4.35$\pm$0.68 \\
HeLMS-39  &     00:29:36.02  &  02:07:13.1 & 2.7659(2) & 478$\pm$20$^*$ & CO(3-2) & 2.65$\pm$0.36 & CO(5-4) & 3.98$\pm$0.55  & &  \\      
HeLMS-40 W1  & 23:53:31.89  &  03:17:19.9 & 3.1445(1) & 766$\pm$42$^*$ & CO(3-2) & 1.26$\pm$0.25 & CO(5-4) & 3.03$\pm$0.42  & & \\       
HeLMS-40 W2  & 23:53:31.58 & 03:17:21.0 & 3.1395(1) & 681$\pm$51 & CO(3-2) & 1.11$\pm$0.30 & CO(5-4) & 1.86$\pm$0.40 & &  \\       
HeLMS-41  & 23:36:33.72  & $-$03:21:20.2 & 2.3353(1) & 557$\pm$61 & CO(3-2) & 2.09$\pm$0.35 & CO(4-3) & 3.22$\pm$0.50  & &   \\     
HeLMS-42  & 23:40:14.29  & $-$07:07:36.5 & 1.9558(7) & 463$\pm$92 & CO(2-1) & 1.90$\pm$0.59 & CO(4-3) & 2.61$\pm$0.72  & &  \\      
HeLMS-43  & 23:34:20.08  & $-$00:34:58.1 & 2.2912(8) & 529$\pm$101 & CO(3-2) & 3.07$\pm$0.99 & CO(4-3) & 2.17$\pm$0.57  & &   \\
HeLMS-44  & 23:14:47.61  & $-$04:56:56.0 & 1.3700(1) & 385$\pm$16$^*$  & CO(2-1) & 1.56$\pm$0.29 & CO(3-2) & 3.27$\pm$0.40  & &  \\      
HeLMS-45  & 00:12:27.07  &  02:08:06.4 & 5.3994(1) & 725$\pm$75$^*$  & CO(5-4) & 6.78$\pm$1.27 & CO(8-7) & 4.22$\pm$0.68  &  [CI](1-0) & 1.15$\pm$0.44 \\
HeLMS-46  & 00:46:22.27  &  07:35:18.2 & 2.5765(2) & 685$\pm$31$^*$ & CO(3-2) & 2.42$\pm$0.37 & CO(4-3) & 3.52$\pm$0.56  & & \\
HeLMS-47 & 23:49:51.73  & $-$03:00:17.5 & 2.2232(1) & 694$\pm$46 & CO(2-1) & 2.15$\pm$0.24 & CO(4-3) & 6.84$\pm$0.68 &  [CI](1-0) & 2.08$\pm$0.59  \\      
HeLMS-48 & 23:28:33.52  & $-$03:14:18.9 & 3.3514(1) & 621$\pm$51 & CO(3-2) & 1.58$\pm$0.32 & CO(4-3) & 2.94$\pm$0.41  & CO(5-4) & 2.94$\pm$0.41   \\
HeLMS-49  & 23:37:21.72  & $-$06:47:42.0 & 2.2154(5) & 531$\pm$111 &          &               & H$_2$O(4$_{23}$-3$_{30}$) & 2.11$\pm$0.63 & [CI](1-0) & 1.89$\pm$0.60 \\
HeLMS-50  & 23:51:01.86  & $-$02:44:23.5 & 2.0531(5) & 1251$\pm$195 & CO(3-2) & 4.41$\pm$0.83 & CO(4-3) & 3.86$\pm$0.36 & &  \\
HeLMS-51  & 23:26:17.61  & $-$02:53:19.6 & 2.1567(5) & 1064$\pm$59$^*$ & CO(4-3) & 5.60$\pm$0.94 &            &          &  [CI](1-0) &  2.99$\pm$0.77 \\
HeLMS-52  &  23:37:27.01 & $-$00:23:41.1 & 2.2092(1) & 344$\pm$25  & CO(3-2) & 3.23$\pm$0.37 & CO(4-3) & 4.26$\pm$0.43  & & \\
HeLMS-54  &  00:27:17.79 &  02:39:46.5 & 2.7070(8) & 299$\pm$69  & CO(3-2) & 0.72$\pm$0.23 &        &          & [CI](1-0) & 0.85$\pm$0.30\\       
HeLMS-55  &  23:28:31.77 & $-$00:40:36.6 & 2.2834(2) & 471$\pm$29$^*$  & CO(3-2) & 4.46$\pm$1.17 & CO(4-3) & 4.74$\pm$0.77 & &  \\
HeLMS-56  &  00:13:25.82 &  04:25:07.3 & 3.3896(3) & 814$\pm$38$^*$  & CO(3-2) & 0.95$\pm$0.18 & CO(4-3) & 1.94$\pm$0.27 & & \\                
HeLMS-57  &  00:35:19.56 &  07:28:03.7 & 1.9817(3) & 762$\pm$106  & CO(2-1) & 2.75$\pm$0.59 & CO(4-3) & 5.15$\pm$1.01 & & \\
\hline
\multicolumn{11}{c}{HerS Sources} \\
\hline
HerS-2 & 01:20:41.65 & $-$00:27:05.8 & 2.0151(1)  & 748$\pm$23$^*$ & CO(2-1) & 6.50$\pm$0.64 & CO(4-3) & 12.78$\pm$1.37 &  &  \\
HerS-3 NE & 01:27:54.06 &  00:49:38.9 & 3.0608(1) & 382$\pm$25  & CO(3-2) & 3.80$\pm$0.33 & CO(5-4) & 4.32$\pm$0.49 & & \\ 
HerS-3 SW & 01:27:54.30 &  00:49:42.6 & 3.0607(3) & 427$\pm$31  & CO(3-2) & 1.45$\pm$0.15 & CO(5-4) & 3.95$\pm$0.48 & & \\
HerS-5    & 01:26:20.60 &  01:29:49.9 & 1.4493(1) & 742$\pm$22$^*$ & CO(2-1) & 2.81$\pm$0.16 & CO(3-2) & 6.00$\pm$0.39 & & \\
HerS-7   & 01:01:33.69 &  00:31:53.7 & 1.9836(5) & 693$\pm$83 & CO(2-1) & 3.91$\pm$0.86 & CO(4-3) & 5.68$\pm$0.91 & & \\
HerS-8   & 01:09:38.94 & $-$01:48:22.9 & 2.2431(2) & 718$\pm$108 & CO(3-2) & 1.14$\pm$0.30 & CO(4-3) & 2.75$\pm$0.56 & &  \\
HerS-9  & 01:09:11.86 & $-$01:17:33.4 & 0.8530(1) & 134$\pm$26 &  & & HCN(3-2) & 0.48$\pm$0.13 & HCO$^+$(3-2) & 0.46$\pm$0.14 \\       
HerS-10  & 01:17:22.40 & 00:56:22.8 & 2.4688(1) & 411$\pm$12$^*$ & CO(3-2) & 2.85$\pm$0.24 &  & &  [CI](1-0)& 1.61$\pm$0.35        \\
HerS-11  & 00:58:47.20 & $-$01:00:16.2 & 4.6618(3) & 531$\pm$86 & CO(4-3) & 1.73$\pm$0.45 & CO(7-6) & 5.08$\pm$1.14 &  & \\       
HerS-12  & 01:25:46.18 & $-$00:11:42.2 & 2.2706(3) & 814$\pm$38$^*$ & CO(3-2) & 5.17$\pm$0.78 & CO(4-3) & 7.45$\pm$1.27 & &  \\       
HerS-13  & 01:25:20.95 &  01:17:24.3 & 2.4759(4) & 564$\pm$60 & CO(3-2) & 4.11$\pm$0.58 &   &      &  [CI](1-0) &  2.87$\pm$0.60  \\
HerS-14  & 01:40:57.35 & $-$01:05:46.8 & 3.3441(1) & 895$\pm$37 & CO(3-2) & 3.33$\pm$0.50 & CO(4-3) &  6.17$\pm$0.71 & CO(5-4) & 8.32$\pm$0.50 \\ 
HerS-15 & 01:21:06.83 & 00:34:55.8 & 2.3019(1) & 672$\pm$19$^*$ & CO(3-2) & 2.50$\pm$0.22 & CO(4-3) & 3.85$\pm$0.38 & &\\
HerS-16 & 02:14:34.48 & 00:59:23.9 & 2.1981(2) & 786$\pm$27$^*$ & CO(4-3) & 6.61$\pm$0.47 &       & & [CI](1-0) & 3.17$\pm$0.50 \\        
HerS-17 & 02:14:02.54 & $-$00:46:11.0 & 3.0182(1) & 793$\pm$18$^*$ & CO(3-2) & 3.94$\pm$0.38 & CO(5-4) & 6.97$\pm$0.46 & & \\        
HerS-18 E       & 01:32:12.31 & 00:17:55.6 & 1.6926(1) & 733$\pm$36 & CO(2-1) & 4.72$\pm$0.45 & CO(3-2) & 7.16$\pm$0.47 & & \\
HerS-18 W & 01:32:11.78 & 00:17:58.2 &  {\it 0.5636(1)} & 364$\pm$75 & {\it CO(2-1)}    & 0.95$\pm$0.25 & & \\
HerS-19 SE & 02:05:29.26 & 00:04:57.0 & {\it 3.1678(1)} & 630$\pm$110 &  {\it CO(3-2)} &   1.62$\pm$0.41 & & &  &  \\
HerS-19 W  & 02:05:28.82 & 00:05:01.2 & {\it  3.1394(1)} & 662$\pm$190 & {\it CO(3-2)} & 1.22$\pm$0.51 &  &   &  &  \\   
HerS-20 & 01:02:46.14 & 01:05:40.1 & 2.0792(1) & 341$\pm$23 & CO(2-1) &  1.95$\pm$0.18 & CO(4-3) & 11.71$\pm$1.91 & & \\ 
\hline
\end{tabular}
\vspace{-0.2cm}
\tablefoot{The uncertainties in the spectroscopic redshifts, $z_{\rm spec}$, given in brackets, correspond to the errors derived from the simultaneous Gaussian fits to the line profiles. Emission lines fitted with double Gaussians are highlighted with an asterisk after the value of $\Delta V$. The redshifts and line identifications that are not robust are in italics (for HerS-18 W and HerS-19 SE \& W). Further details about the line fitting procedure and the derivation of $z_{\rm spec}$ are provided in Appendix~\ref{Appendix: Presentation of Catalogue}.
}
   \label{Appendix_table:emission-lines1}
\end{table*}

\begin{table*}[!htbp]
\tiny
\caption{Spectroscopic redshifts and emission line properties: HerBS sources.}
\begin{tabular}{lcrcccccccc}
\hline
\hline
Source & \multicolumn{2}{c}{Observed Position} & $z_{\rm spec}$ &  \multicolumn{7}{c}{Emission Lines} \\
 & RA & Dec. & & $\Delta V$ & Line & Line Flux &  Line & Line Flux  & Line & Line Flux \\ 
 & \multicolumn{2}{c}{(J2000)} & & $\rm (km \, s^{-1})$ & & $\rm (Jy \, km \, s^{-1})$ & & $\rm (Jy \, km \, s^{-1})$ & & $\rm (Jy \, km \, s^{-1})$ \\ 
\hline
HerBS-38 SE     & 14:46:09.08 & 02:19:19.4 & 2.4775(1) & 674$\pm$19$^*$ & CO(3-2) & 2.19$\pm$0.22 &                               &               & [CI](1-0) & 1.60$\pm$0.19 \\
HerBS-38 W      &  14:46:08.33 & 02:19:29.8 & 2.4158(1) & 873$\pm$67$^*$ & CO(3-2) & 1.66$\pm$0.28       &  CO(4-3) & $<$1.16 & & \\
HerBS-38 NE     &  14:46:09.10 & 02:19:34.1 & 6.5678(1) & 998$\pm$107   &                               &                           & CO(7-6) &  0.74$\pm$0.24 &         H$_2$O(2$_{11}$-2$_{02}$) & 0.82$\pm$0.22 \\
HerBS-46 & 14:45:56.30 & $-$00:48:51.8 & 1.8349(1) & 994$\pm$41$^*$ & CO (2-1) & 1.96$\pm$0.30  & CO(4-3) & 2.94$\pm$0.39 & &  \\ 
HerBS-48 & 12:13:01.53 & $-$00:49:23.2 & 3.1438(2) & 773$\pm$83  & CO(3-2)  & 3.59$\pm$0.69        & CO(5-4) & 5.16$\pm$0.74 &  & \\  
HerBS-50 & 12:03:19.14 & $-$01:12:54.6 & 2.9283(2) &  508$\pm$16$^*$ &  CO(3-2)  & 2.56$\pm$0.25 & CO(5-4) & 4.02$\pm$0.30 &  & \\ 
HerBS-51 &  12:07:09.12  & $-$02:47:01.8 & 2.1827(1) & 296$\pm$10 & CO(3-2) & 1.33$\pm$0.12 & CO(4-3)  & 2.12$\pm$0.10 & [CI](1-0)  & 1.06$\pm$0.27 \\
HerBS-53 E      & 11:51:11.94 & $-$01:26:38.7 & 1.4219(2) & 362$\pm$60 & CO(2-1) & 0.98$\pm$0.23 & CO(3-2) & 1.44$\pm$0.36 &  & \\ 
HerBS-53 W      & 11:51:12.28 & $-$01:26:37.3 & 1.4236(3) & 543$\pm$87 & CO(2-1) & 1.16$\pm$0.30 & CO(3-2) & 2.99$\pm$0.66 & & \\ 
HerBS-61 & 12:01:27.59 & $-$01:40:45.8 & 3.7293(1) & 722$\pm$40$^*$ & CO(3-2)  & 1.50$\pm$0.27  &  CO(4-3) & 2.07$\pm$0.34 & CO(6-5) & 2.95$\pm$0.42 \\ 
HerBS-62 &  12:15:42.81 & $-$00:52:20.8 & 2.5738(1) & 378$\pm$22 & CO(3-2)  & 5.77$\pm$0.60        & CO(4-3) & 7.95$\pm$0.64 &  & \\ 
HerBS-65 & 13:44:22.58  & 23:19:50.1 & 2.6858(5) & 417$\pm$68  & CO(3-2)  & 1.60$\pm$0.48        & CO(5-4) & 4.84$\pm$1.08 &  & \\ 
HerBS-72 &  14:45:12.16 & $-$00:15:10.5 & 3.6380(1) & 541$\pm$124$^*$ & CO(4-3)  & 2.12$\pm$0.22        & CO(6-5) & 1.69$\pm$0.19 &     [CI](1-0) & 0.72$\pm$0.52 \\ 
HerBS-74 & 12:06:00.46 & 00:35:01.1 & 2.5596(3) & 761$\pm$47$^*$ & CO(3-2) & 1.79$\pm$0.38 & CO(4-3) & 1.01$\pm$0.22 & & \\ 
HerBS-76 E & 13:35:34.09 & 34:18:34.8 & 2.3302(1) & 200$\pm$15  & CO(3-2) & 3.01$\pm$0.30 & CO(4-3) & 4.01$\pm$0.45 &  &  \\ 
HerBS-76 W & 13:35:33.83 & 34:18:33.4 & 2.3319(2) & 742$\pm$149 & CO(3-2) & 0.75$\pm$0.12 & CO(4-3) & 1.26$\pm$0.38 &  &  \\ 
HerBS-78  & 14:33:52.56 & 02:04:17.2 & 3.7344(1) & 489$\pm$30  & CO(3-2) & 1.09$\pm$0.17 & CO(4-3) & 2.04$\pm$0.20  & CO(6-5) & 2.05$\pm$0.23 \\ 
HerBS-82  & 12:11:44.88  & 01:06:38.1 & {\it 2.0583(1)}  &      181$\pm$15$^*$  &       {\it CO(4-3)} & 4.07$\pm$0.38               &  & &   & \\
HerBS-83 & 12:18:13.06 & 01:18:43.2 & 3.9438(1) & 742$\pm$114 & CO(4-3) & 1.23$\pm$0.34 & CO(6-5) & 1.62$\pm$0.40 & H$_2$O(2$_{11}$-2$_{02}$) & 0.93$\pm$0.35 \\
HerBS-85 & 11:47:52.85 & $-$00:58:32.0 & 2.8169(1) & 411$\pm$71 & CO(3-2) & 2.25$\pm$0.38 & CO(5-4) & 2.29$\pm$0.47 & [CI](1-0)  &  1.86$\pm$0.55 \\
HerBS-91 E & 09:21:35.82 & 00:01:30.9 & 2.4048(1) & 422$\pm$41 & CO(3-2) & 1.73$\pm$0.27 & CO(4-3) & 1.97$\pm$0.27 & & \\
HerBS-91 C & 09:21:35.55 & 00:01:30.4 & 2.4047(1) & 573$\pm$62 & CO(3-2) & 0.96$\pm$0.19 & CO(4-3) & 1.32$\pm$0.24 &     [CI](1-0) &  1.05$\pm$0.27 \\
HerBS-91 E\&C & 09:21:35.72 & 00:01:30.6 & 2.4048(1) & 573$\pm$62 &  & & & & [CI](1-0) &  1.05$\pm$0.27 \\
HerBS-92 E\&W  & 13:38:08.80 & 25:51:52.9 & 3.2644(2) & 534$\pm$102 & CO(3-2) & 3.06$\pm$0.86 & CO(5-4) & 3.44$\pm$0.86 &  & \\
HerBS-105  & 08:39:31.97 & $-$01:18:00.0 & 2.6684(2) & 520$\pm$29$^*$ & CO(3-2) & 1.28$\pm$0.21 & CO(5-4) & 1.76$\pm$0.34 &  &  \\
HerBS-108  & 08:38:17.43 & $-$00:41:34.4 & 3.7168(1) & 541$\pm$53$^*$  & CO(3-2) &  0.29$\pm$0.10 & CO(4-3) & 2.82$\pm$0.33      & CO(6-5) & 3.28$\pm$0.39   \\
HerBS-109 NW & 13:29:00.35 & 28:19:18.5 & 1.5850(1) & 662$\pm$84 & CO(2-1) & 1.64$\pm$0.29 & CO(3-2) & 2.88$\pm$0.57 &  &   \\
HerBS-109 S & 13:29:00.31 & 28:19:07.5 & 1.5843(3) & 569$\pm$81 & CO(2-1) & 0.93$\pm$0.20 & CO(3-2) & 2.10$\pm$0.42 &  &   \\
HerBS-109 NE & 13:29:00.77 & 28:19:16.3 & 2.8385(2) & 330$\pm$62 & CO(3-2) & 0.70$\pm$0.18 & CO(5-4) & 1.21$\pm$0.37 &  &  \\
HerBS-110  & 14:18:33.24 & 01:02:10.9 & 2.6810(2) & 915$\pm$25$^*$ & CO(3-2) & 2.28$\pm$0.24 & CO(5-4) & 2.95$\pm$0.26 &  &  \\
HerBS-115  & 13:35:38.20 & 26:57:40.1 & 2.3706(1) & 1020$\pm$72  & CO(3-2) & 2.45$\pm$0.29   & CO(4-3) &   3.27$\pm$0.37 & [CI](1-0) & 1.19$\pm$0.36 \\
HerBS-116 E\&W & 12:13:48.05 & 01:08:11.0 & 3.1547(1) & 800$\pm$107 & CO(3-2) & 3.02$\pm$0.57 & CO(5-4) & 4.14$\pm$0.87 &  &    \\
HerBS-124 W & 12:21:58.48 & 00:33:24.9 & 2.2772(1) & 449$\pm$35 & CO(3-2) & 2.26$\pm$0.25 & CO(4-3) & 3.62$\pm$0.45 &  &  \\
HerBS-124 E & 12:21:58.65 & 00:33:25.0 & 2.2781(1) & 630$\pm$175  & CO(3-2) & 0.55$\pm$0.22 & CO(4-3) & 0.85$\pm$0.35 &  &    \\
HerBS-125  & 13:04:32.18 & 29:53:39.6 & 2.5739(1) & 615$\pm$55 & CO(3-2) & 2.29$\pm$0.30 & CO(4-3) & 3.79$\pm$0.50 &  &  \\
HerBS-126  & 14:51:35.27 & $-$01:14:17.3 & 2.5875(3) & 1083$\pm$128 & CO(3-2) & 2.05$\pm$0.33 & CO(4-3) & 0.80$\pm$0.16 & &  \\
HerBS-127  & 13:21:28.76 & 28:20:23.2 & 3.1958(1) & 839$\pm$98$^*$ & CO(3-2) & 2.42$\pm$0.89 & CO(5-4) & 2.53$\pm$1.01 &  &  \\
HerBS-128  & 13:04:14.48 & 30:35:38.2 & 2.0681(2) & 643$\pm$27$^*$ & CO(3-2) & 2.54$\pm$0.32 & CO(4-3) & 2.26$\pm$0.28 &  &   \\
HerBS-129  & 13:00:53.80 & 26:03:00.3 & 3.3074(2) &  645$\pm$19$^*$ & CO(3-2) & 3.13$\pm$0.33 & CO(5-4) & 2.82$\pm$0.28 &  &   \\
HerBS-134  & 13:34:40.42 & 35:31:39.1 & 3.1725(3) & 793$\pm$173 & CO(3-2) & 2.46$\pm$0.83 & CO(5-4) & 2.40$\pm$0.74 &  &  \\
HerBS-136  & 08:53:08.45 & $-$00:57:28.8 & 3.2884(5) & 1042$\pm$88 & CO(3-2) & 1.02$\pm$0.16 & CO(5-4) & 1.77$\pm$0.20 &  &   \\
HerBS-137  & 14:53:37.15 & 00:04:10.2 & 3.0408(3) & 450$\pm$86 & CO(3-2) & 1.45$\pm$0.38 & CO(5-4) & 1.29$\pm$0.43 &  &  \\      
HerBS-140  & 14:21:40.42 & 00:04:46.3 & 2.7799(2) & 770$\pm$26$^*$ & CO(3-2) & 0.69$\pm$0.09 & CO(5-4) & 1.53$\pm$0.16 &  &  \\
HerBS-143  & 14:18:10.10 & $-$00:37:45.5 & 2.2406(1) & 540$\pm$43  & CO(3-2) & 1.43$\pm$0.20 & CO(4-3) & 1.94$\pm$0.27 & [CI](1-0) & 1.00$\pm$0.16  \\
HerBS-147  & 14:34:03.66 & 00:02:30.1 & 3.1150(1) & 1789$\pm$300 & CO(3-2) & 2.00$\pm$0.60 & CO(5-4) & 2.98$\pm$0.68 &  & \\
HerBS-149  & 13:38:27.61 & 31:39:55.6 & 2.6650(1) & 243$\pm$21  & CO(3-2) & 2.29$\pm$0.30 & CO(5-4) & 5.45$\pm$0.65 &  & \\
HerBS-150E & 12:24:59.27 & $-$00:56:51.9 & 3.6732(1) & 780$\pm$128  & CO(4-3) & 1.92$\pm$0.51 & CO(6-5) & 1.03$\pm$0.23 & &  \\
HerBS-150C & 12:24:58.98 & $-$00:56:47.5 & 3.6682(1) & 212$\pm$64 & CO(4-3) &
0.26$\pm$0.10   &         &               & &  \\
HerBS-150W & 12:24:58.70  & $-$00:56:48.7 & 3.6787(1) & 1036$\pm$178  & CO(4-3) & 1.47$\pm$0.32   &         &                     &     & \\
HerBS-153  & 14:42:43.41 & 01:55:04.3 & 3.1501(2)  & 552$\pm$16$^*$   & CO(3-2) & 0.89$\pm$0.19 & CO(5-4) & 3.72$\pm$0.33 &  &  \\
HerBS-157  & 08:49:57.77 & 01:07:10.8 & 1.8964(1)  & 849$\pm$45$^*$ & CO(2-1) & 1.53$\pm$0.33 & CO(4-3) & 1.39$\pm$0.22 &  & \\
HerBS-162 NE    & 14:43:34.73 & $-$00:30:30.5 & 2.4742(4) &  483$\pm$79$^*$  & CO(3-2) & 1.18$\pm$0.27 & CO(4-3) & 0.55$\pm$0.14 &  &  \\
HerBS-162 SW    & 14:43:34.35 & $-$00:30:35.7 & 2.4735(2) & 288$\pm$41$^*$  & CO(3-2) & 1.44$\pm$0.25 & CO(4-3) & 1.27$\pm$0.25 &  &    \\
HerBS-164  & 12:14:16.30 & $-$01:37:03.7 & 2.0126(1) & 644$\pm$97 & CO(2-1) & 1.04$\pm$0.27 & CO(4-3) & 3.25$\pm$0.27 &  &  \\
HerBS-165  & 09:06:13.91 & $-$01:00:41.1 & 2.2251(1) & 445$\pm$67 & CO(3-2) & 0.97$\pm$0.25 & CO(4-3) &  1.49$\pm$0.30 &  &  \\
HerBS-167  & 13:03:41.80 & 31:37:57.9 & 2.2144(7)       & 733$\pm$129$^*$        & CO(3-2) & 1.05$\pm$0.27 & CO(4-3) & 0.97$\pm$0.25 &  &   \\
HerBS-169  & 08:38:59.47 & 02:13:27.4 & 2.6977(1)       & 617$\pm$48     & CO(3-2) & 1.73$\pm$0.18 &                &                     & [CI](1-0) & 0.89$\pm$0.15 \\
HerBS-171  & 08:39:45.21 & 02:10:18.1 & 2.4793(1)       & 382$\pm$56     & CO(3-2) & 0.95$\pm$0.20 & CO(4-3) & 1.17$\pm$0.26 &  &  \\
HerBS-172  &  14:50:40.47 & 00:33:35.8 & 2.9246(4)      & 386$\pm$41     & CO(3-2) & 1.34$\pm$0.22 & CO(5-4) & 0.74$\pm$0.11 &  &  \\
HerBS-175  & 12:19:00.83 & 00:33:26.9 & 3.1575(2)       & 660$\pm$77  & CO(3-2) & 3.48$\pm$0.55 & CO(5-4) & 0.43$\pm$0.09 &  & \\
HerBS-176  & 13:12:22.15 & 27:02:17.8 & 2.9805(4)       & 733$\pm$76     & CO(3-2) & 3.63$\pm$0.53 & CO(5-4) & 5.22$\pm$0.84 &  & \\
HerBS-177  &  11:54:33.72 & 00:50:41.8 & 3.9625(1)      & 606$\pm$53   & CO(4-3) & 3.36$\pm$0.49 & CO(6-5)   & 4.97$\pm$0.75 & H$_2$O(2$_{11}$-2$_{02}$) & 2.28$\pm$0.51 \\
HerBS-179  &  11:55:20.97 & $-$02:13:29.4 & 3.9423(1)   & 391$\pm$39     & CO(4-3) & 2.41$\pm$0.45 & CO(6-5) & 3.71$\pm$0.56 & H$_2$O(2$_{11}$-2$_{02}$) & 0.52$\pm$0.31 \\
\hline
\end{tabular}
\end{table*}

 \addtocounter{table}{-1}

\begin{table*}[!htbp]
\tiny
\caption{{\bf continued} - Spectroscopic redshifts and emission line properties: HerBS sources.}
\begin{tabular}{lcrcccccccc}
\hline
\hline
Source & \multicolumn{2}{c}{Observed Position} & $z_{\rm spec}$ &  \multicolumn{7}{c}{Emission Lines} \\
 & RA & Dec. & & $\Delta V$ & Line & Line Flux &  Line & Line Flux  & Line & Line Flux \\ 
 & \multicolumn{2}{c}{(J2000)} & & $\rm (km \, s^{-1})$ & & $\rm (Jy \, km \, s^{-1})$ & & $\rm (Jy \, km \, s^{-1})$ & & $\rm (Jy \, km \, s^{-1})$ \\ 
\hline
HerBS-180 S     & 13:15:39.55 & 29:22:21.6 & 1.4527(1) & 308$\pm$34     & CO(2-1) & 1.77$\pm$0.26 & CO(3-2) & 1.75$\pm$0.38 &  &  \\
HerBS-183       & 09:04:53.06 & 02:20:16.9 & 1.8910(1) & 687$\pm$21$^*$ & CO(2-1) & 0.89$\pm$0.09 & CO(4-3) & 3.21$\pm$0.32 &  &  \\
HerBS-185       & 09:24:08.92 & $-$00:50:18.1 & 4.3238(1)       & 310$\pm$24  & CO(7-6) & 1.95$\pm$0.40      & H$_2$O(2$_{11}$-2$_{02}$) & 0.85$\pm$0.30 &  & \\ 
--          &             &               &             &             &         &            & [CI](1-0) & 0.51$\pm$0.34 & [CI](2-1) & 1.42$\pm$0.30 \\
HerBS-187 E & 08:37:05.58 & 02:00:31.5 & 1.8285(1) & 591$\pm$76   & CO(2-1)   & 2.09$\pm$0.40 & CO(4-3) & 3.40$\pm$0.63 &  &  \\
HerBS-187 W & 08:37:05.35 & 02:00:32.7 & 1.8274(1) & 403$\pm$78   & CO(2-1)   & 0.60$\pm$0.20 & CO(4-3) & 1.82$\pm$0.47 &  &  \\
HerBS-188 & 08:42:59.98 & 02:50:01.7 & 2.7675(1)  & 1069$\pm$147$^*$ & CO(3-2)   & 1.94$\pm$0.38 & CO(5-4) & 2.54$\pm$0.53 &  & \\
HerBS-190 & 09:04:05.28 & $-$00:33:33.5 & 2.5890(2)  & 784$\pm$77  & CO(3-2)   & 3.17$\pm$0.42 & CO(4-3) & 3.92$\pm$0.67 &  &  \\
HerBS-191 & 12:47:53.28 & 32:24:45.8 & 3.4428(3)  & 761$\pm$36$^*$  & CO(4-3)   & 3.23$\pm$0.55 & CO(5-4) & 2.52$\pm$0.39 &  &  \\
HerBS-193 & 08:53:51.96 & $-$00:08:05.4 & 3.6951(1)  &  680$\pm$30 & CO(3-2)   &  0.71$\pm$0.14 & CO(4-3)   & 1.29$\pm$0.10 & CO(6-5) & 1.54$\pm$0.10    \\
HerBS-194 N & 08:55:21.30 & $-$00:35:56.6 & 2.3335(1) & 467$\pm$84  &  CO(3-2)   & 0.72$\pm$0.21 & CO(4-3) & 1.86$\pm$0.47 &  &  \\
HerBS-194 S & 08:55:21.04 & $-$00:36:11.9 & 2.3316(1) & 538$\pm$140  &  CO(3-2)   & 0.38$\pm$0.14 & CO(4-3) & 0.34$\pm$0.13 &  &  \\
HerBS-197 & 12:20:34.05 & $-$00:38:04.4 & 2.4170(1)  & 195$\pm$10    & CO(3-2)   & 2.59$\pm$0.35 & CO(4-3) & 3.75$\pm$0.28     &                               [CI](1-0) & 1.11$\pm$0.16 \\
HerBS-199 E & 13:33:52.30 & 33:49:13.5 & 1.9248(2) & 438$\pm$66 & CO(2-1) & 1.98$\pm$0.58 & CO(4-3) & 3.05$\pm$0.61 &  &   \\
HerBS-199 W & 13:33:51.45 & 33:49:18.9 & 1.9197(1) & 420$\pm$97 & CO(2-1) & 0.50$\pm$0.45 & CO(4-3) & 2.75$\pm$0.83 &  &  \\
HerBS-201  &  14:11:18.06 & $-$01:06:52.6 & 4.1408(1) & 553$\pm$69 & CO(4-3) & 1.17$\pm$0.25 & CO(7-6) & 1.15$\pm$0.22 &  &  \\
--          &             &               &          &             &        &               & [CI](1-0) & 0.71$\pm$0.21 & [CI](2-1) & 0.50$\pm$0.10 \\
HerBS-202  & 14:33:28.37 & 02:08:09.8 & 2.0224(1) & 411$\pm$21   & CO(2-1) & 1.51$\pm$0.31 & CO(4-3) & 2.66$\pm$0.18 &  &  \\
HerBS-204 W & 13:29:09.21 & 30:09:59.3 & 3.4937(4) & 378$\pm$34  & CO(4-3) & 1.32$\pm$0.16 & CO(5-4) & 2.07$\pm$0.31 &  &   \\
HerBS-204 E & 13:29:09.71 & 30:09:57.1 & 3.4933(2) & 365$\pm$44  & CO(4-3) & 1.01$\pm$0.18 & CO(5-4) & 1.27$\pm$0.23 &  &  \\
HerBS-205 NE &  14:51:32.92 & 02:41:02.8 & 2.9600(1) & 575$\pm$58        & CO(3-2) & 1.26$\pm$0.21 & CO(5-4) & 0.92$\pm$0.13 &  &  \\    
HerBS-205 SE &  14:51:32.91 & 02:40:58.9 & 2.9599(4) & 450$\pm$53        & CO(3-2) & 0.66$\pm$0.13 & CO(5-4) & 0.50$\pm$0.08 &  &  \\
HerBS-205 W & 14:51:32.38 & 02:41:00.5 & 2.9630(1) & 566$\pm$117   & CO(3-2) & 0.27$\pm$0.12 & CO(5-4) & 0.57$\pm$0.16 &  & \\
HerBS-206 &  14:04:21.76 & $-$00:12:17.1 & 2.8122(4)  &  410$\pm$41      & CO(3-2) & 0.54$\pm$0.08 & CO(5-4) & 0.92$\pm$0.13 &  & \\
\hline
\end{tabular}
 \tablefoot{The uncertainties in the spectroscopic redshifts, $z_{\rm spec}$, given in brackets, correspond to the errors derived from the simultaneous Gaussian fits to the line profiles.  Emission lines fitted with double Gaussians are highlighted with an asterisk after the value of $\Delta V$. The redshift and line identifications that are not robust are in italics (in the case of HerBS-82). Further details about the line fitting procedure and the derivation of $z_{\rm spec}$ are provided in Appendix~\ref{Appendix: Presentation of Catalogue}.}
\label{Appendix_table:emission-lines2}
\end{table*}

\begin{table*}[!htbp]
\vspace{3cm}
\tiny
\caption{Spectroscopic redshifts and emission line properties: HerBS sources from \cite{Neri2020}.}
\begin{tabular}{lcrcccccccc}
\hline
\hline
Source & \multicolumn{2}{c}{Observed Position} & $z_{\rm spec}$ &  \multicolumn{7}{c}{Emission Lines} \\
 & RA & Dec. & & $\Delta V$ & Line & Line Flux &  Line & Line Flux  & Line & Line Flux \\ 
 & \multicolumn{2}{c}{(J2000)} & & $\rm (km \, s^{-1})$ & & $\rm (Jy \, km \, s^{-1})$ & & $\rm (Jy \, km \, s^{-1})$ & & $\rm (Jy \, km \, s^{-1})$ \\ 
\hline
HerBS-34        & 13:34:13.87 & 26:04:57.5 & 2.6637(2) & 330$\pm$10$^*$ & CO(3-2) & 2.8$\pm$0.4 & CO(5-4) & 8.4$\pm$0.8 &  & \\
HerBS-43a   & 13:24:18.79 & 32:07:54.4 & 3.2121(1) & 1070$\pm$90$^*$ & CO(4-3) & 5.5$\pm$0.8 & CO(5-4) & 6.7$\pm$0.8 &  & \\
HerBS-43b   & 13:24:19.24 & 32:07:49.2 & 4.0543(7) & 800$\pm$50$^*$ & CO(4-3) & 1.7$\pm$0.3 & CO(5-4) & 1.5$\pm$0.3 & CO(6-5)  & 4.8$\pm$0.9 \\
HerBS-44    & 13:32:55.85 & 34:22:08.4 & 2.9268(2) & 520$\pm$50$^*$ & CO(3-2) & 4.9$\pm$0.5 & CO(5-4) & 12.5$\pm$1.2 & & \\
HerBS-54    & 13:15:40.72 & 26:23:19.6 & 2.4417(3) & 1020$\pm$190 & CO(3-2) & 4.3$\pm$0.4 & CO(4-3) & 8.5$\pm$0.8 & & \\
HerBS-58    & 13:03:33.17 & 24:46:42.3 & 2.0842(1) & 970$\pm$50$^*$ & CO(3-2) & 5.3$\pm$1.0 & CO(4-3) & 5.2$\pm$1.5 & [CI](1-0) & 4.7$\pm$0.5 \\
HerBS-70 E    & 13:01:40.33 & 29:29:16.2 & 2.3077(4) & 770$\pm$50$^*$ & CO(3-2) & 1.8$\pm$0.5 & CO(4-3) & 3.4$\pm$0.3 & [CI](1-0) & 3.5$\pm$0.7 \\
HerBS-70 W    & 13:01:39.31 & 29:29:25.2 & 2.3115(1) & 140$\pm$20 & CO(3-2) & 1.7$\pm$0.3 & CO(4-3) & 2.0$\pm$0.3 & \\
HerBS-79     & 13:14:34.08 & 33:52:20.1 & 2.0782(8) & 870$\pm$70$^*$ & CO(3-2) & 4.1$\pm$0.8 & CO(4-3) & 5.5$\pm$0.5 & & \\
HerBS-89a    & 13:16:11.52 & 28:12:17.7 & 2.9497(1) & 1102$\pm$83$^*$ & CO(3-2) & 4.00$\pm$0.60 & CO(5-4) & 8.40$\pm$0.80 & & \\
HerBS-95 E    & 13:43:42.73 & 26:39:18.0 & 2.9718(3) & 870$\pm$50$^*$ & CO(3-2) & 1.0$\pm$0.1 & CO(5-4) & 3.6$\pm$0.3 & & \\
HerBS-95 W    & 13:43:41.55 & 26:39:22.7 & 2.9729(2) & 540$\pm$30$^*$ & CO(3-2) & 2.4$\pm$0.4 & CO(5-4) & 3.5$\pm$0.3 & & \\
HerBS-113    & 13:12:11.35 & 32:38:37.8 & 2.7870(8) & 900$\pm$200$^*$ & CO(3-2) & 6.1$\pm$1.2 & CO(5-4) & 13.5$\pm$1.4 &  &  \\
HerBS-154    & 13:22:58.11 & 32:50:51.7 & 3.7070(5) & 310$\pm$40 & CO(6-5) & 7.6$\pm$0.7 & [CI](1-0) & 1.3$\pm$0.4 &  $\rm H_2O(2_{11}$-$2_{02})$ & 1.5$\pm$0.3 \\
        \hline
\end{tabular}
\tablefoot{The uncertainties in the spectroscopic redshifts, $z_{\rm spec}$, given in brackets, correspond to the errors derived from Gaussian fits to the line profiles. $\Delta V$ values are the mean linewidths (FWHM) weighted by the peak intensities of the detected CO emission lines. Emission lines fitted with double Gaussians are highlighted with an asterisk after the value of $\Delta V$ - the reader is referred to \citep{Neri2020} for further details.}
   \label{Appendix_table:emission-lines3}
\end{table*}

\begin{figure*}[!ht]
   \centering
\includegraphics[width=0.6\textwidth]{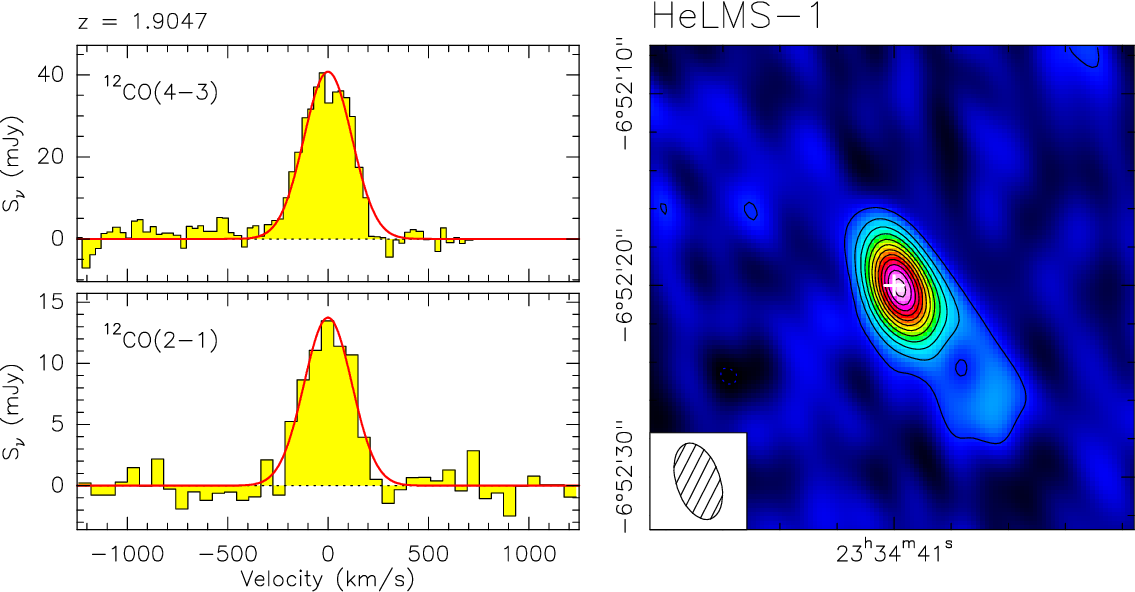} 
\includegraphics[width=0.6\textwidth]{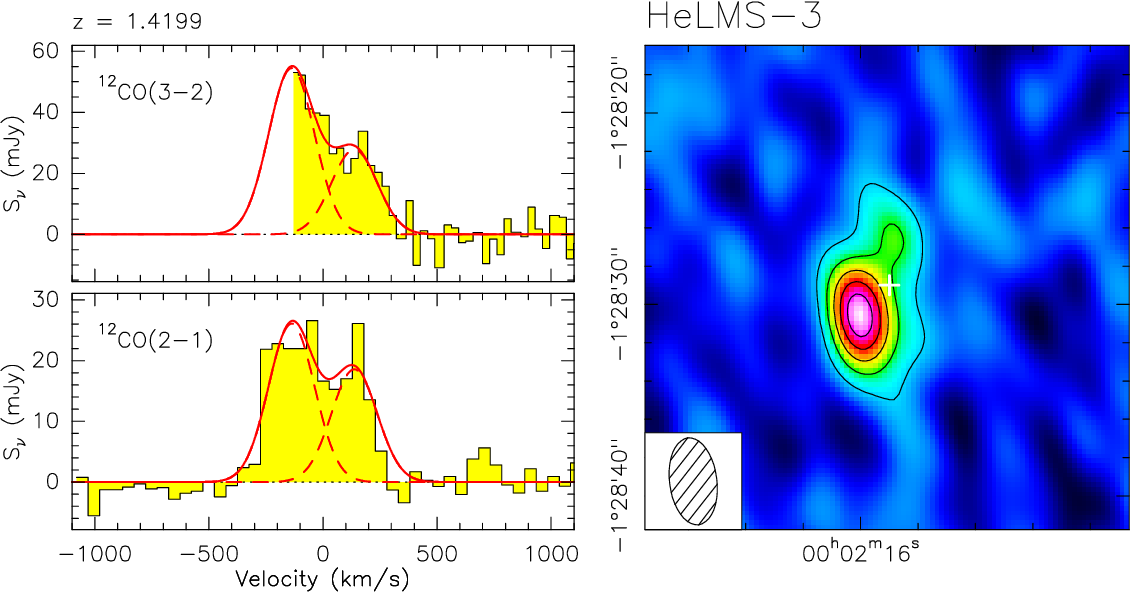}
\includegraphics[width=0.6\textwidth]{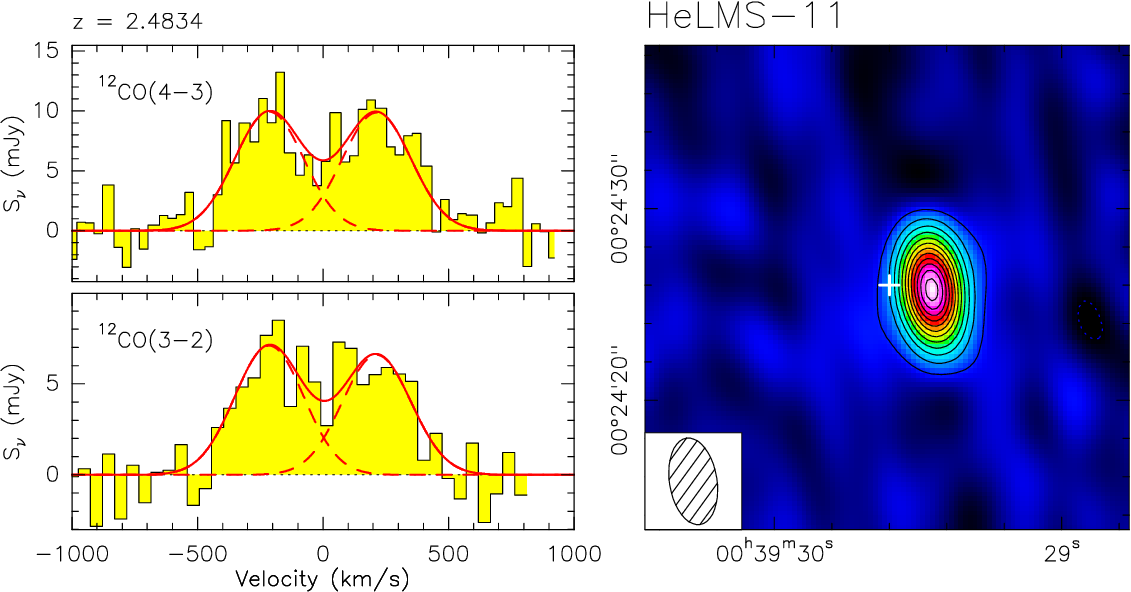}
\includegraphics[width=0.6\textwidth]{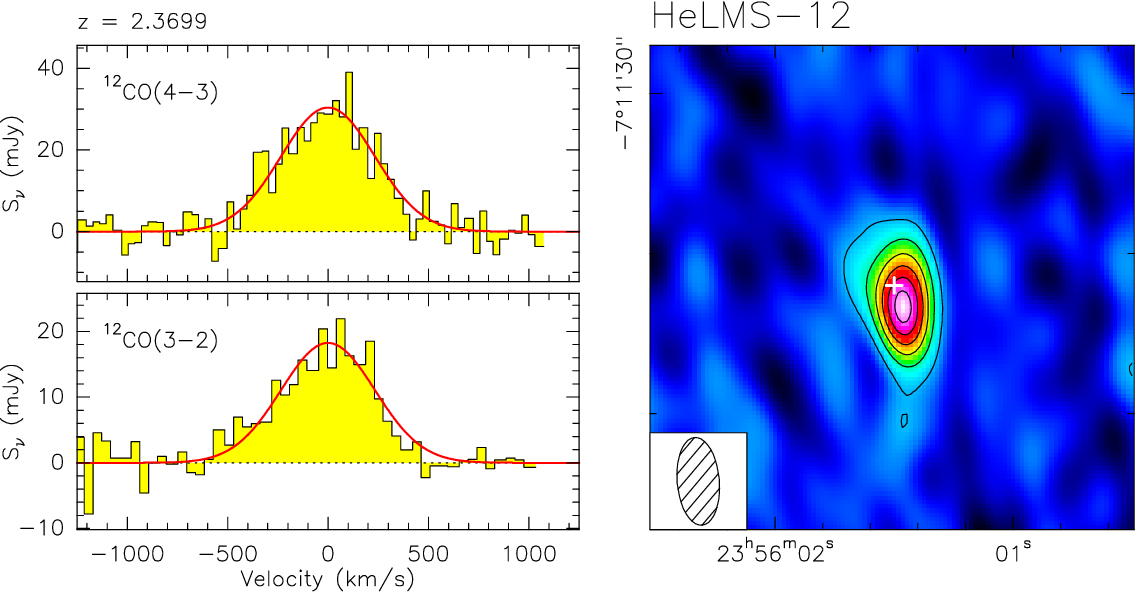}
   \caption{Combined continuum and emission lines maps of the $z$-GAL sources selected from the HeLMS sample and spectra of each of the detected emission lines. Further details are provided in Appendix~\ref{Appendix: Presentation of Catalogue}.}
   \label{figure:spectra_continuum_HeLMS}%
    \end{figure*}

    \addtocounter{figure}{-1}

\begin{figure*}[!ht]
   \centering
\includegraphics[width=0.6\textwidth]{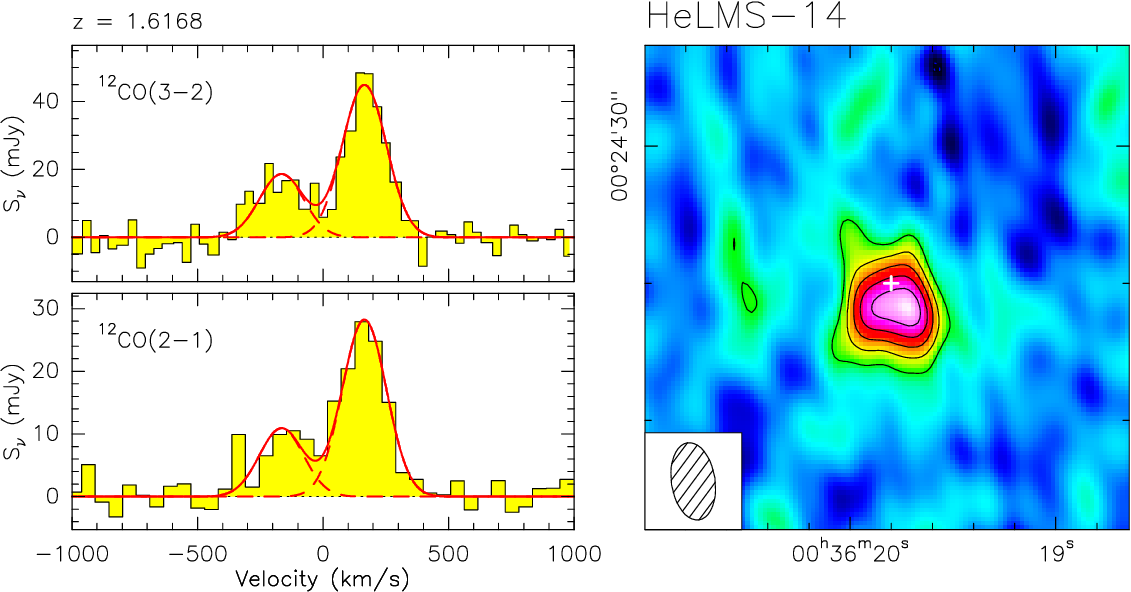}

\vspace{0.8cm}
\includegraphics[width=0.6\textwidth]{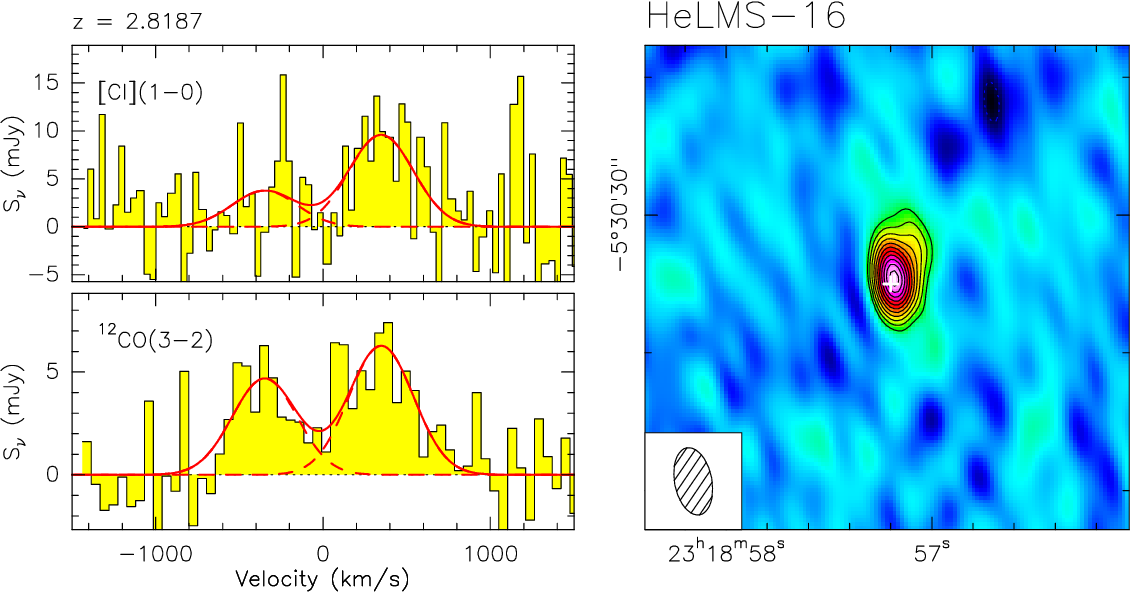}

\vspace{0.8cm}
 \includegraphics[width=0.9\textwidth]{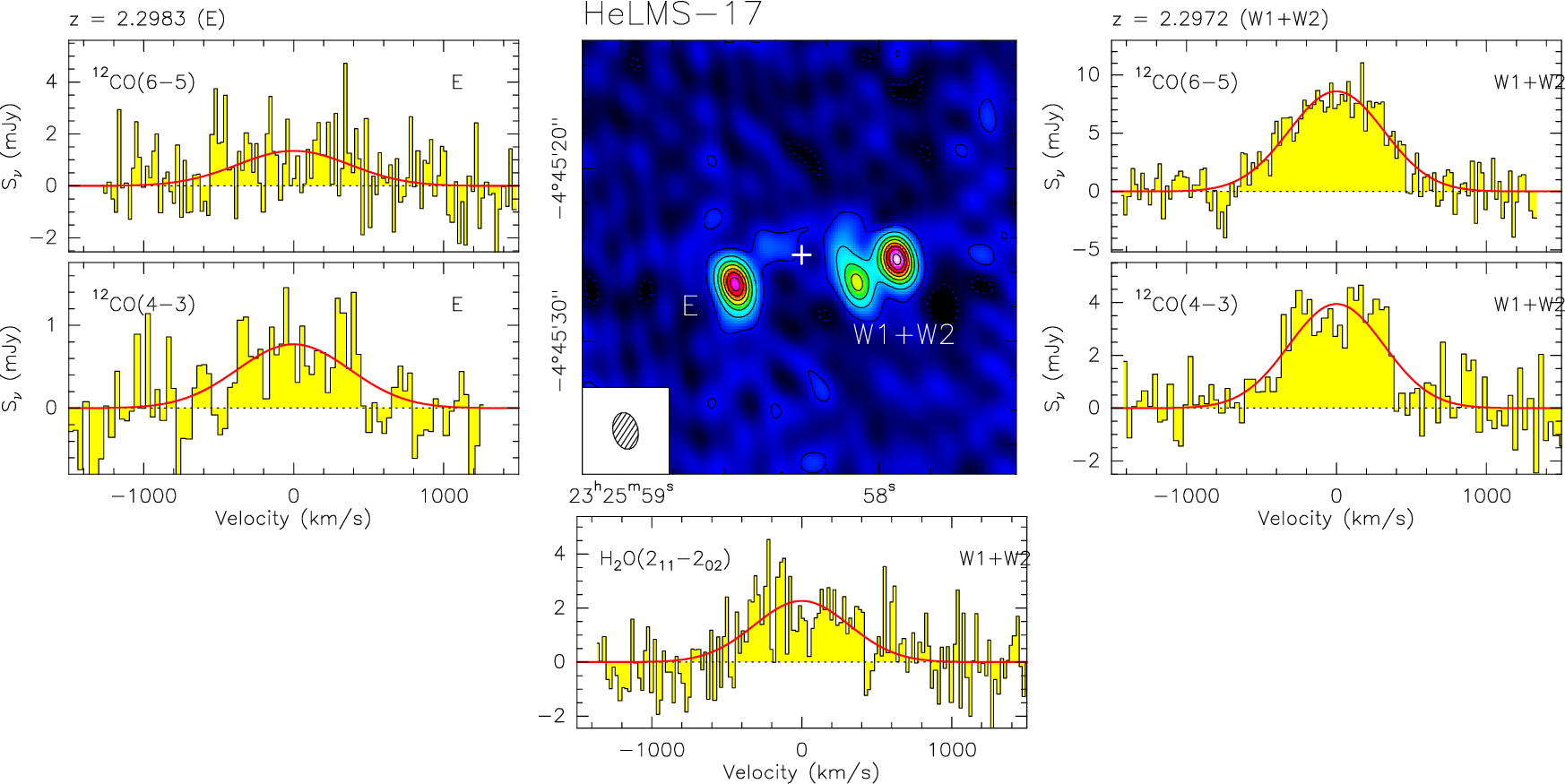}
   \caption{{\bf continued}}
    \end{figure*}

    \addtocounter{figure}{-1}

\begin{figure*}[!ht]
   \centering
 \includegraphics[width=0.9\textwidth]{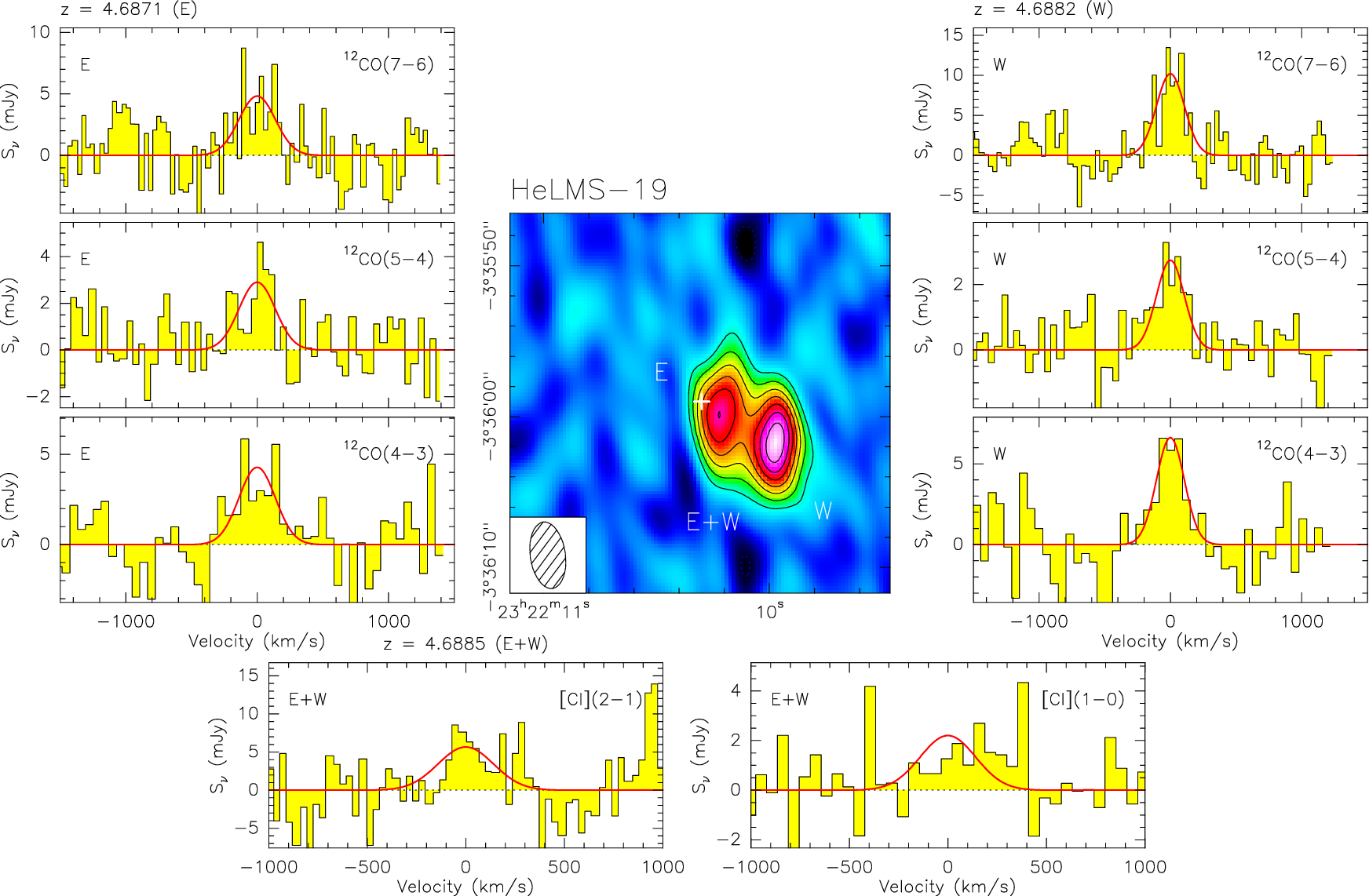}

\vspace{0.8cm}
\includegraphics[width=0.6\textwidth]{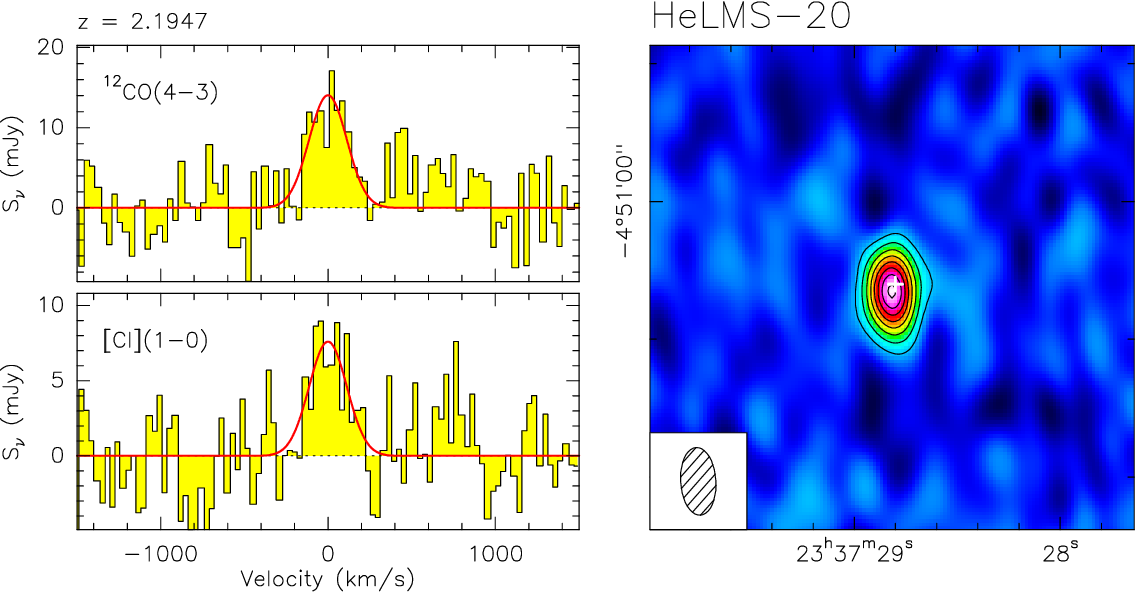}

\vspace{0.8cm}
\includegraphics[width=0.6\textwidth]{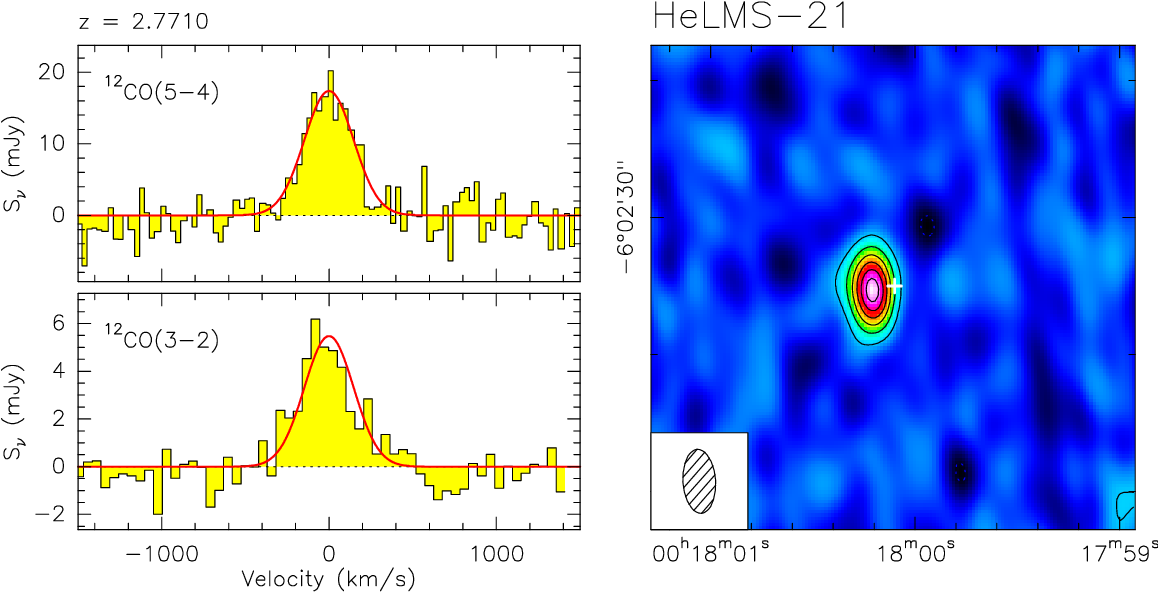}
   \caption{{\bf continued}}
    \end{figure*}

    \addtocounter{figure}{-1}

\begin{figure*}[!ht]
   \centering
\includegraphics[width=0.6\textwidth]{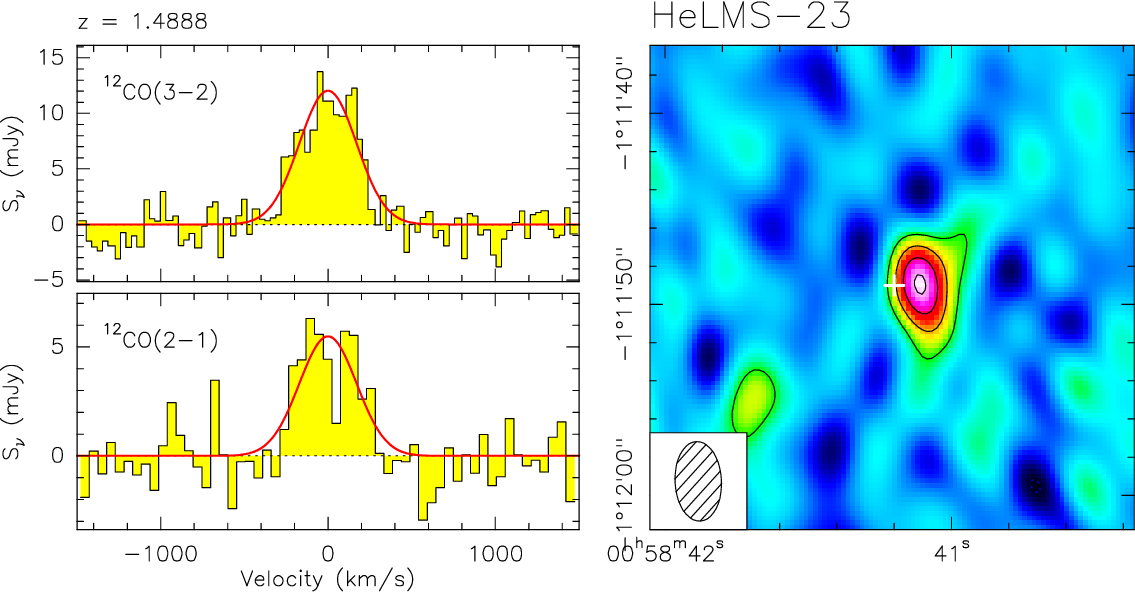}

\vspace{1.5cm}
\includegraphics[width=0.9\textwidth]{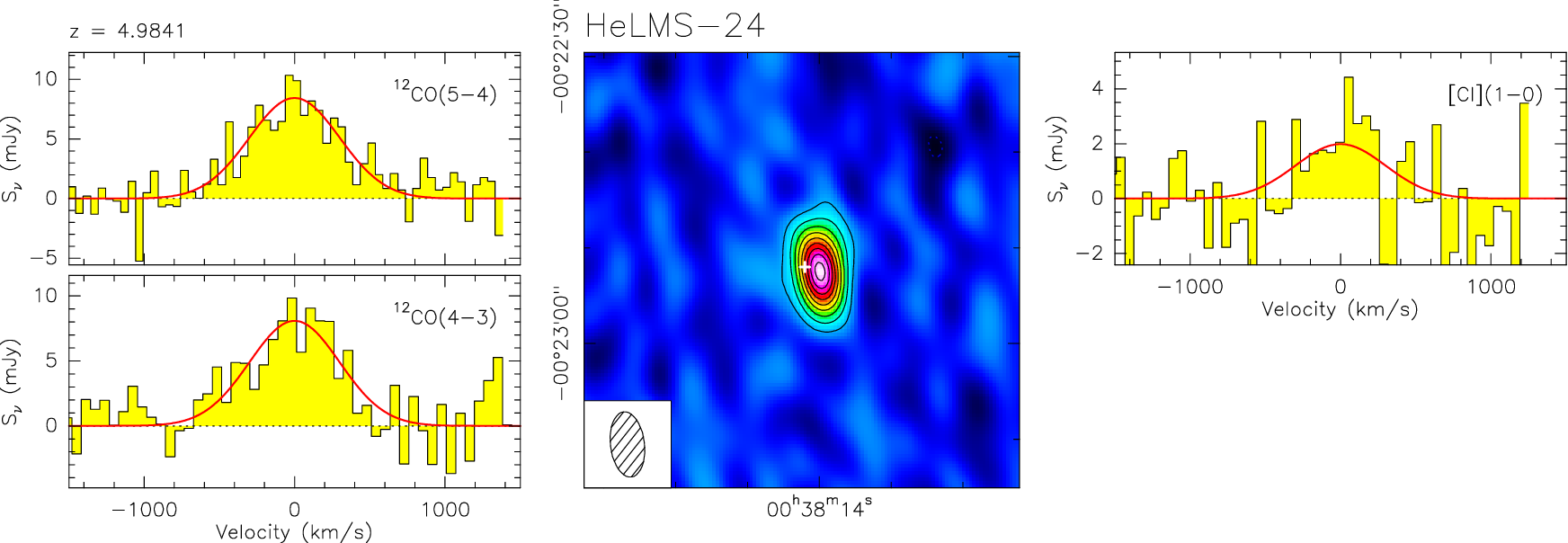}

\vspace{1.5cm}
\includegraphics[width=0.6\textwidth]{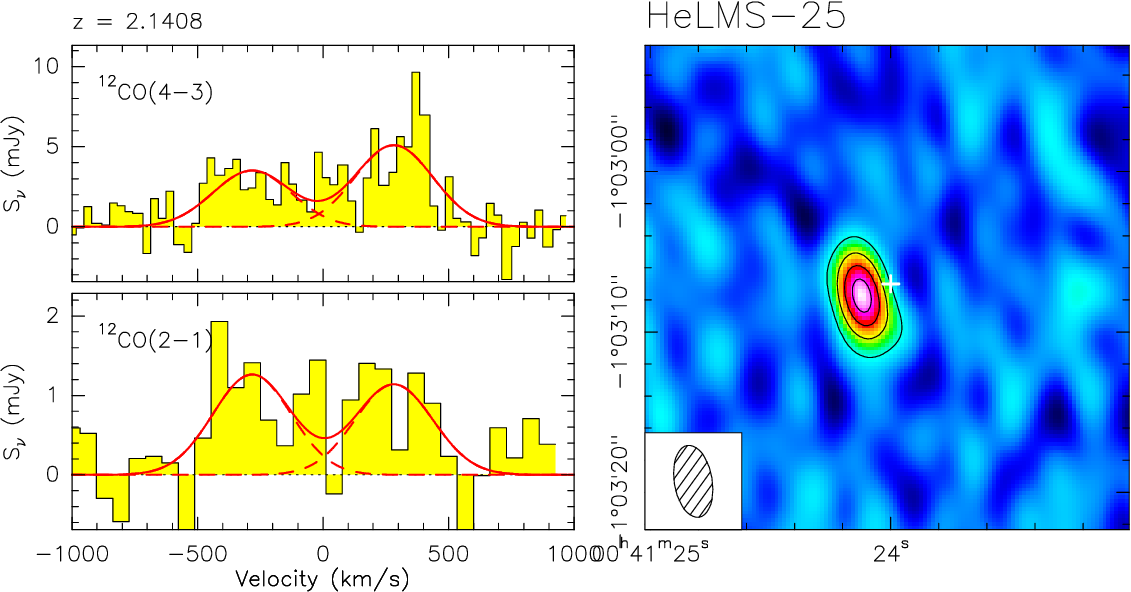}
   \caption{{\bf continued}}
    \end{figure*}
    \addtocounter{figure}{-1}

\begin{figure*}[!ht]
   \centering
\includegraphics[width=0.9\textwidth]{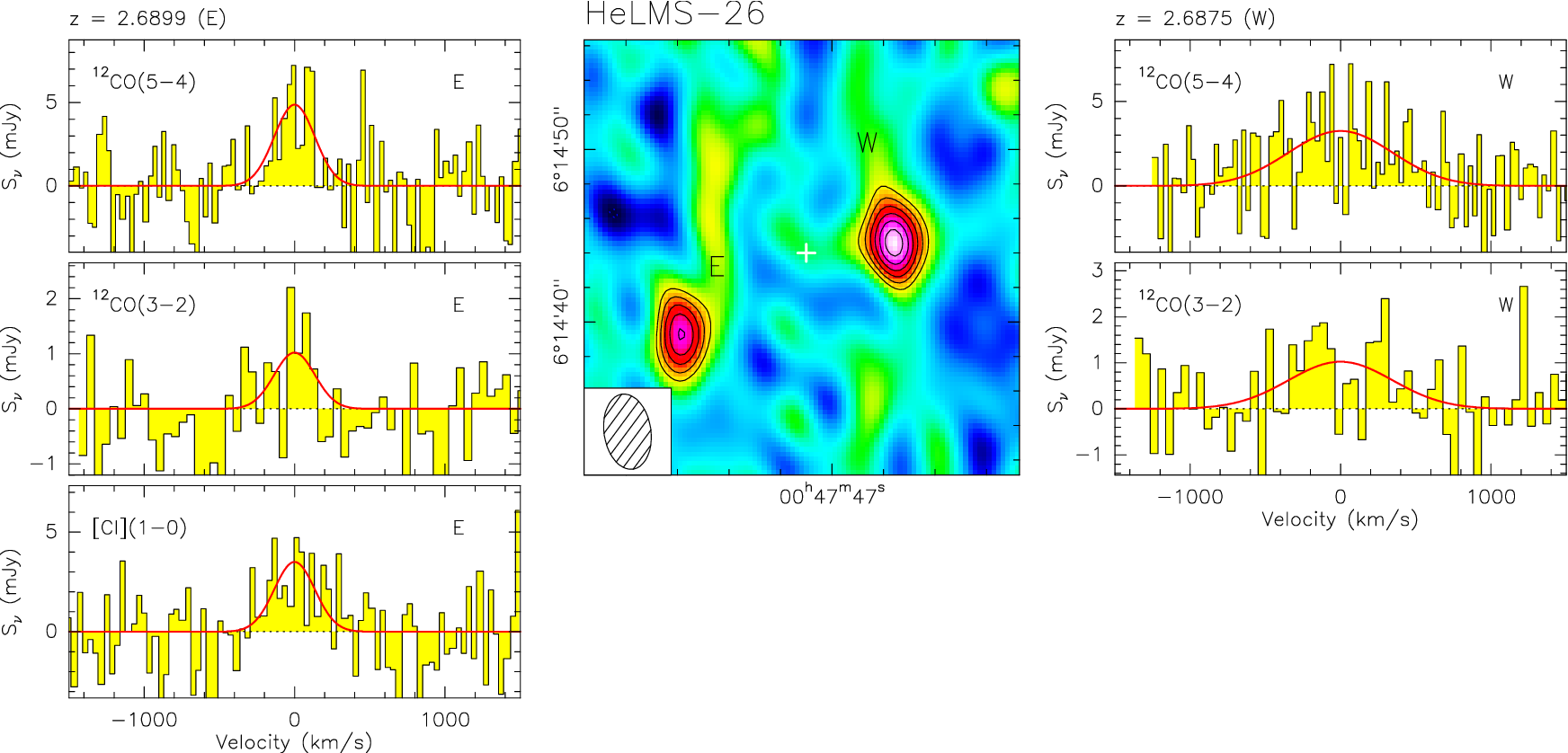}

\vspace{0.8cm}
\includegraphics[width=0.6\textwidth]{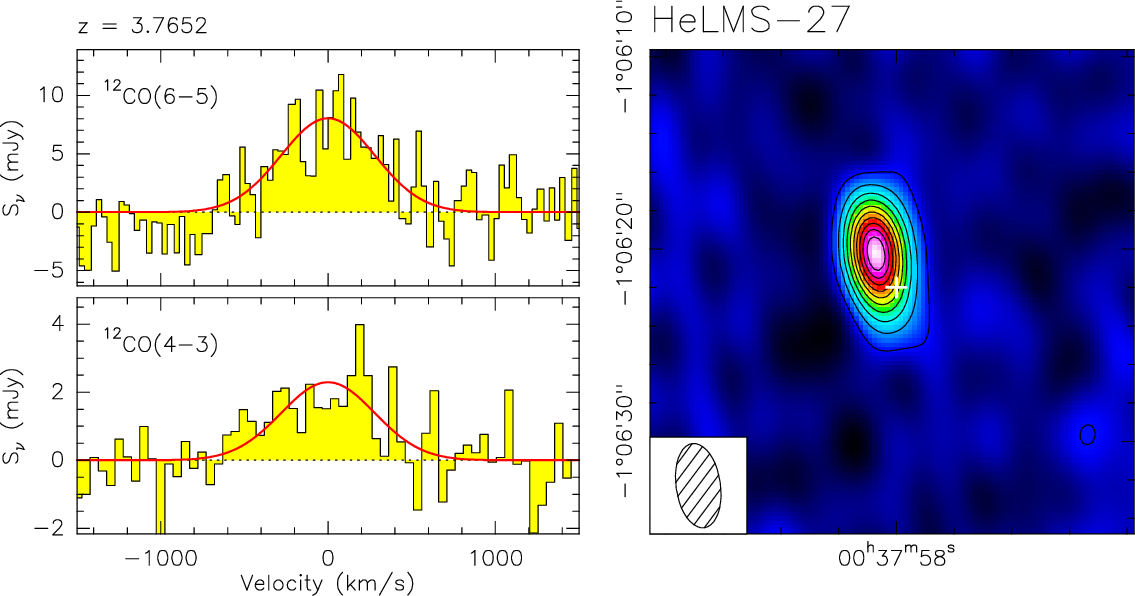}

\vspace{1.2cm}
\includegraphics[width=0.6\textwidth]{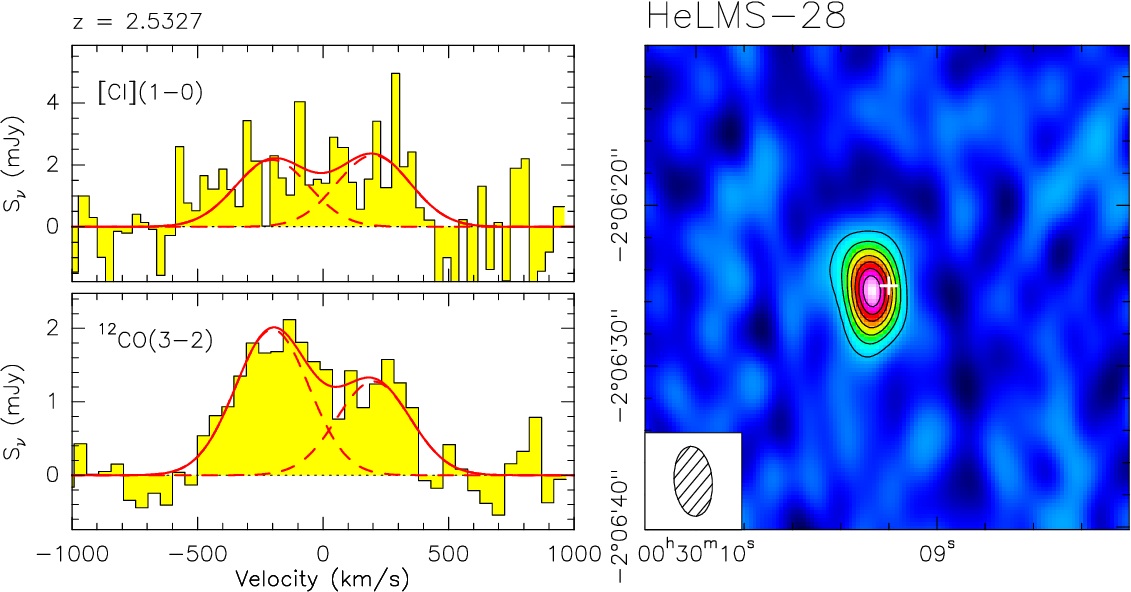}
\caption{{\bf continued}}
    \end{figure*}
 \addtocounter{figure}{-1}
 
\begin{figure*}[!ht]
   \centering
\includegraphics[width=0.6\textwidth]{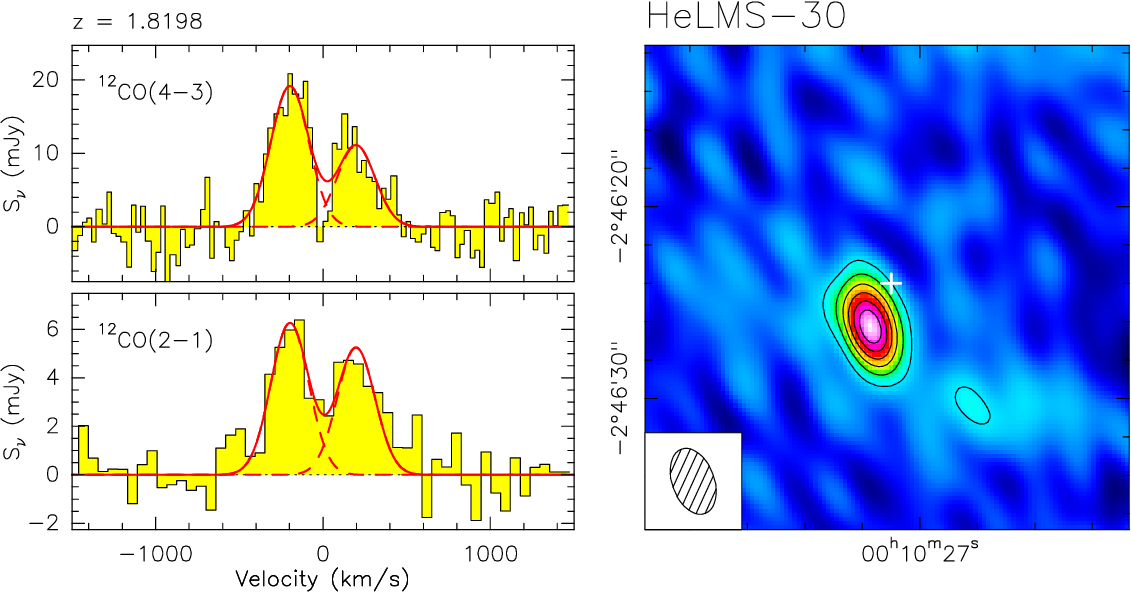}
\includegraphics[width=0.6\textwidth]{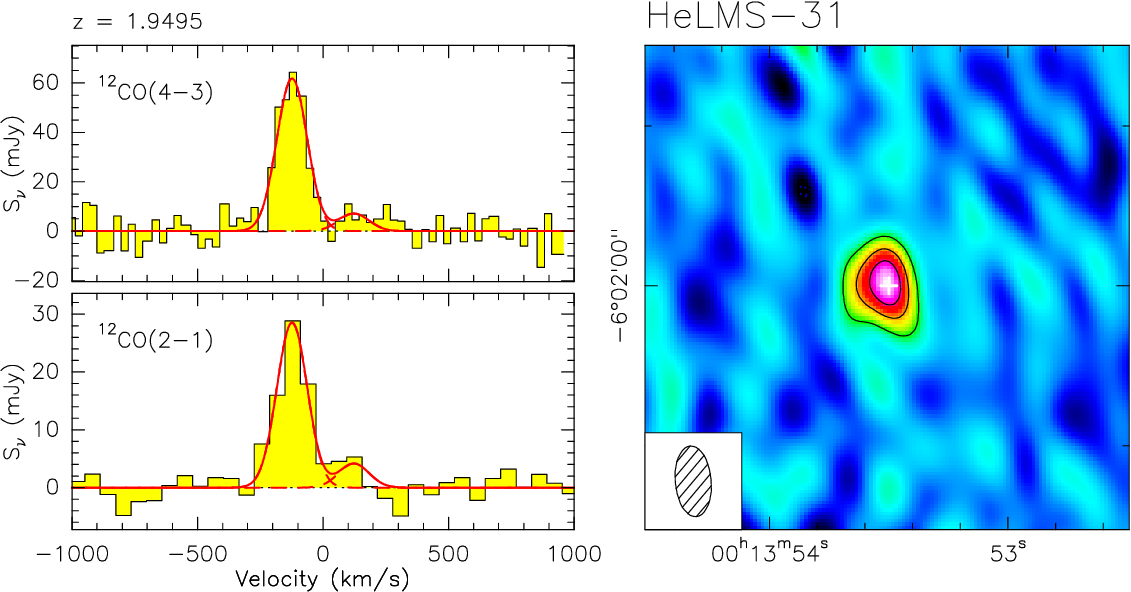}
\includegraphics[width=0.6\textwidth]{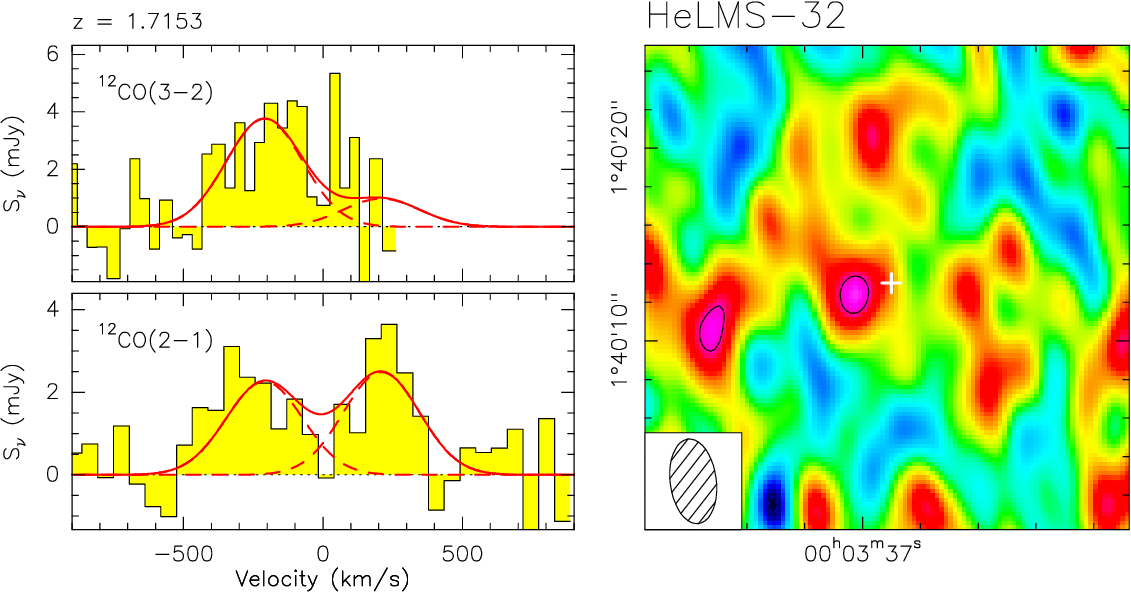}
\includegraphics[width=0.6\textwidth]{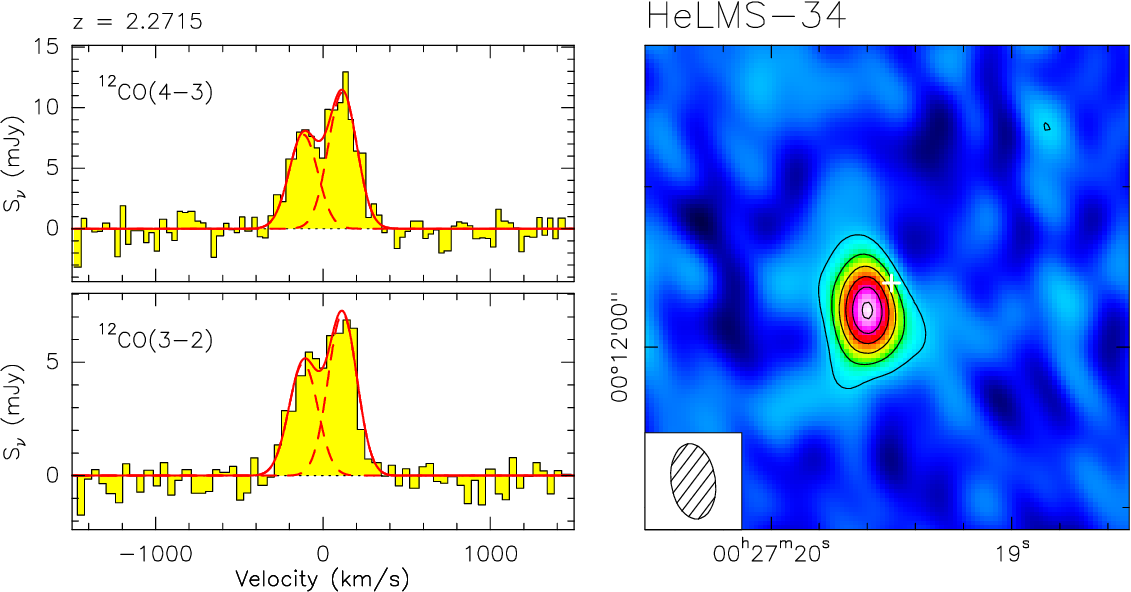} 
\caption{{\bf continued}}
    \end{figure*}
 \addtocounter{figure}{-1}

     \begin{figure*}[!ht]
   \centering
\includegraphics[width=0.6\textwidth]{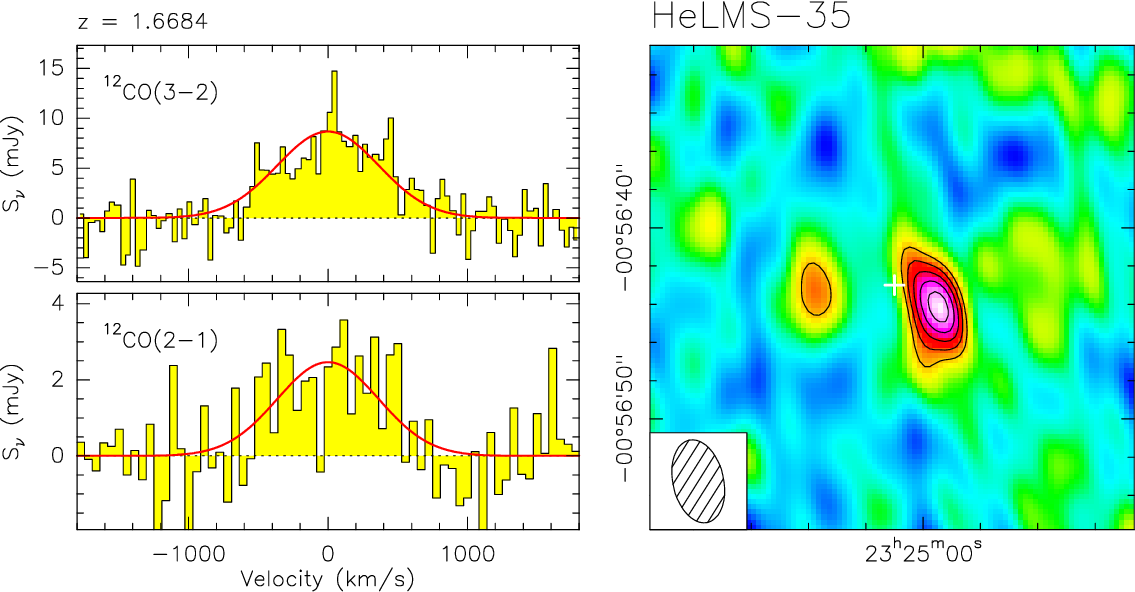}

\vspace{0.4cm}
\includegraphics[width=0.9\textwidth]{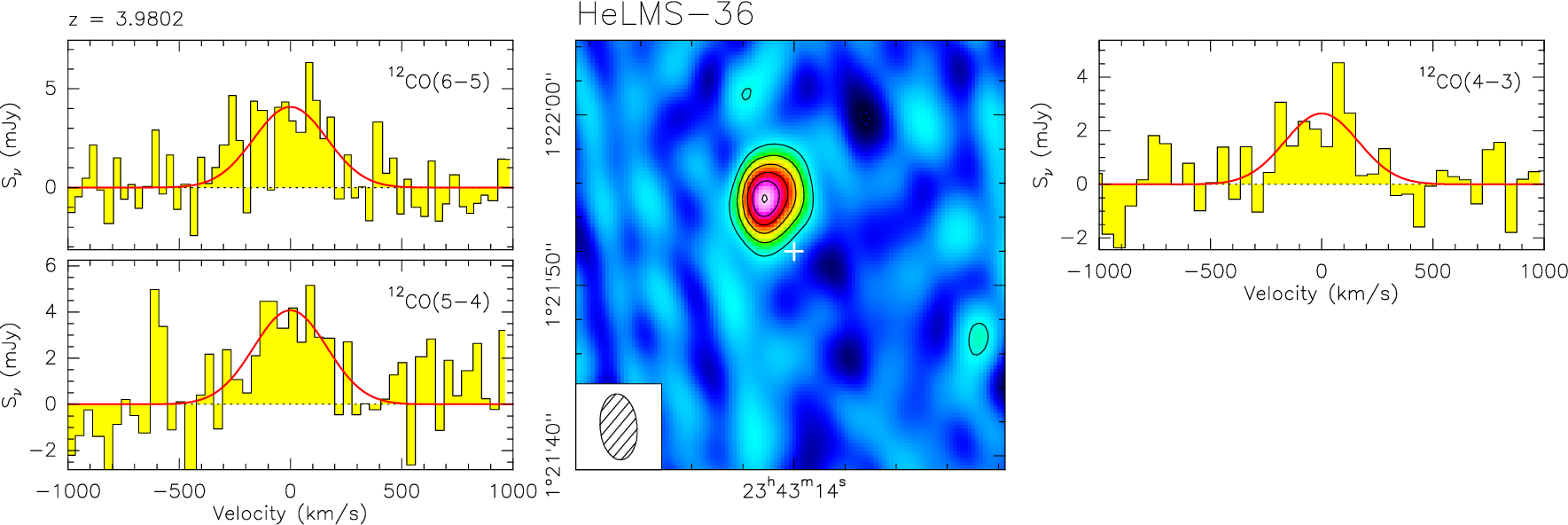}

\vspace{0.4cm}
\includegraphics[width=0.6\textwidth]{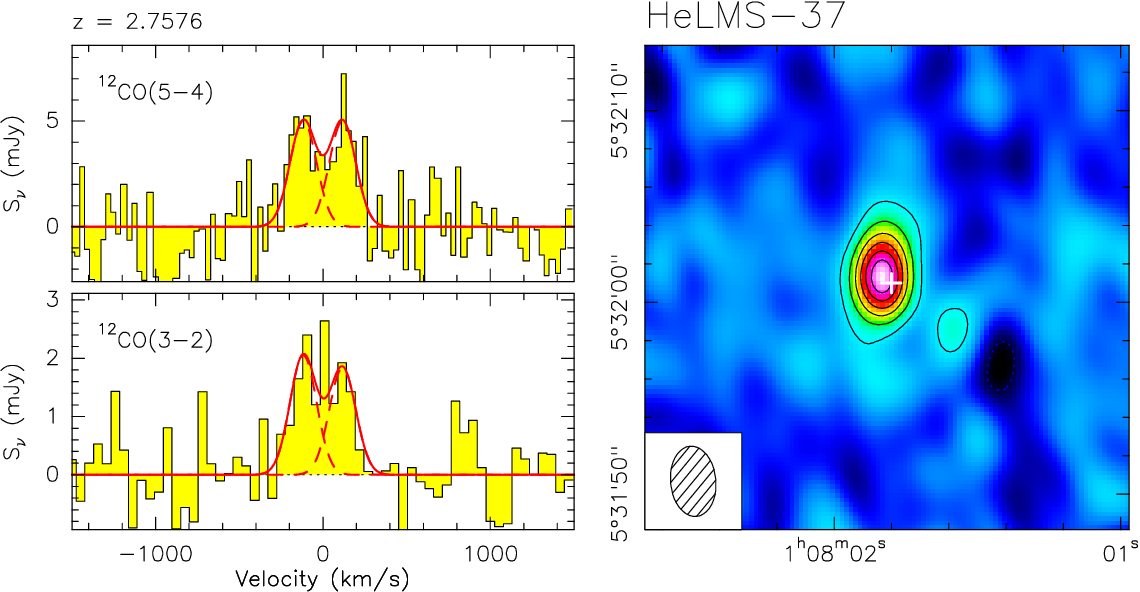}

\vspace{0.4cm}
\includegraphics[width=0.9\textwidth]{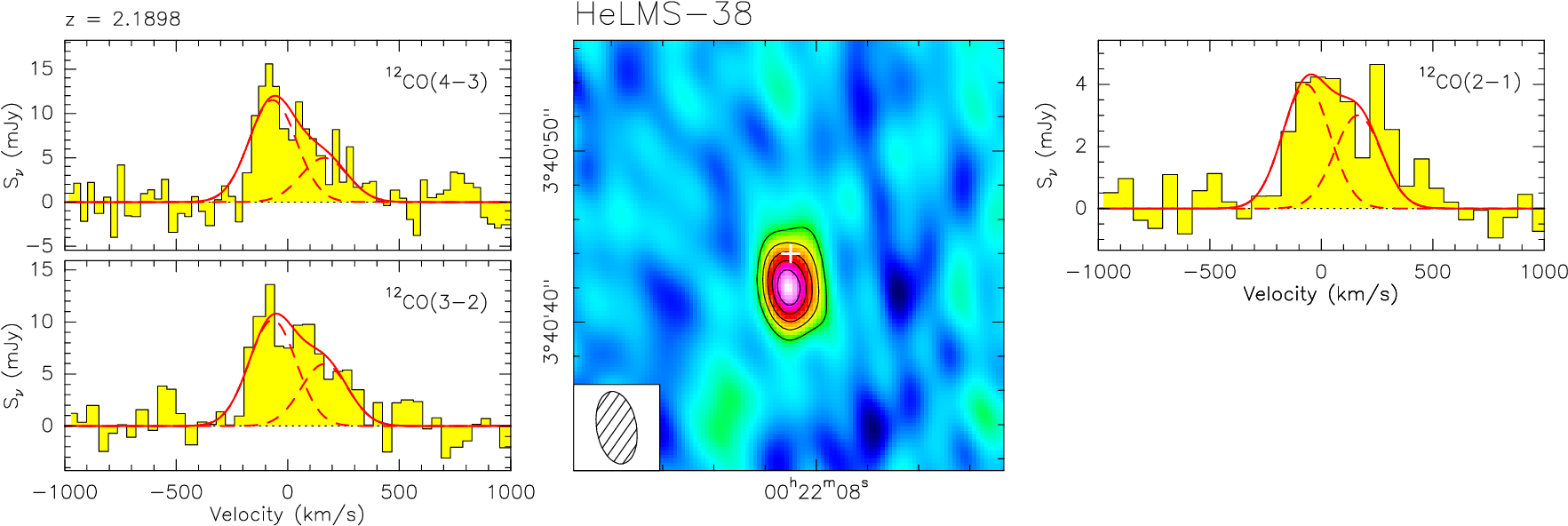}
\caption{{\bf continued}}
    \end{figure*}
  \addtocounter{figure}{-1}
 
 \begin{figure*}[!ht]
   \centering
\includegraphics[width=0.6\textwidth]{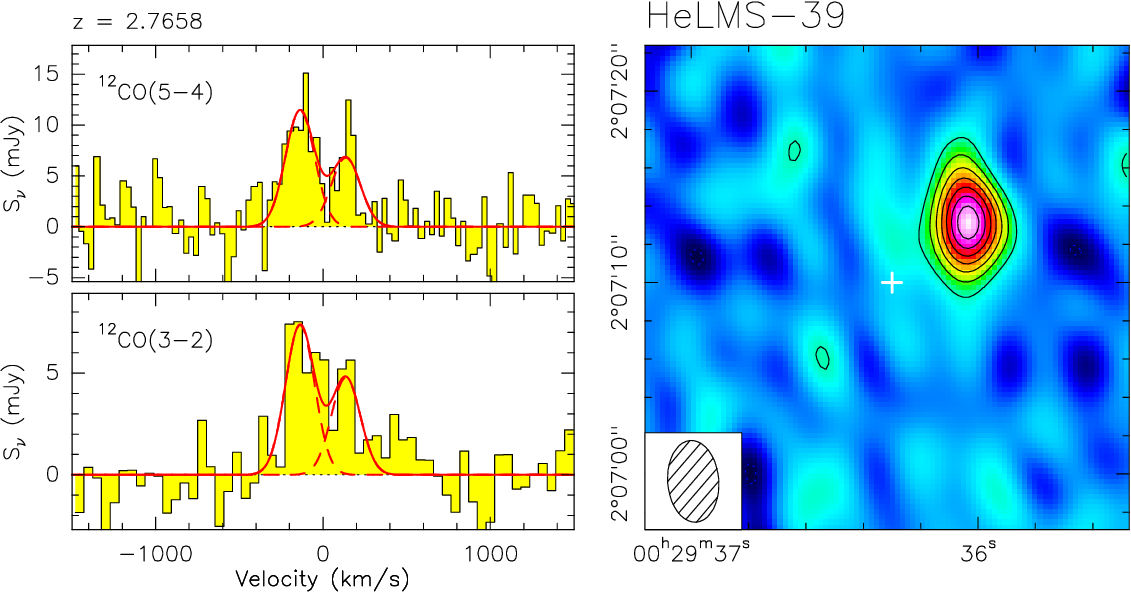}

\vspace{0.4cm}
\includegraphics[width=0.9\textwidth]{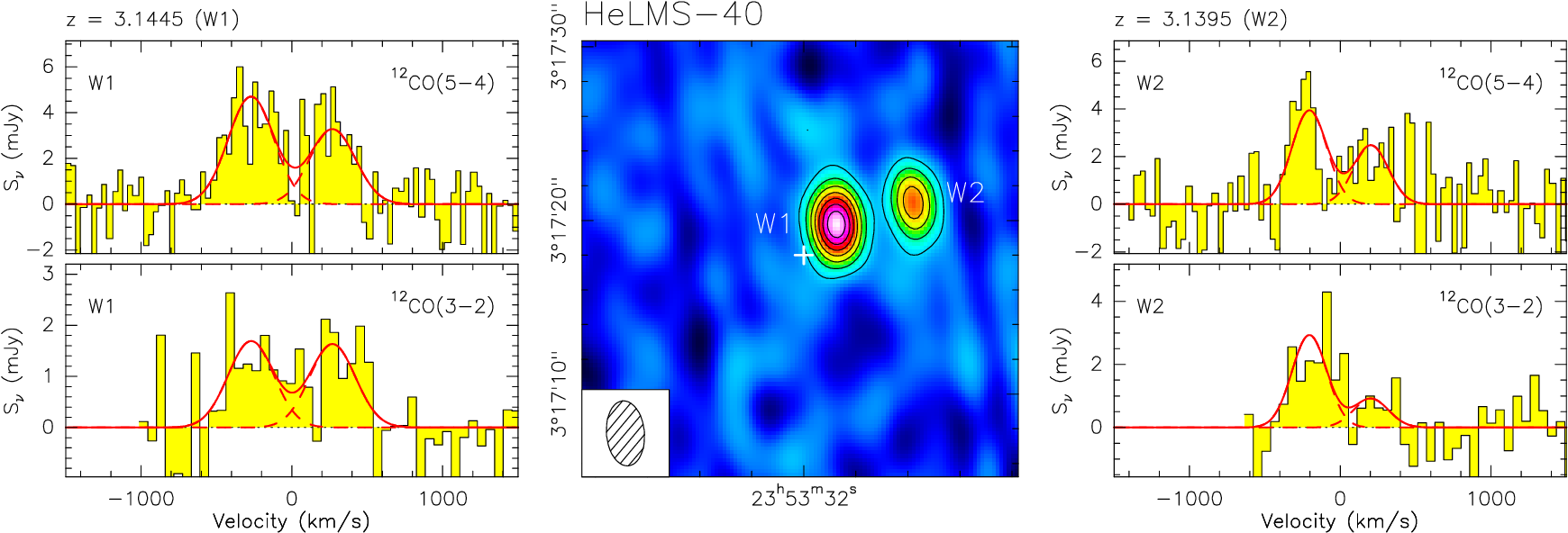}

\vspace{0.4cm}
\includegraphics[width=0.6\textwidth]{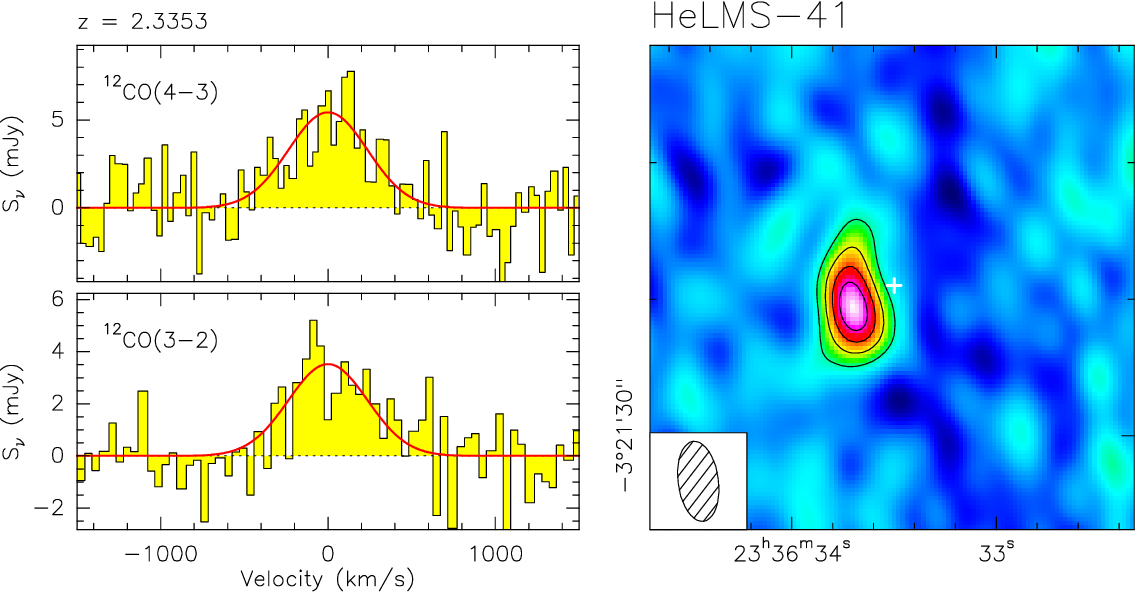}

\vspace{0.4cm}
\includegraphics[width=0.6\textwidth]{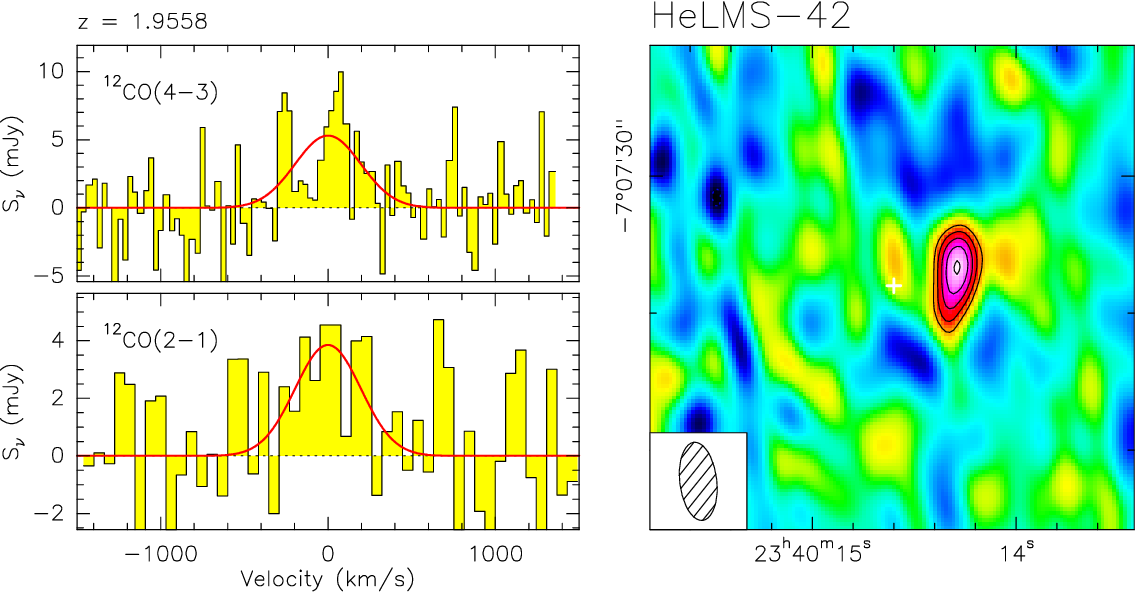}
\caption{{\bf continued}}

    \end{figure*}
 \addtocounter{figure}{-1}
    
\begin{figure*}[!ht]
   \centering
\includegraphics[width=0.6\textwidth]{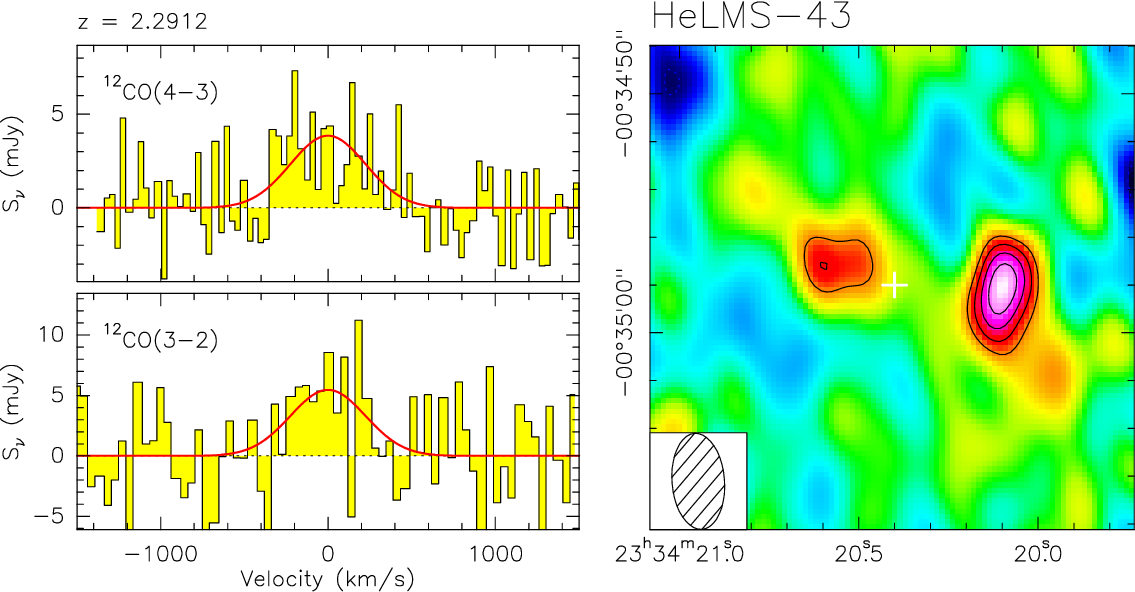}

\vspace{0.4cm}
\includegraphics[width=0.6\textwidth]{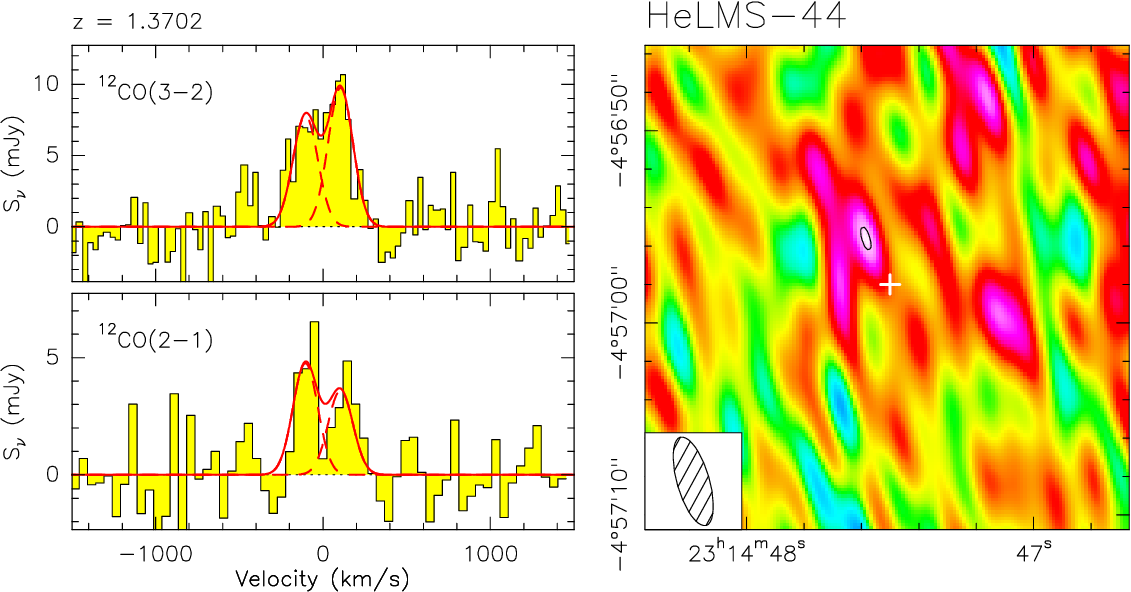}

\vspace{0.4cm}
\includegraphics[width=0.9\textwidth]{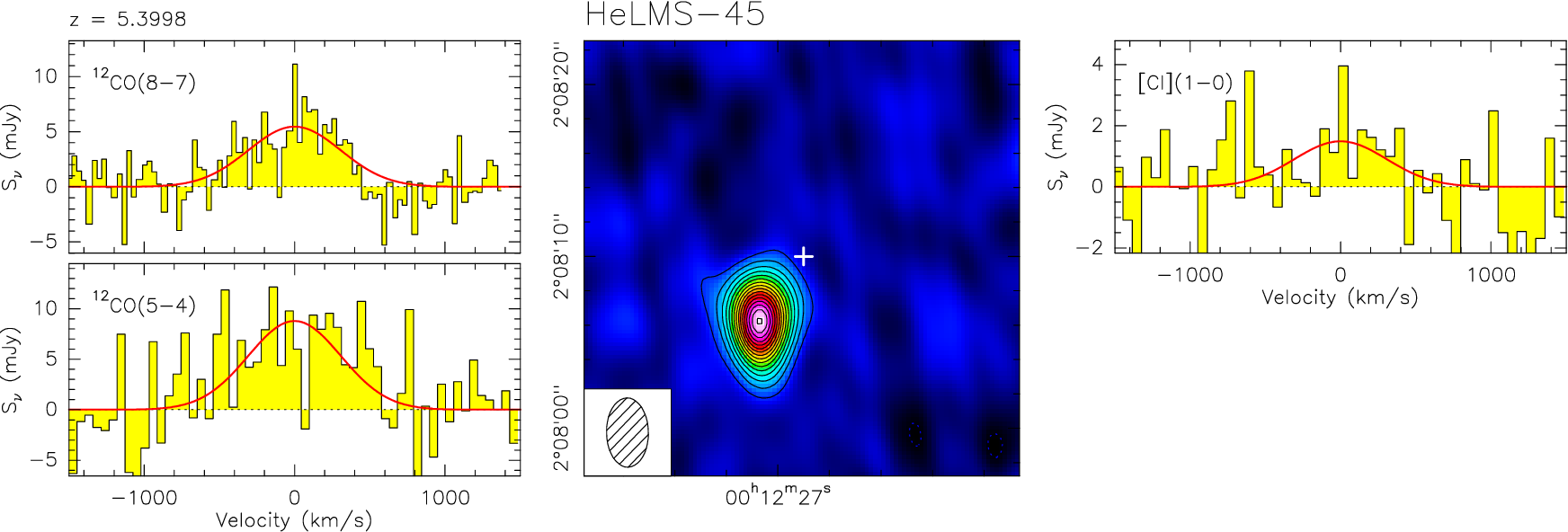}

\vspace{0.4cm}
\includegraphics[width=0.6\textwidth]{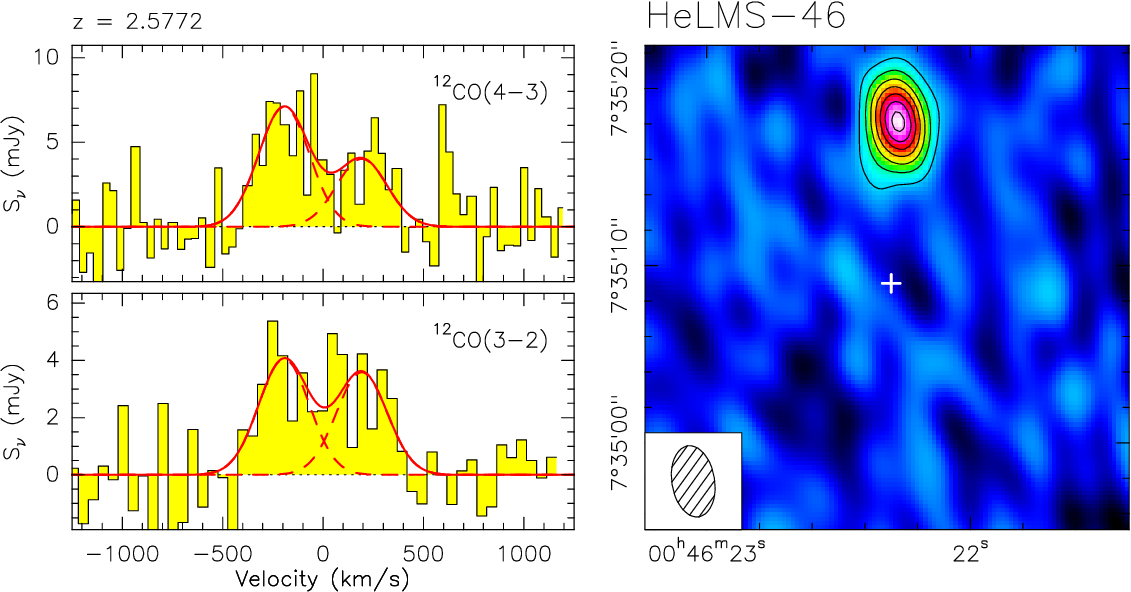}
\caption{{\bf continued}}

    \end{figure*}
 \addtocounter{figure}{-1}
 
    \begin{figure*}[!ht]
   \centering
\includegraphics[width=0.9\textwidth]{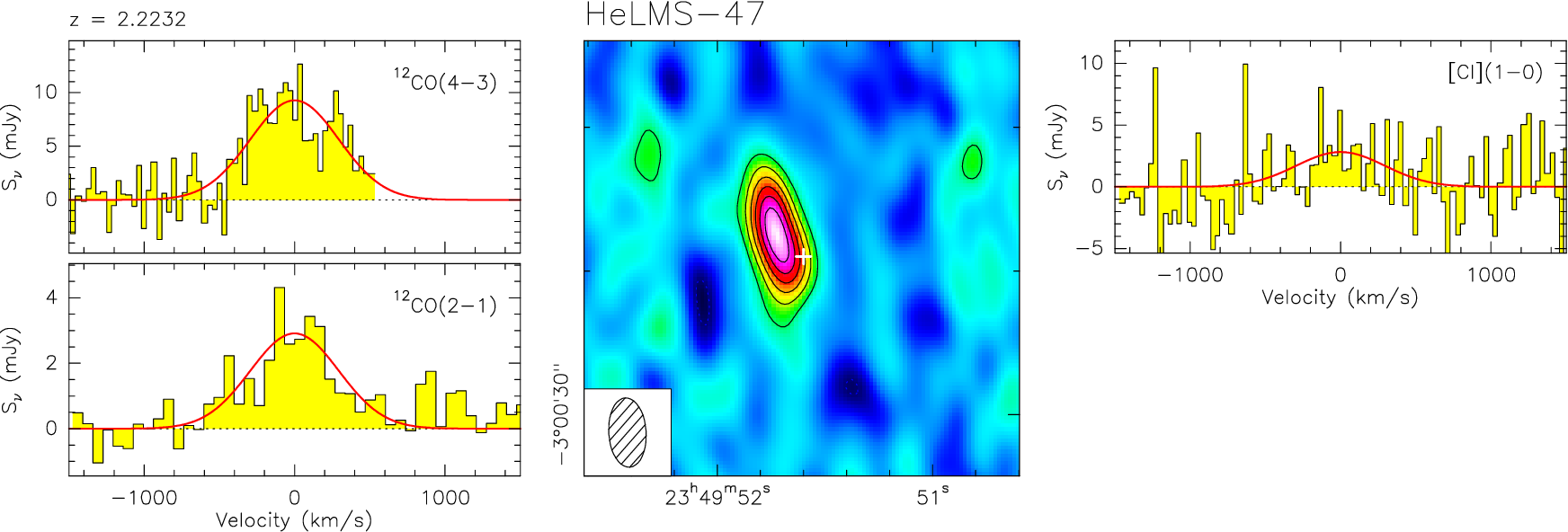}

\vspace{0.4cm}
\includegraphics[width=0.9\textwidth]{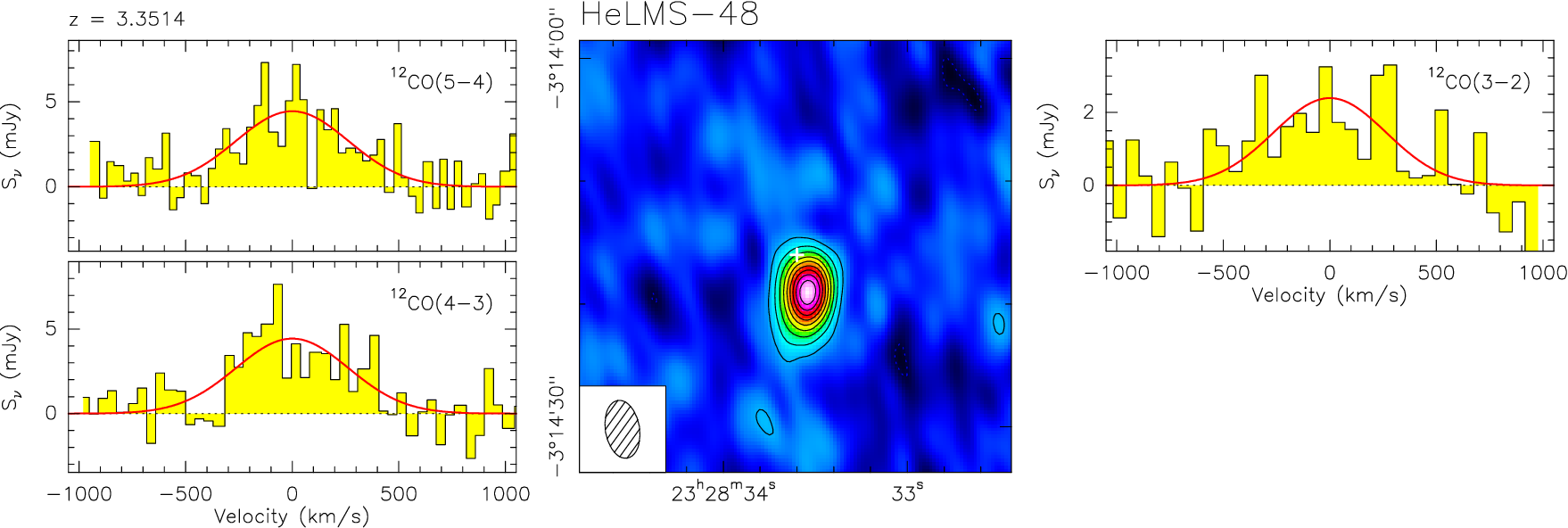}

\vspace{0.4cm}
\includegraphics[width=0.6\textwidth]{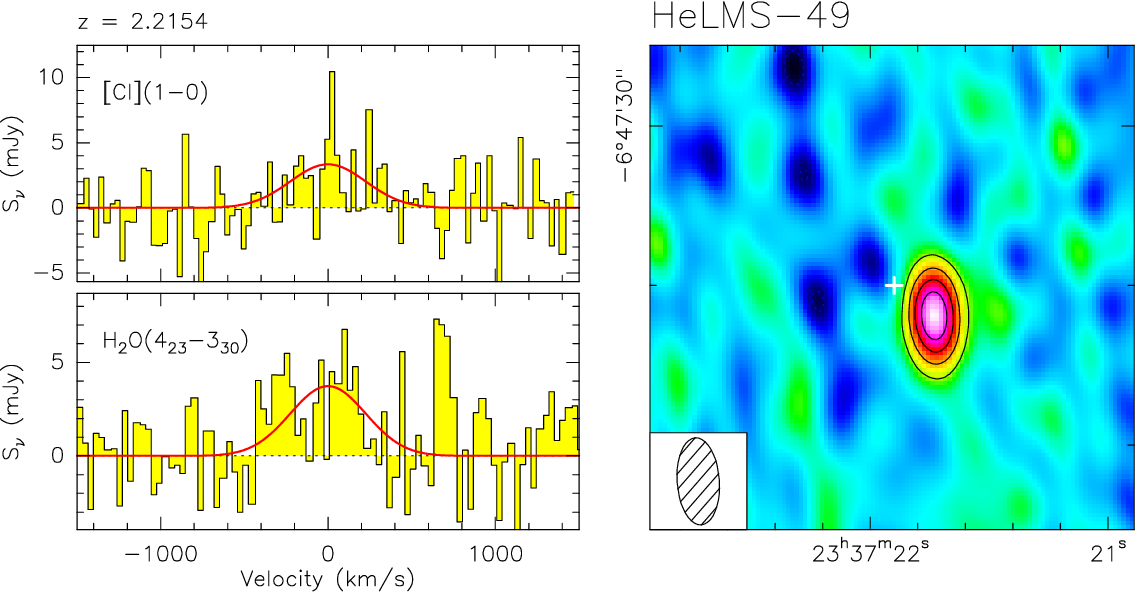}

\vspace{0.4cm}
\includegraphics[width=0.6\textwidth]{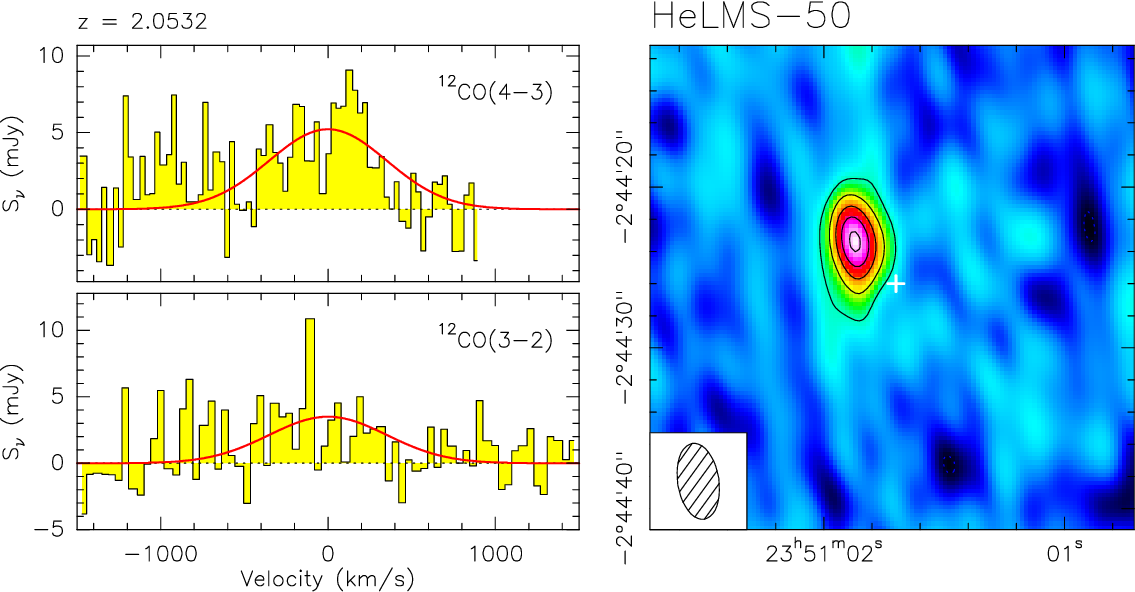}
\caption{{\bf continued}}
    \end{figure*}
 \addtocounter{figure}{-1}
 
\begin{figure*}[!ht]
   \centering
\includegraphics[width=0.6\textwidth]{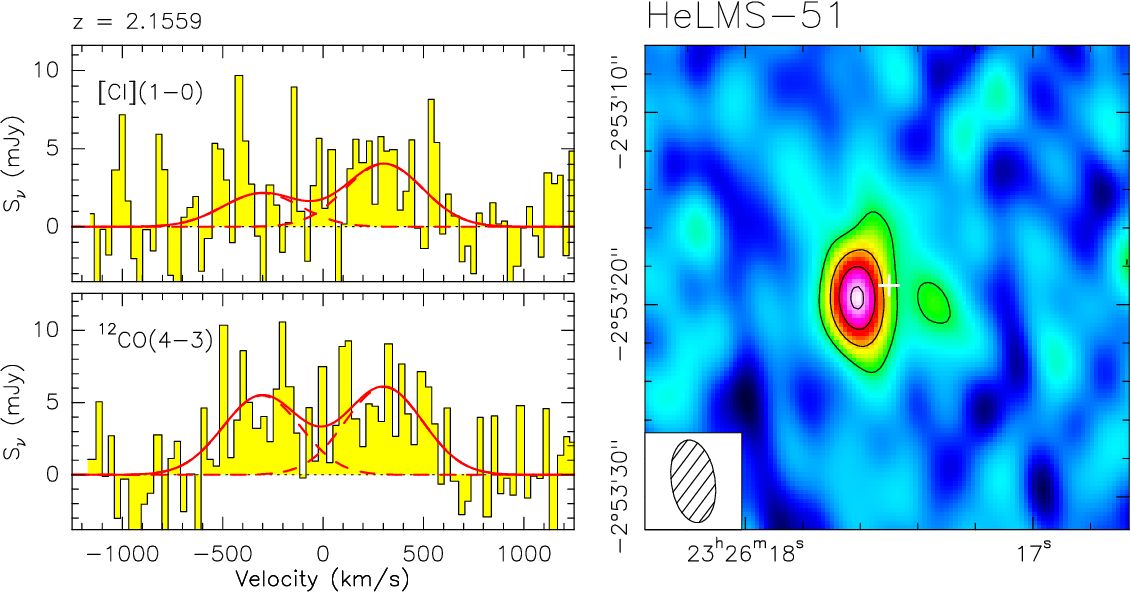}

\vspace{1.4cm}
\includegraphics[width=0.6\textwidth]{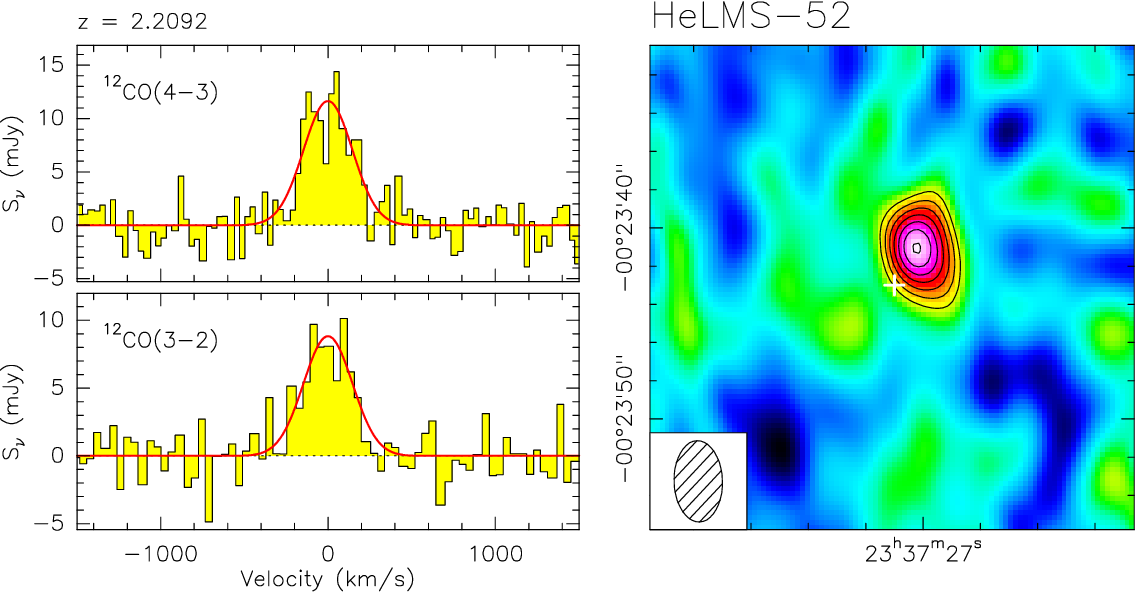}

\vspace{1.4cm}
\includegraphics[width=0.6\textwidth]{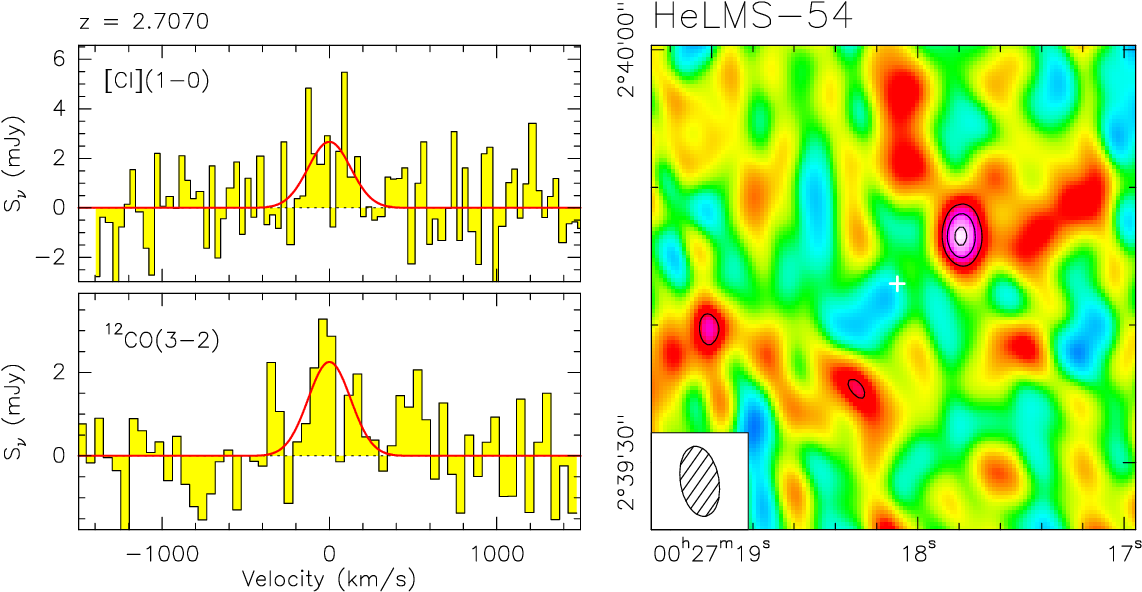}
   \caption{{\bf continued}}
    \end{figure*}

\begin{figure*}[!ht]
   \centering
\includegraphics[width=0.6\textwidth]{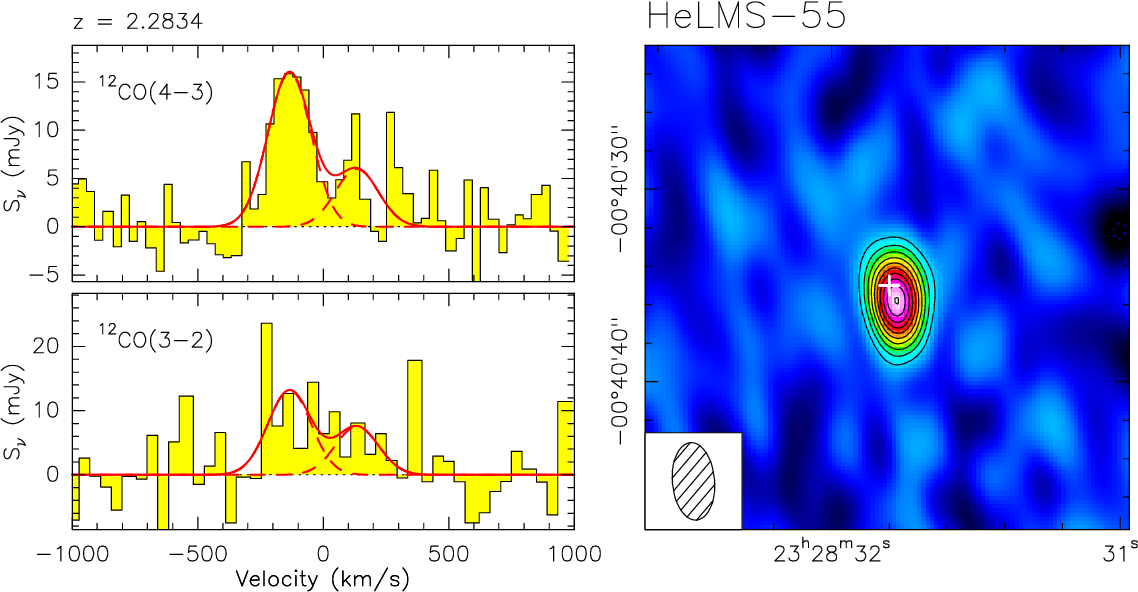}

\vspace{1.4cm}
 \includegraphics[width=0.6\textwidth]{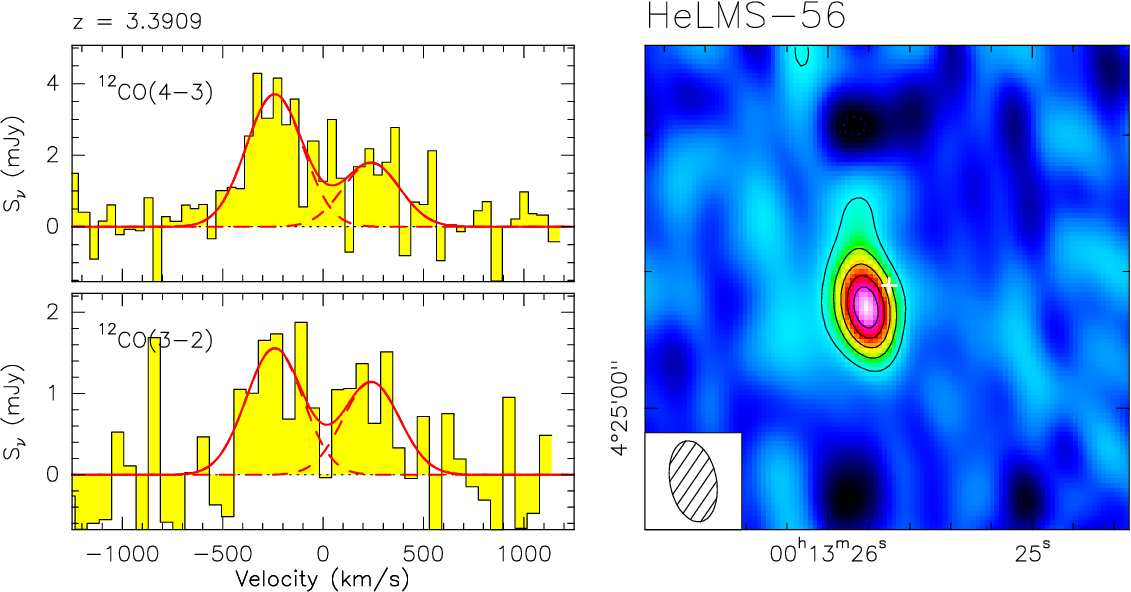}

 \vspace{1.4cm}
\includegraphics[width=0.6\textwidth]{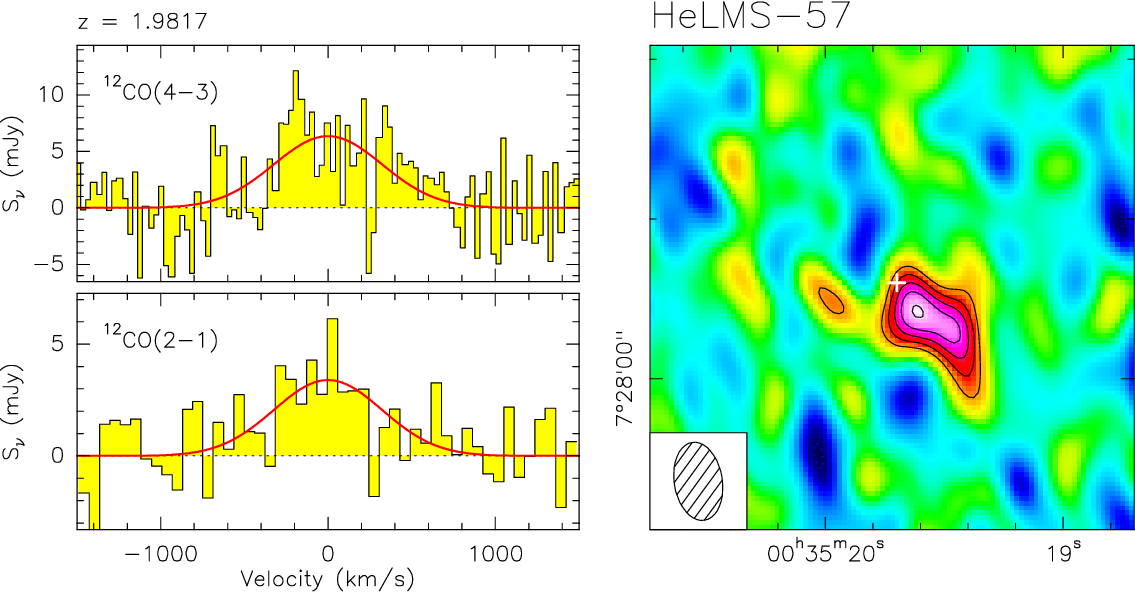}
\caption{{\bf continued}}
    \end{figure*}
 \addtocounter{figure}{-1}
 
\begin{figure*}[!ht]
   \centering
\includegraphics[width=0.6\textwidth]{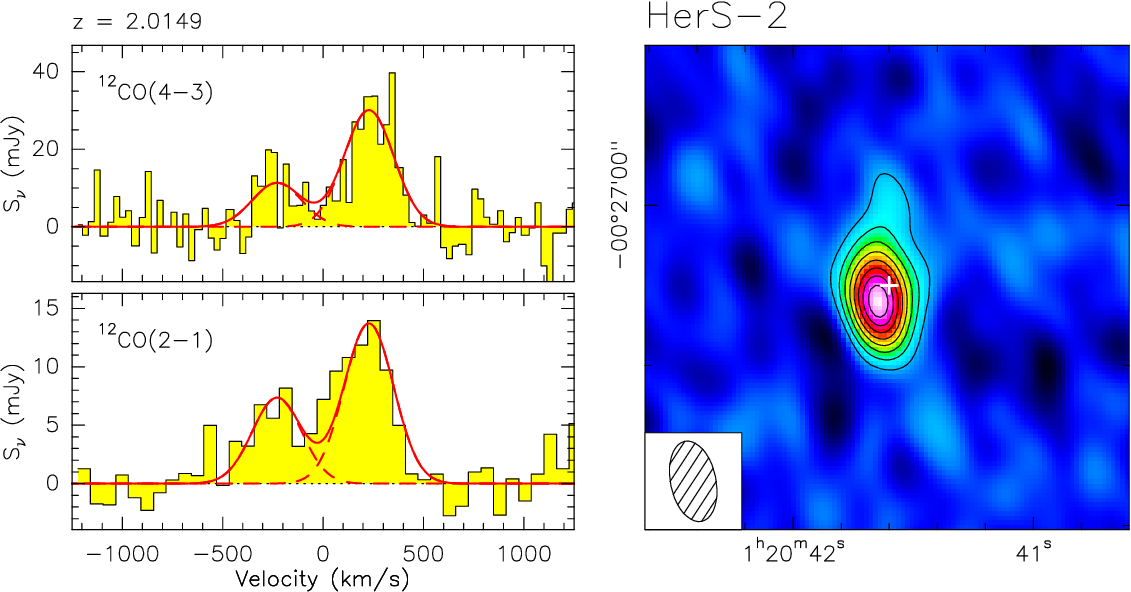}

\vspace{1.4cm}
\includegraphics[width=0.9\textwidth]{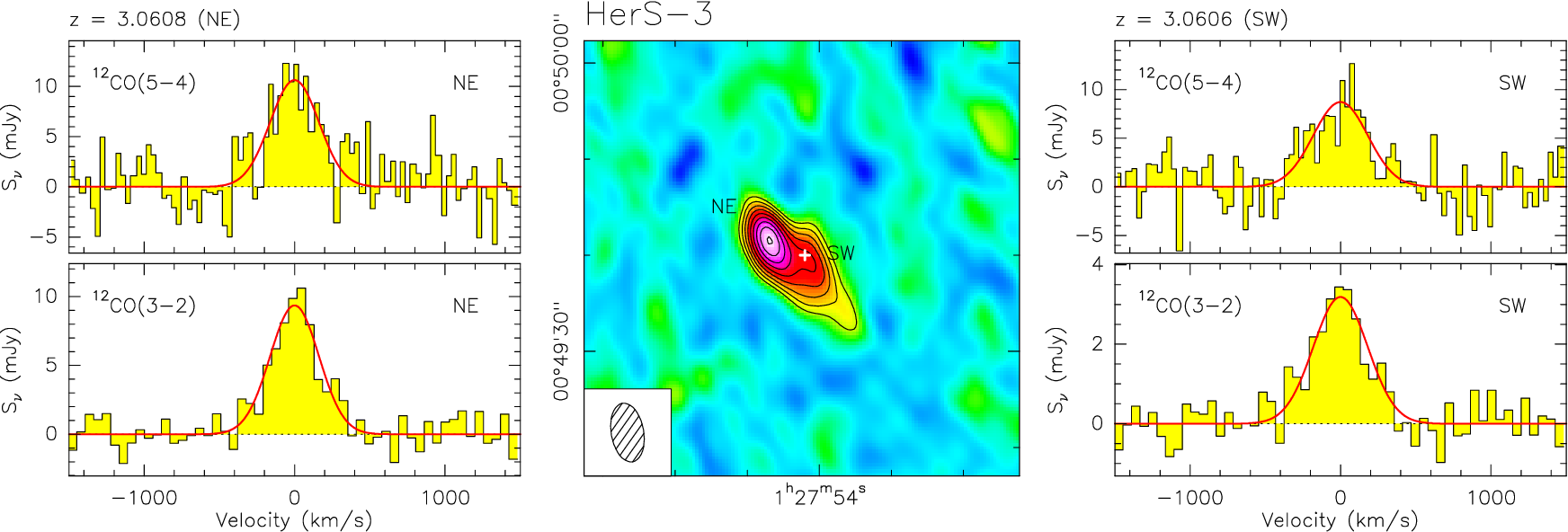}

\vspace{1.4cm}
\includegraphics[width=0.6\textwidth]{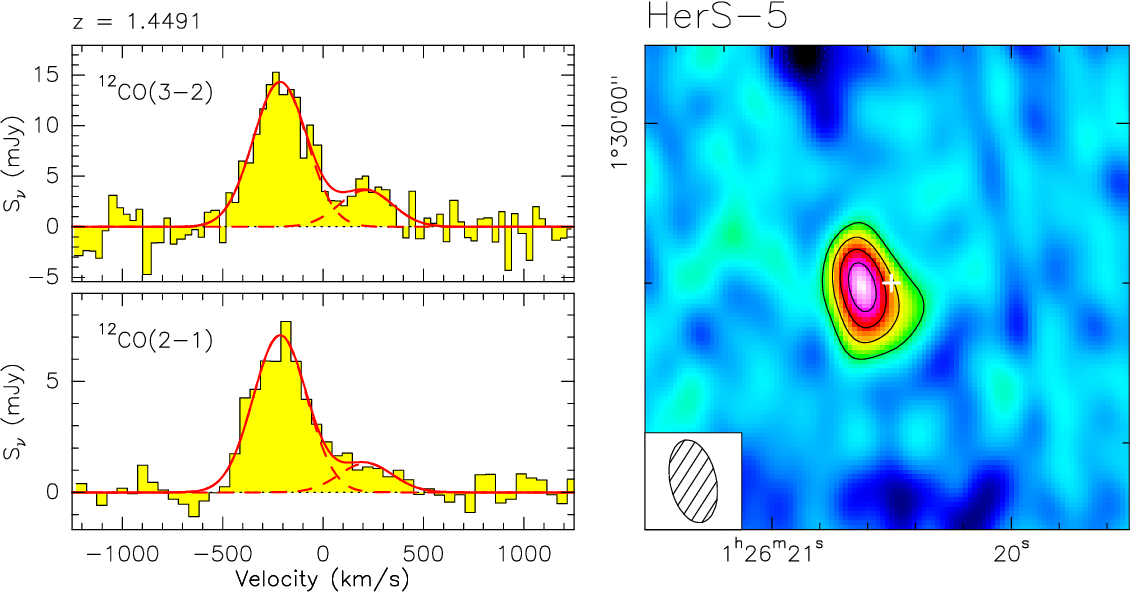}
   \caption{Combined continuum and emission lines maps of the $z$-GAL sources selected from the HerS sample and spectra of each of the detected emission lines. Further details are provided in Appendix~\ref{Appendix: Presentation of Catalogue}.}
\label{figure:spectra_continuum_HerS}%
    \end{figure*}

 \addtocounter{figure}{-1}

\begin{figure*}[!ht]
   \centering
\includegraphics[width=0.6\textwidth]{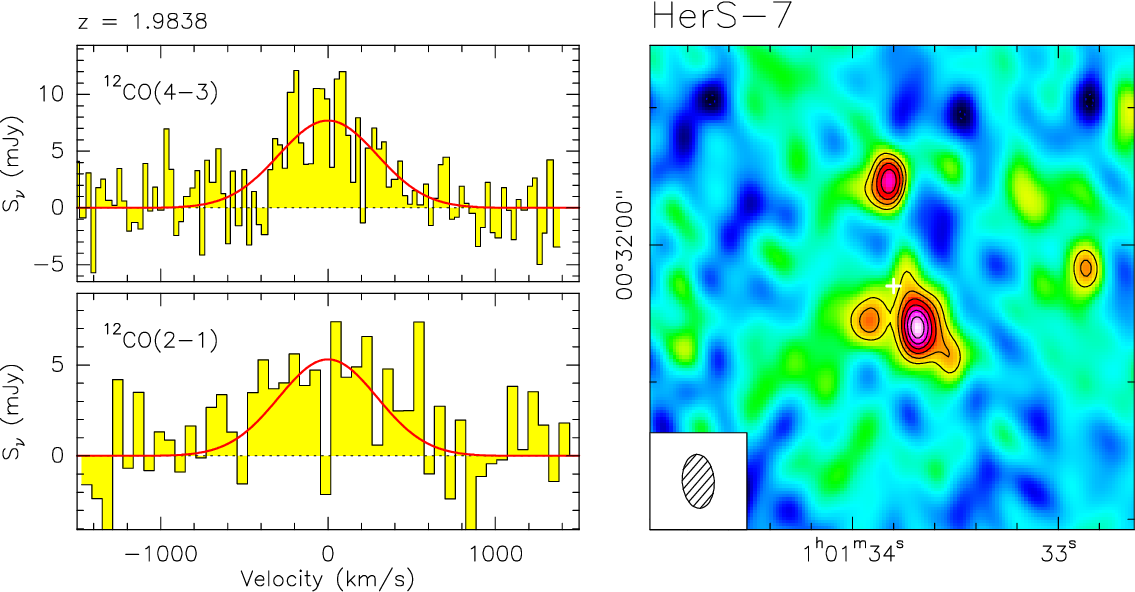}
\includegraphics[width=0.6\textwidth]{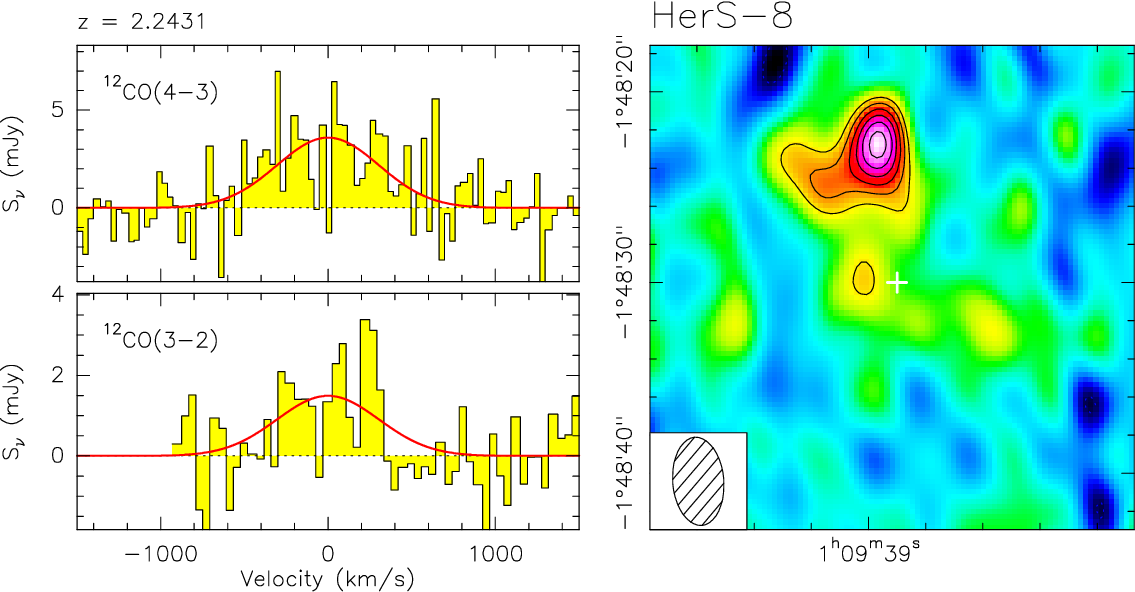}
\includegraphics[width=0.6\textwidth]{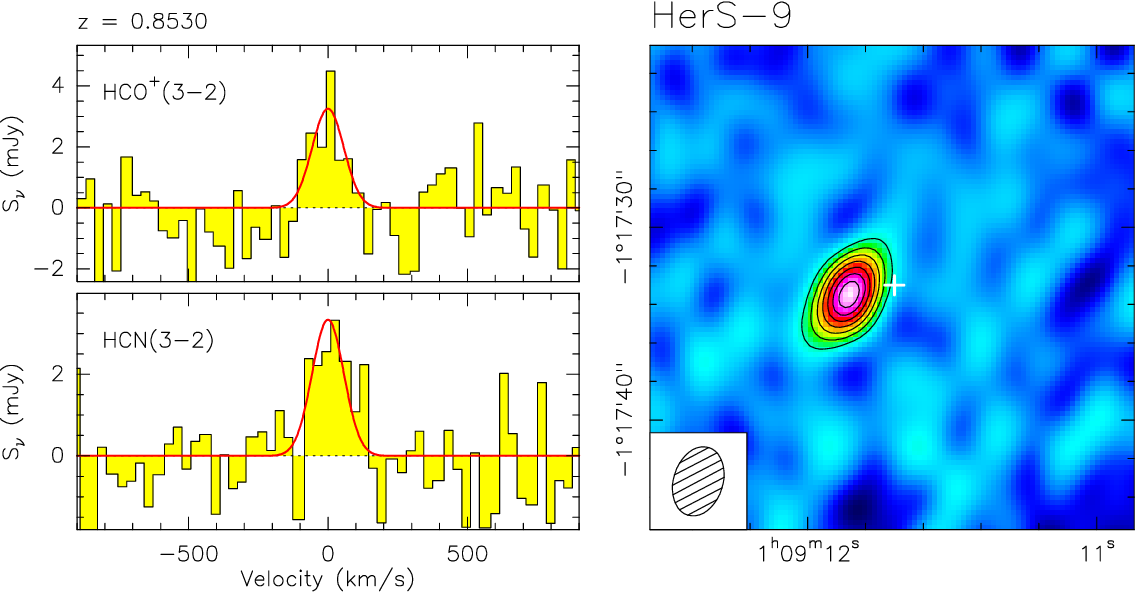}
\includegraphics[width=0.6\textwidth]{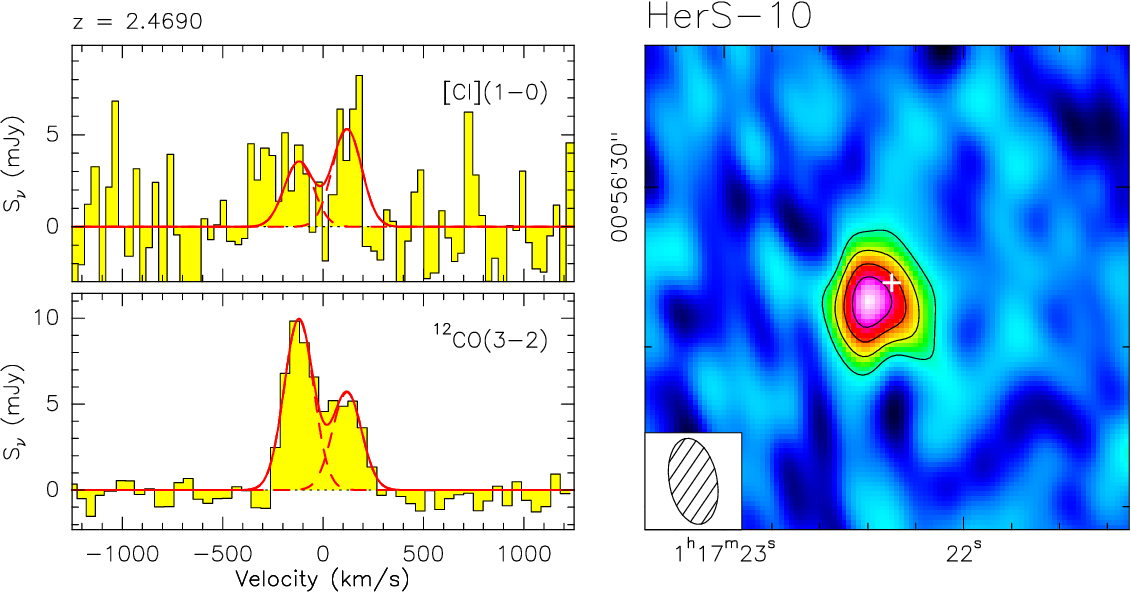}
    \caption{{\bf continued}}
    \end{figure*}

 \addtocounter{figure}{-1}

 \begin{figure*}[!ht]
   \centering
\includegraphics[width=0.6\textwidth]{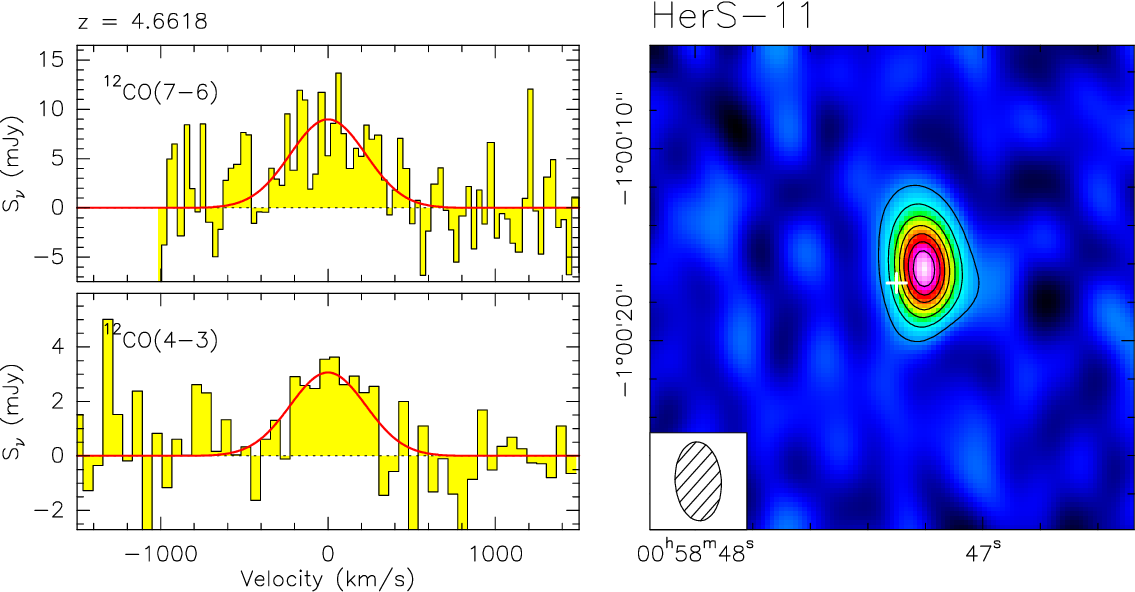}
\includegraphics[width=0.6\textwidth]{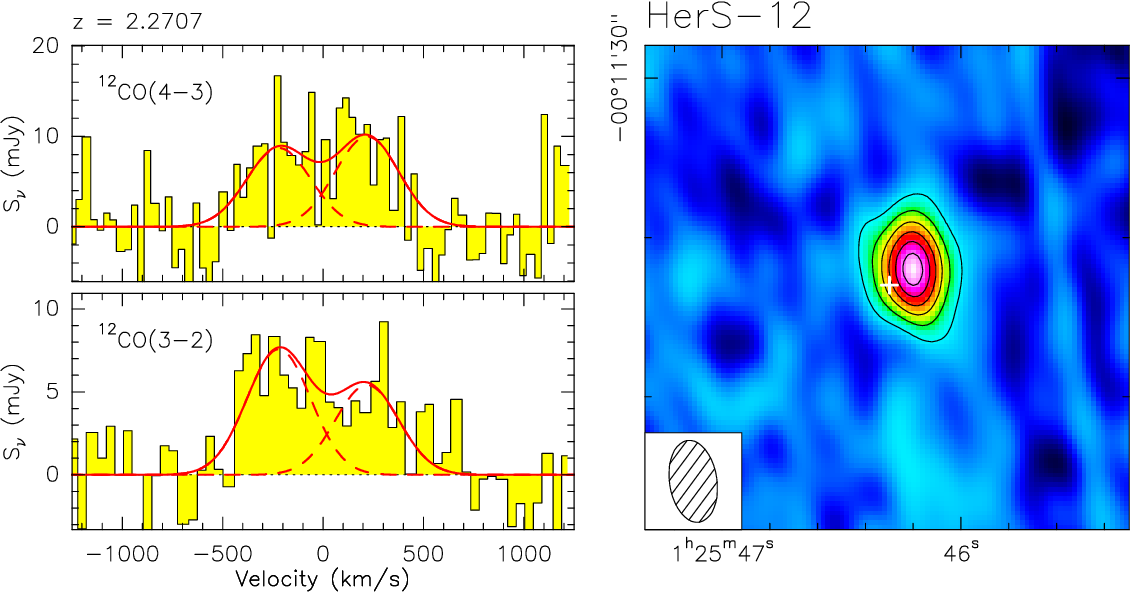}
\includegraphics[width=0.6\textwidth]{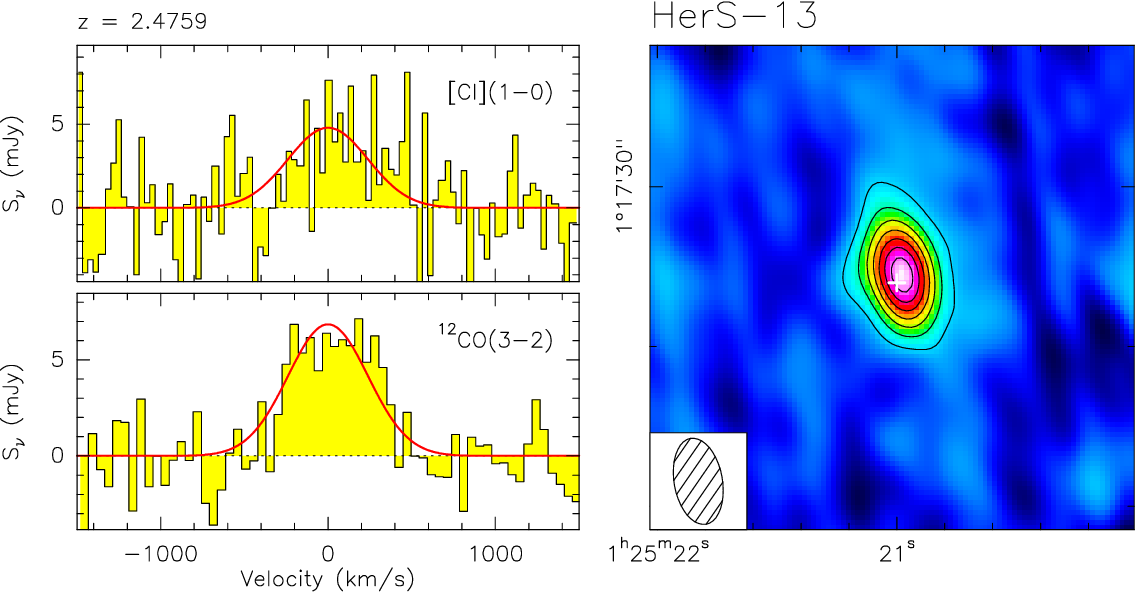}

\vspace{0.4cm}
\includegraphics[width=0.9\textwidth]{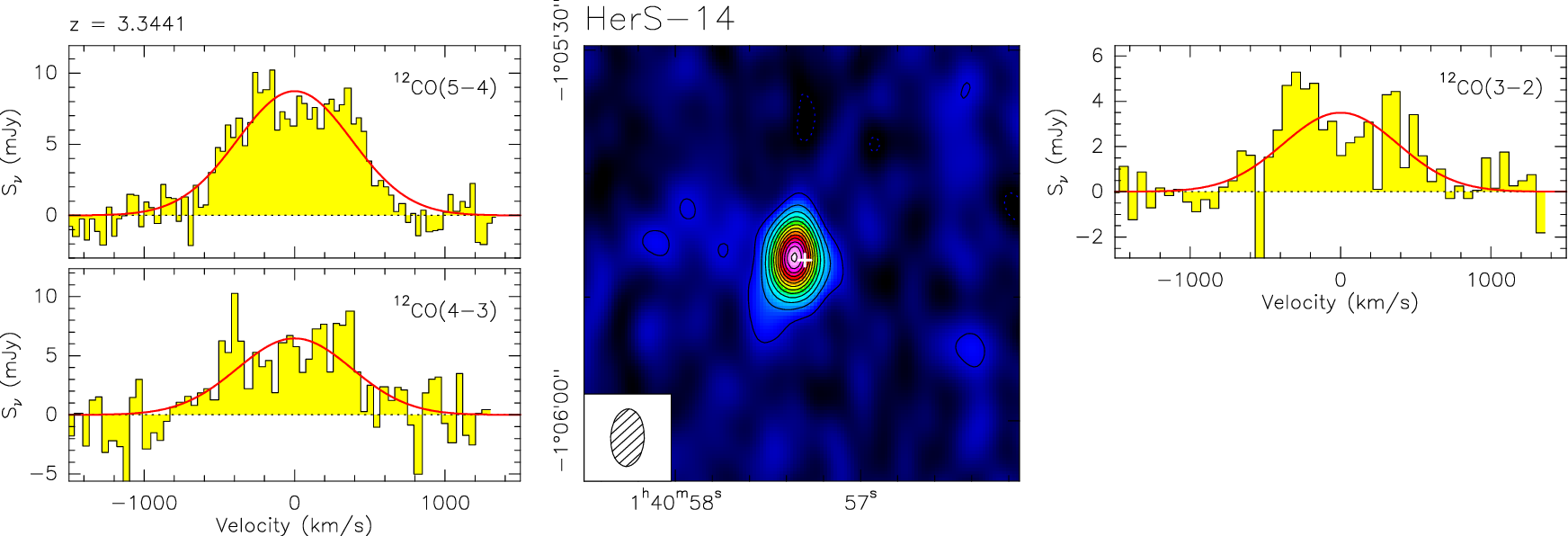} 
    \caption{{\bf continued}}
    \end{figure*}

 \addtocounter{figure}{-1}

 \begin{figure*}[!ht]
   \centering
\includegraphics[width=0.6\textwidth]{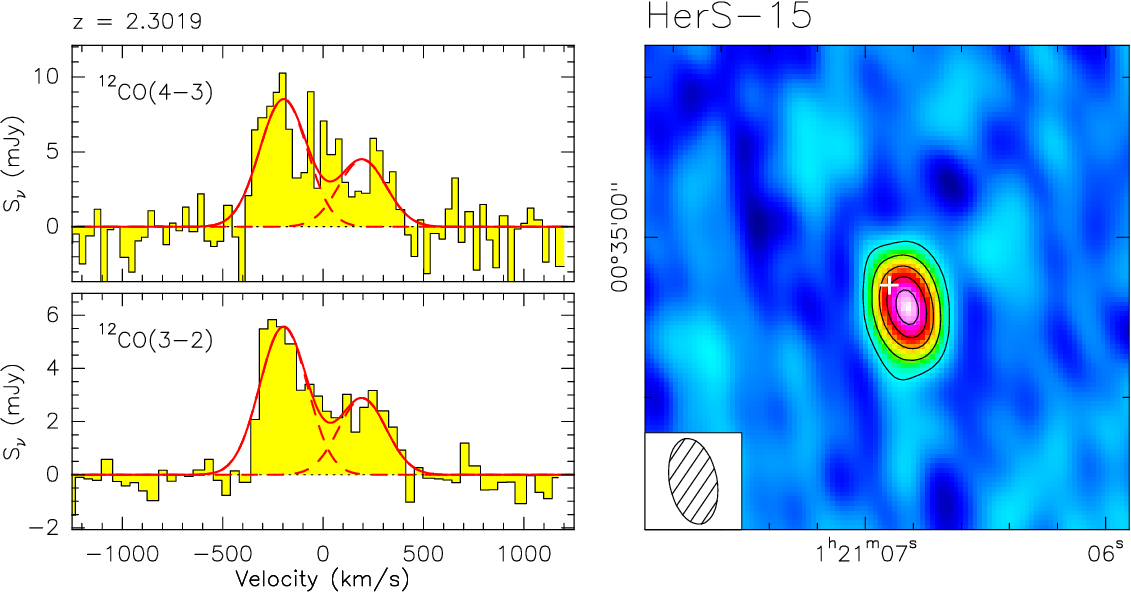}

\vspace{1.2cm}
\includegraphics[width=0.6\textwidth]{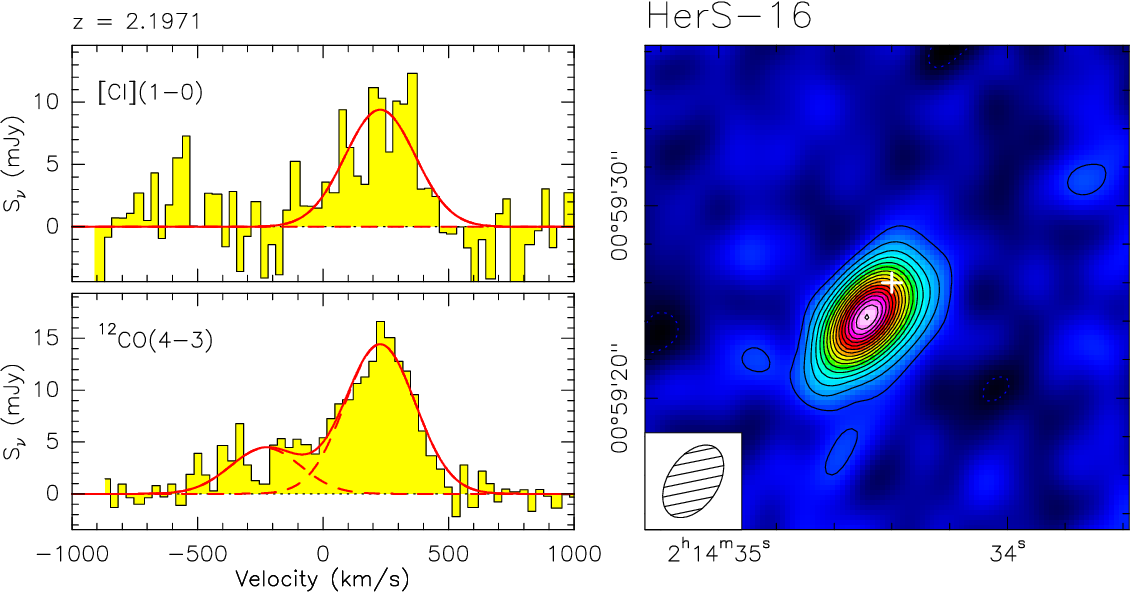}

\vspace{1.2cm}
\includegraphics[width=0.6\textwidth]{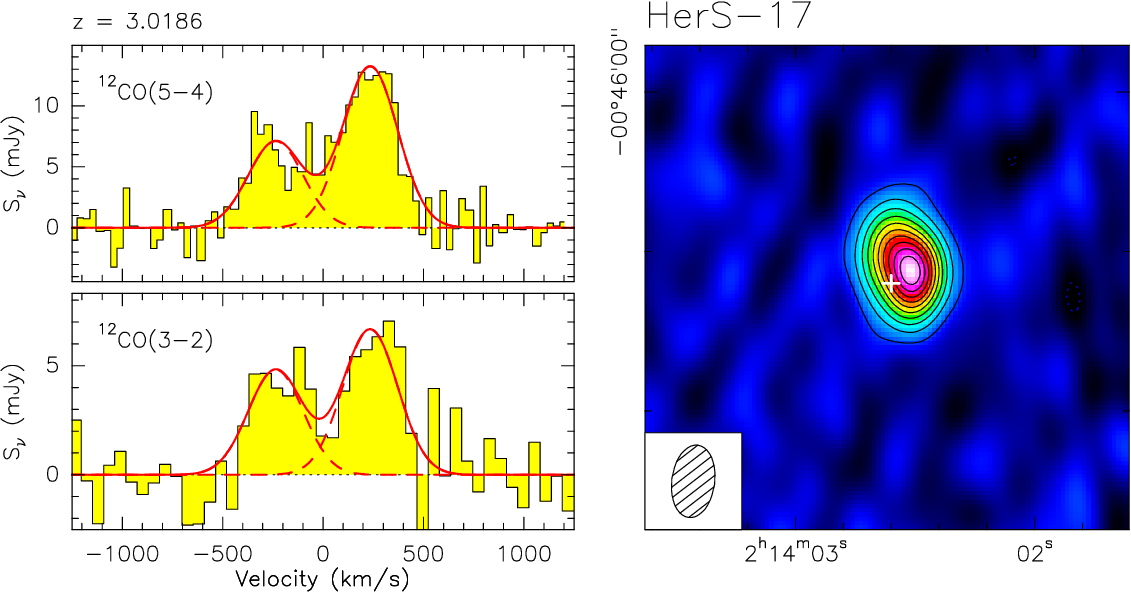}
    \caption{{\bf continued}}
    \end{figure*}
 \addtocounter{figure}{-1}
 
 \begin{figure*}[!ht]
   \centering
\includegraphics[width=0.9\textwidth]{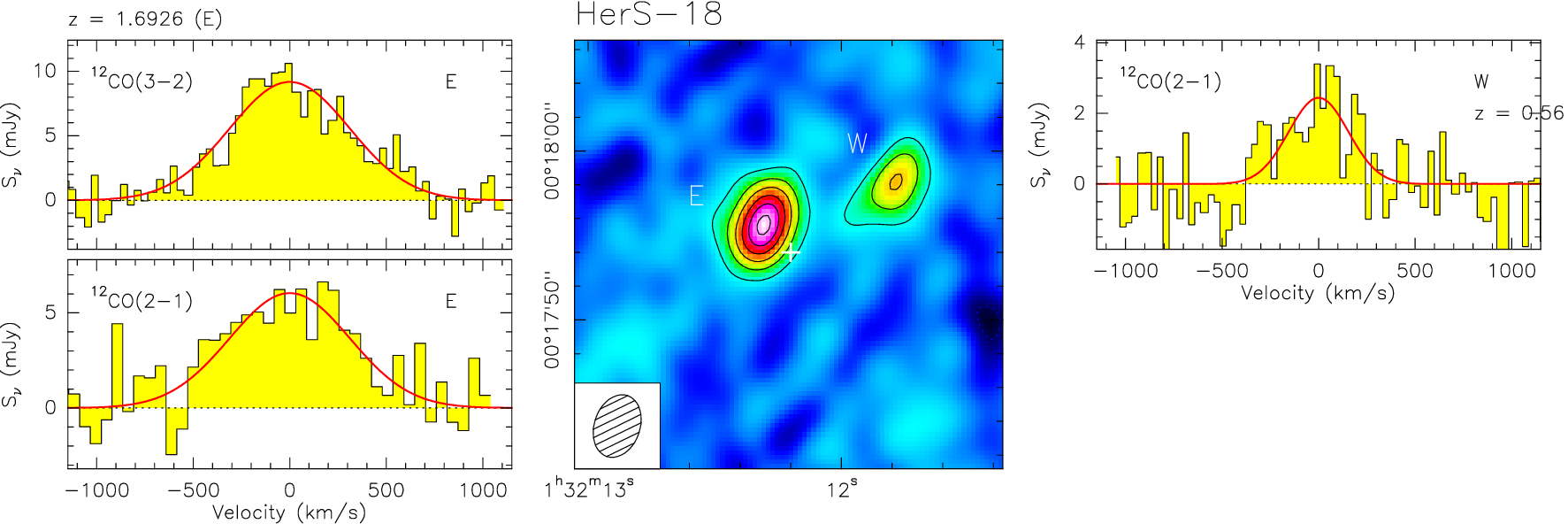}

\vspace{1.2cm}
\includegraphics[width=0.9\textwidth]{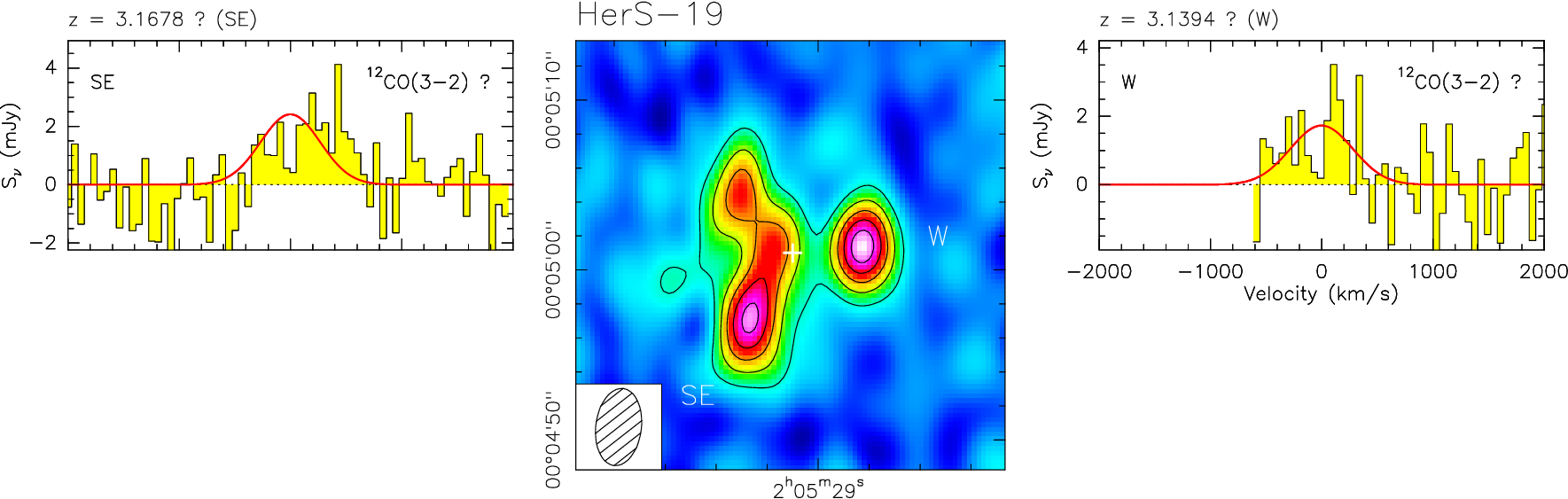}

\vspace{1.2cm}
\includegraphics[width=0.6\textwidth]{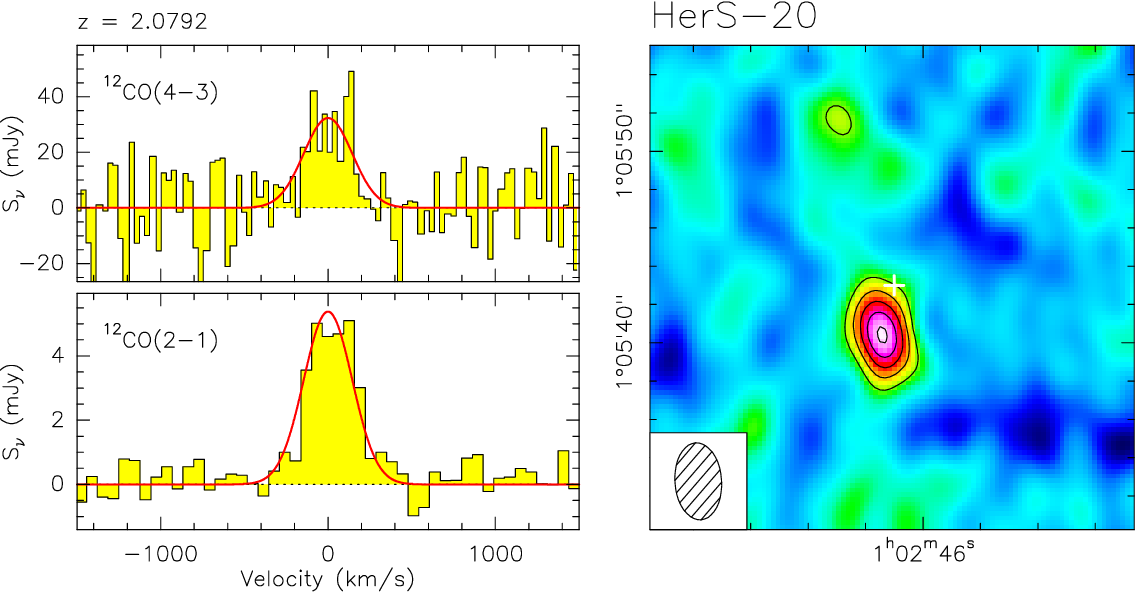}
    \caption{{\bf continued}}
    \end{figure*}

\begin{figure*}[!ht]
   \centering
\includegraphics[width=0.9\textwidth]{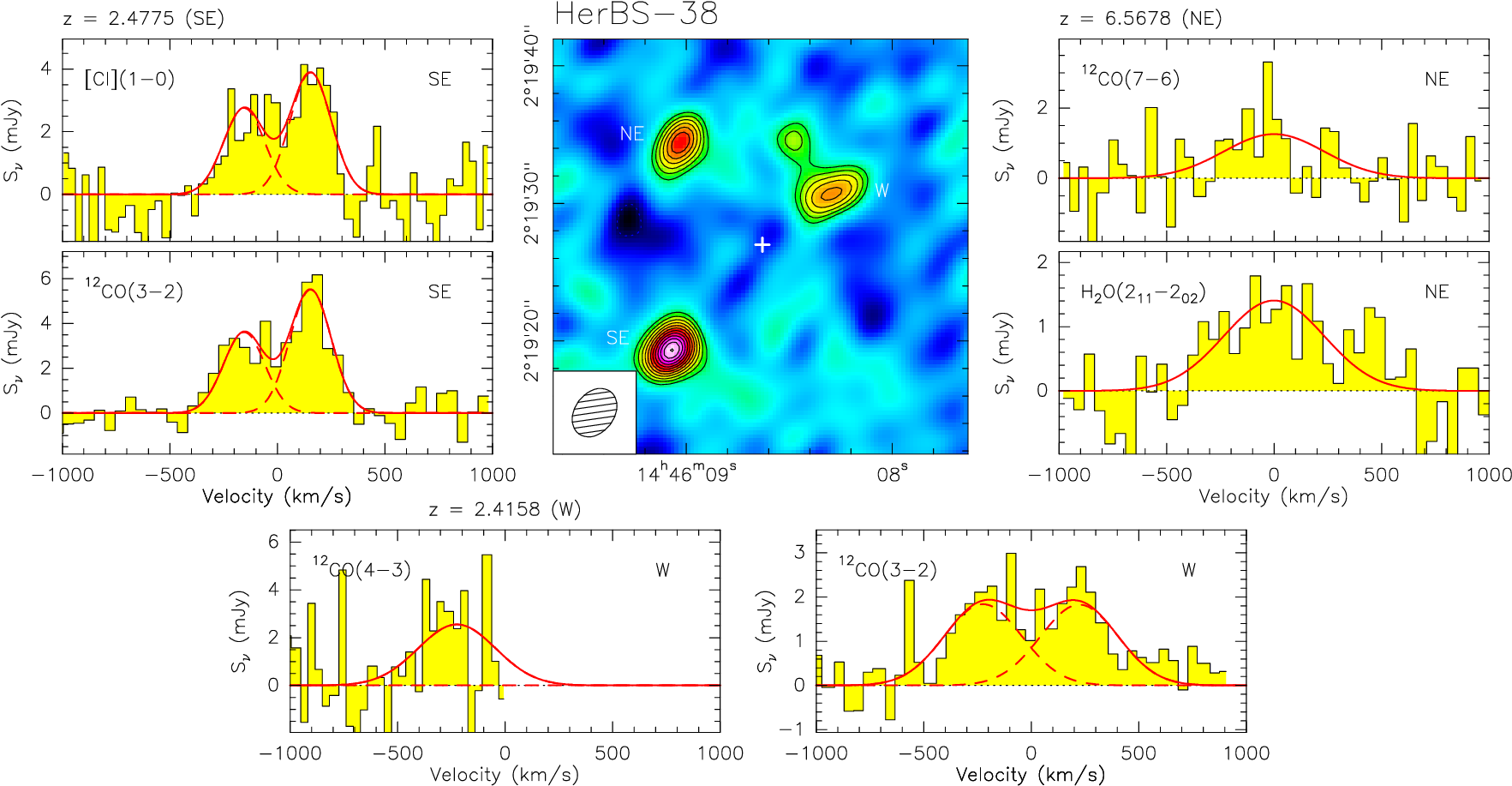}

\vspace{1.0cm}
\includegraphics[width=0.6\textwidth]{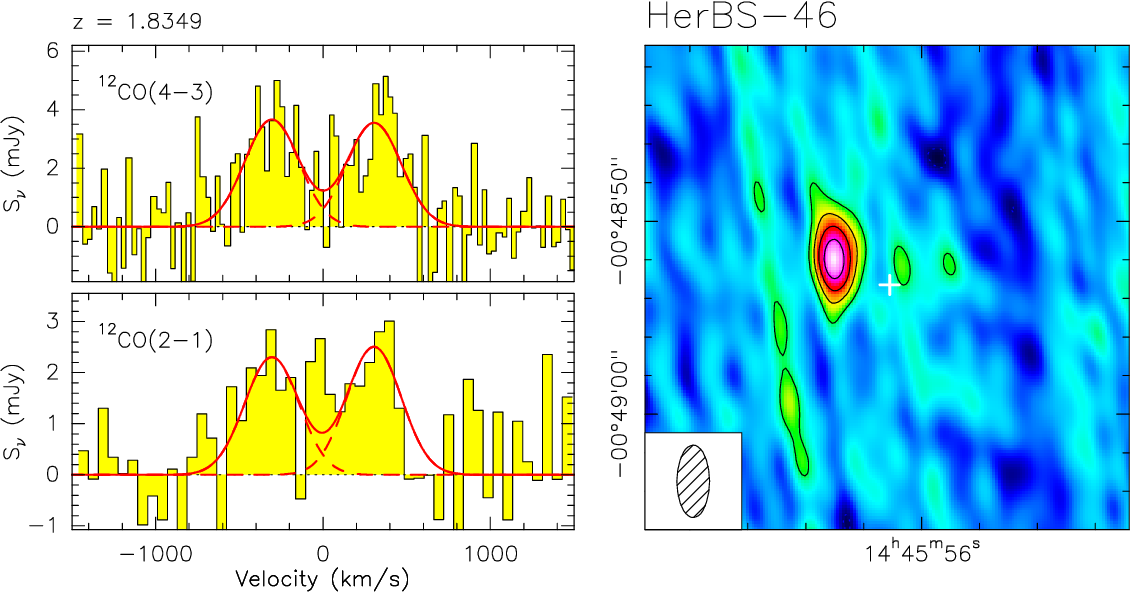}

\vspace{0.5cm}
\includegraphics[width=0.6\textwidth]{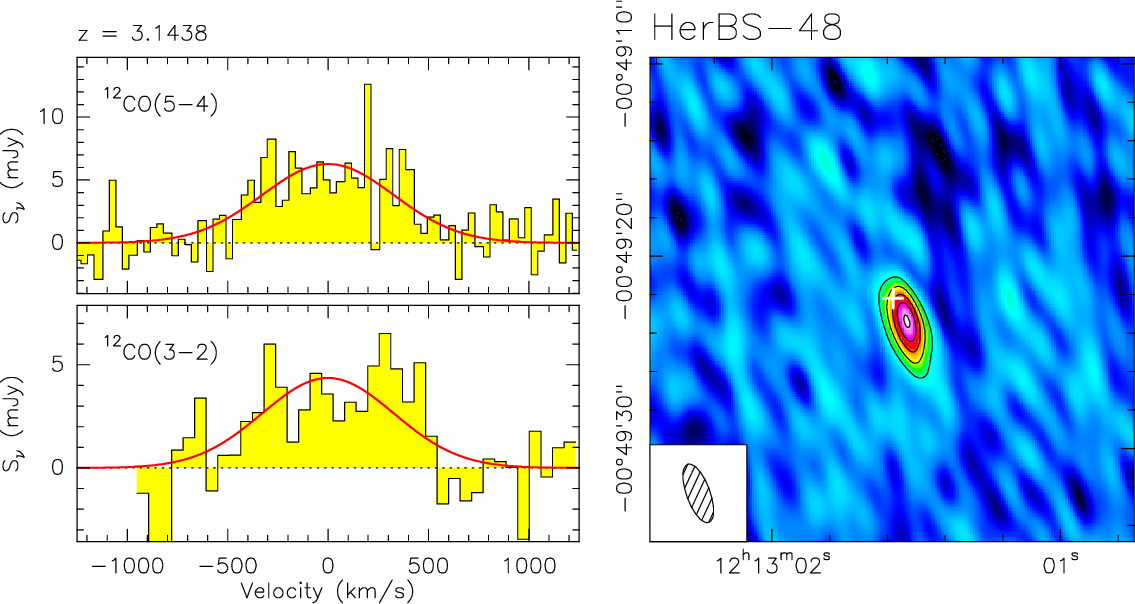}
   \caption{Combined continuum and emission lines maps of the $z$-GAL sources selected from the HerBS sample and spectra of each of the detected emission lines. Further details are provided in Appendix~\ref{Appendix: Presentation of Catalogue}.}
\label{figure:spectra_continuum_HerBS}%
    \end{figure*}
 \addtocounter{figure}{-1}
 
\begin{figure*}[!ht]
   \centering
\includegraphics[width=0.6\textwidth]{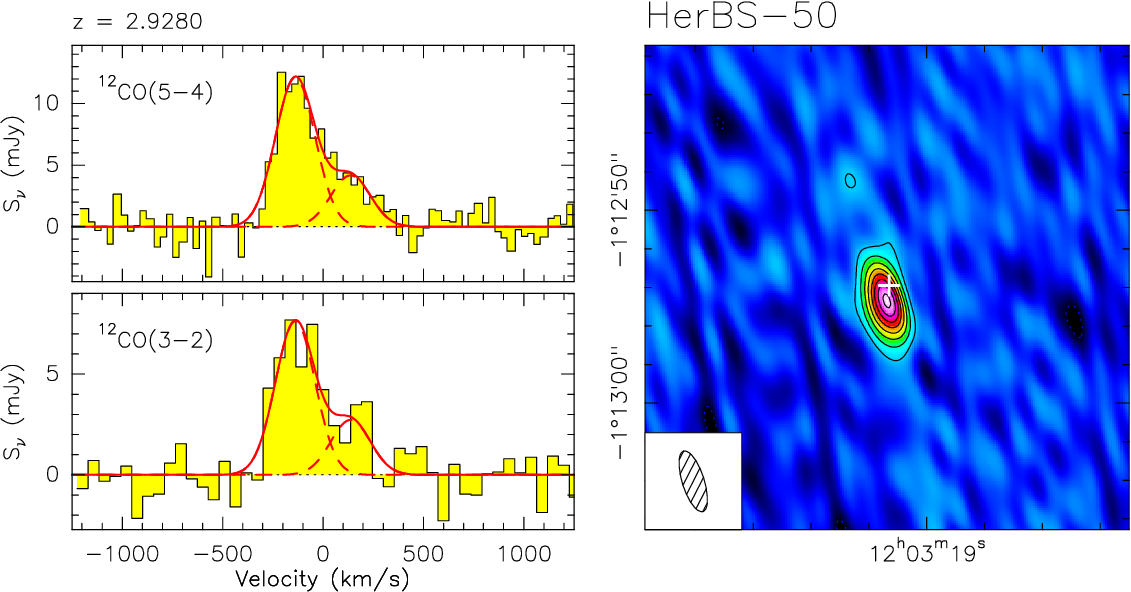}

\vspace{0.3cm}
\includegraphics[width=0.9\textwidth]{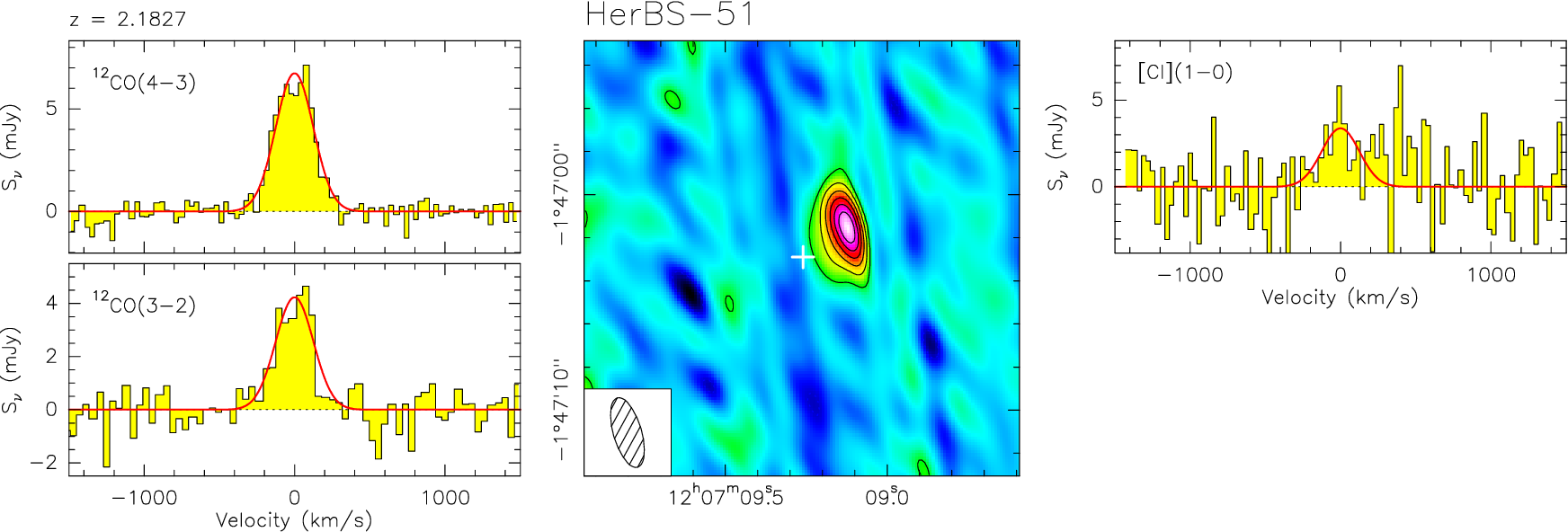}
 \includegraphics[width=0.9\textwidth]{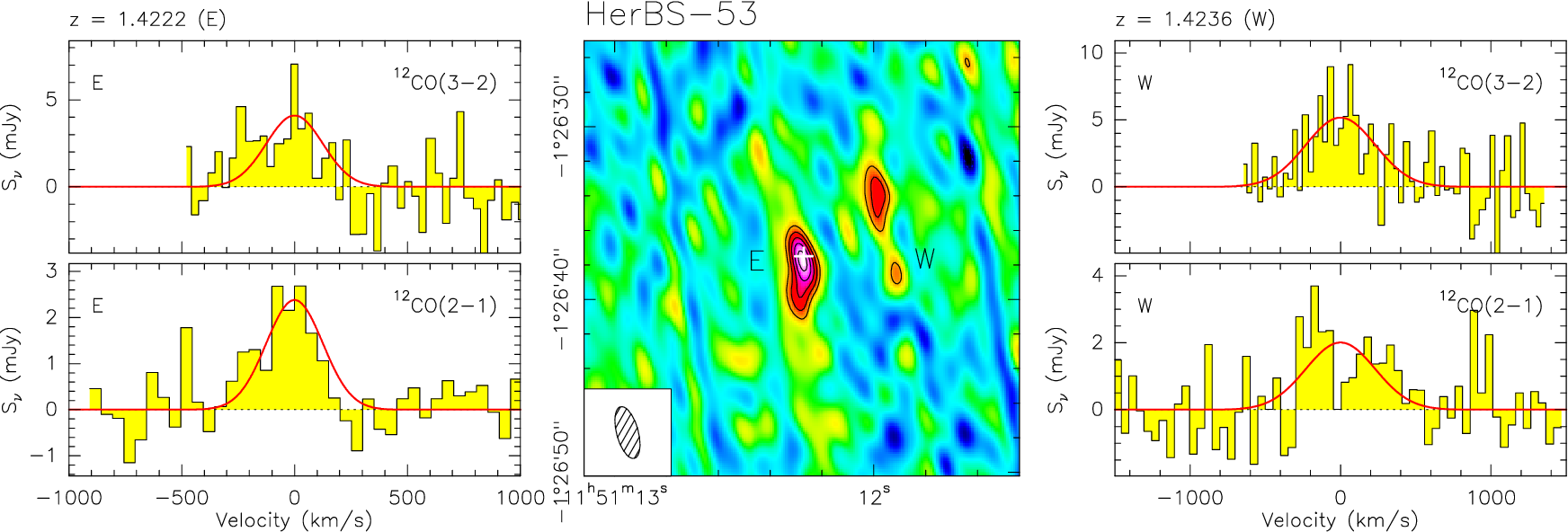}
 \includegraphics[width=0.9\textwidth]{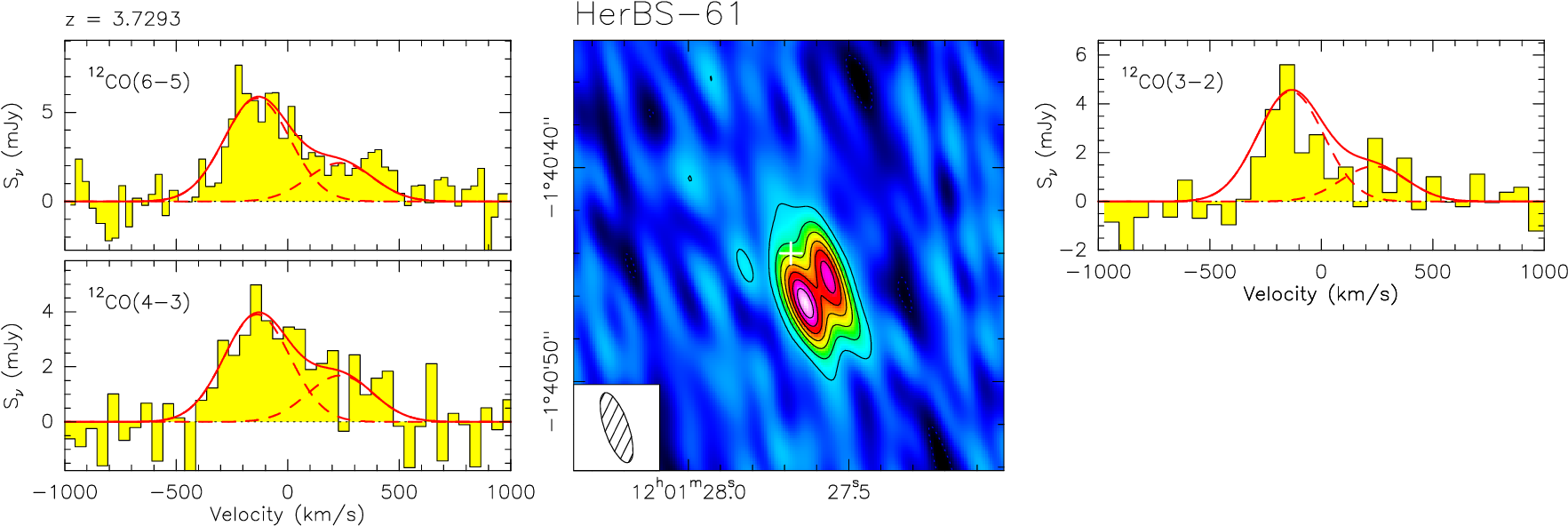}
   \caption{{\bf continued}}
    \end{figure*}
  \addtocounter{figure}{-1}

  \begin{figure*}[!ht]
   \centering
\includegraphics[width=0.6\textwidth]{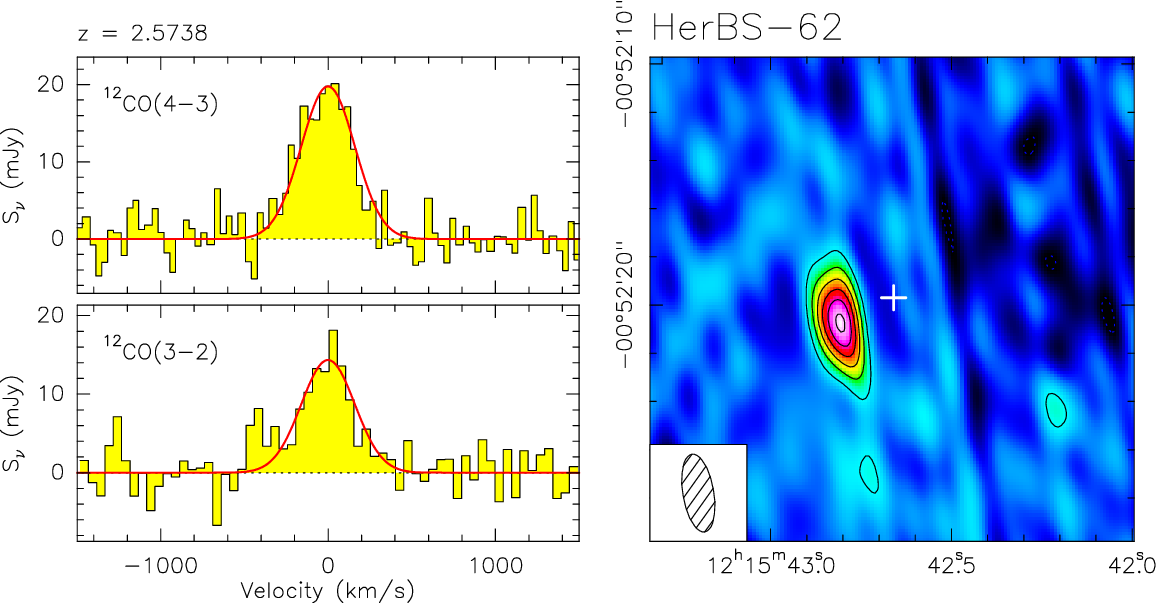}
 \includegraphics[width=0.6\textwidth]{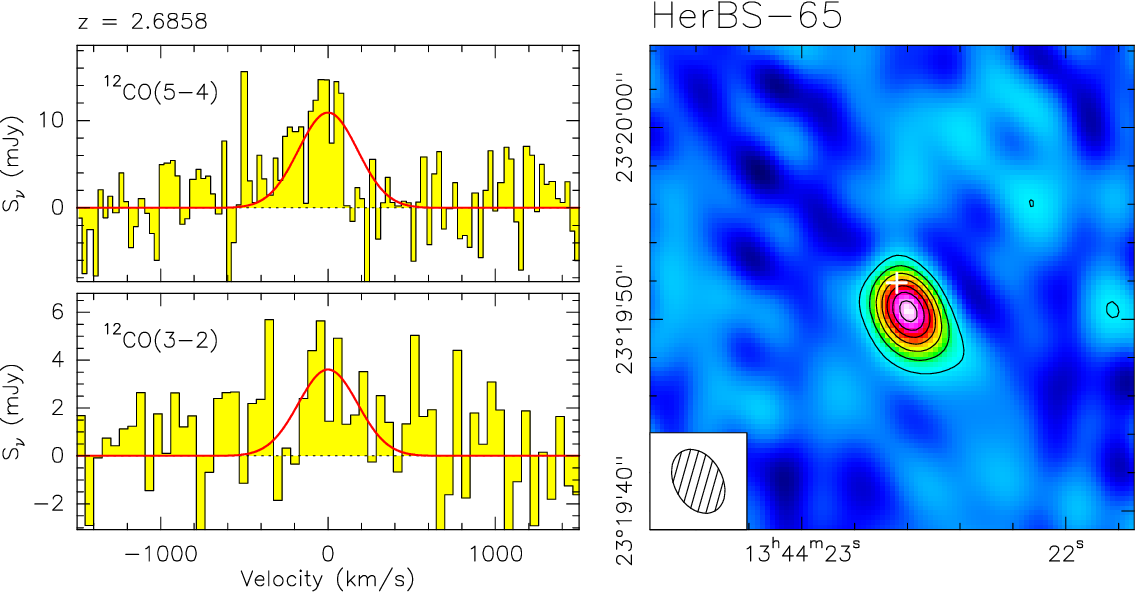}

 \vspace{0.4cm}
 \includegraphics[width=0.9\textwidth]{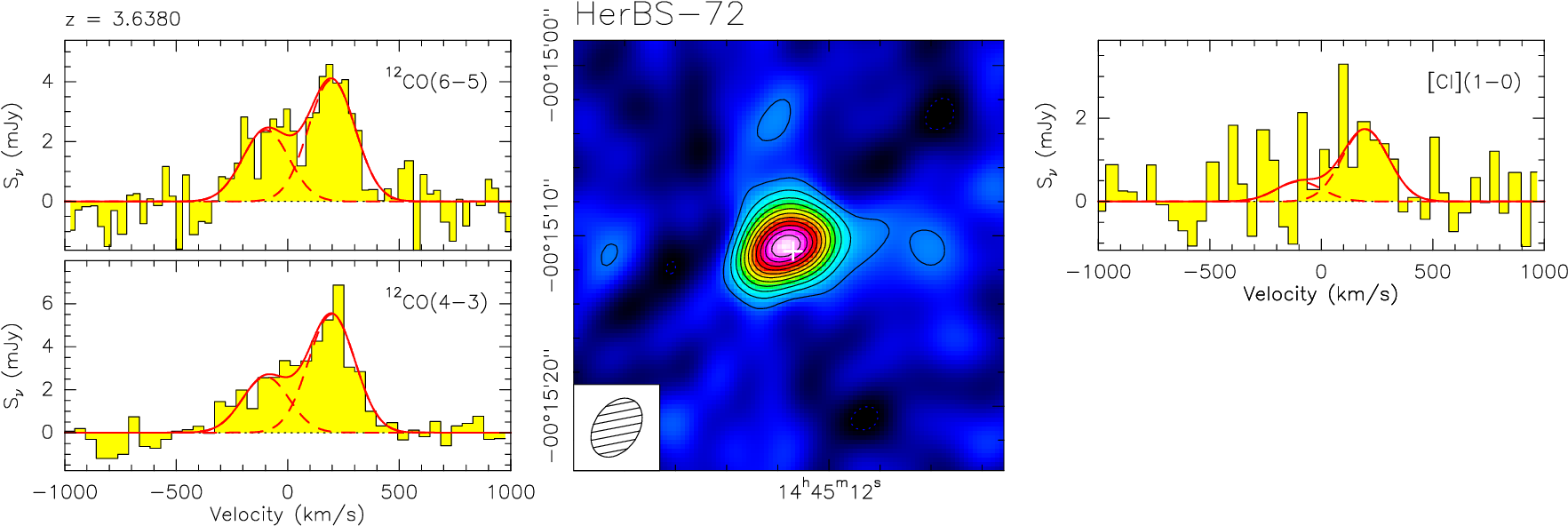}

 \vspace{0.4cm}
 \includegraphics[width=0.6\textwidth]{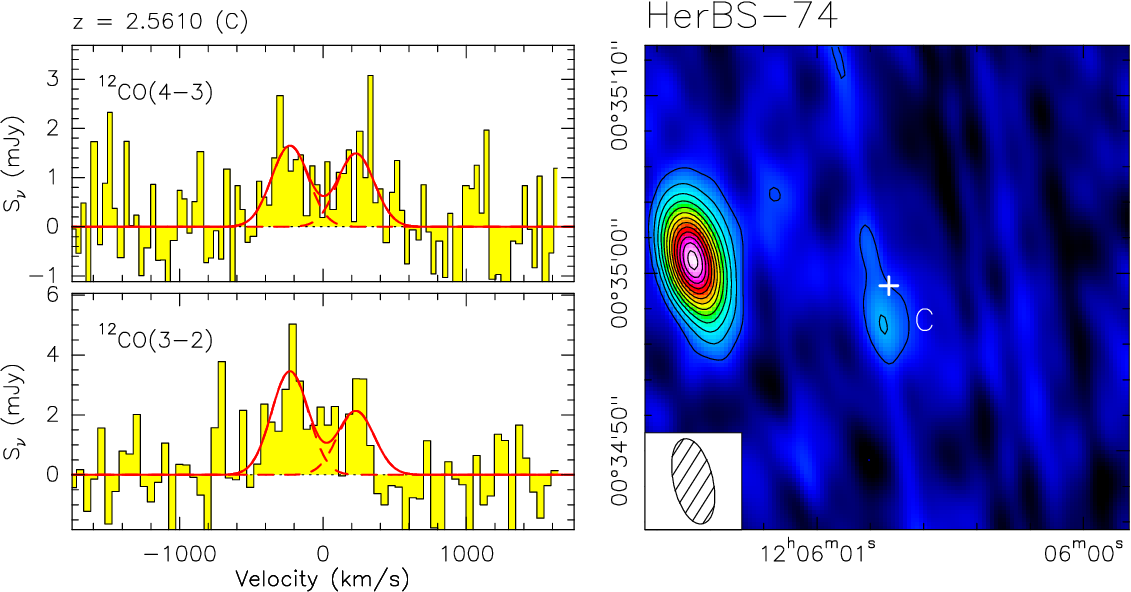}
   \caption{{\bf continued}}
    \end{figure*}
  \addtocounter{figure}{-1}

  \begin{figure*}[!ht]
   \centering
 \includegraphics[width=0.9\textwidth]{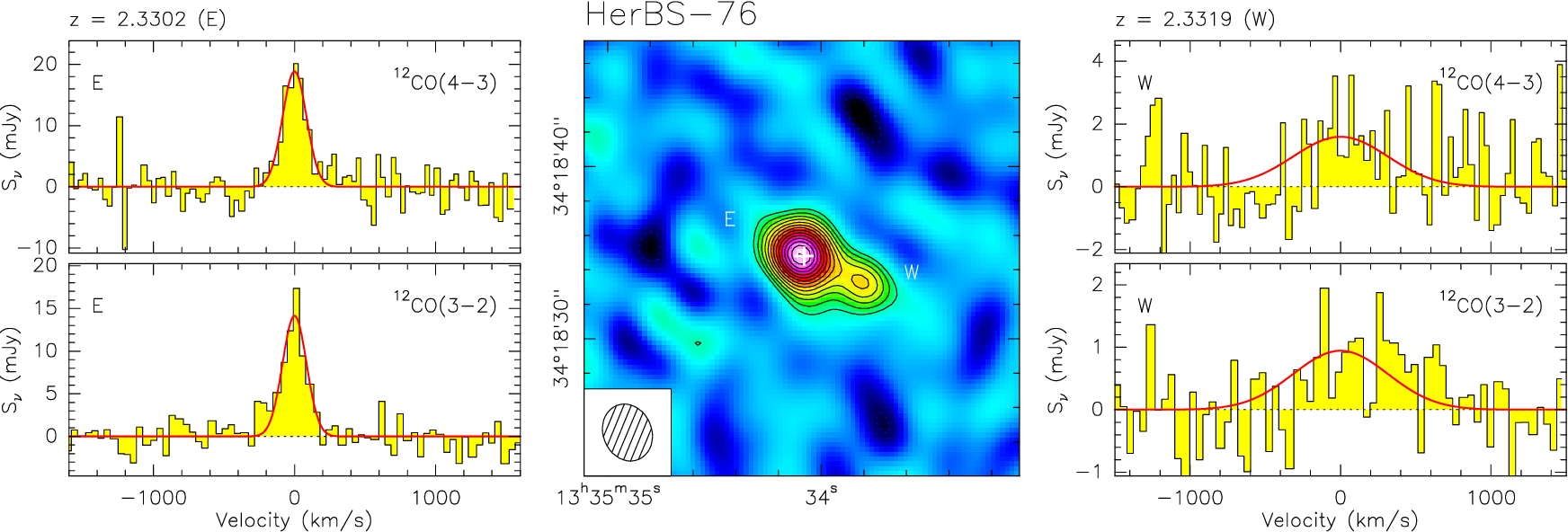}
 \includegraphics[width=0.9\textwidth]{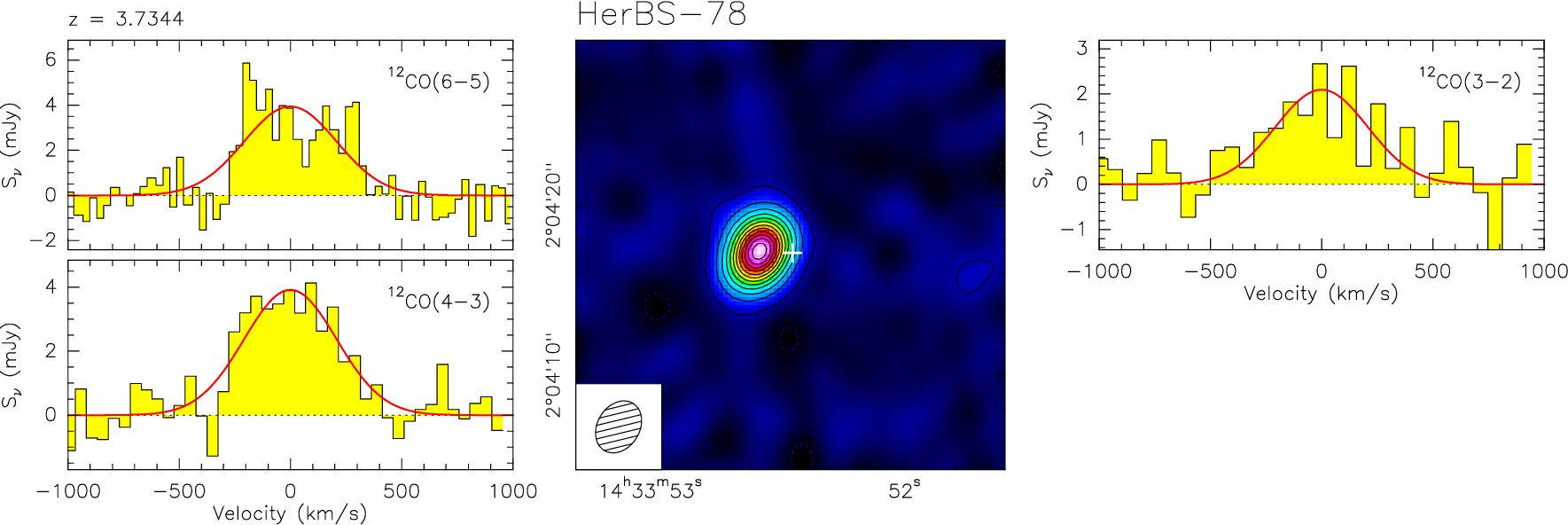}
 \includegraphics[width=0.9\textwidth]{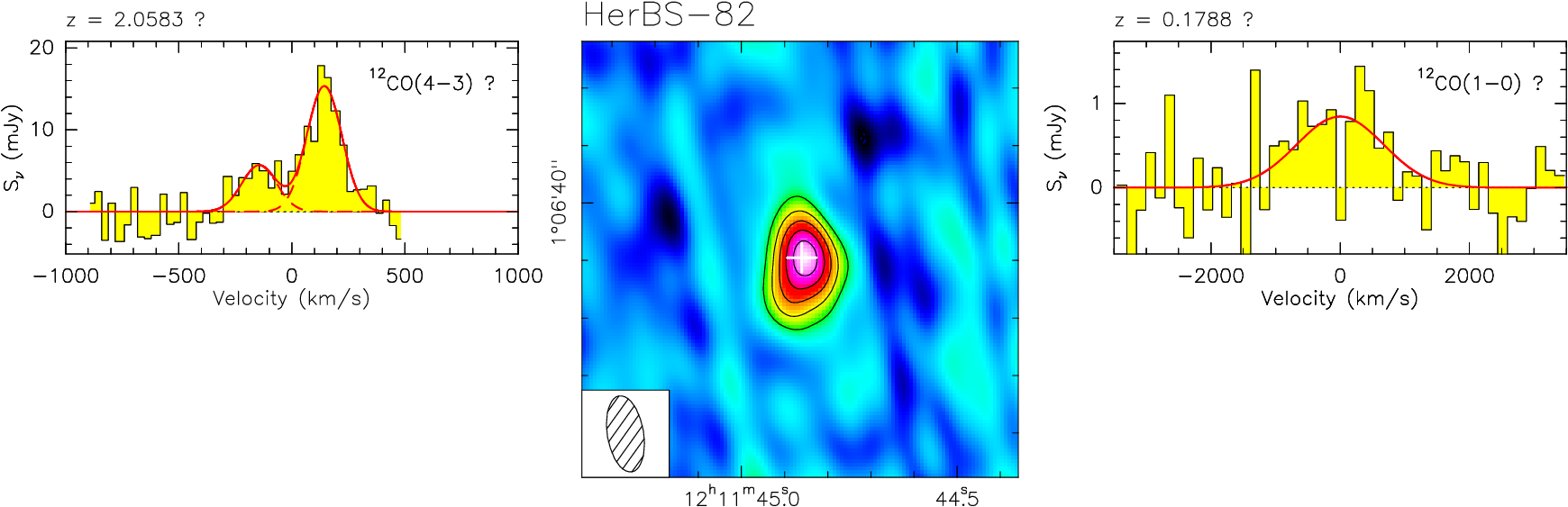}

 \vspace{0.3cm}
 \includegraphics[width=0.9\textwidth]{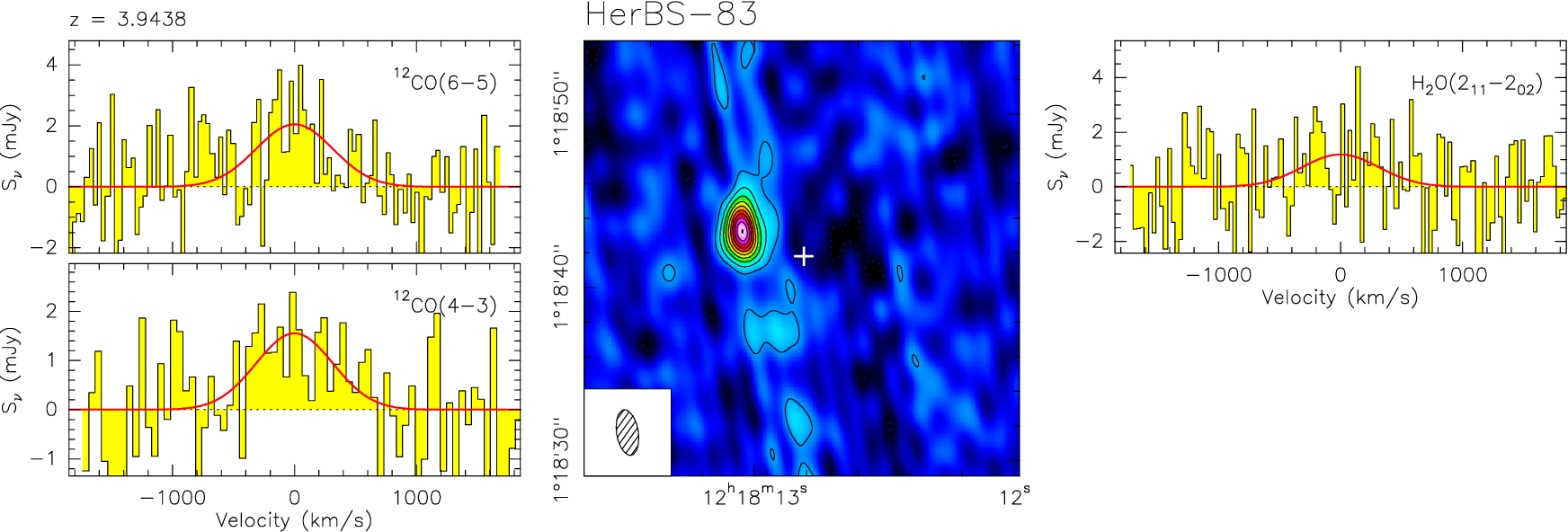}
   \caption{{\bf continued}}
    \end{figure*}
  \addtocounter{figure}{-1}

\begin{figure*}[!ht]
   \centering
 \includegraphics[width=0.9\textwidth]{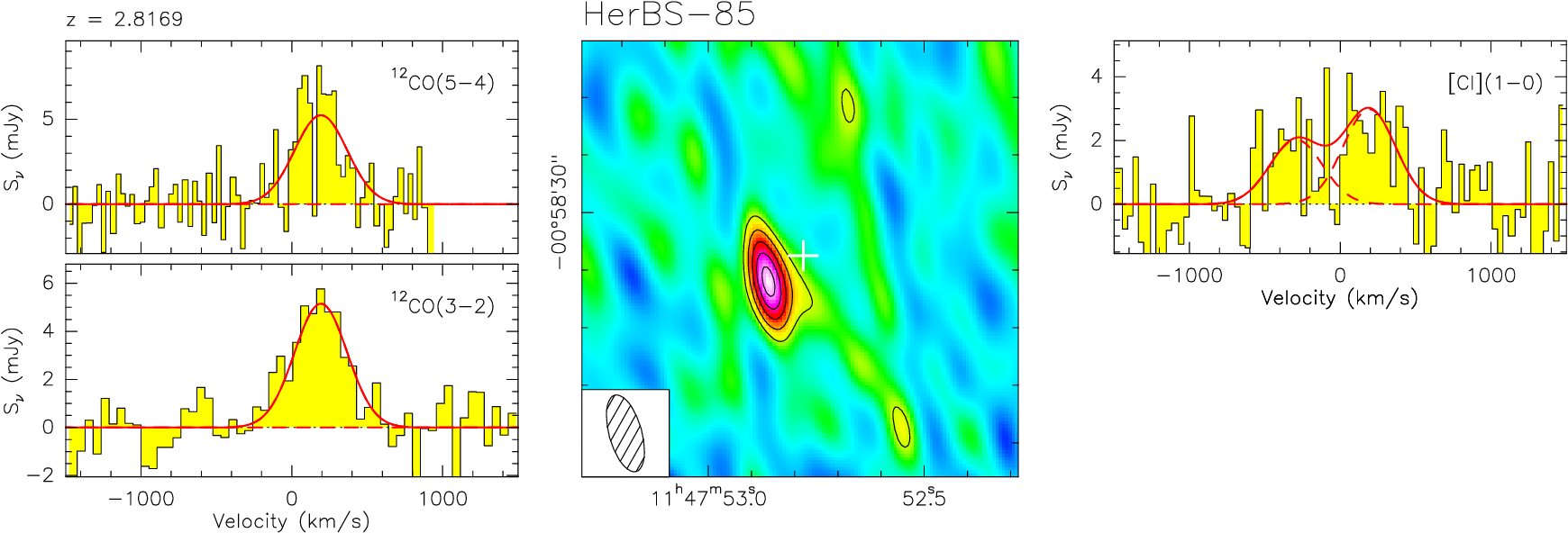}

 \vspace{1.2cm}
\includegraphics[width=0.9\textwidth]{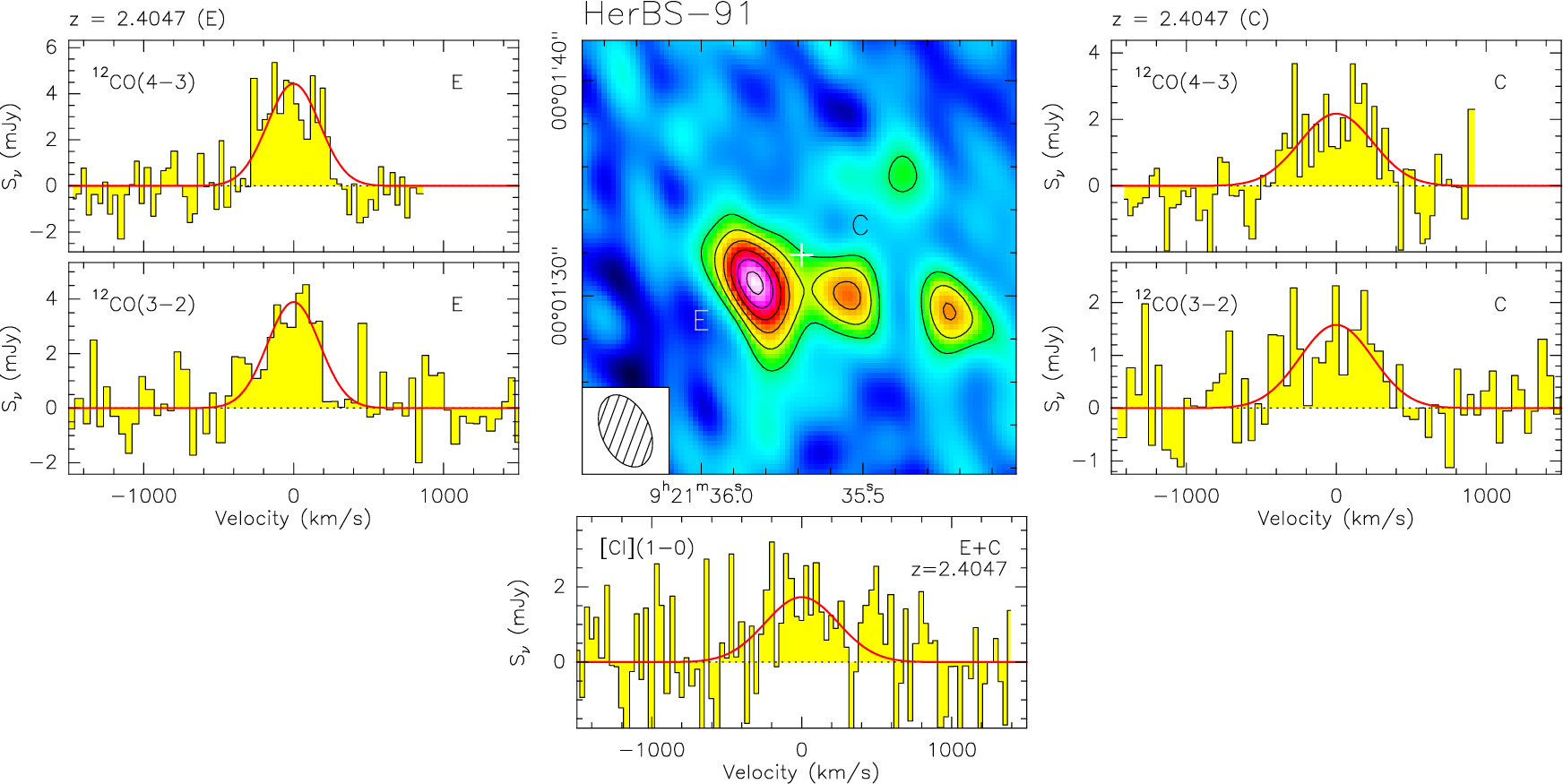}

\vspace{1.2cm}
\includegraphics[width=0.6\textwidth]{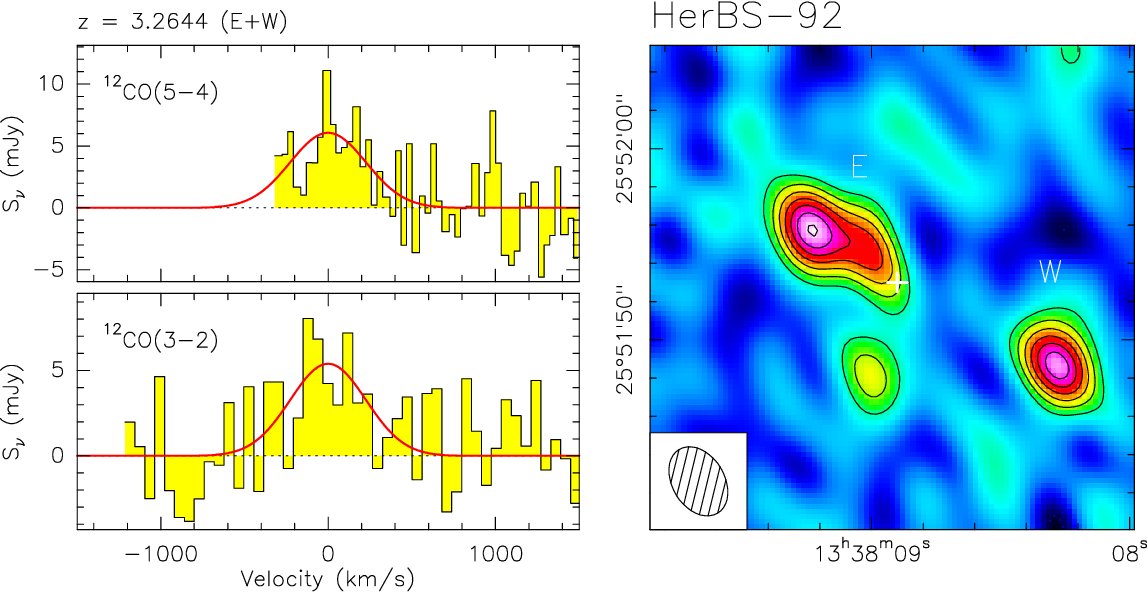}
   \caption{\bf{continued}}

    \end{figure*}
 \addtocounter{figure}{-1}

\begin{figure*}[!ht]
   \centering
\includegraphics[width=0.6\textwidth]{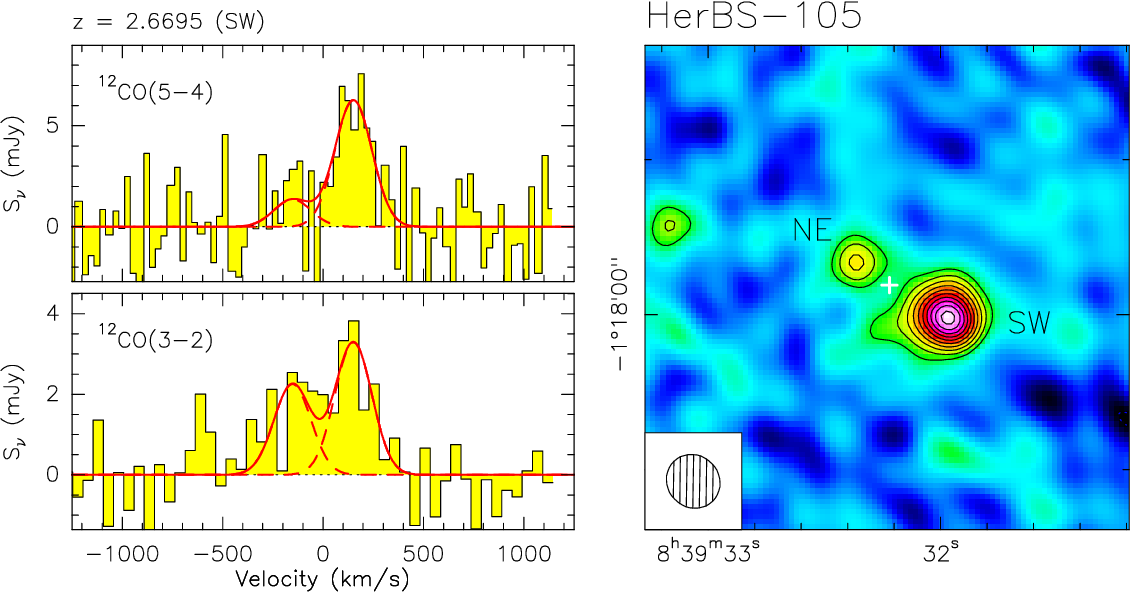}

\vspace{1.0cm}
\includegraphics[width=0.9\textwidth]{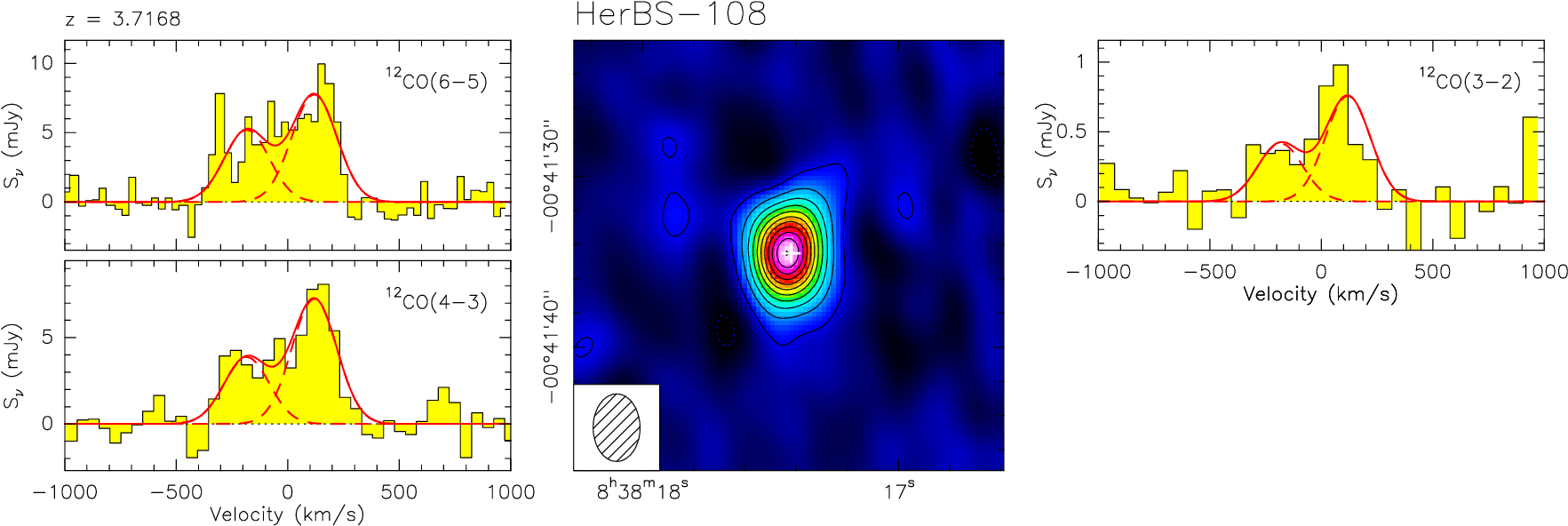}

\vspace{1.0cm}
\includegraphics[width=0.9\textwidth]{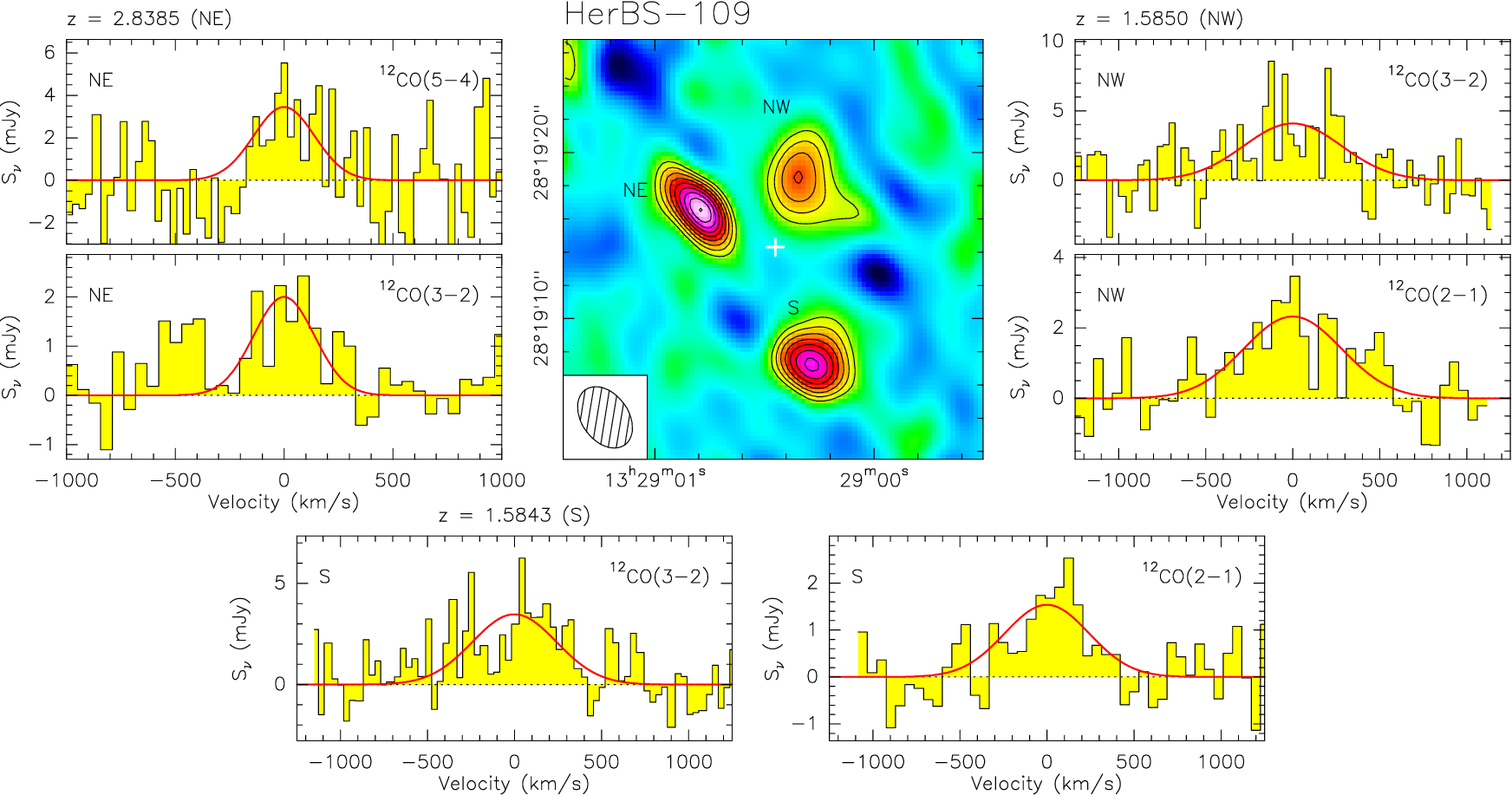}

   \caption{\bf{continued}}
    \end{figure*}
 \addtocounter{figure}{-1}

\begin{figure*}[!ht]
   \centering
\includegraphics[width=0.6\textwidth]{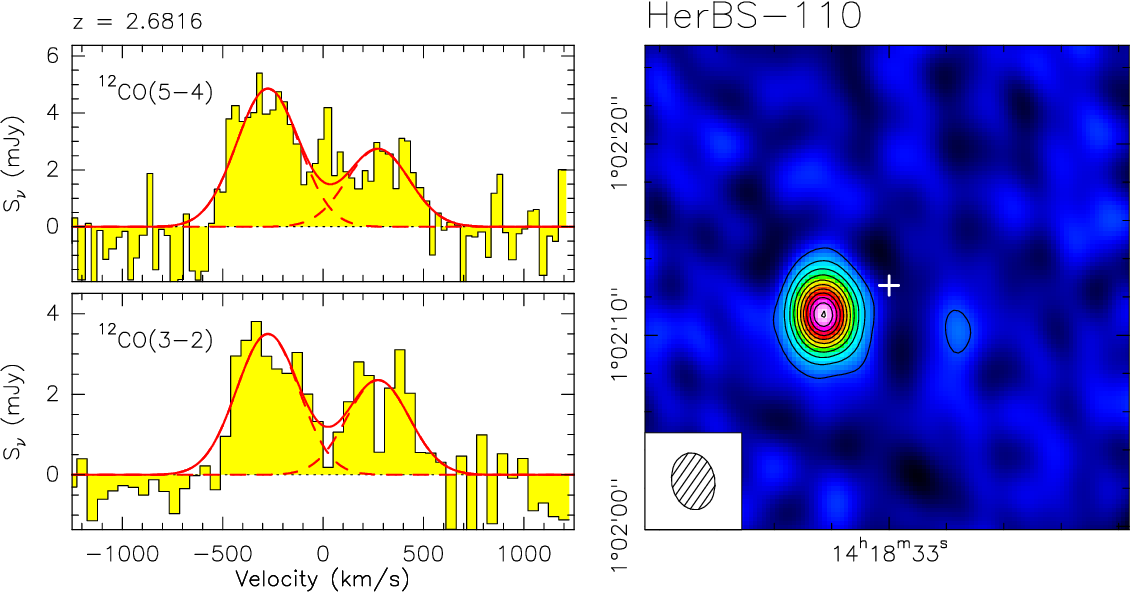}

\vspace{0.4cm}
\includegraphics[width=0.9\textwidth]{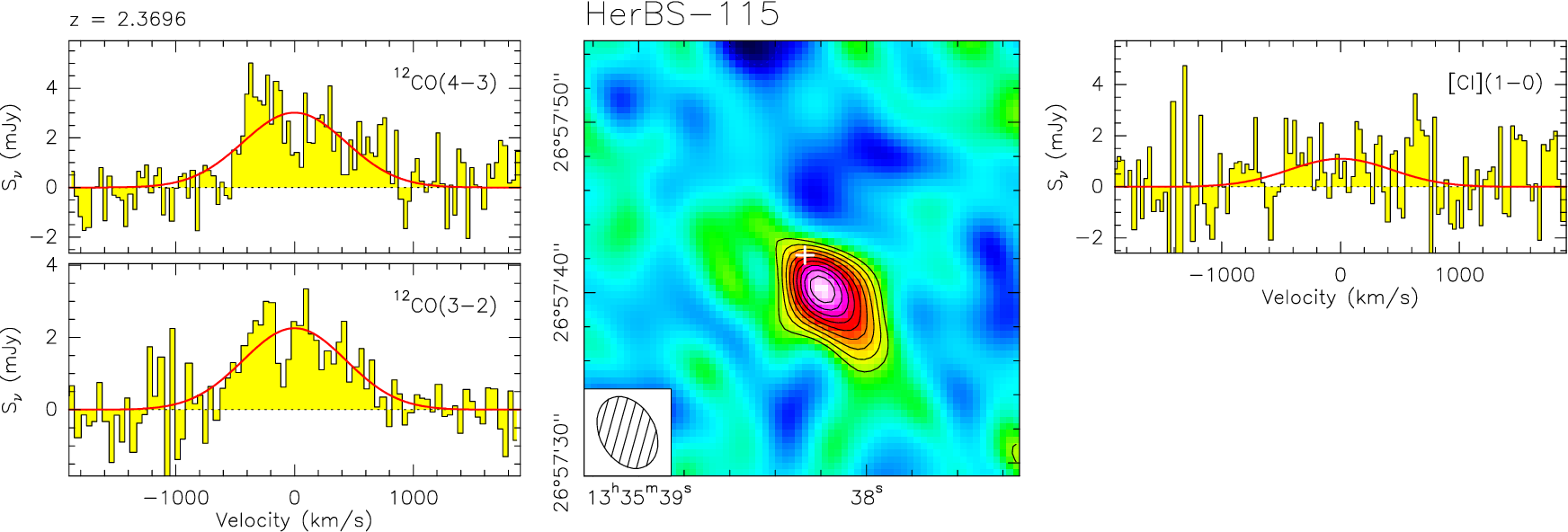}

\vspace{0.4cm}
\includegraphics[width=0.6\textwidth]{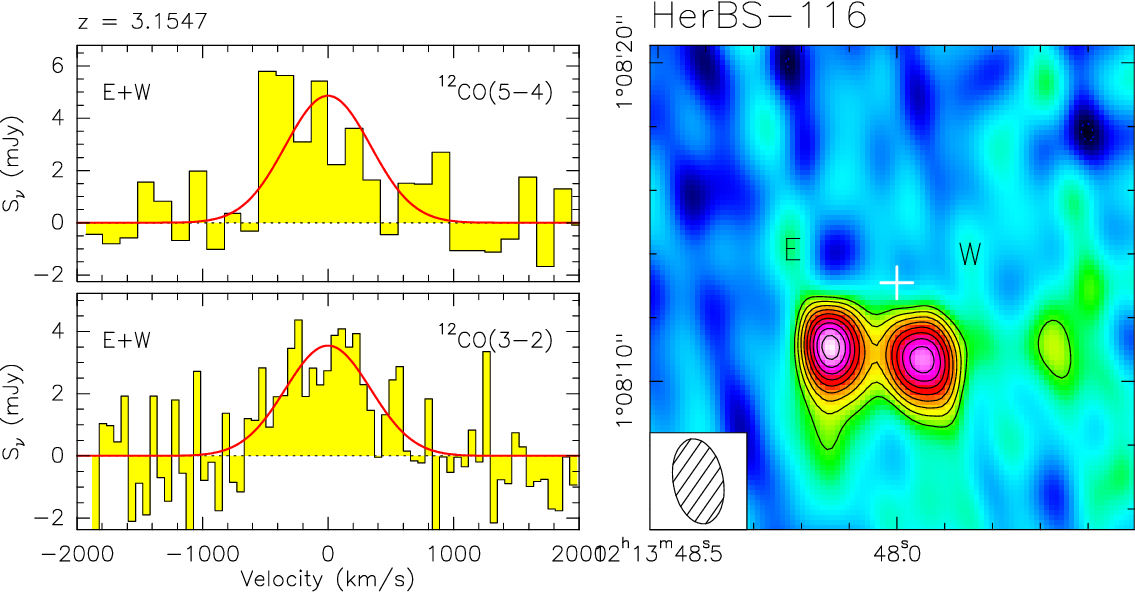}

\vspace{0.4cm}
\includegraphics[width=0.9\textwidth]{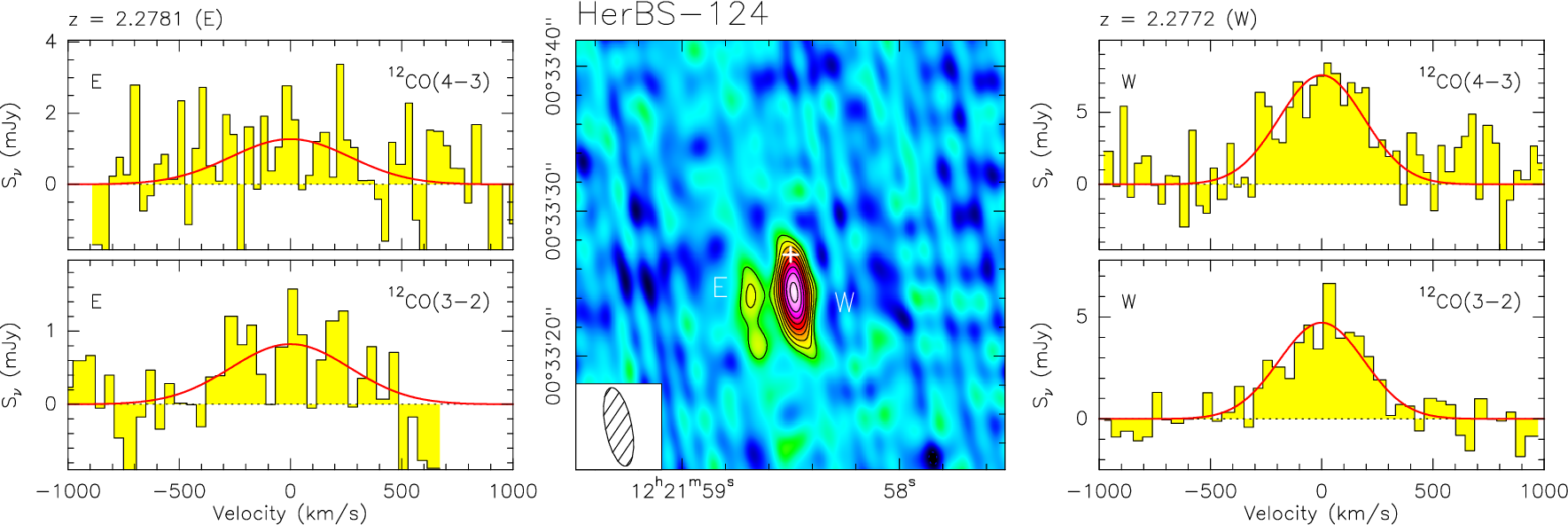}
   \caption{\bf{continued}}
    \end{figure*}
 \addtocounter{figure}{-1}
 
\begin{figure*}[!ht]
   \centering
\includegraphics[width=0.6\textwidth]{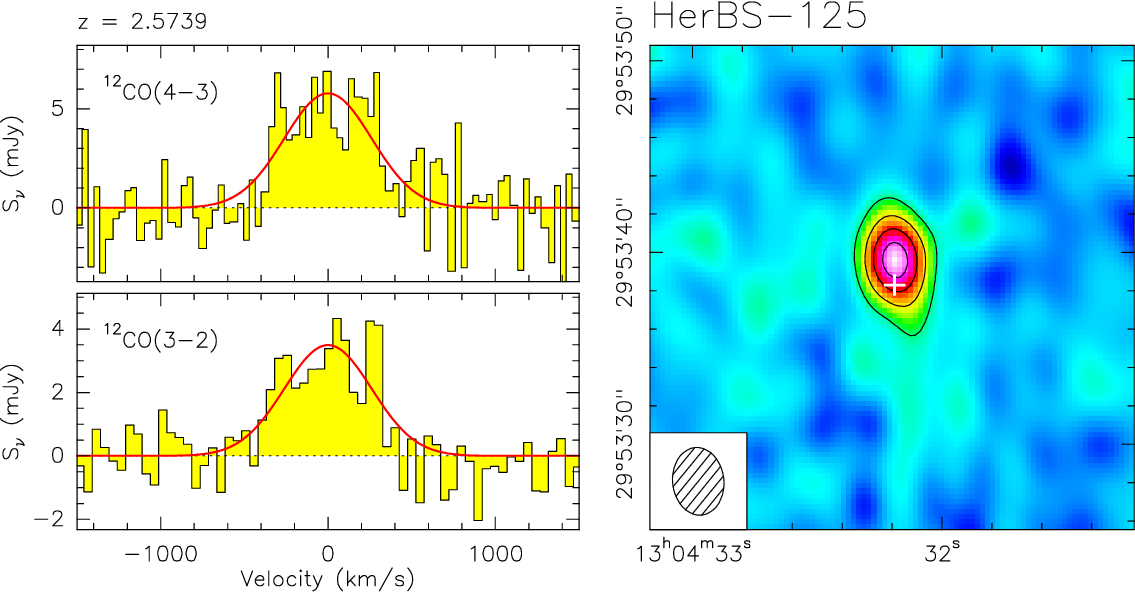}
\includegraphics[width=0.6\textwidth]{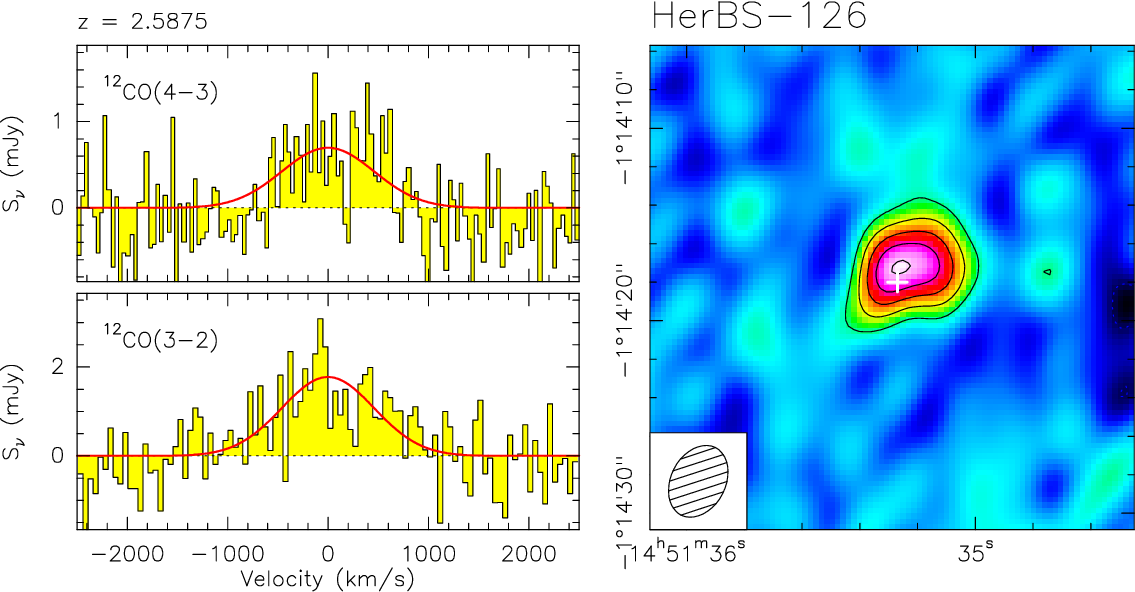}
 \includegraphics[width=0.6\textwidth]{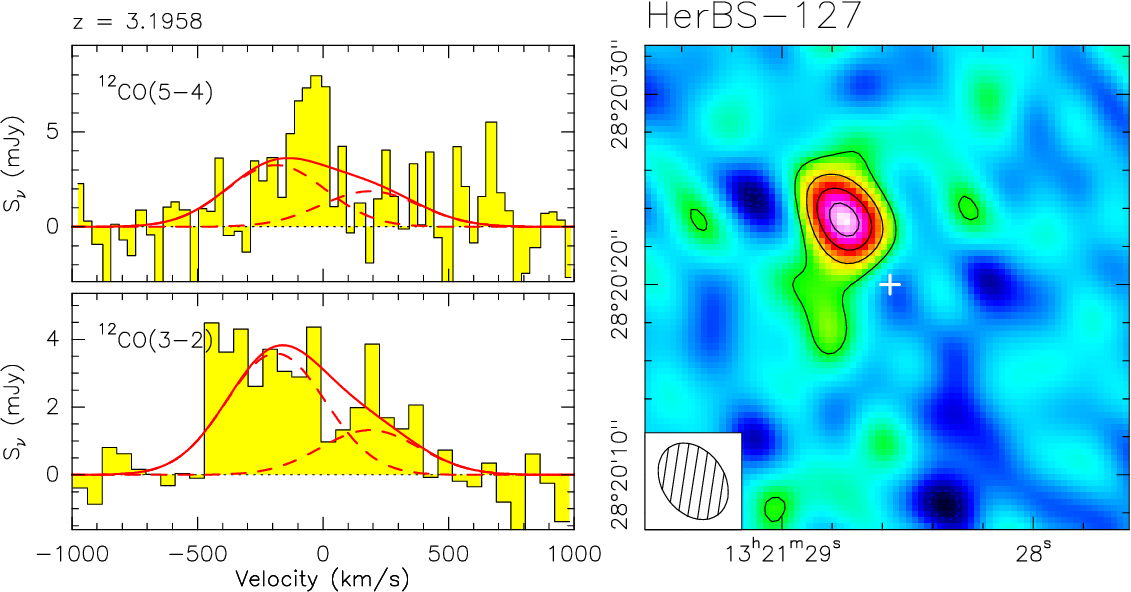}
 \includegraphics[width=0.6\textwidth]{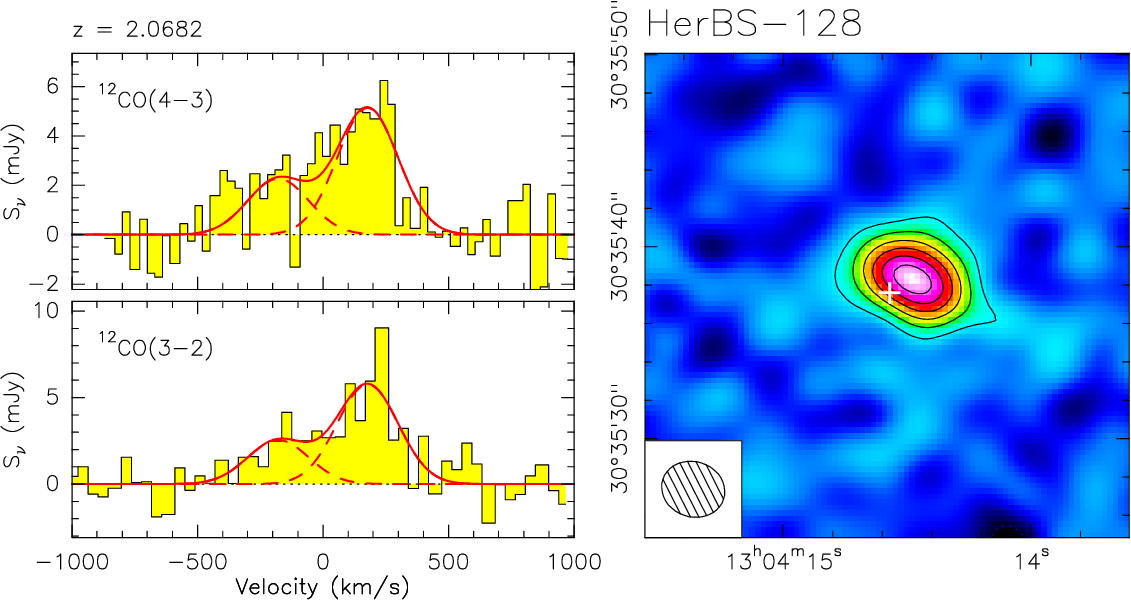}
   \caption{\bf{continued}}
    \end{figure*}
 \addtocounter{figure}{-1}
 
\begin{figure*}[!ht]
   \centering
\includegraphics[width=0.6\textwidth]{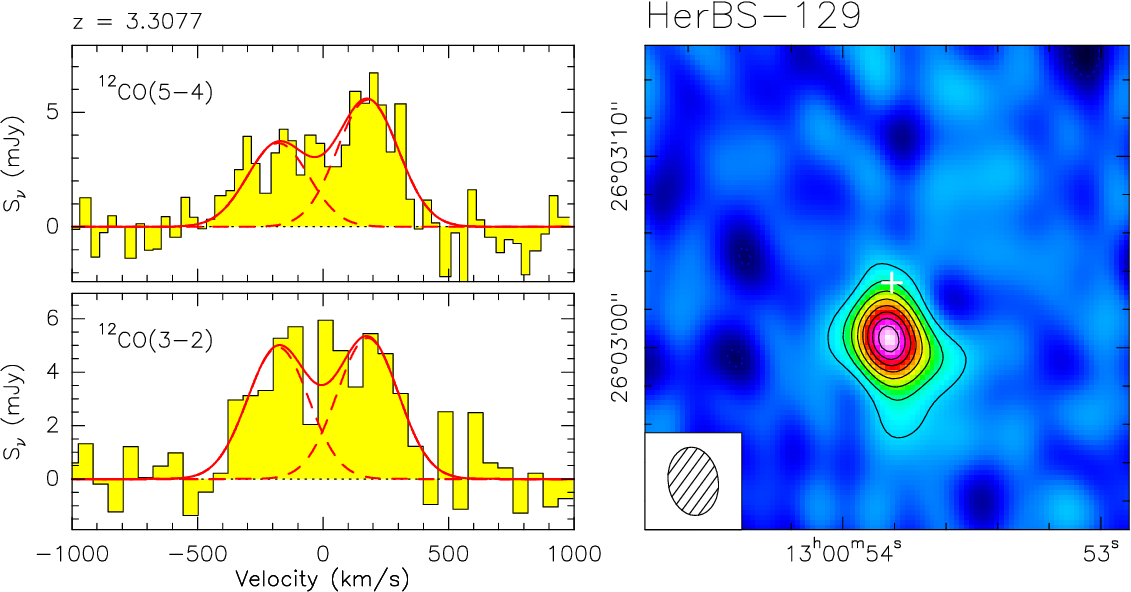} 
\includegraphics[width=0.6\textwidth]{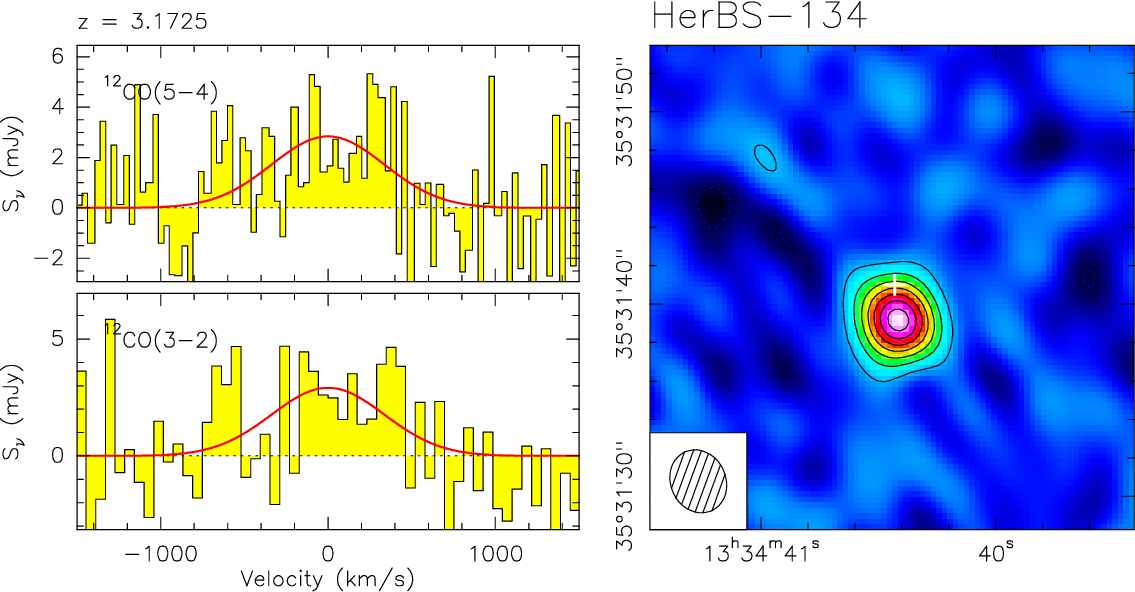}
\includegraphics[width=0.6\textwidth]{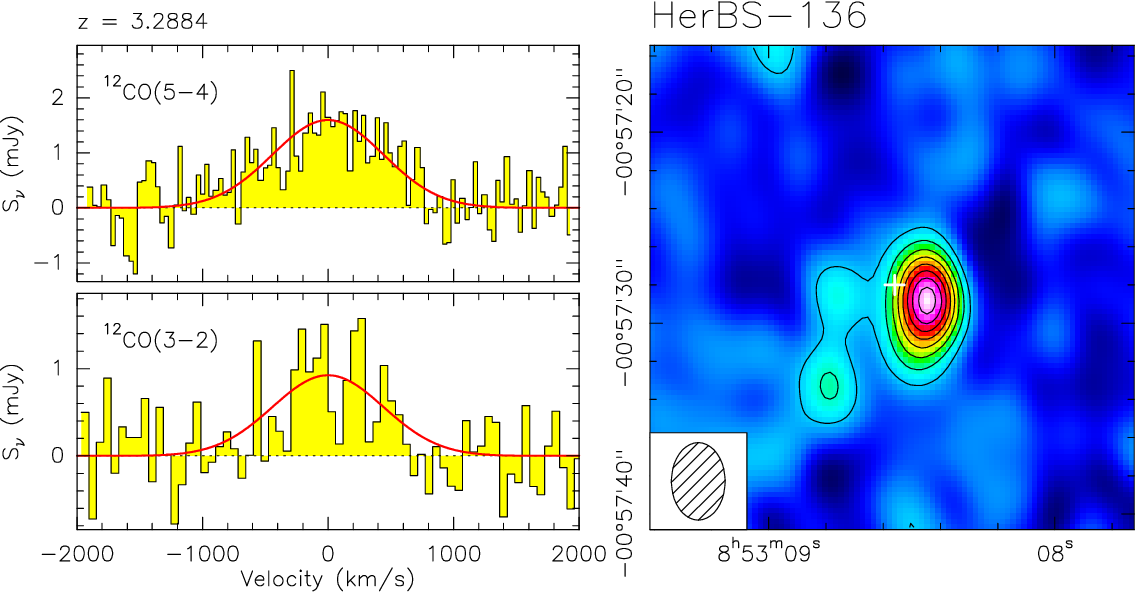}
\includegraphics[width=0.6\textwidth]{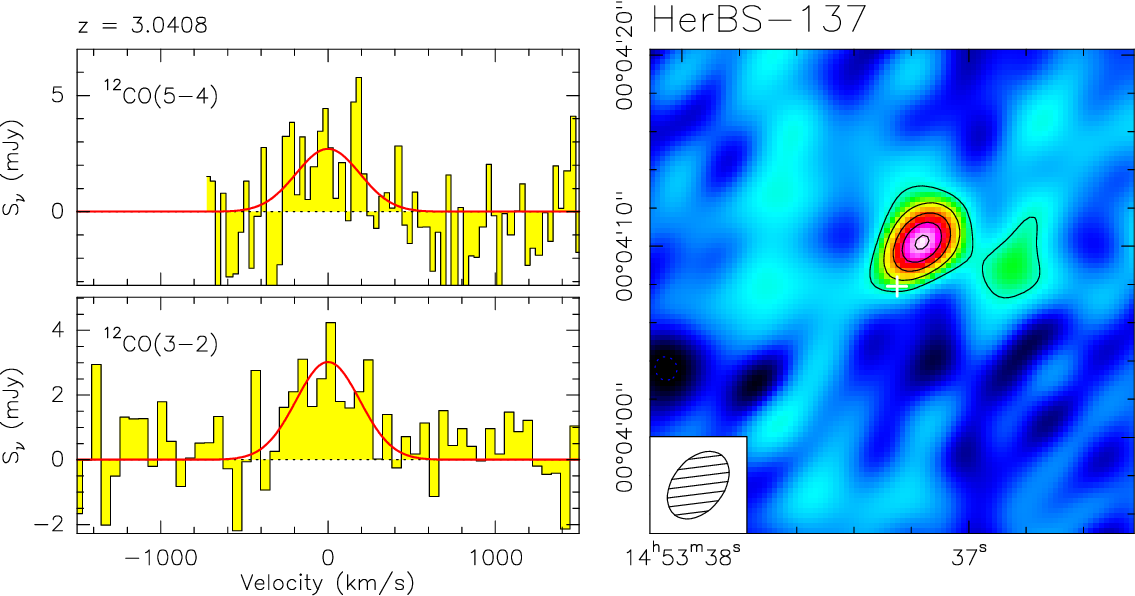}
   \caption{\bf{continued}}
    \end{figure*}
 \addtocounter{figure}{-1}

\begin{figure*}[!ht]
   \centering
\includegraphics[width=0.6\textwidth]{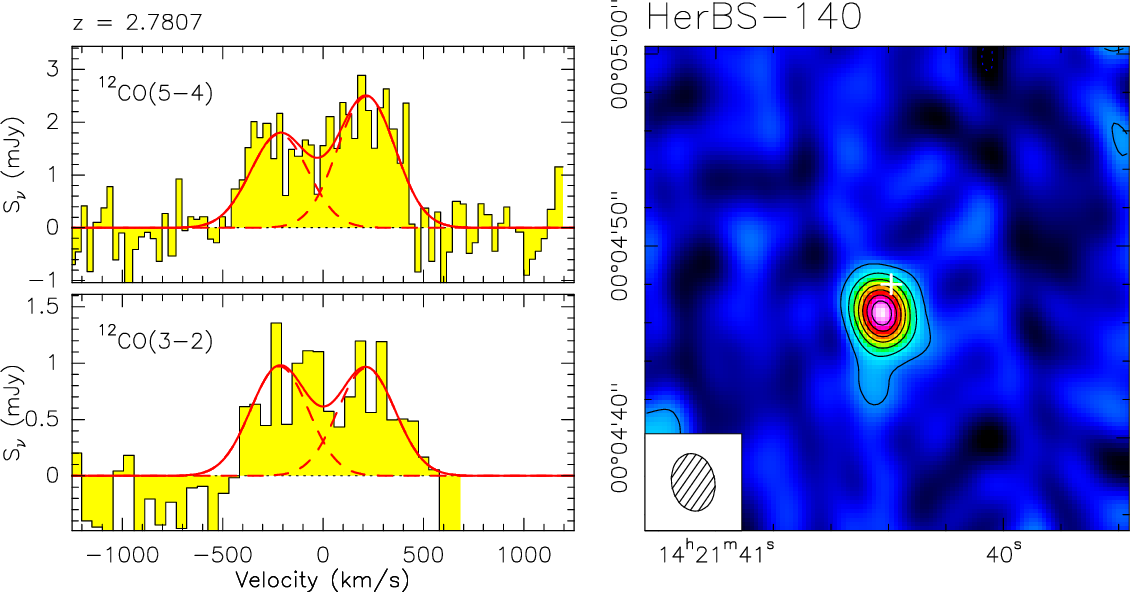}

\vspace{0.4cm}
\includegraphics[width=0.9\textwidth]{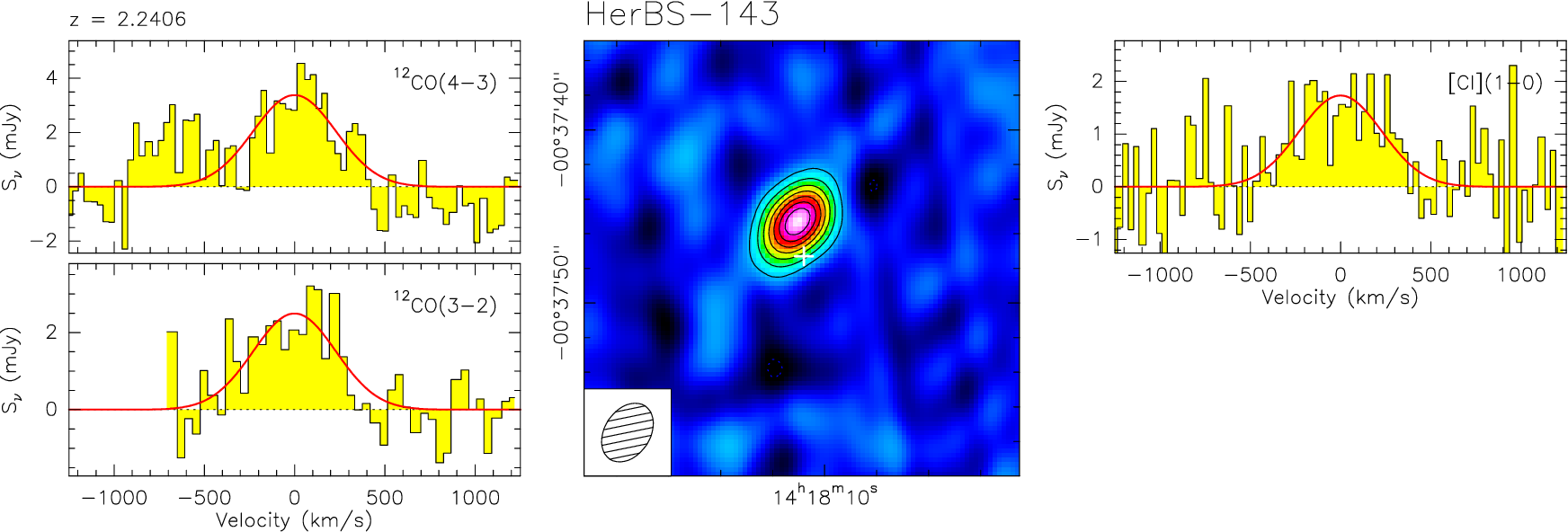}

\vspace{0.4cm}
\includegraphics[width=0.6\textwidth]{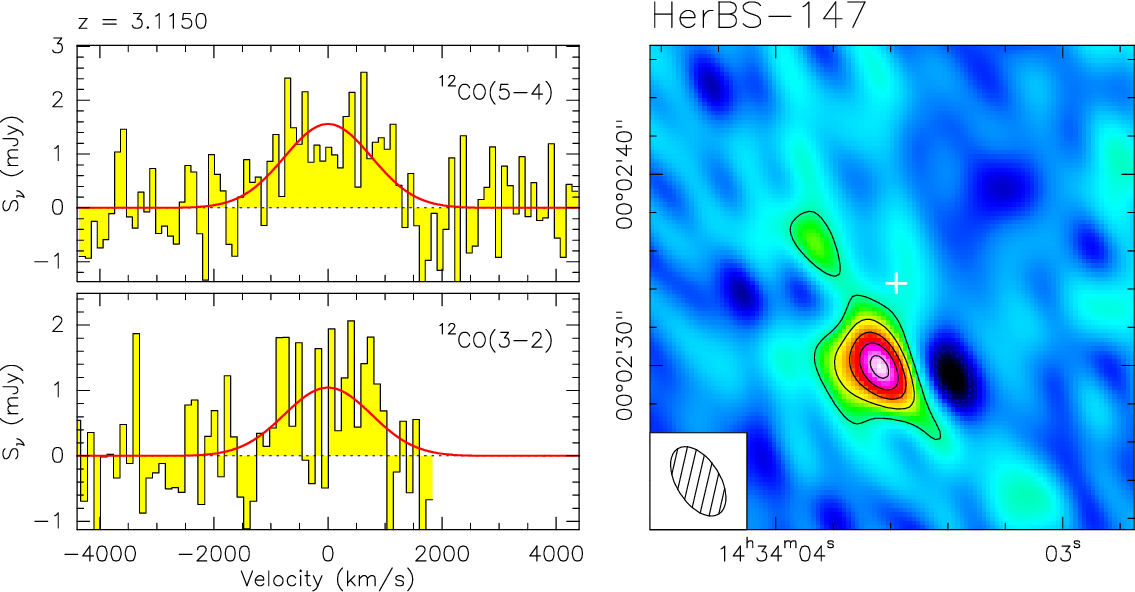}

\vspace{0.4cm}
\includegraphics[width=0.6\textwidth]{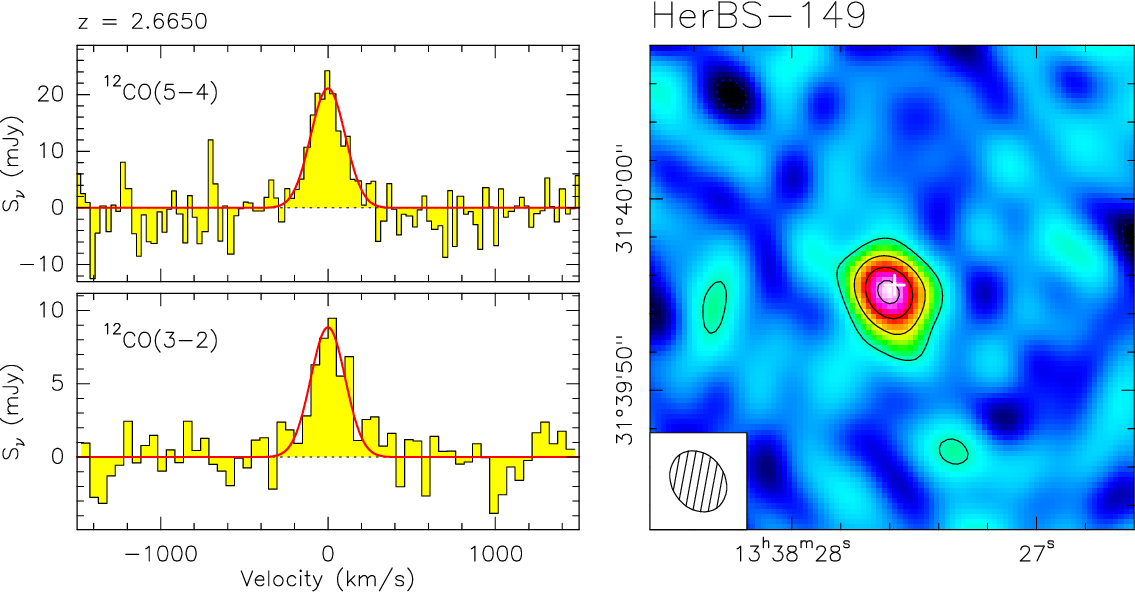}
   \caption{\bf{continued}}
    \end{figure*}
 \addtocounter{figure}{-1}

\begin{figure*}[!ht]
   \centering
\includegraphics[width=0.9\textwidth]{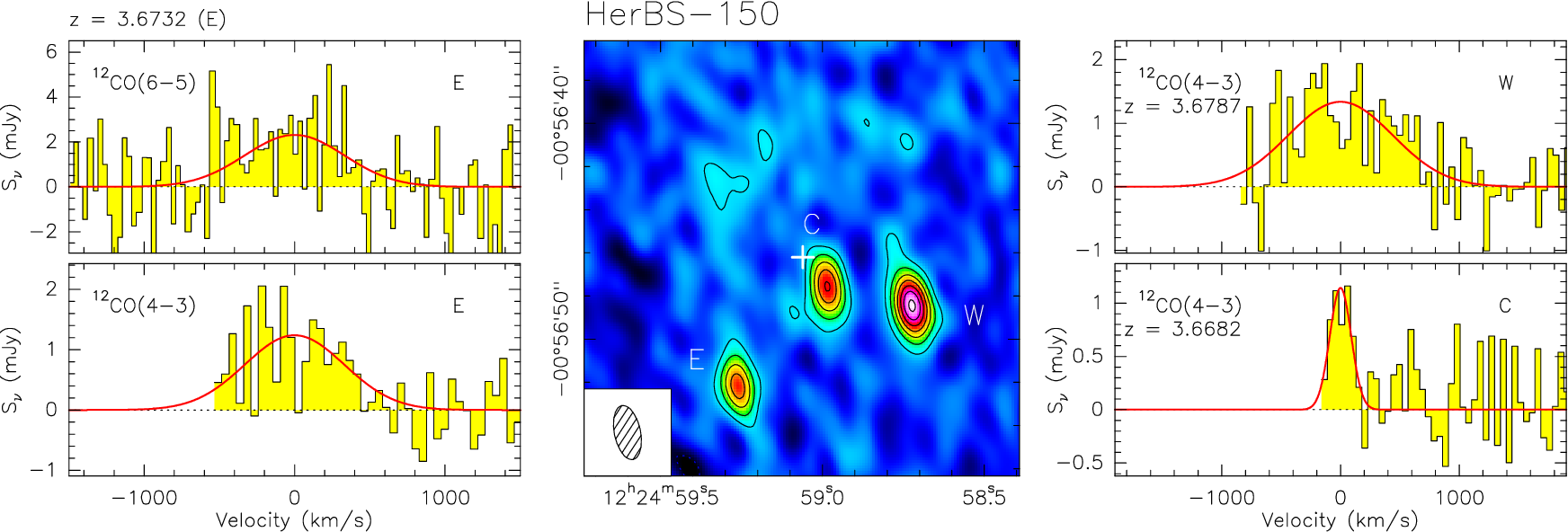}

\vspace{0.4cm}
\includegraphics[width=0.6\textwidth]{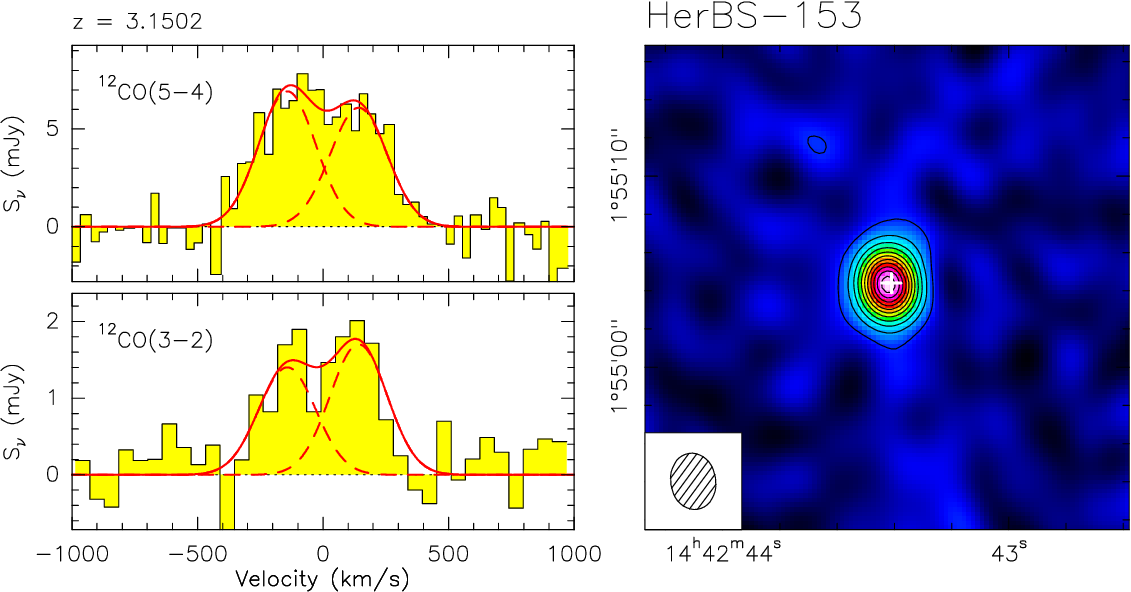}
\includegraphics[width=0.6\textwidth]{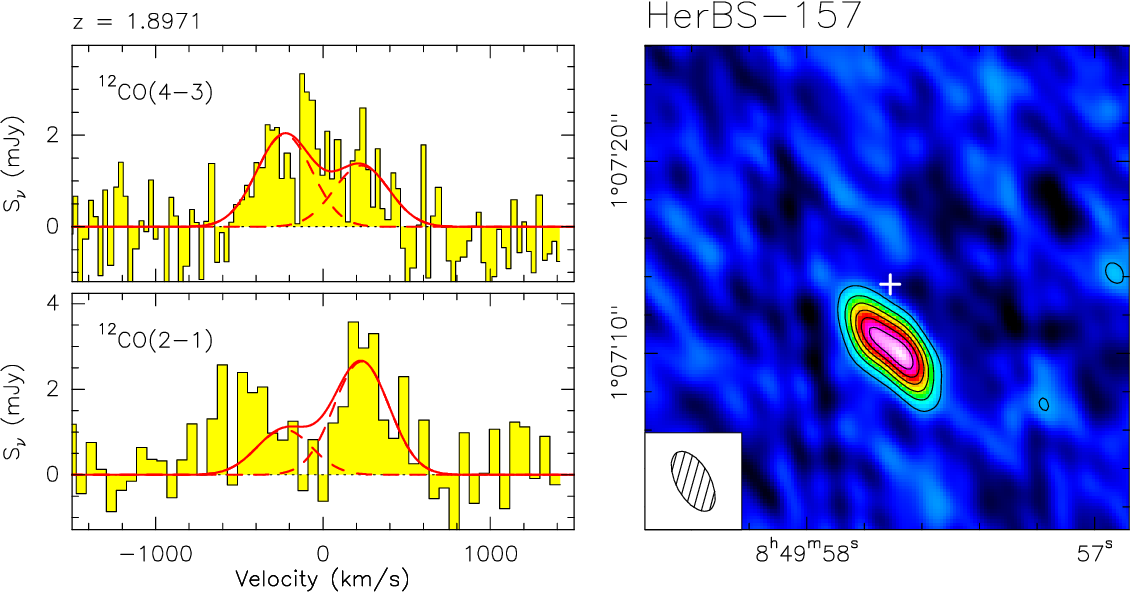}

\vspace{0.6cm}
\includegraphics[width=0.9\textwidth]{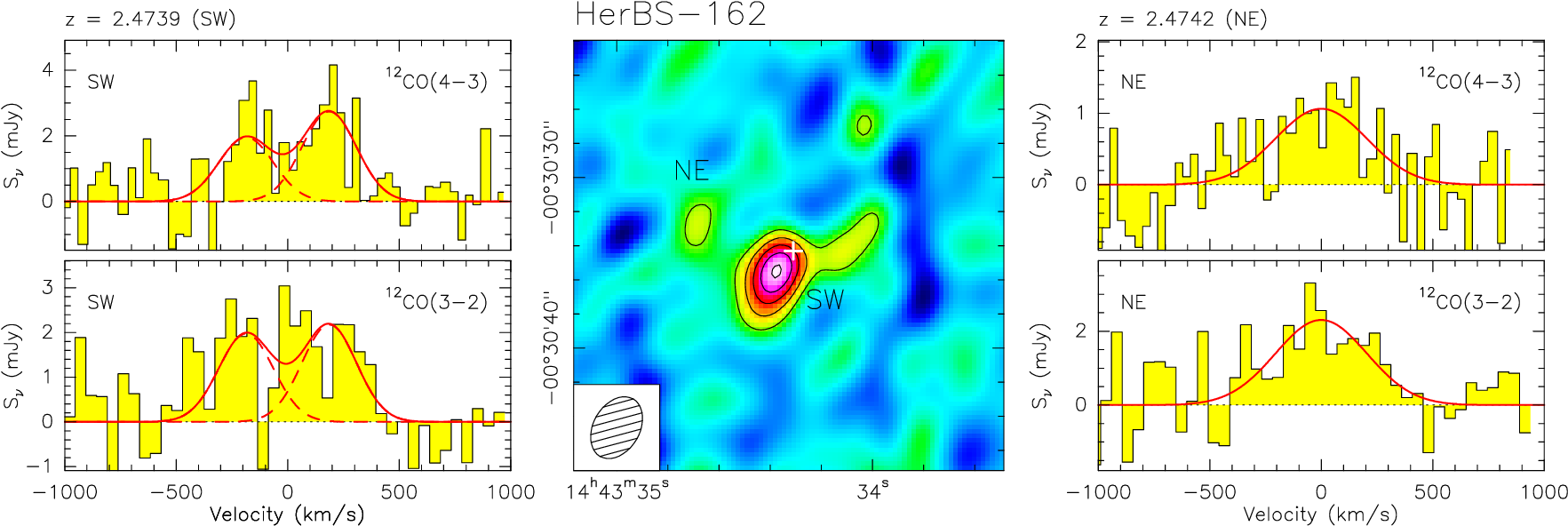}
   \caption{\bf{continued}}
    \end{figure*}
 \addtocounter{figure}{-1}

\begin{figure*}[!ht]
   \centering
\includegraphics[width=0.6\textwidth]{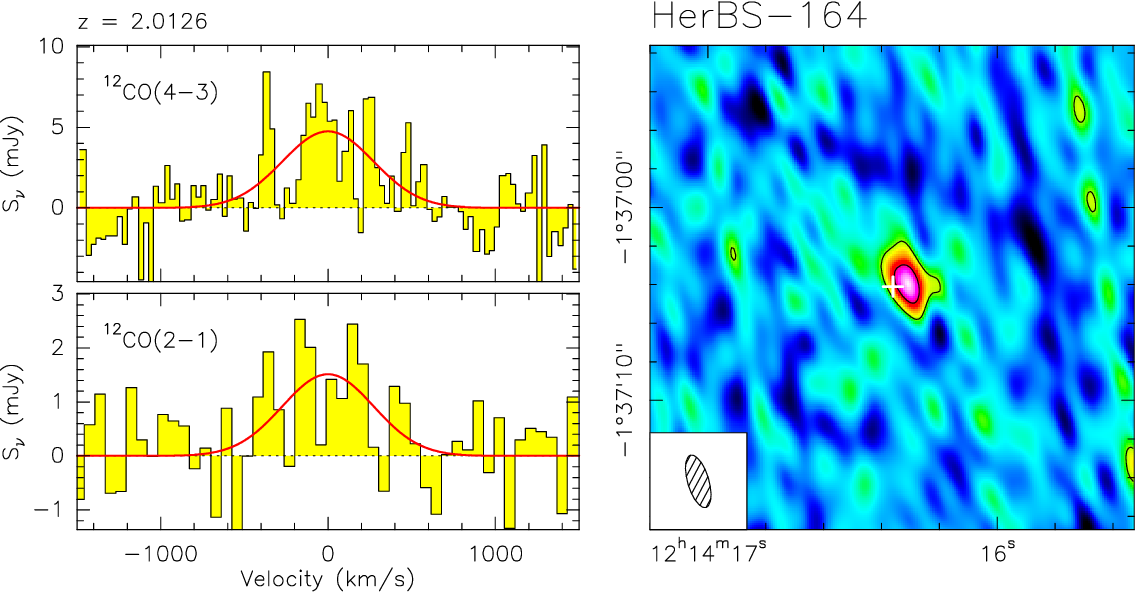}
\includegraphics[width=0.6\textwidth]{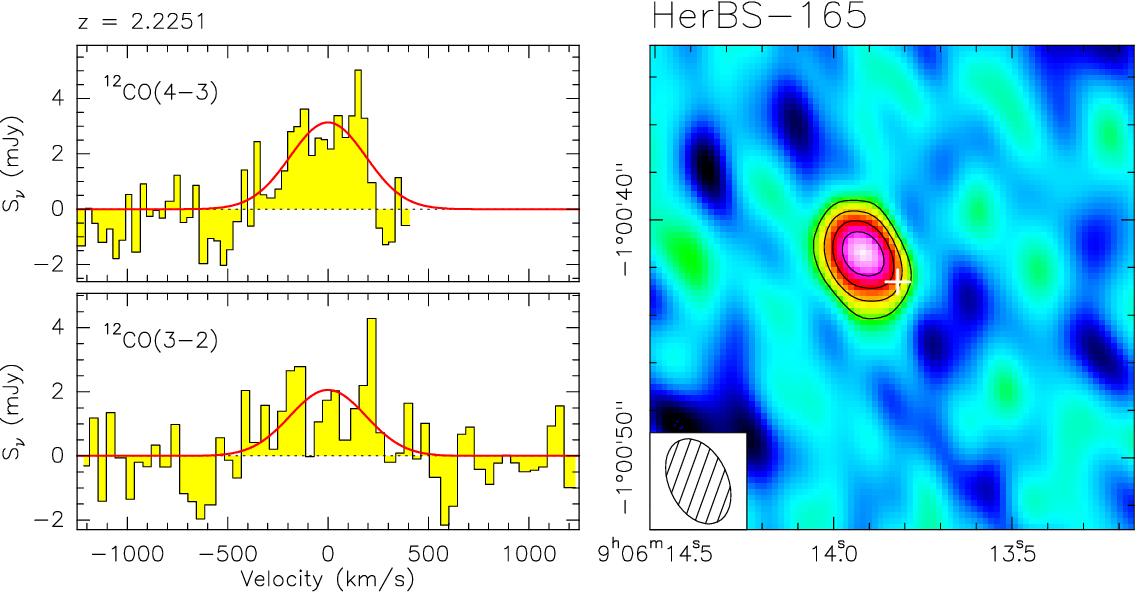}
\includegraphics[width=0.6\textwidth]{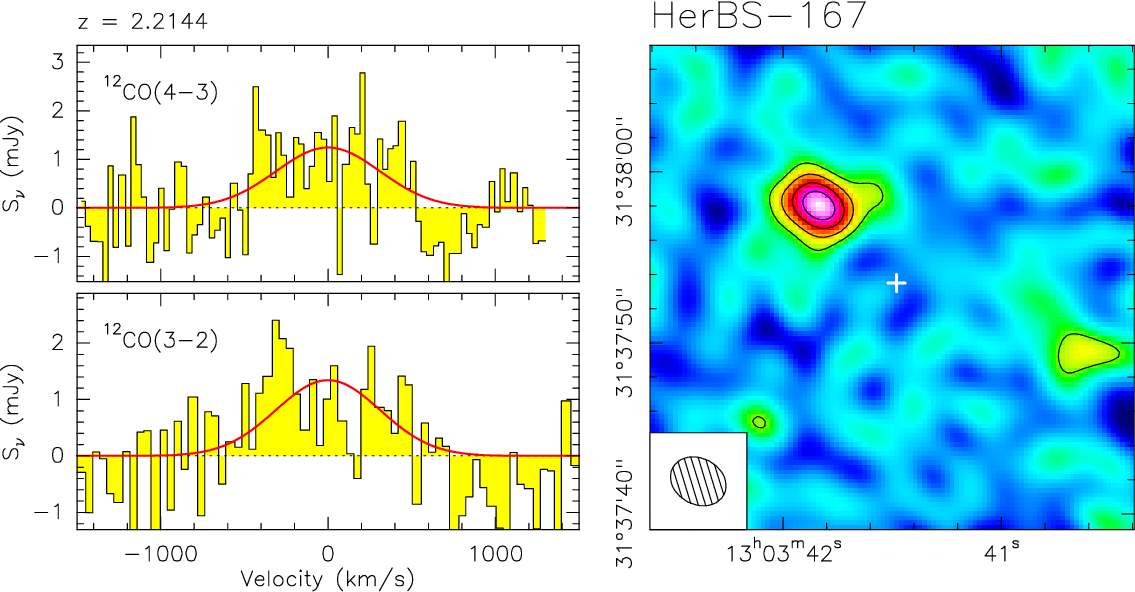}
\includegraphics[width=0.6\textwidth]{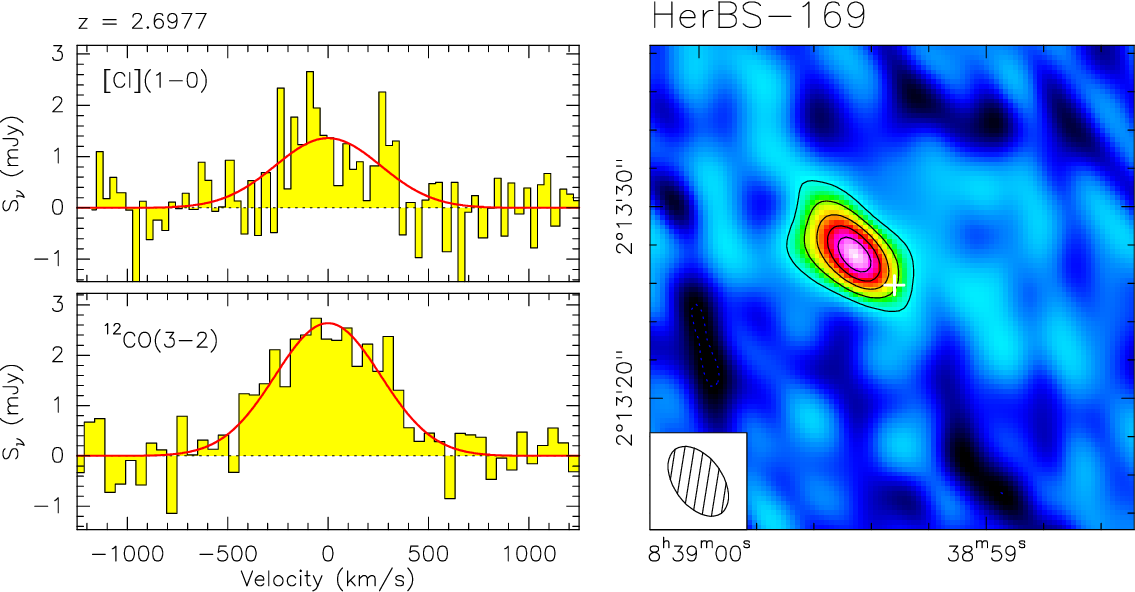}
   \caption{\bf{continued}}
    \end{figure*}
 \addtocounter{figure}{-1}

\begin{figure*}[!ht]
   \centering
\includegraphics[width=0.6\textwidth]{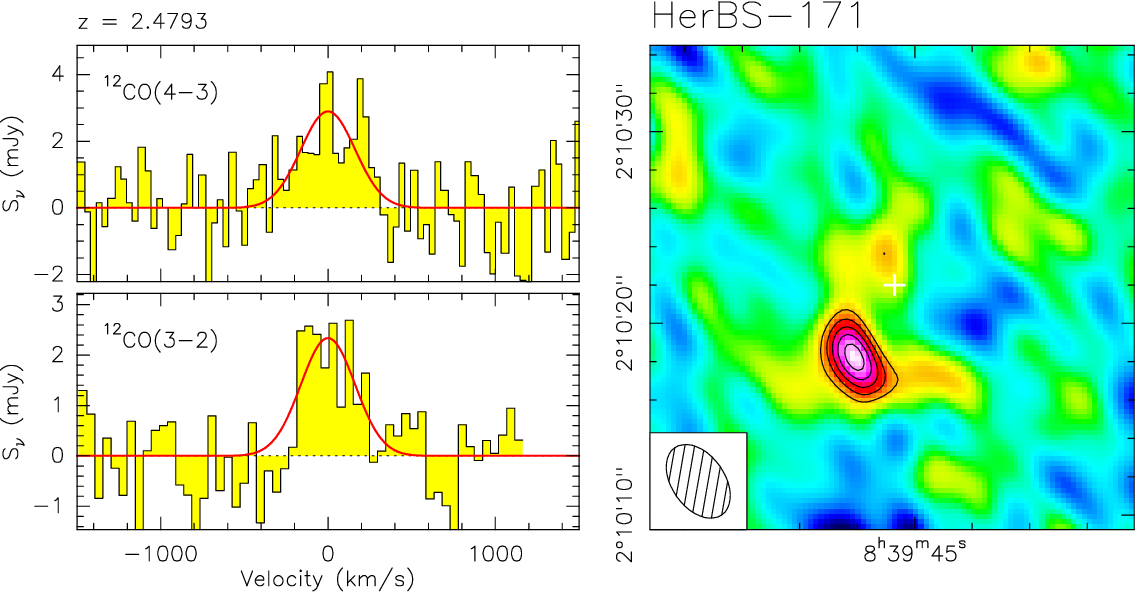}
\includegraphics[width=0.6\textwidth]{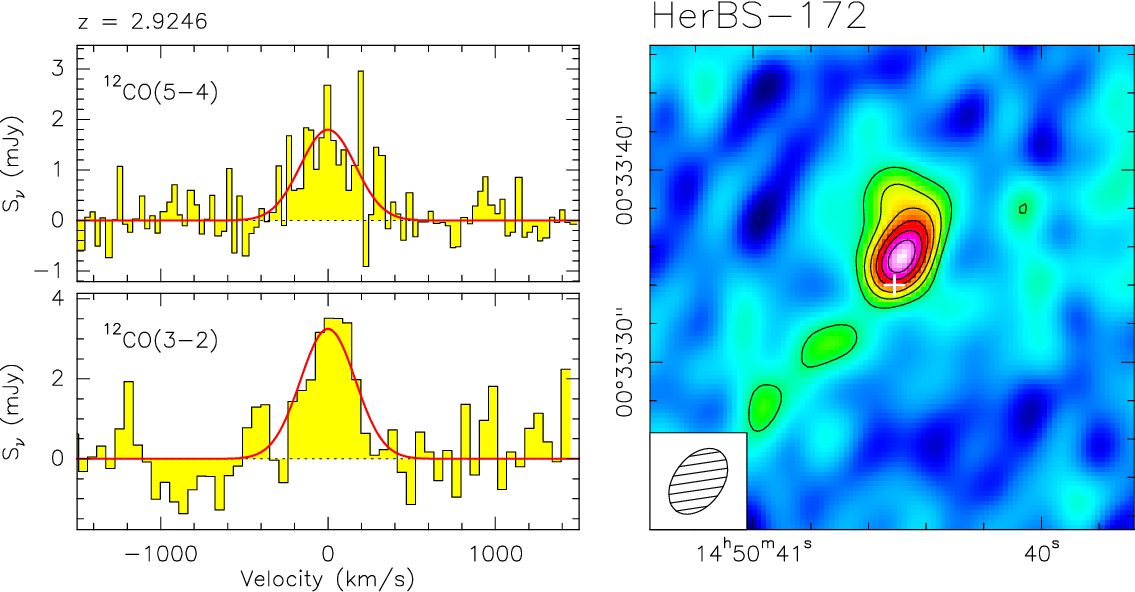}
\includegraphics[width=0.6\textwidth]{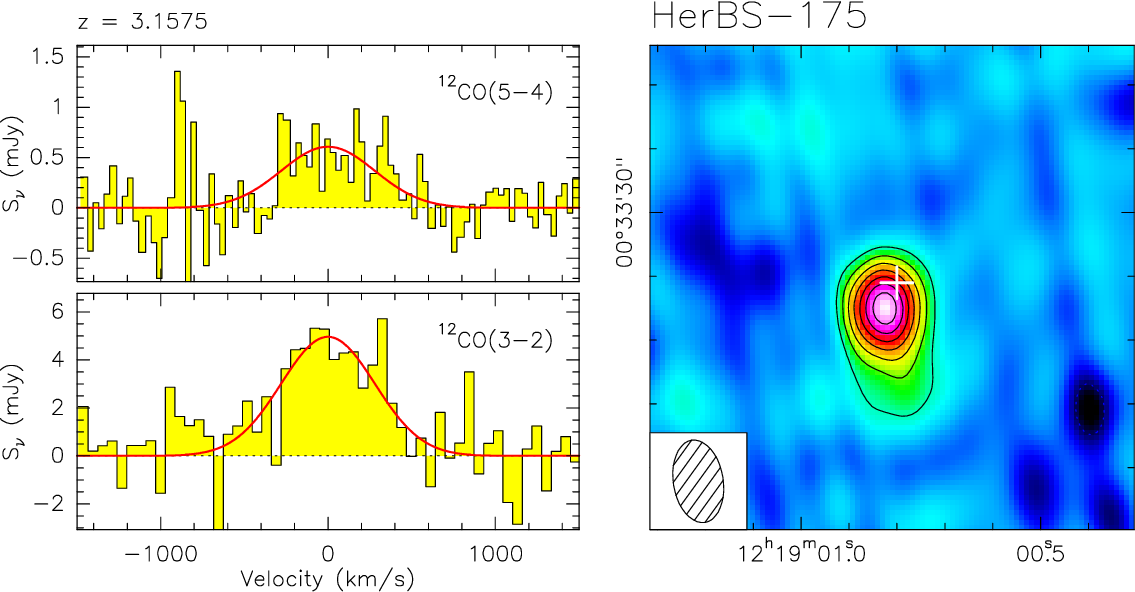}
\includegraphics[width=0.6\textwidth]{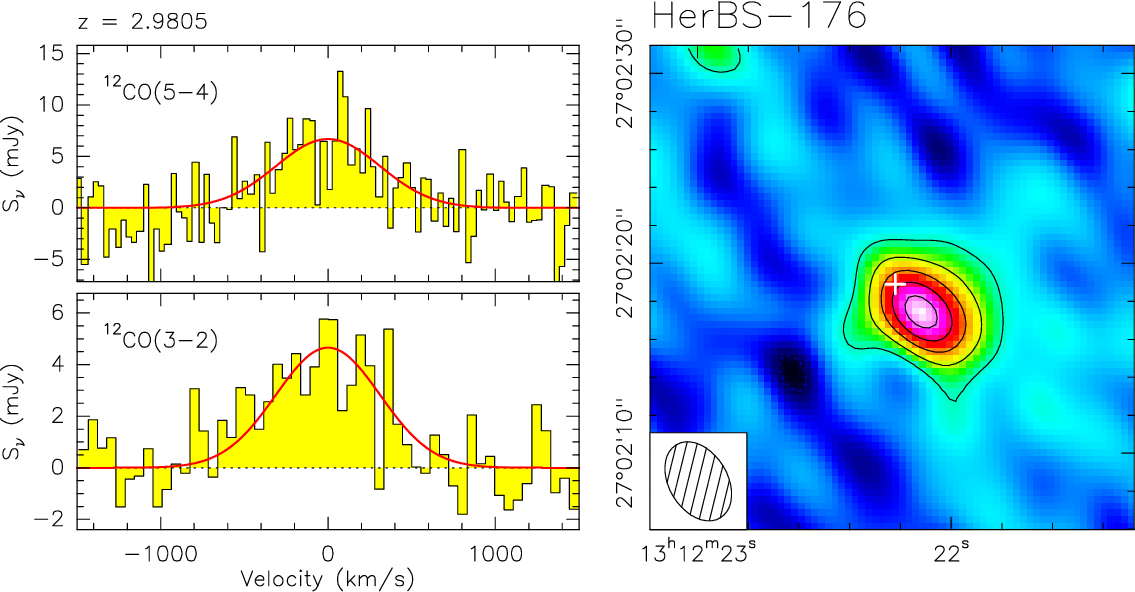} 
   \caption{\bf{continued}}
    \end{figure*}
 \addtocounter{figure}{-1}

\begin{figure*}[!ht]
   \centering
\includegraphics[width=0.9\textwidth]{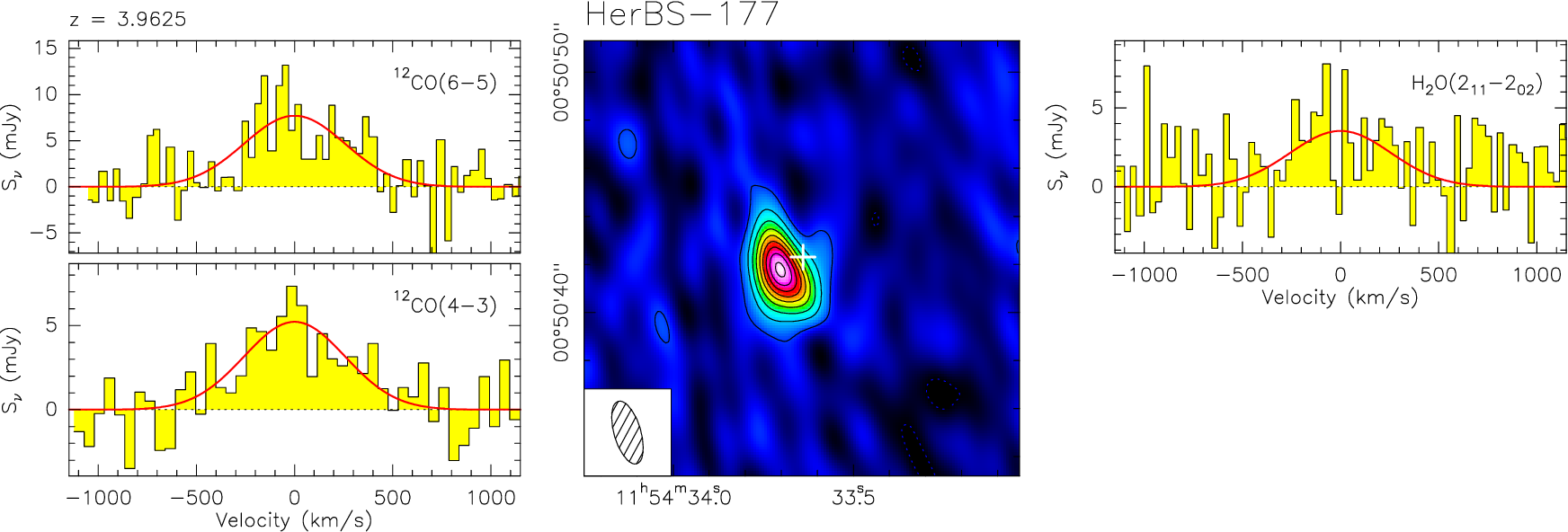}  
\includegraphics[width=0.9\textwidth]{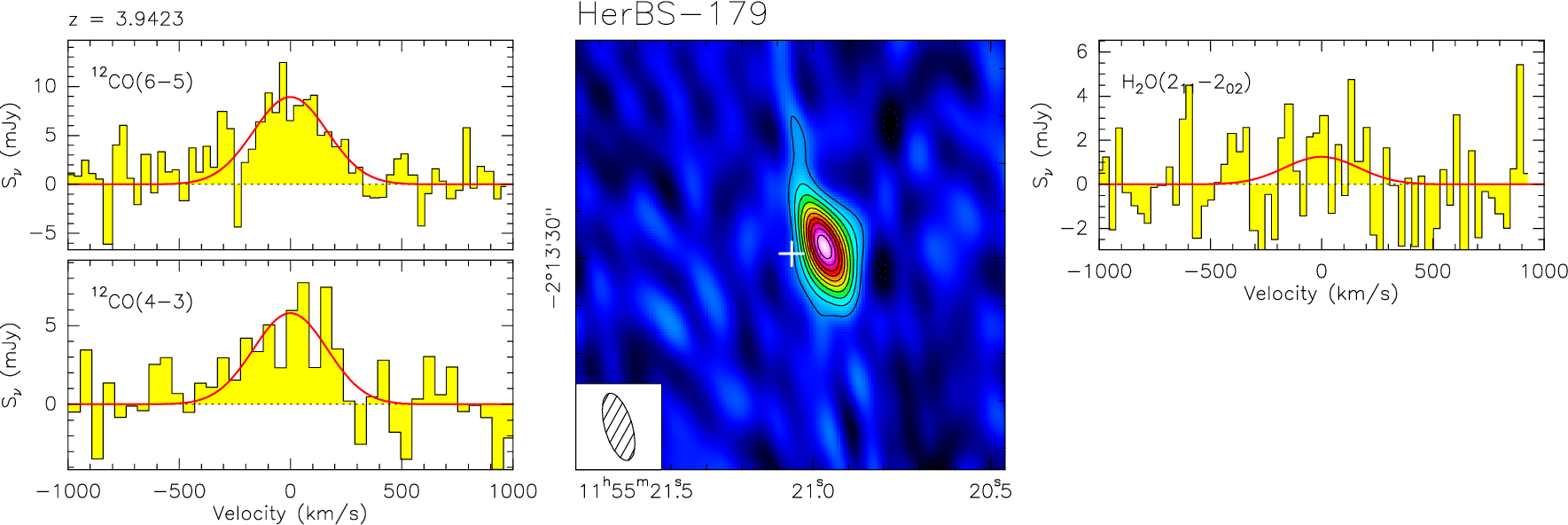}
\includegraphics[width=0.6\textwidth]{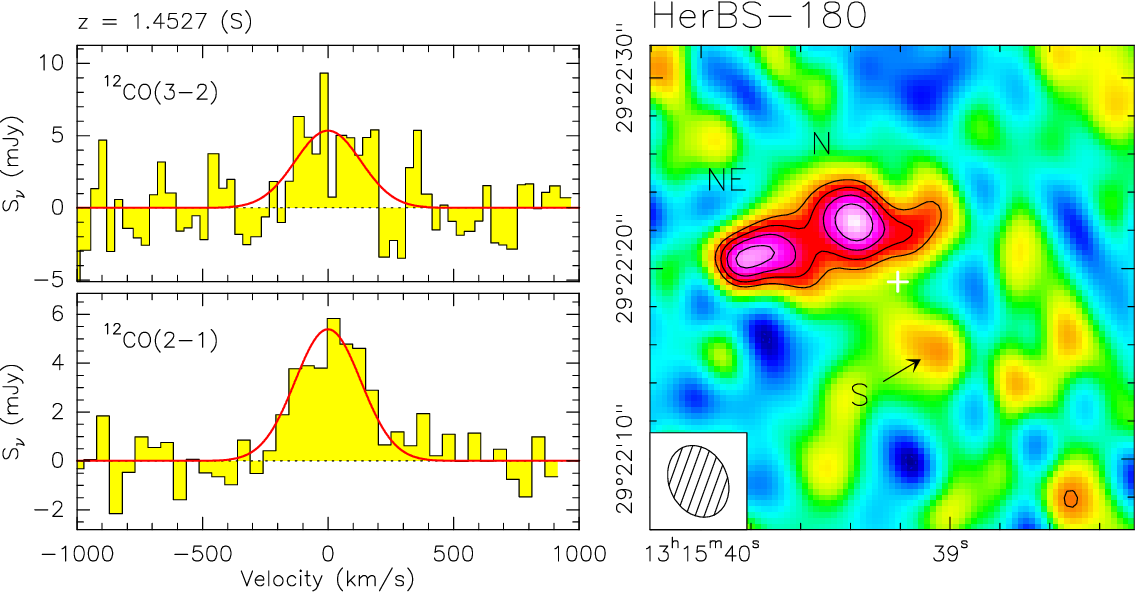}
\includegraphics[width=0.6\textwidth]{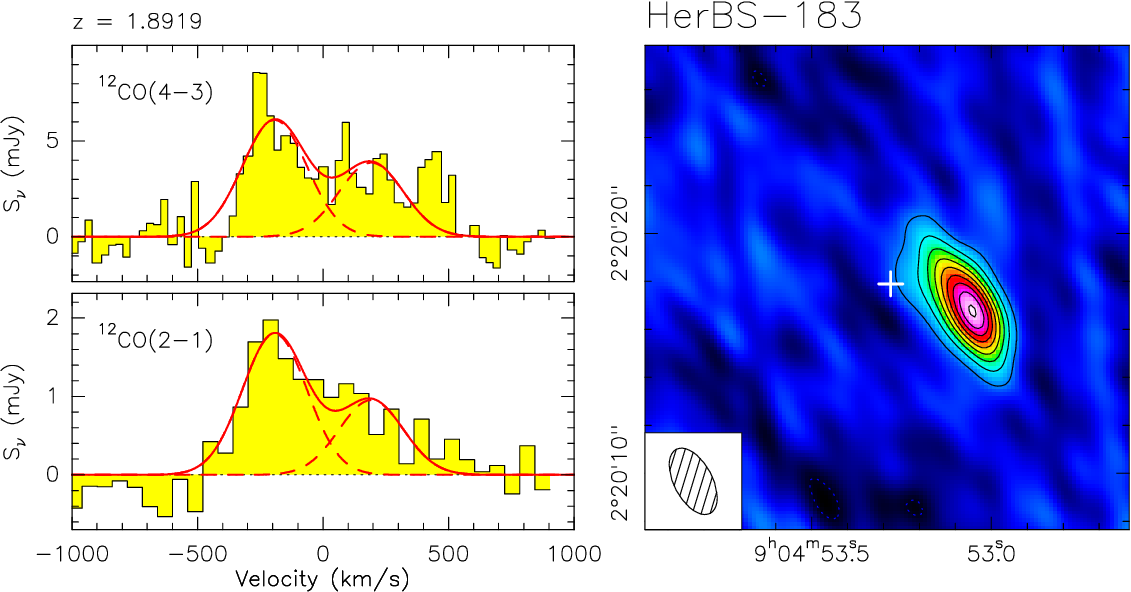}
   \caption{\bf{continued}}
    \end{figure*}
 \addtocounter{figure}{-1}

 \begin{figure*}[!ht]
   \centering
\includegraphics[width=0.9\textwidth]{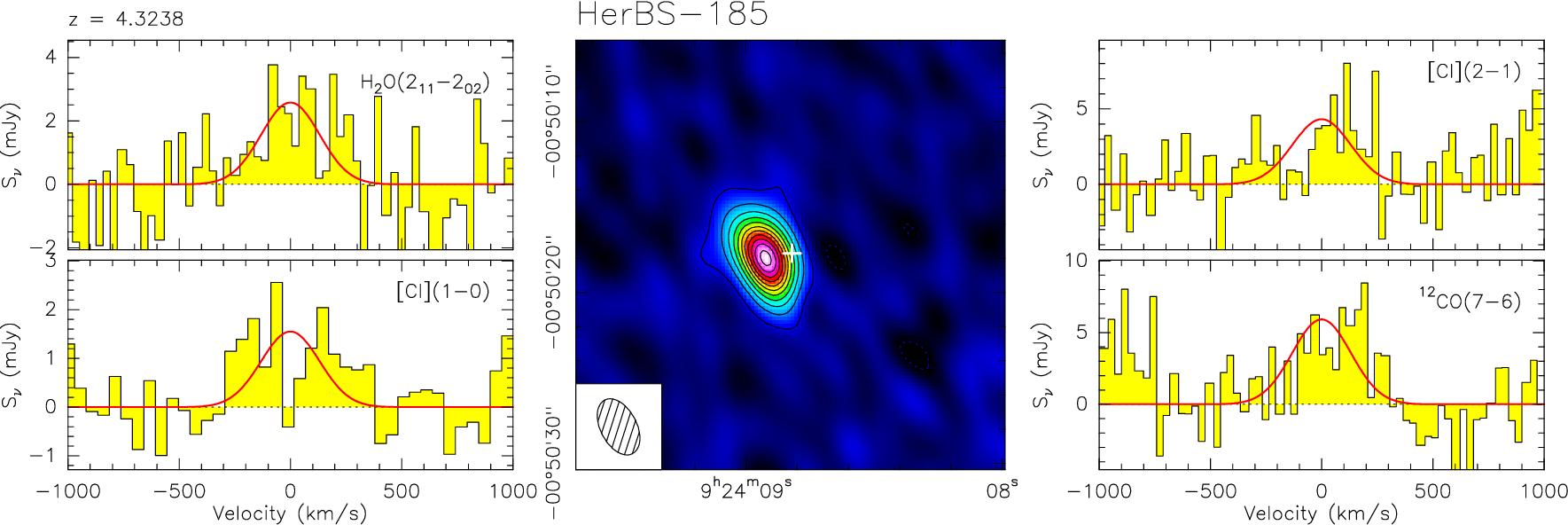}  
\includegraphics[width=0.9\textwidth]{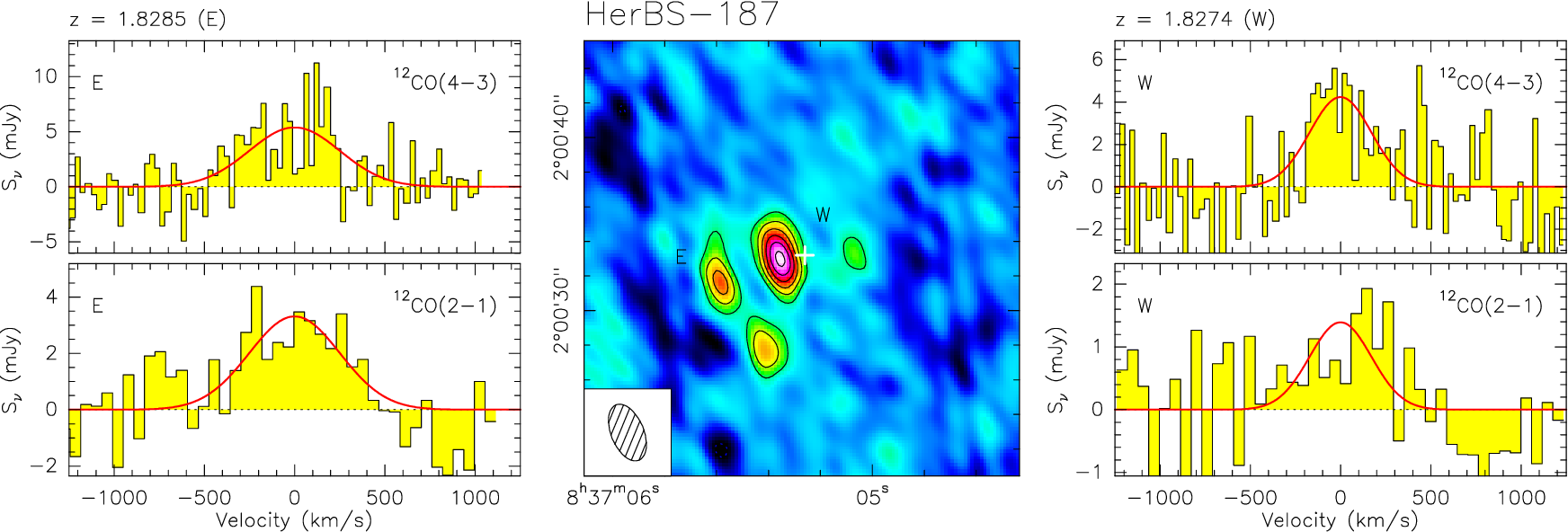}  
\includegraphics[width=0.6\textwidth]{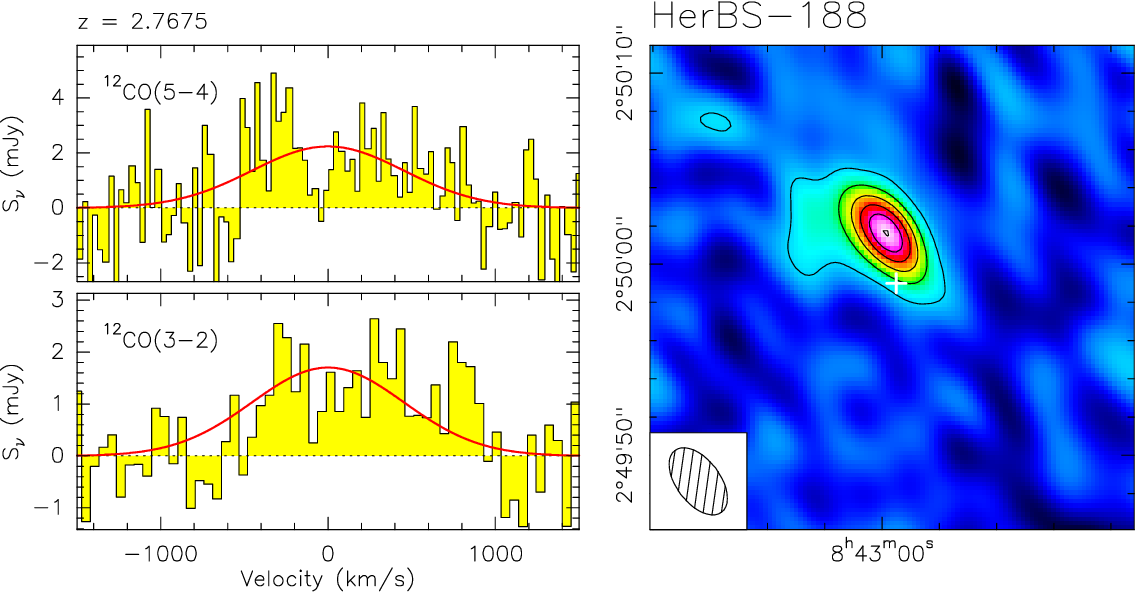}
\includegraphics[width=0.6\textwidth]{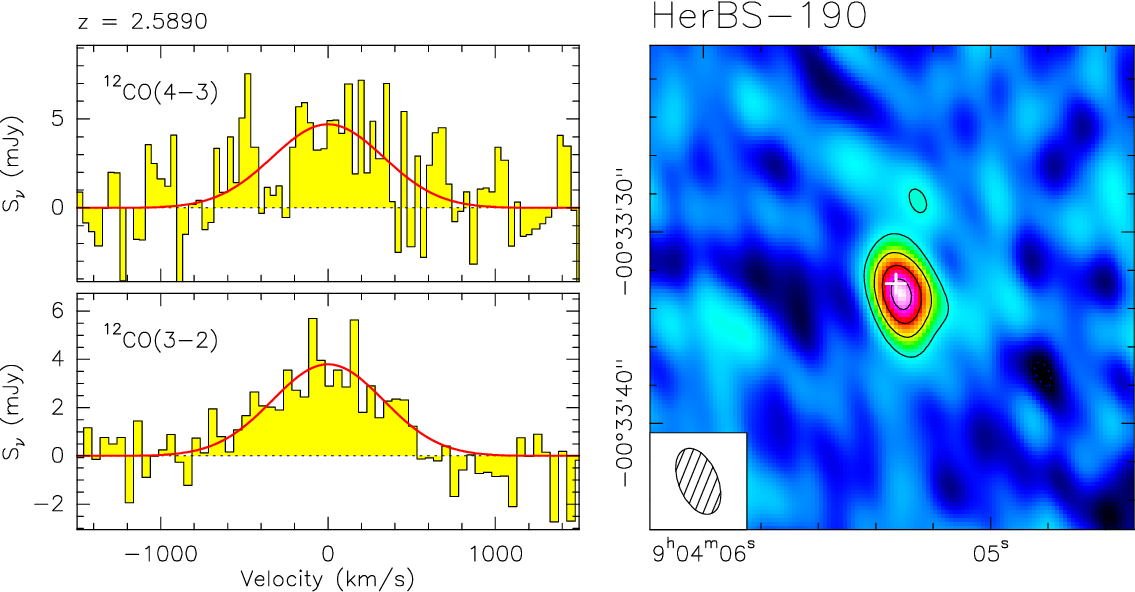}
   \caption{\bf{continued}}
    \end{figure*}
 \addtocounter{figure}{-1}
 
\begin{figure*}[!ht]
   \centering
\includegraphics[width=0.6\textwidth]{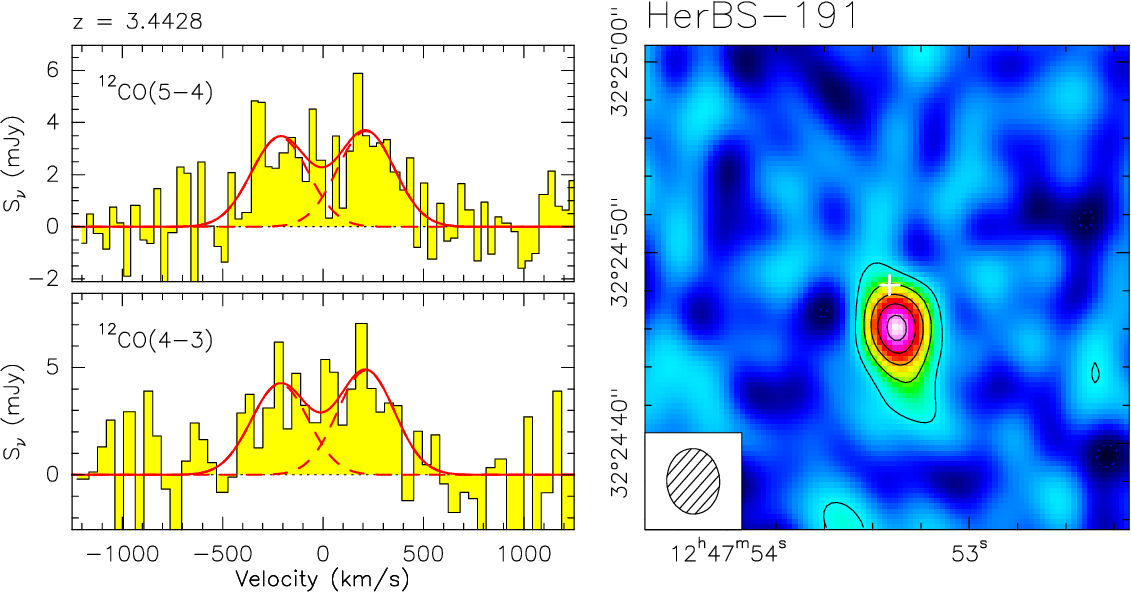}
\includegraphics[width=0.9\textwidth]{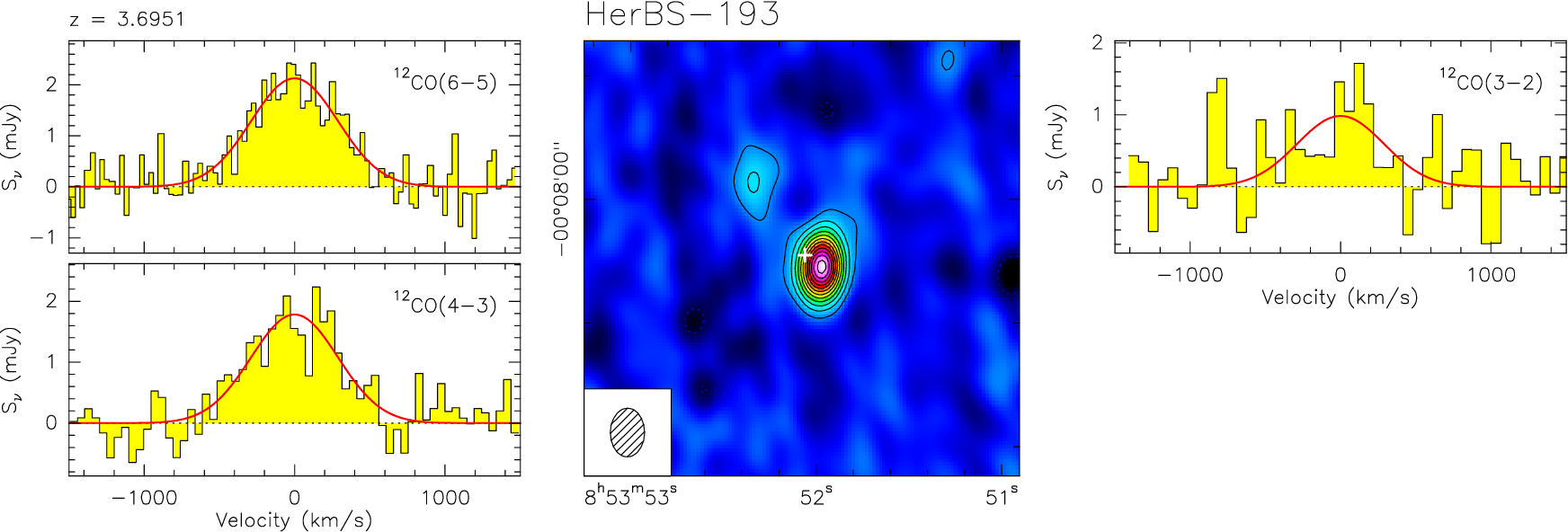}    
\includegraphics[width=0.9\textwidth]{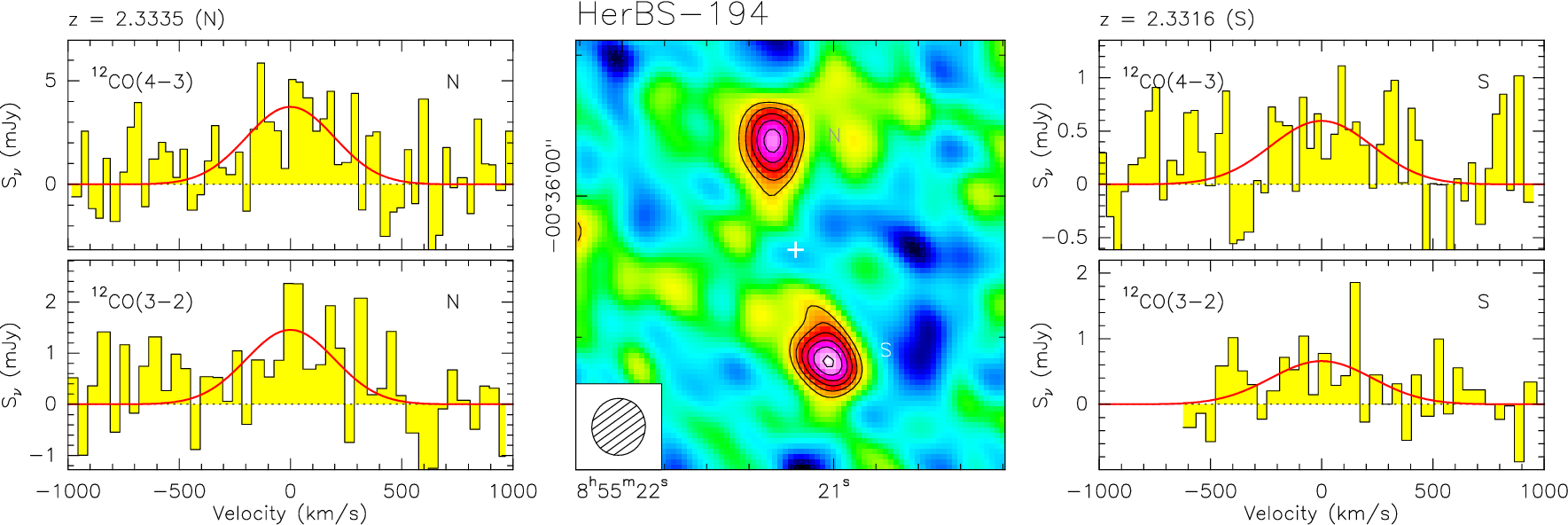}
\includegraphics[width=0.9\textwidth]{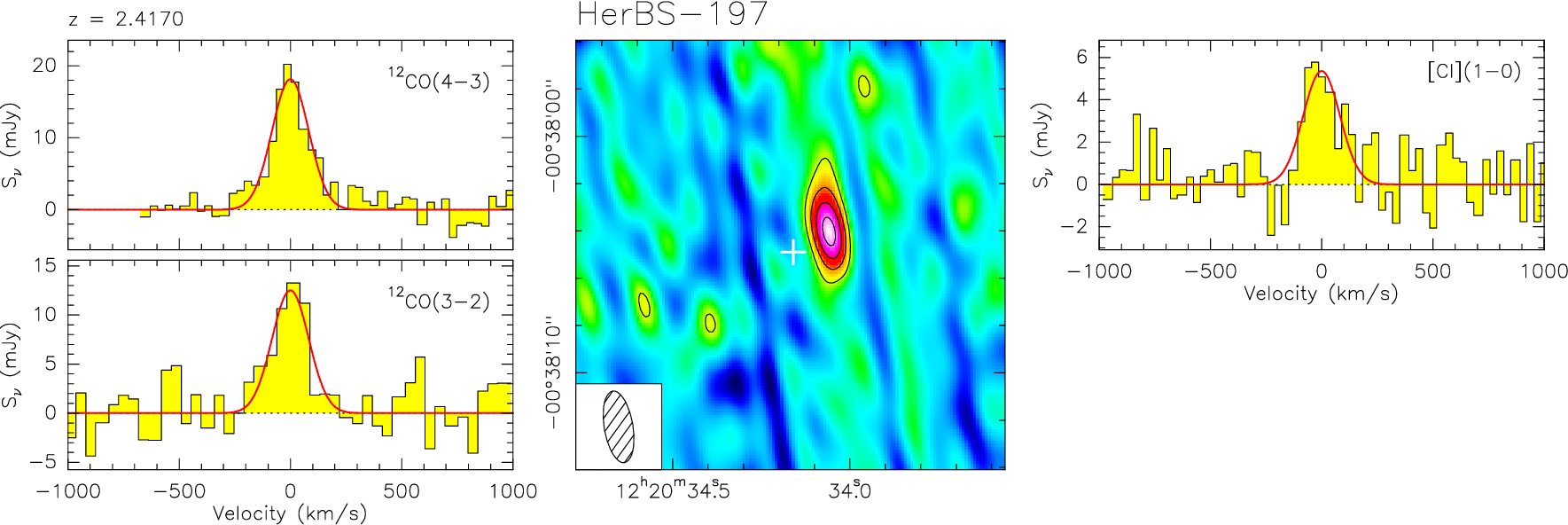}
   \caption{\bf{continued}}
\end{figure*}
\addtocounter{figure}{-1}

\begin{figure*}[!ht]
   \centering
\includegraphics[width=0.9\textwidth]{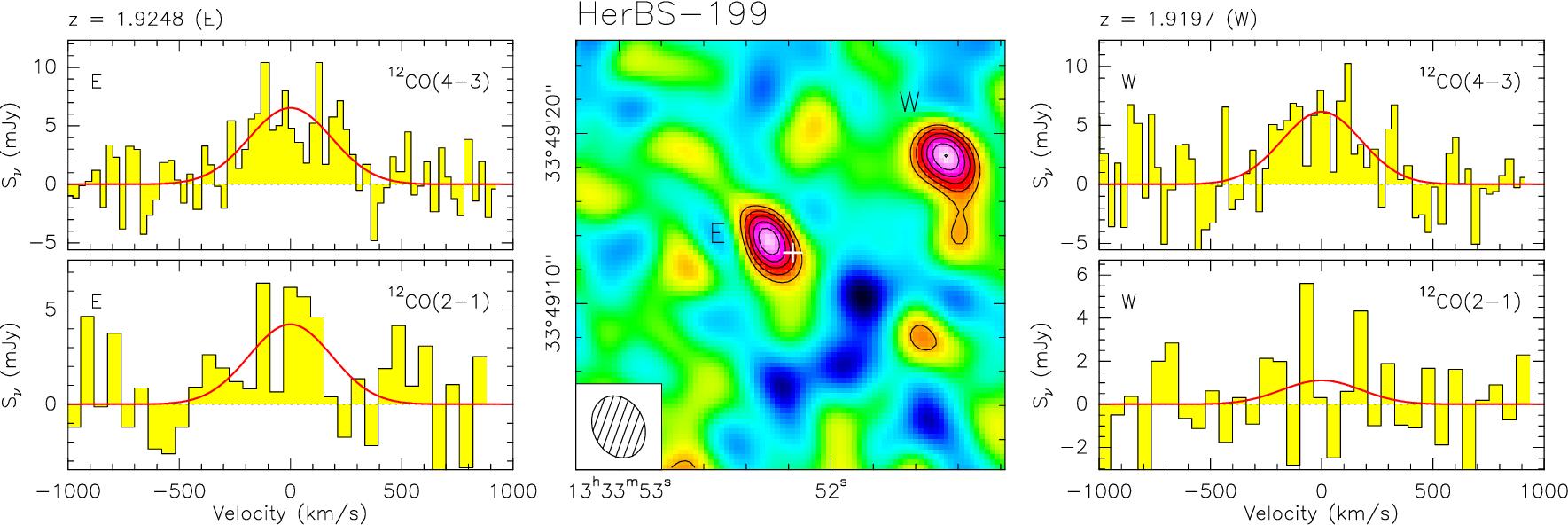}

\vspace{1.3cm}
\includegraphics[width=0.9\textwidth]{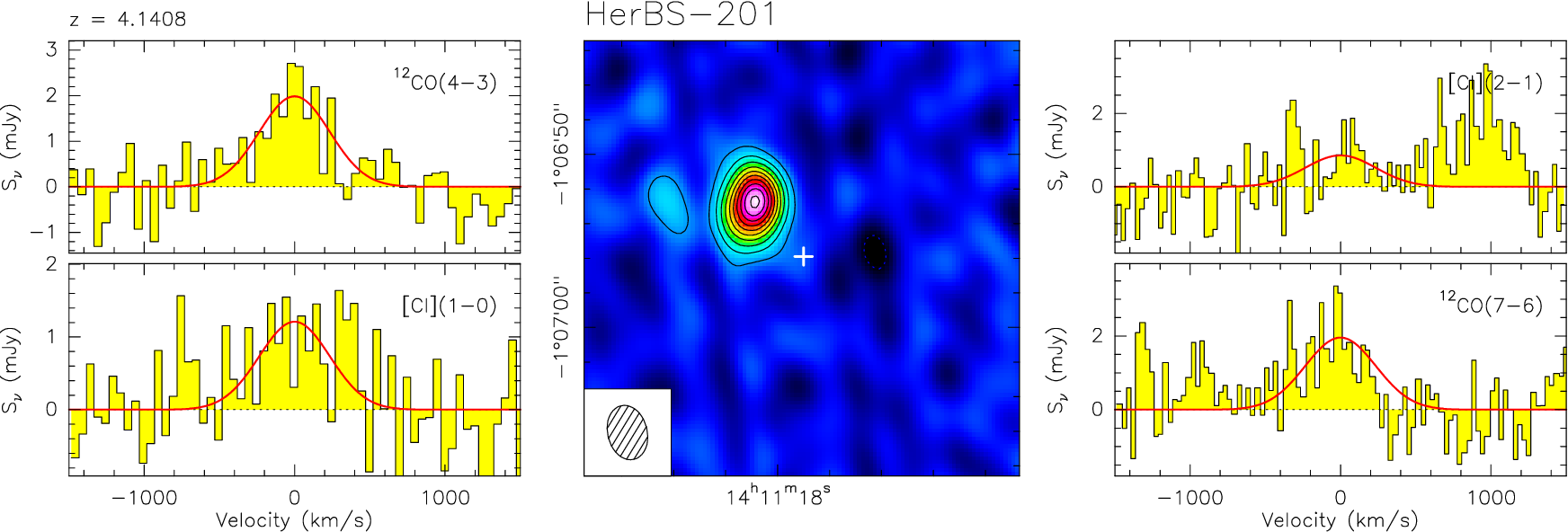}

\vspace{1.3cm}
\includegraphics[width=0.6\textwidth]{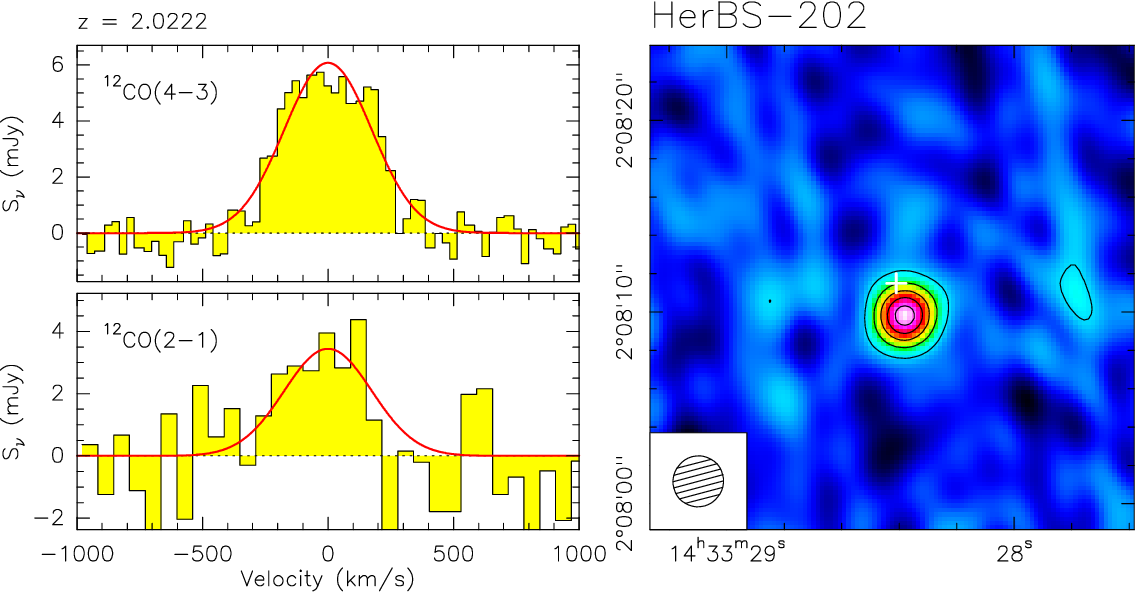}
   \caption{\bf{continued}}
\end{figure*}
\addtocounter{figure}{-1}

\begin{figure*}[!ht]
   \centering
\includegraphics[width=0.9\textwidth]{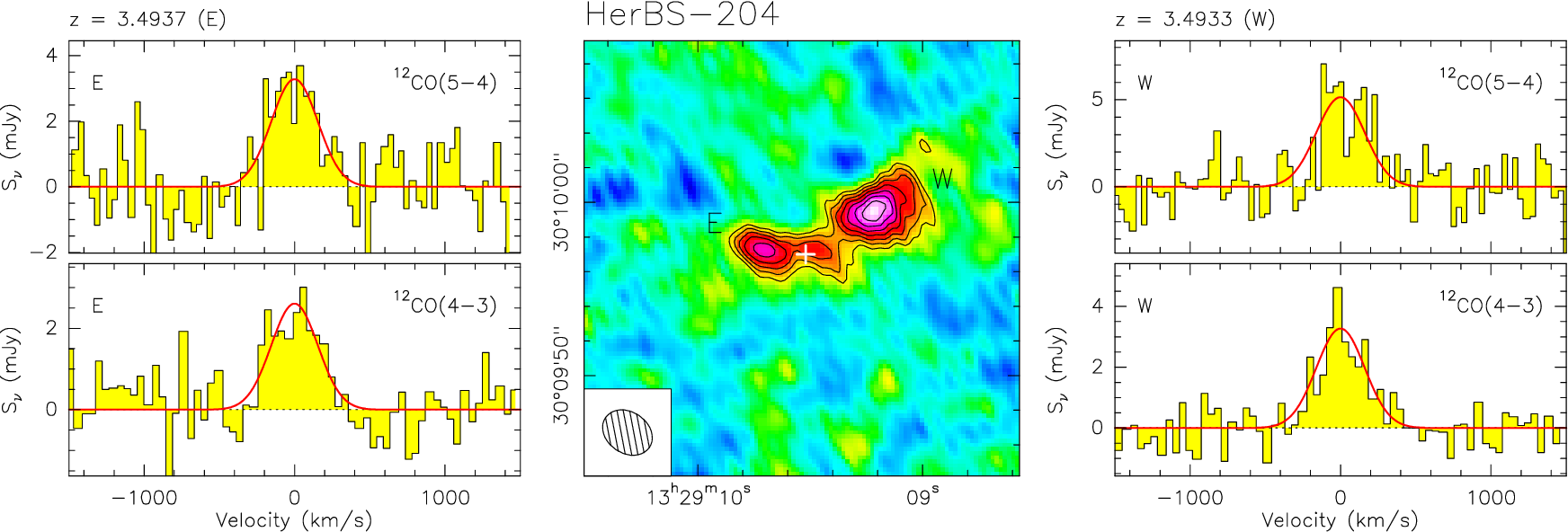}

\vspace{0.8cm}
\includegraphics[width=0.9\textwidth]{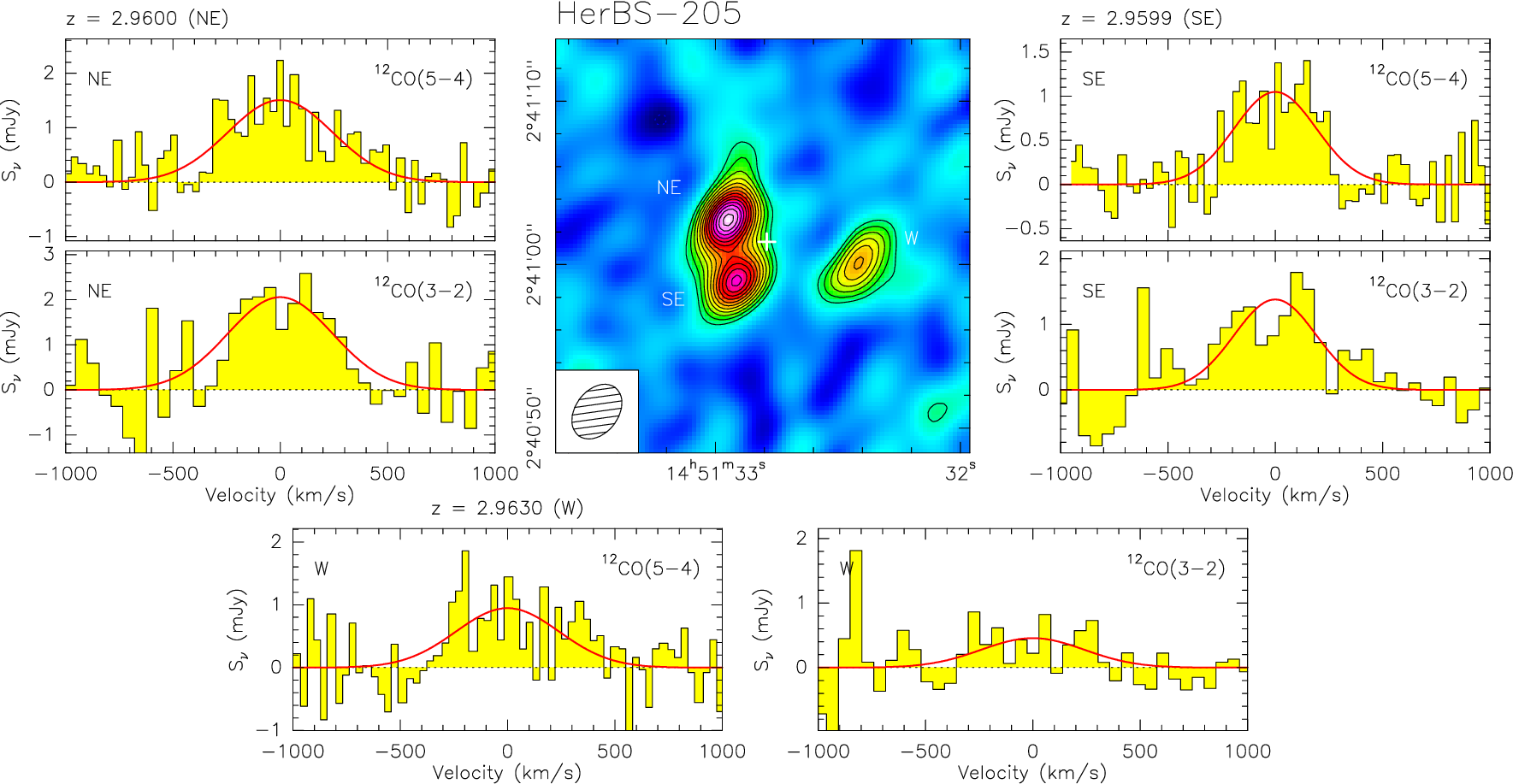}

\vspace{0.8cm}
 \includegraphics[width=0.6\textwidth]{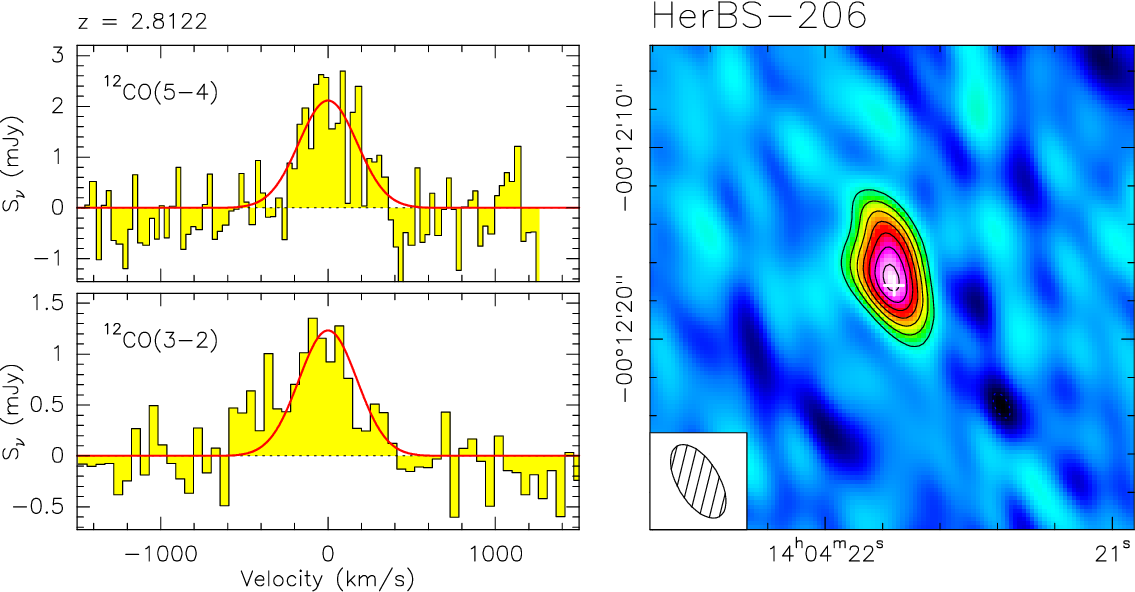}
   \caption{\bf{continued}}
\end{figure*}

\end{appendix}

\end{document}